\documentclass[11pt]{article}

\usepackage{geometry}
\usepackage[inline]{enumitem}
\usepackage[english]{babel}
\usepackage{pifont}%
\newcommand{\xmark}{\ding{55}}%

\newcommand{\bt}{\fontseries{b}\selectfont}
\usepackage[title]{appendix}
\usepackage{ragged2e}
\usepackage{amsmath}
\usepackage{amssymb}
\usepackage{amsthm}
\usepackage{thmtools}
\usepackage{amsfonts}
\usepackage{float}
\usepackage{mathtools}
\usepackage{fancyhdr} 
\usepackage{bm}
\usepackage{natbib}
\usepackage{verbatim}
\usepackage{graphicx}
\usepackage{xcolor}
\usepackage{enumitem}
\usepackage{changepage} %
\usepackage{bbm}
\usepackage{tabularx}
\usepackage{colortbl}
\usepackage{stmaryrd}
\usepackage[refpage]{nomencl}
\usepackage{subfig}
\usepackage{centernot}
\usepackage{marvosym} %
\usepackage{stmaryrd}
\usepackage{etoolbox}
\usepackage{algorithm}
\usepackage{thmtools, thm-restate}
\usepackage{tikz}%
\usetikzlibrary{shapes,backgrounds,intersections}
\usepackage[font = small, labelfont = bf]{caption}

\usepackage{etoolbox}

\usepackage{algpseudocode}%
\usepackage{array,multirow}

\usepackage{booktabs}
\usepackage{pdflscape}
\usepackage{float}
\usepackage{tabu}

\usetikzlibrary{arrows.meta}

\usepackage{authblk}
\usepackage[colorlinks,citecolor=blue,urlcolor=blue,linkcolor=red]{hyperref}
\usepackage{cleveref}

\makeatletter
\def\thmt@refnamewithcomma #1#2#3,#4,#5\@nil{%
  \@xa\def\csname\thmt@envname #1utorefname\endcsname{#3}%
  \ifcsname #2refname\endcsname
    \csname #2refname\expandafter\endcsname\expandafter{\thmt@envname}{#3}{#4}%
  \fi
}

\makeatother

\def\independenT#1#2{\mathrel{\rlap{$#1#2$}\mkern2mu{#1#2}}}

\newcommand\independent{\protect\mathpalette{\protect\independenT}{\perp}}
\def\independenT#1#2{\mathrel{\rlap{$#1#2$}\mkern2mu{#1#2}}}

\renewcommand{\t}{\intercal}

\newcommand{\ep}{\varepsilon}
\renewcommand{\epsilon}{\varepsilon}

		\renewcommand{\i}{\infty}
		
		\renewcommand{\phi}{\varphi}

\newcommand{\Cov}{\mathrm{Cov}}
\newcommand{\Corr}{\mathrm{Corr}}

\newcommand{\tPA}[1]{{\widetilde{\mathrm{PA}}({#1})}}
\newcommand{\PA}[1]{{\mathrm{PA}({#1})}}

\newcommand{\convp}{\stackrel{P}{\longrightarrow}}

\newcommand{\convd}{\stackrel{\mathcal{D}}{\longrightarrow}}

\DeclareMathOperator*{\argmin}{arg\,min}

\DeclareMathOperator*{\plim}{P-lim}

\newcommand{\define}[4]{\expandafter#1\csname#3#4\endcsname{#2{#4}}}
\newcommand{\R}{\mathbb{R}}

\newcommand{\N}{\mathbb{N}}
\forcsvlist{\define{\newcommand}{\mathcal}{c}}{A,B,C,D,E,F,G,H,I,J,K,L,P,R,M,N,U,S,T,W,V,X,Y,Z}
\forcsvlist{\define{\newcommand}{\mathbb}{b}}{A,B,C,D,E,F,G,H,I,J,K,L,P,Q,M,N,Z,T,X,Y}
\forcsvlist{\define{\newcommand}{\mathbf}{f}}{A,B,C,D,E,F,G,H,I,J,K,L,M,N,O,P,Q,R,S,T,U,W,V,X,Y,Z}

\newcommand{\lp}{\left( }
\newcommand{\rp}{\right) }
\newcommand{\lb}{\left\lbrace}
\newcommand{\rb}{\right\rbrace}

\newcommand{\ra}{\rangle}
\newcommand{\la}{\langle}

\crefname{enumi}{}{}

\newtheorem{assumption}{Assumption}

\newtheorem{lemma}{Lemma}

\newtheorem{remark}{Remark}

\renewcommand{\thesection}{\arabic{section}}
\renewcommand{\theequation}{\arabic{section}.\arabic{equation}}
\renewcommand{\thetheorem}{\arabic{section}.\arabic{theorem}}

\renewcommand{\thecorollary}{\arabic{section}.\arabic{corollary}}
\renewcommand{\thelemma}{\arabic{section}.\arabic{lemma}}

\renewcommand{\theremark}{\arabic{section}.\arabic{remark}}

\newenvironment{proofenv}[1] {\textbf{Proof of {#1}:}}{\hfill$\square$\bigskip}

\crefname{assumption}{assumption}{assumptions}
\Crefname{assumption}{Assumption}{Assumptions}

\makenomenclature

\theoremstyle{definition}

\begin{document}
\title{Distributional robustness of K-class\\estimators and the PULSE\footnote{Published in The Econometrics Journal: \url{https://doi.org/10.1093/ectj/utab031}}}
\author{Martin Emil Jakobsen\\Department of Mathematical Sciences\\University of Copenhagen\\m.jakobsen@math.ku.dk \and Jonas Peters\\Department of Mathematical Sciences\\University of Copenhagen\\jonas.peters@math.ku.dk}

 \maketitle

\begin{abstract}\normalsize While causal models are robust in that they are prediction optimal under arbitrarily 
	strong interventions, 
	they may not be optimal when the interventions are bounded.
	We prove that the classical K-class estimator satisfies such optimality by establishing a connection between 
	K-class estimators and anchor regression. 
	This connection 
	further motivates a novel estimator in
	instrumental variable settings 
	that minimizes
	the mean squared prediction
	error subject to the constraint that the estimator lies in an
	asymptotically valid confidence region of the causal coefficient. We call
	this estimator PULSE (p-uncorrelated least squares estimator), relate it to work on invariance,
	show
	that it can be computed efficiently as a data-driven K-class estimator, even though the underlying optimization problem is non-convex, and prove consistency.
	We evaluate 
	the estimators on real data and 
	perform simulation experiments illustrating 
	that PULSE suffers from less variability. There are several settings including 
	weak instrument settings, where it outperforms other estimators. 
\end{abstract}
\textbf{Keywords:} Causality, distributional robustness, instrumental variables

\crefname{section}{section}{sections}
\Crefname{section}{Section}{Sections}
\crefname{appendix}{appendix}{appendices}
\Crefname{appendix}{Appendix}{Appendices}

\section{Introduction}
Learning causal parameters from data has been 
a key challenge in many scientific fields 
and has been a long-studied problem in 
econometrics
\citep[e.g.][]{	goldberger1972structural, koopmans1953studies,wold1954causality}.
Many years after the groundbreaking work by
\citet{Fisher35} and \citet{Peirce83},
causality plays again an increasingly important role 
in machine learning and statistics,
two research areas that
are most often considered part of mathematics or computer science 
\citep[e.g.,][]{Imbens2015, Pearl2009, Peters2017,Spirtes2000}.
 Even though current research in mathematics and computer science on the one hand and econometrics on the other, does not develop independently, we believe that there is a lot of potential for more fruitful interaction between these two fields.
Differences in the 
language have emerged, which can 
make 
communication difficult, but
the target of inference,
the underlying principles, and 
the methodology 
in both fields are closely related.
This paper 
establishes
a link between two developments 
in these fields:
K-class estimation 
which aims at 
estimation of causal parameters with good 
statistical properties 
and 
invariance principles
that are used to build methods that 
are robust with respect to distributional shifts. 
This connection allows us to prove distributional robustness guarantees for K-class estimators and motivates a new estimator, PULSE.
We summarize our main results in Section~\ref{sec:summary}.

\subsection{Related work}
Given causal background knowledge, 
causal parameters 
can be estimated 
when taking into account 
confounding effects between treatment and outcome.
Several related 
techniques have been suggested to tackle that problem,
including 
variable adjustment \citep[][]{Pearl2009}, 
propensity score matching \citep[][]{Rosenbaum83},
inverse probability weighting \citep[][]{Horvitz1952} or 
G-computation \citep[][]{Robins1986}.

If some of the relevant variables have not been observed, 
one may instead 
use exogenous variation in the data to 
infer causal parameters, 
e.g., in the setting of instrumental variables
\citep[e.g.,][]{angrist1995identification,Newey2013,Wang2016,wright1928tariff}. 
Limited information estimators leverage instrumental variables 
to conduct single equation inference. 
An example of such methods is the two-stage 
least squares estimators (TSLS) developed by \citet{theil1953repeated}. 
Instead of minimizing the 
residual sum of squares as 
done by the
ordinary least square (OLS) estimator, 
the TSLS
minimizes
the sample-covariance between the instruments and regression residuals. 
TSLS estimators are consistent, but
are known to have suboptimal finite sample properties, e.g., they only have moments op to the degree of  over-identification \citep[][]{mariano1972existence}.  
\citet{Kadane1971} shows that under suitable conditions, the mean squared error of TSLS might even be larger than the one of OLS 
if the sample size is small (more precisely and using the notation introduced below, if $0\leq n-q\leq 2(3-(q_2-d_1))$, where $q_2-d_1$ is the degree of overidentification). This result is another indication that under certain conditions, it might be beneficial to use the OLS for regularization.
Another method of inferring causal parameters in structural equation models is the limited information maximum likelihood (LIML) estimator due to \citet{anderson1949estimation}. %
\citet{theil1958economic} introduced K-class estimators, which contain OLS, TSLS and the LIML estimator as special cases. This class of estimators is parametrized by a deterministic or stochastic parameter $\kappa\in[0,\i)$ that depends on the observational data. Under mild regularity conditions a member of this class is consistent and asymptotically normally distributed if 
$(\kappa-1)$ and 
$\sqrt{n}(\kappa-1)$ 
converge, respectively,
to zero in probability when 
$n$ tends to infinity; see, e.g.\ \citet{MARIANO1975},  \citet{mariano2001simultaneous}. While the LIML does not have moments of any order, it shares the same asymptotic normal distribution with TSLS.
Based on simulation studies,
\citet{anderson_hildenbrand_1983} argues that,
in many 
practically relevant
cases,
the normal approximation to a 
finite-sample estimator 
is inadequate for TSLS but 
a useful approximation in the case of LIML.
Using Monte Carlo simulations, 
\citet{hahn2004estimation} 
recommend that the no-moment estimator LIML should not be used in weak instrument situations, where Fuller estimators have a substantially smaller MSE.
The Fuller estimators
\citep[][]{fuller1977some} form
a subclass of the K-class estimators based on a modification to the LIML, which fixes the no-moment problem while maintaining consistency and asymptotic normality. 
\citet{Kiviet2020} proposes a modification to the OLS estimator that makes use of explicit knowledge of the partial correlation between the covariates and the unobserved noise in $Y$. 
\citet{AndrewsArmstrong2017} propose an unbiased estimator 
that is based on knowledge of the sign of the first stage regression and the variance the reduced form errors and that is less dispersed than TSLS, for example. 
\citet{judge2012minimum} consider an affine combination of the OLS and TSLS estimators, which, again, yields
a modification in the space of estimators.
We prove that our proposed estimator, PULSE, can also be written as a data driven K-class estimator. 
As such, it minimizes a convex combination of the OLS and TSLS loss functions and can, in general, not be written as a convex combination of the estimators.

All of the above methods exploit 
background knowledge, e.g., 
in form of exogeneity of some of the variables.  
If no such background knowledge is available, 
it may still be possible,
under additional assumptions, 
to infer the
causal structure, e.g., represented by a graph, from observational (or observational and interventional) data. 
This problem is sometimes referred to as 
causal discovery.
Constraint-based methods 
assume that the underlying distribution is Markov and faithful with respect to the causal graph and perform 
conditional independence tests to infer (parts of) the graph; see, e.g.\ \citet{Spirtes2000}.
Score-based methods 
assume a certain statistical model and optimize (penalized) likelihood scores; see, e.g.\ \citet{Chickering2002}.
Some methods exploit a simple form of causal assignments, such as 
additive noise
(e.g.,  \citealp{Peters2014jmlr}, and \citealp{Shimizu2006})
and others are based on exploiting invariance statements \citep[e.g.,][]{Peters2016jrssb, Meinshausen2016}. 
Many of such methods assume causal sufficiency, i.e., 
that all causally relevant variables have been observed, but some versions exist that
allow for hidden variables; see, e.g.\ \citet{Claassen2013} and \citet{Spirtes1995}.

Recent works in the fields of machine learning and computational statistics \citep[e.g.][]{HeinzeDeml17,Pfister2019pnas,Scholkopf2012}
investigate whether
causal ideas can help to make machine learning
methods more robust.
The reasoning is that causal models are robust against any intervention in the 
following sense. 
Consider a target 
or response
variable $Y$ and
covariates $X_1, \ldots, X_p$. 
If we regress $Y$ on the set 
$X_S$, 
$S \subseteq \{1, \ldots, p\}$,
of direct causes,
then this regression function 
$x \mapsto E[Y|X_S = x]$
does not change 
when intervening on any of the covariates (which is sometimes referred to as `invariance').  
This statement 
can be proved using the 
local Markov property \citep[][]{Lauritzen1996}, 
for example,
but the underlying fundamental principle 
has been discussed already several decades ago;
most prominently
using the terms 
`autonomy' or `modularity' (\citealp{Haavelmo1944}, and \citealp{Aldrich1989}).
As a result, causal models 
of the form 
$x \mapsto E[Y|X_S = x]$
may perform well
in prediction tasks, 
where, in the test distribution,
the covariates have been intervened on.
If, however, training and
test distributions
coincide, a model 
focusing only on prediction and the estimand 
$x \mapsto E[Y|X = x]$
may outperform a causal approach.

The two models described above (OLS and the causal model) formally solve a minimax problem on
distributional robustness. 
Consider therefore an 
acyclic linear structural equation model (SEM)  over $(Y,X)$ with observational distribution $F$. Details on SEMs and interventions can be found in
\Cref{sec:simsem}.
Assume that the assignment for $Y$ equals
$Y = \gamma_0^\t X + \ep_Y$
for some $\gamma_0\in \R^{d}$. The
variables corresponding to non-zero entries
in $\gamma_0^\t X$
are called
the parents of $Y$, and
$\ep_Y$ is assumed to be independent of these parents.
Then, the 
mean squared prediction error 
when considering the observational
distribution is not necessarily minimized by
$\gamma_0$, that is, in general, we have
$\gamma_0 \neq \gamma_{\text{OLS}} := \argmin_{\gamma} E_F\left[ (Y-\gamma^\t X)^2\right]$. 
Intuitively, 
we may improve the prediction of $Y$
by including other variables than the parents of $Y$, such as its descendants.
When considering distributional robustness, we 
are interested in finding a $\gamma$ that minimizes
the worst case expected squared prediction error
over a class of distributions, $\mathcal{F}$, 
that is, 
\begin{align}
	\label{eq:optimF}
	\argmin_\gamma \sup_{F\in \cF} E_F\left[ (Y-\gamma^\t X)^2\right].
\end{align}
If we observe data from 
all different distributions in $\mathcal{F}$ 
(and know which data point comes from which distribution), we can 
tackle 
this optimization 
directly
\citep[][]{MeinshausenMiniMax}.
But estimators of~\Cref{eq:optimF}
may be available even
if we do not observe data 
from each distribution in $F$.
The true causal coefficient $\gamma_0$, for example, 
minimizes \Cref{eq:optimF}
when $\cF$ is the set of all possible 
(hard)
interventions on $X$ \citep[e.g.,][]{rojas2018invariant}. 
The OLS solution is optimal when $\mathcal{F}$ only contains the training distribution.
In this sense,
the OLS solution and the true causal coefficient 
constitutes the end points of a spectrum of
estimators that are prediction optimal under a 
certain class of distributions.

Intuitively, models trading off causality and predictability may perform
well in situations, where the test distribution is only moderately
different from the training distribution. 
Anchor regression by \citet{AnchorRegression}, see Section~\ref{sec:AR} for details, is 
one approach formalizing this intuition
in a linear setup.
Similarly to an instrumental variable setting, 
one assumes the existence of exogenous variables that are called $A$ (for anchor)
which may or may not act directly on the target $Y$.
The proposed estimator minimizes a convex
combination of the residual sum of squares and the TSLS loss function
and is shown to be prediction optimal 
in the sense of
\Cref{eq:optimF} 
for a 
class $\mathcal{F}$
containing interventions on the covariates up to 
a certain strength; this strength depends on a regularization parameter: the weight that is used in the convex combination of anchor regression. %
Other approaches \citep[][]{Pfister2019stab, rojas2018invariant,Magliacane2018} search over different subsets $S$ and
aim to choose sets that are both invariant and predictive.

\subsection{Summary and contributions} \label{sec:summary}

This paper contains two main contributions: A distributional robustness property of K-class estimators with fixed $\kappa$-parameter and a novel estimator for causal coefficients called the p-uncorrelated least squares estimator (PULSE). The following two sections 
summarize our contributions.
\subsubsection{Distributional robustness of K-class estimators.}
\label{sec:intro:robustness}

In \Cref{SEC:ROBUSTNESS} we show
that anchor regression is closely related to 
K-class estimators. 
In particular,
we prove that for a restricted subclass of models K-class estimators
can be written as 
anchor regression estimators.
For this subclass, this directly implies a distributional
robustness property of K-class estimators.
We then 
prove a similar robustness
property for general
K-class estimators with a fixed penalty parameter, 
and show that these properties hold even if 
the model is misspecified. 

Consider a possibly cyclic linear SEM over the variables
$(Y,X,H,A)$ of the form
\begin{align*} 
	\begin{bmatrix}
		Y & X^\t & H^\t \end{bmatrix} 
	:=  \begin{bmatrix}
		Y &X^\t  & H^\t  \end{bmatrix} B   +  A^\t M+ \ep^\t,
\end{align*}
subject to regularity conditions that ensure that the distribution of $(Y,X,H,A)$ is well-defined.
Here, B and M are constant matrices, the random vectors $A$ and $\ep$ are defined on a common probability space $(\Omega, \cF,P)$, $Y$ is the endogenous target for the single equation inference, $X$ are the observed endogenous variables, $H$ are hidden endogenous variables and $A$ are exogenous variables independent from the unobserved noise innovations $\ep$. 

SEMs allow for the notion of interventions, i.e., modeling external manipulations of the system. In this work, we are only concerned with interventions on the exogenous variables $A$ of the form $\text{do}(A:=v)$. 
Because $A$ is exogeneous, these interventions can be defined as follows: they change the distribution of $A$ to that of a random vector $v$. The
interventional distribution of the variables $(Y,X,H,A)$ under the intervention $\text{do}(A:=v)$ is given by the simultaneous distribution of $(X_v, Y_v,H_v,v)$ generated by the SEM 
\begin{align*}
	\begin{bmatrix}
		Y_v & X_v^\t & H_v^\t \end{bmatrix} := \begin{bmatrix}
		Y_v & X_v^\t & H_v^\t \end{bmatrix} B   + v^\t M+ \ep.
\end{align*}
Thus, the intervention does not change any of the original structural assignments of the endogenous variables. 
Instead, the change in the distribution of the exogeneous variable propagates through the system. We henceforth let $E^{\mathrm{do}(A:=v)}$ denote the expectation with respect to the interventional distribution of the system under the intervention $\mathrm{do}(A:=v)$. More details on interventions can be found \Cref{sec:simsem}

Let $(\fY,\fX,\fH,\fA)$ consist of $n$ row-wise independent and identically distributed copies of the random vector $(Y,X,H,A)$ and consider the single equation of interest
\begin{align*} 
	\fY = \fX \gamma_0 + \fA \beta_0  + \fH \eta_0 + \bm{\ep}_Y = \fX \gamma_0 + \fA \beta_0  + \tilde{\fU}_Y.
\end{align*}
The K-class estimator with parameter $\kappa$ using non-sample information that only $\fZ_* \subset [\fX \, \, \fA]$ have non-zero coefficients in the target equation of interest is given by
\begin{align*}
	\hat{\alpha}_{\text{K}}^n(\kappa) =  (\fZ_*^\t (I-\kappa P_\fA^\perp)\fZ_*)^{-1} \fZ_*^\t(I-\kappa P_\fA^\perp)\fY,
\end{align*}
where $P_\fA^\perp$ is the projection onto the orthogonal complement of the column space of $\fA$. 
For a fixed $\kappa \in [0,1)$ K-class estimators can be represented by a penalized regression problem $\hat{\alpha}_{\text{K}}^n(\kappa)  = \argmin_{\alpha} l_{\mathrm{OLS}}^n(\alpha)+ \kappa/(1-\kappa)l_{\mathrm{IV}}^n(\alpha)$, where $l_{\mathrm{OLS}}^n$ and $l_{\mathrm{IV}}^n$ are the empirical OLS and TSLS loss functions, respectively. This representation and the ideas of \cite{AnchorRegression} allow us to prove that K-class estimator converges to a coefficient that is minimax optimal when considering all distributions induced by a certain set of interventions of $A$. More specifically, we show that for a fixed $\kappa$ and regardless of identifiability,
\begin{align*}
	\hat{\alpha}_{\text{K}}^n(\kappa) \overset{P}{\underset{n\to\i }{\longrightarrow}} \argmin_{\alpha}\sup_{v\in C(\kappa)}  E^{\mathrm{do}(A:=v)}\left[ (Y - \alpha^\t Z_*   )^2 \right],
\end{align*}
where $
C(\kappa) := \{ v:\Omega \to \R^{q}:  \mathrm{Cov}(v,\ep)=0, \,   E    [ vv^\t ]  \preceq \frac{1}{1-\kappa} E[AA^\t] \}$. 
The argmin on the right-hand side 
minimizes the worst case prediction error when considering 
interventions up to a certain strength (measured by the set $C(\kappa)$). 
This objective becomes relevant when we consider a response variable with several covariates and
aim to minimize the mean squared prediction error of future realizations of the system of interest that do not follow the training distribution. 
The above result says that 
if the new realizations correspond to (unknown)
interventions on the exogenous variables that are of bounded strength, K-class estimators with fixed $\kappa\in (0,1)$ 
minimize the worst case prediction performance and, in particular, outperform the true causal parameter and the least squares solution (see also \Cref{fig:DistRobustness} in \Cref{app:DistributionalRobustness}).
For $\kappa$ approaching one, we recover the guarantee of the causal solution and for $\kappa$ approaching zero, the set of distributions contains the training distribution.
The above minimax property therefore adds to the discussion 
whether 
non-consistent
K-class estimators with 
penalty parameter not 
converging to 
one
can be useful; see, e.g.\ \cite{dhrymes1974}.

\subsubsection{The PULSE estimator} \label{sec:summarypuls}
\Cref{SEC:PULSE} contains the second main contribution in this work. We
propose a novel data driven K-class estimator for causal coefficients,
which we call the p-uncorrelated least square estimator (PULSE). 
As above, we consider
a single endogenous target 
in an SEM (or simultaneous equation model)
and aim to predict it 
from observed predictors that are with a priori (non-sample) information known to be either endogenous or exogenous. The PULSE estimator can be written in several equivalent forms. It can, first, be seen as a data-driven K-class estimator 
	\begin{align*}
		\hat{\alpha}^n_{\mathrm{K}}( \lambda^\star_n/(1+\lambda^\star_n)  ) = \argmin_{\alpha} l_{\mathrm{OLS}}^n(\alpha) + \lambda^\star_n l_{\mathrm{IV}}^n(\alpha) ,
	\end{align*}
	where
	\begin{align*}
		\lambda^\star_n
		:= \inf \left\{ \lambda > 0\,:\,  \begin{tabular}{c}
			\text{testing }\text{Corr}$(A,Y-Z\hat{\alpha}_{\text{K}}^n(\lambda/(1+\lambda)))=0$ \\
			\text{yields a }p\text{-value }$ \geq p_{\min}$
		\end{tabular}
		\right\},
	\end{align*}
	for some pre-specified level of the hypothesis test $p_{\min}\in(0,1)$. 
	In words, the PULSE estimator outputs the K-class estimator closest to the OLS while maintaining a non-rejected test of uncorrelatedness.
	In principle, PULSE can be used with any testing procedure.
	The choice of test, however, may influence the difficulty of the 
	resulting optimization problem. In this paper, we investigate PULSE in connection with a specific class of hypothesis tests that, for example, contain the test of \cite{anderson1949estimation}.  For these hypothesis tests we develop an efficient 
	and 
	provably correct optimization method, that is based on binary line search and quadratic programming.

	We show that our estimator can, second, 
	be written as the solution to a constrained optimization problem. To that end, define the primal problems
	\begin{align*} 
		\hat{\alpha}_{\text{Pr}}^n(t) :=  \begin{array}{ll}
			\argmin_{\alpha} & l_{\mathrm{OLS}}^n(\alpha)  \\
			\text{subject to} & l_{\mathrm{IV}}^n(\alpha)\leq t.
		\end{array}
	\end{align*}
	For the choice $t^\star_n := \sup\{t\,:\, \text{testing } \mathrm{Corr}(A,Y-Z\hat{\alpha}_{\text{Pr}}^n(t))=0$ $\text{yields a}$ $p\text{-value} \geq p_{\min}\}$,
	we provide a detailed analysis proving that  
	$\hat{\alpha}^n_{\mathrm{K}}
	(\lambda^\star_n/(1+\lambda^\star_n) ) 
	= \hat{\alpha}_{\text{Pr}}^n(t^\star_n)$. %

	For the 
	testing procedure proposed in this paper, we show that, third, 
	PULSE can be written as 
	\begin{align*} 
		\begin{array}{ll}
			\text{argmin}_\alpha & l_{\mathrm{OLS}}^n(\alpha;\fY,\fZ)  \\
			\text{subject to} & \alpha \in \cA_n(1-p_{\min}),
		\end{array}
	\end{align*}
	where $\cA_n(1-p_{\min})$ is the non-convex acceptance region for our test of uncorrelatedness.

	This third formulation allows for a simple interpretation of our estimator: among all coefficients (not restricted to K-class estimators) that do not yield a rejection of uncorrelatedness, we choose the one that yields the best prediction. 
	If the acceptance region is empty it outputs a warning indicating a possible model misspecification or an assumption violation to the user (in that case, one can formally output another estimator such as TSLS or Fuller, yielding PULSE well-defined).

	In the just-identified setup, the TSLS estimator solves a 
	normal equation which is equivalent to setting a sample covariance between the instruments and the resulting prediction residuals to zero; it then corresponds to $t=0$. 
	For this (and the over-identified) setting, we prove that PULSE is a consistent estimator for the causal coefficient. 
	
	The TSLS does not have a finite variance if there is insufficient degree of overidentification, for example. 
	In particular for weak instruments, this usually comes with poor finite sample performance. In such cases, however, the acceptance region of uncorrelatedness is usually large. This yields a weak constraint in the optimization problem and the PULSE will be closer to the OLS, which in certain settings suffers from less variability \citep[see, e.g.,][]{hahn2004estimation,HAHN05}. 
	In simulations we indeed see that, similarly to other data-driven K-class estimators that are pulled towards the OLS, such as Fuller estimators, the PULSE comes with beneficial finite sample properties compared to TSLS and LIML.

	Unlike other estimators such as LIML or the classical TSLS, the PULSE is well-defined in under-identified settings, too. Here, its objective is still to find the best predictive solution among all parameters that do not reject uncorrelatedness. 
	Uncorrelatedness to the exogeneous variable is sometimes referred to as invariance. 
	The idea of choosing the best predictive among all invariant models has been investigated in several works \citep[e.g.][]{Pfister2019stab, rojas2018invariant, Magliacane2018}  with the motivation to 
	find models that generalize well (in particular, with respect to interventions on the exogenous variables). 
	Existing methods, however, focus on selecting subsets of variables and then consider least squares regression of the response variable onto the full subset. 
	PULSE can recover such type of solutions if they are indeed optimal. 
	But it also allows to search over coefficients that are different from least squares regression for sets of variables. 
	Consequently, PULSE allows us to find solutions in situations, where the above methods would not find any invariant subsets, which may often be the case if there are hidden variables (see \Cref{app:underidentifiedexperiment} for an example).

	We show in a simulation study that there are several settings in which 
	PULSE outperforms existing estimators both in terms of MSE ordering and several one-dimensional scalarizations of the MSE. More specifically, we show that PULSE can outperform the TSLS and Fuller estimators in weak instrument situations, for example, where Fuller estimators are known to
	have good MSE properties; see, e.g.\ \citet{hahn2004estimation} and \citet{stock2002survey}.

	Implementation of  PULSE and code for experiments (R) are available on GitHub.\footnote{
		\url{https://github.com/MartinEmilJakobsen/PULSE}}

	\section{Robustness properties of K-class estimators} \label{SEC:ROBUSTNESS}
	\setcounter{equation}{0}
	In this section 
	we consider
	K-class 
	estimators (\citealp{theil1958economic}, and \citealp{nagar1959bias})
	and show a connection with anchor regression  of \citet{AnchorRegression}. %
	In \Cref{sec:KclassInIVModelCoincidesWithAr} we establish the connection 
	in models where we use \textit{a priori} information that there are no included exogenous variables in the target equation of interest. 
	In \Cref{sec:KclassAsPenalizedRegression} we then show that general
	K-class estimators can be written as the solution to a penalized regression problem. In \Cref{sec:intervrobustnessofKclass} we utilize this representation and the ideas of \citet{AnchorRegression} to prove a distributional robustness guarantee of general K-class estimators with fixed $\kappa\in[0,1)$,
	even under model misspecification and non-identifiability. 	Proofs of results in this section can be found in \Cref{sec:RobustnessProofs}.
	
	\subsection{Setup and assumptions} \label{sec:SetupAndAssumptionsRobustness}
	Denote the random vectors $Y\in \R, X\in \R^{d}, A\in \R^{q},H\in \R^r$  and $\ep\in \R^{d+1+r}$ 
	by the target, endogenous regressor, anchors, hidden and noise variables, respectively.
	Let further %
	$(Y,X,H)$ be generated by the possibly cyclic  structural equation model (SEM)
	\begin{align} \label{ARModel}
		\begin{bmatrix}
			Y & X^\t & H^\t \end{bmatrix} 
		:=  \begin{bmatrix}
			Y &X^\t  & H^\t  \end{bmatrix} B   +  A^\t M+ \ep^\t,
	\end{align}
	for some random vectors $\ep\independent A$ and constant matrices $B$ and $M$.  Let $(\fY,\fX,\fH,\fA)$ consist of $n\geq \min\{d,q\}$ row-wise independent and identically distributed copies of the random vector $(Y,X,H,A)$. Solving for the endogenous variables we get the structural and reduced form equations $[\,\fY\,\, \fX\,\, \fH\,]\, \Gamma = \fA M + \bm{\ep}$ and $[\,\fY\,\, \fX\,\, \fH\,] = \fA \Pi + \bm{\ep} \Gamma^{-1}$, 
	where $\Gamma := I-B$ and  $\Pi := M \Gamma^{-1}$.  Assume without loss of generality that $\Gamma$ has a unity diagonal, such that the target equation of interest is given by
	\begin{align} \label{eq:StructuralEquationOfInterest}
		\fY = \fX \gamma_0 + \fA \beta_0  + \fH \eta_0 + \bm{\ep}_Y = %
		\fZ \alpha_0 + \tilde{\fU}_Y,
	\end{align}
	where $(1, -\gamma_0,- \eta_0)\in \R^{(1+d+r)}$, $ \beta_0\in \R^q$ and $\bm{\ep}_Y$  are the first columns of $\Gamma$, $M$ and $\bm{\ep}\in \R^n$ respectively, $\fZ := [
	\fX \, \, \fA]$, $\alpha_0 = (\gamma_0, \beta_0)\in \R^{d+q}$ and $\tilde{\fU}_Y := \fH\eta_0 + \bm{\ep}_Y$.  
	
	The possible dependence between the noise $\tilde \fU_Y$ and the endogenous variables, i.e., the influence by hidden variables, generally,
	renders the standard OLS approach for estimating $\alpha_0$ inconsistent. 
	Instead, one can make use of the components in $A$ that have vanishing coefficient in 
	\Cref{eq:StructuralEquationOfInterest} 
	for consistent estimation.  In the remainder of this work, we disregard any \textit{a priori} (non-sample) 
	information not concerning the target equation.
	The question of identifiability of $\alpha_0$ has been studied extensively \citep[][]{frisch38,Haavelmo1944,measuringtheequationsystems} and more recent overviews can be found in, e.g., 
	\citet{Didelez2010}, \citet{fisher1966identification}, and \citet{greene2003econometric}.

	We will use the following assumptions concerning the structure of the SEM:

	\begin{assumption}[Global assumptions] \label{ass:global} 
		\begin{enumerate*}[label=(\alph*),ref=\ref{ass:global}.(\alph*)] \justifying
			\item 
			$(Y,X,H,A)$ is generated in accordance with the SEM in \Cref{ARModel}; 
			\label{ass:linearSEM}
			\item $\rho(B)<1$ where $\rho(B)$ is the spectral radius of $B$;  \label{ass:SpectralRadiusOfBLessThanOne}
			\item $\ep$ has jointly independent marginals $\ep_1, \ldots, \ep_{d+1+r}$;\label{ass:epIndependentMarginals}
			\item $A$ and $\ep$ are independent; \label{ass:AindepEp}
			\item No variable in $Y$, $X$ and $H$ is an ancestor of $A$, that is, $A$ is exogenous;
			\label{ass:Aexogenous}
			\item $E[\|\ep\|^2_2]$, $E[\|A\|^2_2] < \i$; \label{ass:SecondMomentEp}
			\label{ass:SecondMomentA}
			\item $E[\ep] =0$. \label{ass:EpMeanZero}
			\item $\text{Var}(A) \succ 0$, i.e., the variance matrix of $A$ is positive definite; \label{ass:VarianceOfAPositiveDefinite}
			\item $\fA^\t \fA$ is almost surely of full rank;  \label{ass:AtAfullRank}
		\end{enumerate*}
	\end{assumption}
	\begin{assumption}[Finite sample  assumptions] \label{ass:finiteass}
		\begin{enumerate*}[label=(\alph*),ref=\ref{ass:finiteass}.(\alph*)]
			\item 	$\fZ_*^\t \fZ_*$ is almost surely of full rank;  \label{ass:ZtZfullRank}
			\item $\fA^\t \fZ_*$ is almost surely of full column rank. \label{ass:AtZfullRank}
			\item $\fX^\t \fX$ is almost surely of full rank; \label{ass:XtXfullRank}
		\end{enumerate*}
	\end{assumption}
	\begin{assumption}[Population assumptions] \label{ass:popass}
		\begin{enumerate*}[label=(\alph*),ref=\ref{ass:popass}.(\alph*)]
			\item $\text{Var}(Z_*) \succ 0$, i.e., the variance matrix of $Z_*$ is positive definite;\label{ass:VarianceOfZPositiveDefinite}
			\item 	$E[AZ_*^\t]$ is of full column rank. \label{ass:EAZtFullColumnRank}
		\end{enumerate*}
	\end{assumption}
	We will henceforth assume that \Cref{ass:global} always holds. This assumption ensure that the SEM and that the TSLS objectives are well-defined.
	In the above assumptions, $Z_*$ and $\fZ_*$ are 
	generic placeholders for a subset of endogenous and exogenous variables from 
	$[
	X^\t \, \,  A^\t
	]^\t$ 
	and
	$[
	\fX \, \,  \fA
	]$, respectively,
	which should be clear from the context in which they are used.  Both Assumption \Cref{ass:AtAfullRank} and Assumption \Cref{ass:XtXfullRank} hold if $X$ and $A$ have density with respect to Lebesgue measure, which in turn is guaranteed by Assumption \Cref{ass:AindepEp} if $A$ and $\ep$ have density with respect to Lebesgue measure.
	Assumption \Cref{ass:VarianceOfAPositiveDefinite} and \Cref{ass:AtAfullRank} implies that the instrumental variable objective functions introduced below is almost surely well-defined and Assumption \Cref{ass:XtXfullRank} yields that the ordinary least square solution is almost surely well-defined. Assumption \Cref{ass:SecondMomentA,ass:SecondMomentEp} implies that $Y,X$ and $H$ all have finite second moments. %
	For Assumption \Cref{ass:EAZtFullColumnRank} and \Cref{ass:AtZfullRank} it is necessary that $q\geq \mathrm{dim}(Z_*)$, i.e., that the setup must be just- or over-identified; see \Cref{sec:SetupAndAssumptionsPULSE} below.
	\subsection{Distributional robustness of anchor regression} \label{sec:AR}
	\citet{AnchorRegression} proposes a method, called anchor regression,
	for predicting 
	the endogenous target variable $Y$ 
	from the endogenous variables $X$. %
	The collection of exogenous variables $A$,
	called anchors, are not included in that prediction model. 
	Anchor regression trades off predictability and invariance
	by considering a convex combination of 
	the ordinary least square (OLS) loss function and
	the two-stage least square (IV) loss function using the anchors as instruments.
	More formally, we define 
	\begin{align} \label{eq:lossPop}
		l_{\mathrm{OLS}}(\gamma ;Y,X) &:= E(Y-\gamma^\t X)^2, \\
		l_{\mathrm{IV}}(\gamma ;Y,X,A) %
		&:=E(A(Y-\gamma^\t X))^\t   E(A A^\t)^{-1} E(A(Y- \gamma^\t X)), \notag \\ \label{eq:lossEmp}
		l^n_{\mathrm{OLS}}(\gamma;\fY,\fX) &:=  n^{-1} (\fY-\fX \gamma)^\t (\fY-\fX \gamma),\\
		l^n_{\mathrm{IV}}(\gamma;\fY,\fX, \fA) &:=n^{-1} (\mathbf{Y}-\mathbf{X}\gamma)^\t  P_\fA (\mathbf{Y}-\mathbf{X}\gamma),
	\end{align}
	as the population and finite sample versions of the loss functions.
	$P_\fA = \fA(\fA^\t \fA)^{-1}\fA^\t$
	is the orthogonal projection onto the column space of $\fA$. 
	To simplify notation, we omit the dependence on $Y$, $X$, $A$, $\fA$, $\fX$ or $\fY$ when they are clear from a given context.
	For a penalty parameter $\lambda> -1$, 
	the anchor regression coefficients are defined as
	\begin{align} %
		\gamma_{\mathrm{AR}}(\lambda)&:= %
		\argmin_{\gamma\in \R^d} \{l_{\text{OLS}}(\gamma)+\lambda l_{\text{IV}}(\gamma) \},	\quad
		\label{eq:EmpiricalARestimator}
		\hat{\gamma}_{\mathrm{AR}}^n(\lambda):= %
		\argmin_{\gamma\in \R^d} \{l_{\text{OLS}}^n(\gamma)+\lambda l_{\text{IV}}^n (\gamma)\}.
	\end{align}
	The estimator $\hat{\gamma}_{\text{AR}}^n(\lambda)$ consistently estimates the population estimand $\gamma_{\text{AR}}(\lambda)$
	and
	minimizes prediction error while simultaneously penalizing a transformed sample covariance between the anchors and the resulting prediction residuals. 
	Unlike the TSLS estimator, for example, the anchor regression estimator is 
	almost surely well-defined under the rank condition of Assumption \Cref{ass:XtXfullRank}, even if the model is under-identified, that is, there are less exogenous than endogenous variables. The solution to the empirical minimization problem of anchor regression is given by
	\begin{align} \label{eq:ARsolution}
		\hat{\gamma}_{\text{AR}}^n(\lambda) = [\fX^\t (I+\lambda P_\fA)\fX]^{-1}\fX^\t(I+\lambda P_\fA)\fY,
	\end{align}
	which follows from solving the normal equation of \Cref{eq:EmpiricalARestimator}.

	The motivation of anchor regression
	is not to %
	infer a causal parameter.
	Instead, for a fixed penalty parameter 
	$\lambda$,
	the estimator is shown to
	possess a distributional or interventional robustness 
	property: the estimator is 
	optimal when predicting under interventions on the exogenous 
	variables that are below a certain intervention strength. 
	By Theorem 1 of \citet{AnchorRegression} it holds that $$
	\gamma_{\text{AR}}(\lambda) = \argmin_{\gamma\in \R^d} \sup_{v\in C(\lambda)} E^{\text{do}(A:=v)}\left[ \lp Y-\gamma^\t X \rp^2	\right],$$ 
	where $
	C(\lambda) := \lb  v:\Omega \to \R^q : \text{Cov}(v,\ep)=0, E(v v^\t) \preceq (\lambda+1) E(A A^\t )  \rb.$ 
	
	\subsection{Distributional robustness of K-class estimators} \label{sec:RobustnessOfKclass}
	We now introduce the limited information estimators known as K-class estimators  (\citealp{theil1958economic}, and \citealp{nagar1959bias}) used
	for single equation inference. 
	Suppose that we are given non-sample information about which components of $\gamma_0$ and $\beta_0$, of \Cref{eq:StructuralEquationOfInterest}, are zero. 
	We can then 
	partition $\fX = [
	\fX_* \, \, \fX_{-*}
	] \in \R^{n\times(d_1+d_2)}$, $\fA = [\fA_{*} \, \,  \fA_{-*}]\in \R^{n\times(q_1+q_2)}$ and $\fZ = [
	\fZ_* \, \, \fZ_{-*}
	] = [
	\fX_{*} \, \,  \fA_* \, \,  \fX_{-*}  \, \,  \fA_{-*}
	]$
	with $ \fZ \in \R^{n\times((d_1+q_1)+(d_2+q_2))}$, where $\fX_{-*}$ and $\fA_{-*}$ corresponds to the variables for which 
	our non-sample information states that 
	the components of $\gamma_0$ and $\beta_0$ are zero, respectively. 
	We call the variables corresponding to $\fA_*$ included exogenous variables. 
	Similarly, we write
	$\gamma_{0} =(\gamma_{0,*},\gamma_{0,-*}) $, $\beta_0 = (\beta_{0,*},\beta_{0,-*})$ and  $\alpha_0 = (
	\alpha_{0,*}, \alpha_{0,- *}) =(\gamma_{0,*},\beta_{0,*},\gamma_{0,-*},\beta_{0,-*})$.  The structural equation of interest then reduces to $
	\fY = \fX_* \gamma_{0,*} + \fX_{-*} \gamma_{0,-*} +  \fA_* \beta_{0,*} +\fA_{-*} \beta_{0,-*} + \tilde{\fU}_Y = \fZ_* \alpha_{0,*} + \fU_Y$, 
	where $\fU_Y = \fX_{-*} \gamma_{0,-*} + \fA_{-*} \beta_{0,-*}+ \fH\eta_0 + \bm{\ep}_Y$.
	In the case that the non-sample information is indeed correct, we have that $\fU_Y = \tilde{\fU}_Y= \fH\eta_0 + \bm{\ep}_Y$. When well-defined, the K-class estimator with 
	parameter $\kappa\in \R$ for a simultaneous estimation of $\alpha_{0,*}$ is given by
	
	\begin{align} \label{eq:KclassEstimatorInSingleLine}
		\hat{\alpha}_{\text{K}}^n(\kappa;\fY,\fZ_*,\fA) =  (\fZ_*^\t (I-\kappa P_\fA^\perp)\fZ_*)^{-1} \fZ_*^\t(I-\kappa P_\fA^\perp)\fY,
	\end{align}
	where 
	$I-\kappa P_\fA^\perp = I- \kappa(I-P_\fA) = (1-\kappa)I + \kappa P_\fA$. %
	
	Comparing \Cref{eq:ARsolution,eq:KclassEstimatorInSingleLine}
	suggests a close connection 
	between anchor regression 
	and K-class estimators for inference of structural equations with no included exogenous variables. In the following subsections, we establish this connection and subsequently extend the distributional robustness
	property to general K-class estimators.

	\subsubsection{K-class estimators in models with no included exogenous variables}

	\label{sec:KclassInIVModelCoincidesWithAr}
	Assume that,
	in addition to \Cref{ass:global}, we have the 
	non-sample information that $\beta_0=0$, that is, no exogenous variable in $A$ directly affects the target variable $Y$. %
	By direct comparison we see that the K-class estimator for $\kappa<1$ coincides with the anchor regression estimator with penalty parameter $\lambda =\kappa/(1-\kappa)$, i.e., $\hat{\gamma}_{\text{K}}^n(\kappa)=  \gamma_{\mathrm{AR}}^n \lp\frac{\kappa }{1-\kappa} \rp$.
	Equivalently, we have $\gamma_{\mathrm{AR}}^n \lp \lambda \rp =  \gamma_{\text{K}}^n \lp\lambda /(1+\lambda) \rp$ for any  $\lambda >-1$. As such, the K-class estimator, for a fixed $\kappa$,  inherits the following distributional robustness property:
	\begin{align} \label{eq:robIV}
		\gamma_{\text{K}}(\kappa) &=  \gamma_{\mathrm{AR}} \lp\frac{\kappa }{1-\kappa} \rp = \argmin_{\gamma\in \R^d} \sup_{v\in C(\kappa/(1-\kappa))} E^{\mathrm{do}(A:=v)} \left[ \lp Y-\gamma^\t X\rp^2 \right],
	\end{align}
	where $
	C(\kappa/(1-\kappa)) = \{  v:\Omega \to \R^q : \text{Cov}(v,\ep)=0, E[v v^\t] \preceq \frac{1}{1-\kappa} E[A A^\t]  \}$. 
	This statement holds by Theorem 1 of \citet{AnchorRegression}. 
	
	In an identifiable model with $P \lim_{n\to \i}\kappa =1$ we have that $\hat{\gamma}^n_{\text{K}}(\kappa)$
	consistently estimates the causal parameter; see  e.g.\ \citet{mariano2001simultaneous}. 
	For such
	a choice of $\kappa$, the robustness above is just a weaker version of what the causal coefficient can guarantee. 
	However, the above result in~\Cref{eq:robIV}
	establishes a robustness property for fixed $\kappa <1$, even in cases where the model is not identifiable. 
	Furthermore,
	since we did not use that the non-sample information that $\beta_0=0$ was true, 
	the robustness property is resilient to model misspecification in terms of excluding included exogenous variables from the target equation which generally also breaks identifiability. %

	\subsubsection{The K-class estimators as penalized regression estimators} \label{sec:KclassAsPenalizedRegression}
	We now show that 
	general K-class estimators 
	can be written as
	solutions to penalized regression problems. 
	The first appearance of such a representation is, to the best of our knowledge, due to \citet{mcdonald1977k} building upon previous work of \citet{basmann1960finite,basmann1960asymptotic}. Their representation, however, 
	concerns only the endogenous part $\gamma$.
	We require a slightly different statement and
	will show that the entire K-class estimator of $\alpha_{0,*}$, i.e., the simultaneous estimation of $\gamma_{0,*}$ and $\beta_{0,*}$,
	can be written as a penalized regression problem. 
	Let therefore $l_{\mathrm{IV}}(\alpha;\fY,\fZ_*,\fA)$, $l_{\mathrm{IV}}^n(\alpha;\fY,\fZ_*,\fA)$ and $l_{\mathrm{OLS}}(\alpha;\fY,\fZ_*)$, $l_{\mathrm{OLS}}^n(\alpha;\fY,\fZ_*)$ denote the population and empirical TSLS and OLS loss functions as defined in \Crefrange{eq:lossPop}{eq:lossEmp}. That is, the TSLS loss function for regressing $\fY$ on the included endogenous and exogenous variables $\fZ_*$ using the exogeneity of $\fA$ and $\fA_{-*}$ as instruments and the OLS loss function for regressing $\fY$ on $\fZ_*$. 
	We define the K-class population and finite-sample loss functions as 
	an affine combination of the two loss functions above. That is, 
	\begin{align} \label{KclassLossFunctionPop}
		l_{\mathrm{K}}(\alpha;\kappa,Y,Z_*,A)&= (1-\kappa)l_{\mathrm{OLS}}(\alpha;Y,Z_*) + \kappa l_{\mathrm{IV}}(\alpha;Y,Z_*,A), \\ \label{KclassLossFunctionEmp}
		l_{\mathrm{K}}^n(\alpha;\kappa,\fY,\fZ_*,\fA)&= (1-\kappa)l_{\mathrm{OLS}}^n(\alpha;\fY,\fZ_*) + \kappa l_{\mathrm{IV}}^n(\alpha;\fY,\fZ_*,\fA). 
	\end{align}
	\begin{restatable}[]{proposition}{PenalizedKClassSolutionUniqueAndExists}
		\label{lm:PenalizedKClassSolutionUniqueAndExists}%
		Consider one of the following scenarios: 1) $\kappa <1$ and Assumption \Cref{ass:ZtZfullRank} holds, or 2) $\kappa = 1$ and Assumption \Cref{ass:AtZfullRank} holds.
		The estimator minimizing the empirical loss function of \Cref{KclassLossFunctionEmp} is almost surely well-defined  and coincides with the K-class estimator of \Cref{eq:KclassEstimatorInSingleLine}. That is, it almost surely holds that
		\begin{align} \label{eq:kclassminim}
			\hat{\alpha}_{\mathrm{K}}^n(\kappa;\fY,\fZ_*,\fA) =\argmin_{\alpha\in \R^{d_1+q_1}} l_{\mathrm{K}}^n(\alpha;\kappa,\fY,\fZ_*,\fA).
		\end{align}
	\end{restatable}
	
	Assuming $\kappa\not =1$, we can rewrite \Cref{eq:kclassminim} to
	\begin{align}
		\label{eq:KclassLossFunctionAsPenalizedOLS}
		\hat{\alpha}_{\text{K}}^n(\kappa;\fY,\fZ_*,\fA) = \argmin_{\alpha\in \R^{d_1+q_1}} \{  l^n_{\mathrm{OLS}}(\alpha;\fY,\fZ_*) + \frac{\kappa}{1-\kappa} l^n_{\mathrm{IV}}(\alpha;\fY,\fZ_*,\fA) \}.
	\end{align}
	Thus, 
	K-class estimators
	seek to minimize the ordinary least squares loss for regressing $\fY$ on  $\fZ_*$, while simultaneously penalizing the strength of a transform on the sample covariance between  %
	the prediction residuals and  %
	collection of %
	exogenous variables $\fA$.

	In the following section, we 
	consider a population version of the above quantity.
	If we replace the finite sample \Cref{ass:finiteass}
	with the 
	corresponding population \Cref{ass:popass}, we get that the minimization estimator of the empirical loss function of \Cref{KclassLossFunctionEmp} is asymptotically well-defined. Furthermore, 
	we now prove that
	whenever the population assumptions are satisfied, then,
	for any fixed $\kappa \in [0,1]$, 
	$\hat{\alpha}_{\text{K}}^n(\kappa;\fY,\fZ_*,\fA)$ 
	converges in probability towards the population K-class estimand.
	
	\begin{restatable}[]{proposition}{PopulationPenalizedKClassSolutionUniqueAndExists}
		\label{lm:PopulationPenalizedKClassSolutionUniqueAndExists}%
		Consider one of the following scenarios: 1) $\kappa \in [0,1)$ and Assumption  \Cref{ass:VarianceOfZPositiveDefinite} holds, or 2) $\kappa = 1$ and Assumption  \Cref{ass:EAZtFullColumnRank} holds.
		It holds that  $(\hat{\alpha}_{\text{K}}^n(\kappa;\fY,\fZ_*,\fA))_{n\geq 1}$ is an asymptotically well-defined sequence of estimators. %
		Furthermore, the sequence consistently estimates the well-defined population K-class estimand. That is,

		$$
		\hat{\alpha}_{\mathrm{K}}^n(\kappa;\fY,\fZ_*,\fA) \overset{P}{\underset{n\to\i }{\longrightarrow}} \alpha_{\mathrm{K}}(\kappa;Y, Z_*,A) := \argmin_{\alpha\in \R^{d_1+q_1}}l_{\mathrm{K}}(\alpha;\kappa,Y,Z_*,A).
		$$
	\end{restatable}

	\subsubsection{Distributional robustness of general K-class estimators}
	\label{sec:intervrobustnessofKclass}
	We are now able to prove that the 
	general K-class estimator
	possesses a robustness property similar to the 
	statements above.
	It is prediction optimal under a set of interventions, now 
	including interventions on all
	exogenous $A$ up to a certain strength. 
	\begin{restatable}[]{theorem}{TheoremIntRobustKclass}
		\label{sthm:TheoremIntRobustKclas}%
		Let \Cref{ass:global} hold. 
		For any fixed $\kappa \in[0,1)$ and $Z_*=(X_*, A_*)$ with $X_*\subseteq X$ and $A_* \subseteq A$, we have, whenever the population K-class estimand is well-defined, that 
		\begin{align*}
			\alpha_{\mathrm{K}}(\kappa;Y,Z_*,A)&= \argmin_{\alpha\in \R^{d_1+q_1}}\sup_{v\in C(\kappa)}  E^{\mathrm{do}(A:=v)}\left[ (Y - \alpha^\t Z_*   )^2 \right], %
		\end{align*}
		where $
		C(\kappa) := \lb v:\Omega \to \R^{q}:  \mathrm{Cov}(v,\ep)=0, \,   E    [ vv^\t ]  \preceq \frac{1}{1-\kappa} E[AA^\t] \rb$.
	\end{restatable}
	Here, $E^{\text{do}(A:=v)}$ denotes the expectation with respect to the distribution entailed under the intervention $\text{do}(A:=v)$ 
	(see Section~\ref{sec:intro:robustness} and \Cref{sec:simsem})
	and $(\Omega,\cF,P)$ is the common background probability space on which $A$ and $\ep$ are defined.

	In words, 
	among all linear prediction methods of $Y$ using $Z_*$ as predictors, 
	the 
	K-class estimator 
	with parameter $\kappa$
	has the lowest possible worst case mean squared prediction  error when considering 
	all interventions on the exogenous variables $A$ contained in $C(\kappa)$. 
	As $\kappa$ approaches one, the
	estimator is prediction optimal under 
	a class of arbitrarily strong
	interventions %
	in the direction of the variance of $A$. (Here, $\kappa$ is arbitrary but fixed; the statement does not cover data-driven choices of $\kappa$, such as LIML or Fuller.)
	The above result is a consequence of the relation between anchor regression and 
	K-class estimators.
	The special case $A_* = \emptyset$ 
	is a consequence of Theorem~1 by \citet{AnchorRegression}. 
	Our 
	proof 
	follows similar arguments but 
	additionally allows for $A_* \not = \emptyset$.
	
	The property in Theorem~\ref{sthm:TheoremIntRobustKclas}
	has a decision-theoretic interpretation (see \citet{Chamberlain2007} for an application of decision theory in IV models based on another loss function).
	Consider a response $Y$, covariates $Z_*$ and a distribution (specified by $\theta$) over $(Y, Z_*)$,
	and the squared loss $\ell(Y, Z, \alpha) := (Y-\alpha^\top Z_*)^2$.
	Then, assuming finite variances, for each distribution the risk 
	$E_{\theta}[(Y-\alpha^\top Z_*)^2]$ is minimized by the (population) OLS solution
	$\alpha=\alpha_\theta := \mathrm{cov}_\theta(Z_*)^{-1}\mathrm{cov}_\theta(Z_*,Y)$.
	In the setting of Theorem~\ref{sthm:TheoremIntRobustKclas}, we are given a distribution over $(Y,Z_*)$, specified by $\theta$,  but we are 
	interested in minimizing the risk
	$E_{\theta, v}[(Y-\alpha_{\theta}^\top Z_*)^2]$ 
	for another 
	distribution that is induced by an intervention and specified by $(\theta, v)$. 
	The above result states that the K-class estimator minimizes a worst-case risk when considering all $v \in C(\kappa)$.

	Theorem~\ref{sthm:TheoremIntRobustKclas}
	makes use of the language of SEMs in that it yields the notion of interventions.\footnote{In particular, we have not considered the SEM as a model for counterfactual statements.}
	As such,  the result can be rephrased using other
	causal frameworks. The crucial assumptions are the exogeneity of $A$ and the linearity of the system. 
	Furthermore, the result is robust with respect to 
	several types of model misspecifications that breaks identifiability of $\alpha_0$, such as excluding included endogenous or exogenous predictors or the existence of
	latent variables; see \Cref{rm:ModelMispecification} in \Cref{app:AddRemarks}.

	\section{The p-uncorrelated least square estimator} \label{SEC:PULSE}
	\setcounter{equation}{0}
	We now introduce 
	the p-uncorrelated least square estimator (PULSE). 
	As discussed in \Cref{sec:summary}, PULSE allows for different representations. In this section we start with the third representation and show the equivalence of the other representations afterwards.

	Consider predicting the target $Y$ from endogenous and possibly exogenous regressors $Z$. Let therefore
	$\cH_0(\alpha)$ denote the hypothesis that the prediction residuals using $\alpha$ as a regression coefficient is simultaneously uncorrelated with every exogenous variable, that is, $\cH_0(\alpha) : \text{Corr}(A,Y-\alpha^\t Z) =0$.
	This hypothesis is in some models under certain conditions equivalent to the hypothesis that $\alpha$ is the true causal coefficient. One of these conditions is the rank condition \Cref{ass:EAZtfullrank} introduced below, also known as the rank condition for identification; \citet{wooldridge2010econometric}.

	The two-stage least square (TSLS) estimator exploits the equivalence between the causal coefficient and the zero correlation between the instruments and the regression residuals. Here, one minimizes
	a sample covariance between the instruments and the regression residuals: we can write
	$l^n_{\mathrm{IV}}(\alpha;\fY,\fZ, \fA) 
	= \| \widehat{\text{Cov}}_n(A,Y-\alpha^\t Z) \|^2_{(n^{-1}\fA^\t \fA)^{-1}}$ when $A$ is mean zero.\footnote{$\|\cdot\|_{(n^{-1}\fA^\t \fA)^{-1}}$ is the norm induced by the inner product 
		$\la x,y\ra = x^\t (n^{-1}\fA^\t \fA)^{-1} y$.}
	In the just-identified setup 
	the TSLS estimator yields a sample covariance that is exactly zero and is known to be 
	unstable, in that it has no moments of any order. 
	Intuitively, the constraint of vanishing 
	sample covariance 
	may be too strong. 
		
		Let $T(\alpha;\fY,\fZ,\fA)$ be a finite sample test statistic for testing  the hypothesis $\cH_0(\alpha)$ 
		and let $\text{p-value}(T(\alpha;\fY,\fZ,\fA))$ denote the p-value associated with the test of $\cH_0(\alpha)$.
		We then define the p-uncorrelated least square estimator (PULSE) as
		\begin{align} \label{eq:PULSEfirstEQ}
			\hat{\alpha}^n_{\mathrm{PULSE}}(p_{\min}) = \begin{array}{ll}
				\text{argmin}_\alpha & l_{\mathrm{OLS}}^n(\alpha;\fY,\fZ)  \\
				\text{subject to} & \text{p-value}(T(\alpha;\fY,\fZ,\fA)) \geq p_{\min},
			\end{array}
		\end{align}
		where  $p_{\min}$ is a pre-specified level of the hypothesis test.  In words, we aim to minimize the mean squared prediction error
		among all coefficients which yield a
		p-value for testing $\cH_0(\alpha)$ 
		that does not fall below some pre-specified level-threshold $p_{\min} \in (0,1)$, such as $p_{\min}= 0.05$. 
		That is, the minimization is constrained to the acceptance 
		region of the test, i.e., a confidence region for the causal coefficient in the identified setup. 
		Among these coefficient, 
		we choose the solution that is `closest' to the OLS
		solution.\footnote{  
			Here, closeness is measured in the
			OLS distance:
			We define the OLS norm via
			$\|\alpha\|_{\text{OLS}}^2 := 
			l_{\mathrm{OLS}}^n(\alpha + \hat{\alpha}^n_{\mathrm{OLS}})
			- 
			l_{\mathrm{OLS}}^n(\hat{\alpha}^n_{\mathrm{OLS}})
			= \alpha^\top \mathbf{Z}^T \mathbf{Z} \alpha$, where $\hat{\alpha}^n_{\mathrm{OLS}}$ is the OLS estimator.
			This defines a norm (rather than a semi-norm) 
			if $\mathbf{Z}^T \mathbf{Z}$ is
			non-degenerate. Minimizing 
			$l_{\mathrm{OLS}}^n(\alpha)=\|\fY-\fZ\alpha\|_2^2 = (\alpha-\hat{\alpha}^n_{\mathrm{OLS}})^\t \fZ^\t \fZ (\alpha- \hat{\alpha}^n_{\mathrm{OLS}})+\|\fY-\fZ \hat{\alpha}^n_{\mathrm{OLS}}\|_2^2$ is equivalent 
			to 
			minimizing
			$\|\alpha - \hat{\alpha}^n_{\mathrm{OLS}}\|_{\text{OLS}}^2$.
		}
		
		Thus, PULSE allows for an intuitive interpretation. We will see in the experimental section that it has good finite sample performance, in particular for weak instruments.
		Unlike other estimators, such as LIML, the above estimator is well-defined in the under-identified setup, too.\footnote{
			The PULSE estimator is defined for finite samples, but the 
			following deliberation may help to build intuition:
			In an under-identified IV setting, minimizing $l_{\mathrm{OLS}}(\gamma)$ under the constraint that $l_{\mathrm{IV}}(\gamma) = 0$, 
			can be seen as choosing,
			under all causal models compatible with the distribution,
			the model with the least amount confounding -- when using 
			$E(Y-\gamma^\top X)^2 - E(Y-\gamma_{\mathrm{OLS}}^\top X)^2$
			as a measure for confounding.
		} 
		In such cases, PULSE extends on existing literature that aims to trade-off predictability and invariance but that so far has been restricted to search over subsets of variables 
		(see Sections~\ref{sec:summarypuls} and \Cref{app:underidentifiedexperiment}).
		To maintain consistency of the estimator the chosen test must have asymptotic power of one.

		In this paper,
		we propose a class of significance tests, that contains, e.g., the Anderson-Rubin test \citep[][]{anderson1949estimation}.
		While the objective function in \Cref{eq:PULSEfirstEQ}
		is quadratic in $\alpha$, the resulting constraint is, 
		in general, non-convex. 
		In Section~\ref{sec:DualPULSE}, we develop a computationally efficient procedure that 
		provably solves the optimization problem at low computational cost. 
		Other choices of tests are possible,  
		too, but may result in even harder optimization problems.

		In \Cref{sec:SetupAndAssumptionsPULSE}, we briefly introduce the setup and assumptions.
		In \Cref{sec:VanishCorr}, we specify a class of asymptotically consistent tests for $\cH_0(\alpha)$. 
		In \Cref{sec:PULSEdefi} we formally define the PULSE estimator.
		In \Cref{sec:PrimalPULSE}, we show that the PULSE estimator is well-defined by proving that it is equivalent to a solvable convex quadratically constrained quadratic program which we denote by the primal PULSE. %
		In \Cref{sec:DualPULSE}, we utilize duality theory and derive an alternative representation which we denote by the dual PULSE. 
		This representation 
		yields a computationally feasible algorithm 
		and shows that the PULSE estimator is a K-class estimator
		with a 
		data-driven %
		$\kappa$. 
		Proofs of results in this section can be found in \Cref{sec:RemainingProofsOfSecPULSE} unless stated otherwise.

		\subsection{Setup and assumptions}\label{sec:SetupAndAssumptionsPULSE}
		In the following sections we again let $(\fY,\fX,\fH,\fA)$ consist of $n\geq \min\{d,q\}$ row-wise independent and identically distributed copies of $(Y,X,H,A)$ generated in accordance with the SEM in \Cref{ARModel}.	The structural equation of interest is  $Y = \gamma^\t_0 X + \eta^\t_0  H + \beta^\t_0 A +  \ep_{Y}$.  
		Assume that we have some non-sample information about which $d_2=d-d_1$ and $q_2=q-q_1$ coefficients of $\gamma_0$ and $\beta_0$, respectively, are zero. As in \Cref{SEC:ROBUSTNESS}, 
		we let the subscript $*$ denote the variables and coefficients that are non-zero according to the non-sample information but to simplify notation, we 
		drop the $*$ subscript from $Z$, $\fZ$ and $\alpha_0$; that is, we write
		$Z =[X_*^\t \; A_*^\t]^\t \in \R^{d_1+q_1}$, 
		$\fZ=[
		\fX_* \, \, \fA_*
		]\in \R^{n\times(d_1+q_1)}$ and $\alpha_0 :=(\gamma_{0,*}^\t,\beta_{0,*}^\t)^\t:\in \R^{d_1+q_1}$.  That is, $Y = \alpha^\t_{0} Z + U_Y$,
		where $U_Y = \alpha_{0,-*}^\t Z_{-*}+\eta^\t_0  H + \ep_Y$. If the non-sample information is true, then $U_Y = \eta^\t_0  H + \ep_Y$.

		We define a setup as being under- just- and over-identified by the degree of over-identification $q_2-d_1$ being negative, equal to zero and positive, respectively. That is, the number of excluded exogenous variables $A_{-*}$ being less, equal or larger than the number of included endogenous variables $X_{*}$ in the target equation.

		We assume that the global assumptions of \Cref{ass:global} 
		from \Cref{sec:SetupAndAssumptionsRobustness}  still hold. 
		Furthermore, we will %
		make use of the following situational assumptions
		\begin{assumption} \label{ass:AIndepUYandMeanZeroA}
			\begin{enumerate*}[label=(\alph*),ref=\ref{ass:AIndepUYandMeanZeroA}.(\alph*)]
				\item $A \independent U_Y$; \label{ass:AIndepUy}
				\item 	$E[A]=0$.  \label{ass:MeanZeroA}
			\end{enumerate*}
		\end{assumption}
		\begin{assumption}
			$\ep$ has non-degenerate marginals.\label{ass:NonDegenYNoise}
		\end{assumption}
		\begin{assumption} \label{ass:ZtZfullrankandAtZfullrank}
			\begin{enumerate*}[label=(\alph*),ref=\ref{ass:ZtZfullrankandAtZfullrank}.(\alph*)]
				\item 	$\fZ^\t \fZ$ is of full rank; \label{ass:ZtZfullrank}
				\item 	$\fA^\t \fZ$ is of full rank. \label{ass:AtZfullrank}
			\end{enumerate*}
		\end{assumption}
		\begin{assumption}
			$[\fZ \, \, \fY]$ is of full column rank. \label{ass:ZYfullcolrank}
		\end{assumption}
		\begin{assumption}
			$E[AZ^\t]$ is of full rank. \label{ass:EAZtfullrank}
		\end{assumption}

		Assumption \Cref{ass:AIndepUy} holds if our non-sample information is true, and the instrument set $A$ is independent of all unobserved endogenous variables $H_i$ which directly affect the target $Y$. This holds, for example, if the latent variables are source nodes, that is, they have no parents in the causal graph of the corresponding SEM.
		Assumption \Cref{ass:MeanZeroA} can be achieved by centering the data. Strictly speaking, this introduces a weak dependence structure in the
		observations, which is commonly ignored. Alternatively, one can perform sample splitting. For more details on this assumption and the possibility of relaxing it, see \Cref{rm:AssumptionMeanZero}. 
		Assumption \Cref{ass:ZtZfullrank} ensures that K-class estimators for $\kappa < 1$ are
		well-defined, regardless of the over-identification degree. 
		In the under-identified setup, Assumption \Cref{ass:AtZfullrank} yields that there exists a subspace of solutions minimizing $l_{\text{IV}}^n(\alpha)$. In the just- and over-identified setup this assumption ensures
		that $l_{\text{IV}}^n(\alpha)$ has a unique minimizer 
		given by the two-stage least squares estimator $\hat{\alpha}_{\text{TSLS}}^n := (\fZ^\t P_\fA \fZ)^{-1} \fZ^\t P_\fA \fY$. 
		\Cref{ass:ZYfullcolrank} is used to ensure that the ordinary least square objective function $l_{\text{OLS}}^n(\alpha;\fY,\fZ)$ is strictly positive, such that division by this function is always well-defined.  \Cref{ass:NonDegenYNoise,ass:EAZtfullrank} ensure that various limiting arguments are valid. In the just- and over-identified setup \Cref{ass:EAZtfullrank} is known as the rank condition for identification.%

		\subsection{Testing for vanishing correlation} \label{sec:VanishCorr}
		
		We now introduce a class of tests for the null hypothesis 
		$\cH_0(\alpha) : \text{Corr}(A,Y-Z\alpha) =0$
		that have point-wise asymptotic level and 
		pointwise asymptotic power. 
		These tests
		will allow us to define the corresponding PULSE estimator. 
		When \Cref{ass:ZYfullcolrank} holds we can define  $T_n^c:\R^{d_1+q_1} \to \R$  by
		\begin{align*}
			T_n^c(\alpha) := c(n) \frac{l_{\text{IV}}^n(\alpha)}{l_{\text{OLS}}^n(\alpha)} =  
			c(n) \frac{\|P_\fA (\fY- \fZ \alpha) \|_2^2}{\|\fY- \fZ \alpha\|_2^2},
		\end{align*}
		where $c(n)$ is a function that will typically scale linearly in $n$.
		Let us denote the $1-p$ quantile of the central Chi-Squared distribution with $q$ degrees of freedom by
		$Q_{\chi^2_{q}}(1-p)$.
		By %
		standard limiting theory
		we can test $\cH_0(\alpha)$ in the following manner. 
		\begin{restatable}[Level and power of the test]{lemma}{TheoremTestingVanishingCorr}
			\label{prop:TestingVanishingCorr}
			Let \Cref{ass:AIndepUYandMeanZeroA,ass:NonDegenYNoise,ass:ZYfullcolrank} hold
			and assume that $c(n) \sim n$ as $n\to\i$. For any $p\in (0,1)$ 
			and any fixed $\alpha$, the statistical test rejecting the null hypothesis $\cH_0(\alpha)$ if $T_n^c(\alpha) > Q_{\chi^2_{q}}(1-p),$  has point-wise 
			asymptotic level $p$ and point-wise asymptotic power of 1 against all alternatives as  $n \rightarrow \infty$.
		\end{restatable}
		
		\begin{remark} \label{rm:AssumptionMeanZero}
			\textnormal{Assumption \Cref{ass:MeanZeroA}, $E[A]=0$, is important for the test statistic to be asymptotic pivotal under the null hypothesis, 
				that is, 
				to ensure that the asymptotic distribution of $T_n^c(\alpha)$ 
				does not depend on the 
				model parameters except for $q$. 
				We can drop this assumption 
				if we change the null hypothesis to $\cH_0(\alpha):E[A(Y-Z^\t\alpha)]=0$ and add the assumption that $E[U_Y]=0$. 	Furthermore, if we are in the just- or over-identified setup and \Cref{ass:EAZtfullrank} holds, both of these hypotheses are under their respective assumptions equivalent to $\tilde{\cH}_0(\alpha): \alpha=\alpha_0$. That is, the test in \Cref{prop:TestingVanishingCorr} becomes an asymptotically consistent test for the causal coefficient.}
		\end{remark}
		Depending on the choice of $c(n)$,
		this class contains several tests, some of which are well known. 
		With $c(n) = n-q+Q_{\chi^2_q}(1-p_{\min})$,
		for example, one
		recovers a test that is equivalent to the 
		asymptotic version of the 
		Anderson-Rubin test  (\citealp{anderson1950asymptotic}).
		We make this connection precise 
		in \Cref{rm:ConnectionToAndersonRubinCI}
		in \Cref{app:AddRemarks}. 
		The Anderson-Rubin test is robust to weak instruments in the sense that the limiting distribution of the test-statistic under the null-hypothesis is not affected by weak instrument asymptotics; see, e.g.\ \citet{staiger1997instrumental}  and \citet{stock2002survey}.\footnote{Weak instrument asymptotics is a model scheme where the instrument strength tends to zero at a rate of $n^{-1/2}$, i.e., the reduced form structural equation for the endogenous variables is given by $\fX = \fA n^{-1/2} \Pi_X +\bm{\ep} \Gamma^{-1}_X$.} For weak instruments, the confidence region may be unbounded with large probability; see \cite{dufour1997some}.
		\cite{moreira2009tests} show that the test suffers 
		from loss of power in the over-identified setting.

		To simplify notation, we will from now on work with the choice 
		$c(n) = n$
		and define the acceptance region with level $p_{\min}\in(0,1)$
		as
		$\cA_n(1-p_{\min}) := \{\alpha \in \R^{d_1+q_1}: T_n(\alpha) \leq Q_{\chi^2_{q}}(1-p_{\min})\}$,
		where $T_n(\alpha)$ corresponds to the choice $c(n) = n$.

		\subsection{The PULSE estimator} \label{sec:PULSEdefi}
		For any level $p_{\min}\in(0,1)$, we formally define the PULSE estimator of \Cref{eq:PULSEfirstEQ} by letting the feasible 
		set
		be given by the acceptance region $\cA_n(1-p_{\min})$ of $\cH_0(\alpha)$ using the test of \Cref{prop:TestingVanishingCorr}. That is, we consider
		\begin{align}  \label{eq:PULSE}
			\hat{\alpha}^n_{\mathrm{PULSE}}(p_{\min}) := \begin{array}{ll}
				\argmin_\alpha & l_{\mathrm{OLS}}^n(\alpha)  \\
				\text{subject to} & T_n(\alpha) \leq Q_{\chi^2_{q}}(1-p_{\min}).
			\end{array}
		\end{align}
		In general, 
		this 
		is a non-convex optimization problem (\citealp{boyd2004convex})
		as the constraint function is non-convex,
		see the blue contours in 
		\Cref{fig:LevelsetsTestAndOLS}(left).
		From \Cref{fig:LevelsetsTestAndOLS}(right) we 
		see that 
		in the given example the problem nevertheless has a unique and well-defined solution: 
		the smallest level-set of $l_{\text{OLS}}^n$ with a non-empty intersection of the acceptance region $\{\alpha : T_n(\alpha) \leq Q_{\chi^2_{q}}(1-p_{\min})\}$ intersects with the latter region in a unique point. 
		In \Cref{sec:PrimalPULSE}, we prove that this is not a coincidence: 
		\Cref{eq:PULSE}
		has a unique solution that coincides with the solution of a strictly convex, quadratically constrained quadratic program (QCQP) with a data-dependent constraint bound. 
		In \Cref{sec:DualPULSE}, we further derive 
		an equivalent Lagrangian dual problem.
		This has two important implications.
		(1) It allows us to construct a 
		computationally efficient procedure to compute a solution of the non-convex problem above, and (2), 
		it shows that the PULSE estimator can be written as K-class estimators.

		Estimators with similar constraints albeit different optimization objective have been studied by \cite{gautier2011high}. In \Cref{rm:Pretest} in \Cref{app:AddRemarks} we briefly discuss the connection to pre-test estimators. 
		Furthermore, any method for inverting the test, see, e.g., \cite{davidson2014confidence}, yields a valid confidence set including the proposed point estimator (given that the method outputs the point estimator when the acceptance region is empty).

		\subsection{Primal representation of PULSE} 
		\label{sec:PrimalPULSE}
		We now derive a
		QCQP representation of the PULSE problem, 
		which we call the primal PULSE. 
		For all  $t\geq 0$ define the 
		empirical primal minimization problem (Primal.$t.n$) by
		\begin{align}  \label{PR.t.n}
			\begin{array}{ll}
				\text{minimize}_\alpha & l_{\mathrm{OLS}}^n(\alpha;\fY,\fZ)  \\
				\text{subject to} & l_{\mathrm{IV}}^n(\alpha;\fY,\fZ,\fA)\leq t.
			\end{array}
		\end{align}
		We drop the dependence of $\fY$, $\fZ$ and $\fA$ and refer to the objective and constraint functions as $l_{\text{OLS}}^n(\alpha)$ and $l_{\text{IV}}^n(\alpha)$. The following lemma shows that under suitable assumptions these problems are solvable, strictly convex QCQP problems satisfying Slater's condition.

		\begin{restatable}[Unique solvability of the primal]{lemma}{LemmaPrimalUniqueSolution}		\label{lm:PrimalUniqueSolAndSlatersConditions}
			Let \Cref{ass:ZtZfullrankandAtZfullrank} hold. It holds that $\alpha \mapsto l_{\mathrm{OLS}}^n(\alpha)$ and $\alpha \mapsto l_{\mathrm{IV}}^n(\alpha)$ are strictly convex and convex, respectively. Furthermore, for any $t > \inf_{\alpha}l_{\mathrm{IV}}^n(\alpha)$ it holds that the constrained minimization problem (Primal$.t.n$) has a unique solution and satisfies Slater's condition. In the under- and just-identified setup the constraint bound requirement	
			is equivalent to $t>0$ and in the over-identified setup to $t> l_{\mathrm{IV}}^n(\hat{\alpha}^n_{\mathrm{TSLS}})$, where $\hat{\alpha}^n_{\mathrm{TSLS}}= (\fZ^\t P_\fA \fZ)^{-1}\fZ^\t P_\fA \fY $.
		\end{restatable}
		
\begin{figure}[t] 
			\centering
			\includegraphics[width=\linewidth-0pt]{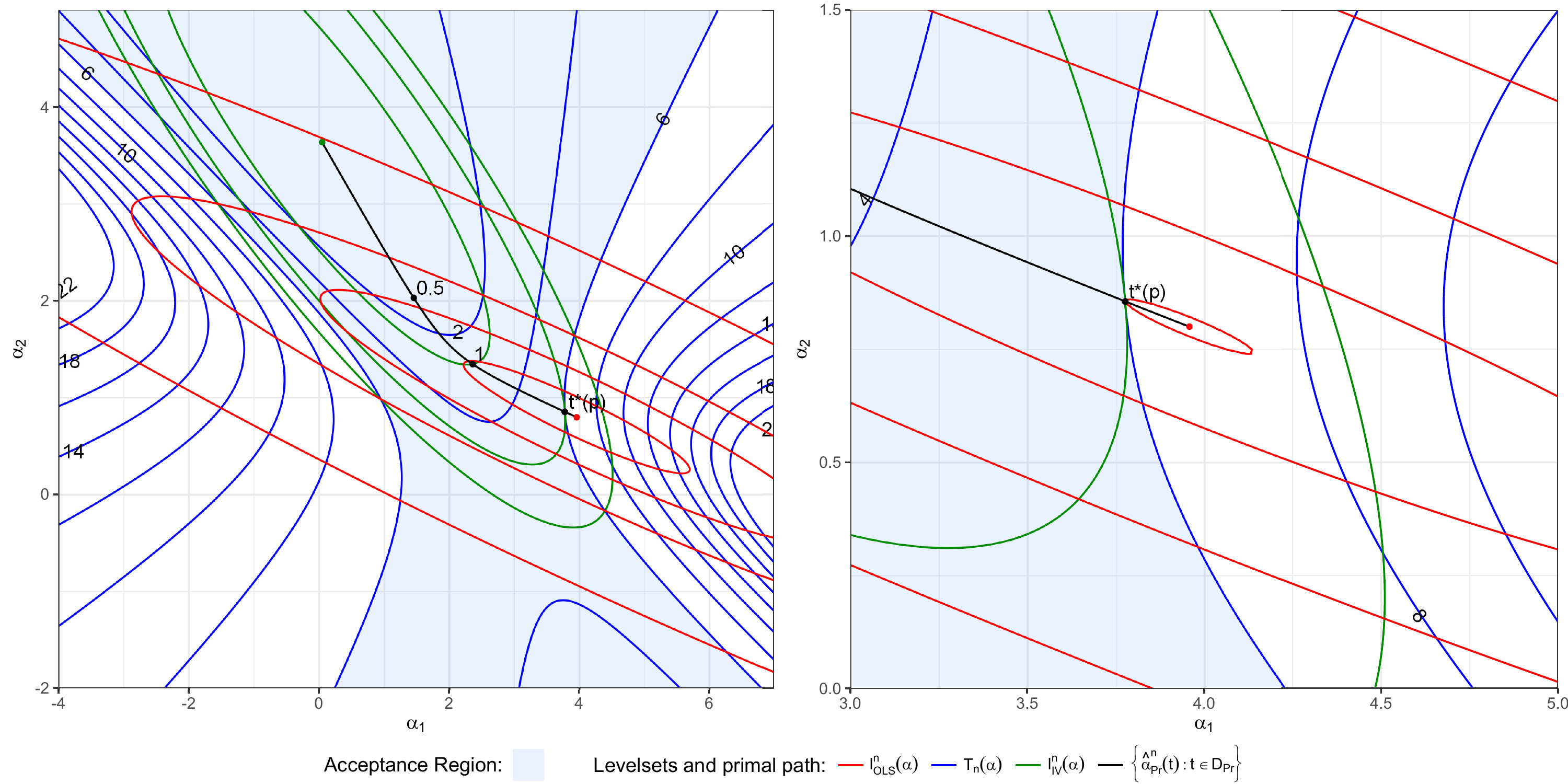}
			\caption{ Illustrations of the level sets of $l_{\mathrm{OLS}}^n$ (red contours), the proposed test-statistic $T_n$ (blue contours) and $l_{\text{IV}}^n$ (green contours) in a just-identified setup. The example is generated with a two dimensional anchor $A = (A_1,A_2)$, one of which is included, and one included endogenous variable $X$, i.e., $Y = \alpha_1 X + \alpha_2 A_1 + H+ \ep_Y$ with $(\alpha_1,\alpha_2)=(1,1)$. Both illustrations 
				show level sets from the same setup, but they use different scales. 
				The black text denotes the level of the test-statistic contours. In this setup, the PULSE constraint bound, the rejection threshold of the test with $p_{\min}=0.05$, is $Q_{\chi^2_{2}}(0.95) \approx 5.99$. The blue level sets of $T_n$ are 
				non-convex. The sublevel set of the test, corresponding to the acceptance region, is illustrated by the blue area. In the right plot, we see that the smallest level set of $l_{\text{OLS}}^n$ that has a non-empty intersection with the $Q_{\chi^2_{q}}(1-p_{\min})$-sublevel set of $T_n$ is a singleton (black dot, $t^*(p)$). This shows that 
				in this example
				the PULSE problem is solvable and has a unique solution. The $l_{\text{IV}}^n$ level set that intersects this singleton   is exactly the $t_n^\star(p_{\min})$-level set of $l_{\text{IV}}^n$, illustrating the statement of \Cref{thm:pPULSESolvesPULSE} in that the primal PULSE 
				with that choice of $t$	
				solves the PULSE problem. The black line visualizes the solutions	
				$\{\hat{\alpha}_{\text{Pr}}^n(t)  : t \in D_{\text{Pr} }\}$. The black points and corresponding text labels indicates which constraint bound $t$ yields the specific point. In general, the class of primal solutions 
				does not coincide with the class of convex combinations of the OLS and the TSLS estimators.} %
			\label{fig:LevelsetsTestAndOLSAndIV} \label{fig:LevelsetsTestAndOLS}
		\end{figure}

		We restrict the constraint bounds to $D_{\text{Pr}}:=(\inf_\alpha l_{\mathrm{IV}}^n(\alpha), l_{\text{IV}}^n(\hat{\alpha}_{\text{OLS}}^n)]$. Considering $t$ that are larger than 
		$\inf_\alpha l_{\mathrm{IV}}^n(\alpha)$
		ensures that the problem 
		(Primal$.t.n$) is uniquely solvable and
		furthermore that Slater's condition is satisfied (see Lemma~\ref{lm:PrimalUniqueSolAndSlatersConditions} above). 
		Slater's condition 
		will play a role 
		in \Cref{sec:DualPULSE} when establishing a sufficiently strong connection with its corresponding dual problem for which we can 
		derive a (semi-)closed form solution.
		Constraint bounds greater than or  equal to $l_{\text{IV}}^n(\hat{\alpha}_{\text{OLS}}^n)$ yield identical solutions. 
		Whenever %
		well-defined, let $\hat{\alpha}_{\text{Pr}}^n:D_{\text{Pr}}\to \R^{d_1+q_1}$ denote the constrained minimization estimator given by the solution to the (Primal$.t.n$) problem
		\begin{align} \label{eq:PrimalProblemSolutionDef}
			\hat{\alpha}_{\text{Pr}}^n(t) :=  \begin{array}{ll}
				\argmin_{\alpha} & l_{\mathrm{OLS}}^n(\alpha)  \\
				\text{subject to} & l_{\mathrm{IV}}^n(\alpha)\leq t.
			\end{array}
		\end{align}
		We now prove that 
		for a specific choice of $t$, the
		PULSE and the primal PULSE yield the same solutions. 
		Define
		$t_n^\star(p_{\min})$ as the data-dependent constraint bound given by
		\begin{align} \label{eq.Def.t.star.p}
			t_n^\star(p_{\min}) := \sup \{ t \in ( \inf_{\alpha}l_{\mathrm{IV}}^n(\alpha),l_{\mathrm{IV}}^n(\hat{\alpha}_{\mathrm{OLS}}^n) ] : T_n(\hat{\alpha}_{\mathrm{Pr}}^n(t))\leq Q_{\chi^2_{q}}(1-p_{\min})\}.
		\end{align}
		If $t^\star_n(p_{\min})>-\i$ or equivalently $t^\star_n(p_{\min})\in D_{\text{Pr}}$ we define the primal PULSE problem and its solution by (Primal$.t^\star_n(p_{\min}).n$) and $\hat{\alpha}_{\mathrm{Pr}}^n(t_n^\star(p_{\min}))$. 
		The following theorem yields conditions for when the solutions to the primal PULSE and PULSE problems coincide.
		\begin{restatable}[Primal representation of PULSE]{theorem}{pPULSESolvesPULSE}
			\label{thm:pPULSESolvesPULSE} 
			Let $p_{\min}\in(0,1)$ and \Cref{ass:ZtZfullrankandAtZfullrank,ass:ZYfullcolrank} hold and assume that  $t_n^\star(p_{\min}) >-\i$. If
			$T_n(\hat{\alpha}_{\mathrm{Pr}}^n(t_n^\star(p_{\min})))\leq Q_{\chi^2_{q}}(1-p_{\min})$, then the PULSE problem has a unique solution given by the primal PULSE solution. That is, %
			$
			\hat{\alpha}^n_{\mathrm{PULSE}}(p_{\min}) = \hat{\alpha}_{\mathrm{Pr}}^n(t_n^\star(p_{\min})).$
		\end{restatable}
		We show that $t_n^\star(p_{\min}) >-\i$ is a sufficient condition for $T_n(\hat{\alpha}_{\mathrm{Pr}}^n(t_n^\star(p_{\min})))\leq Q_{\chi^2_{q}}(1-p_{\min})$ in the proof of \Cref{thm:PULSEpPULSEdPULSEEequivalent}. The sufficiency of $t_n^\star(p_{\min}) >-\i$ is postponed to the latter proof as it easily follows
		from the dual representation.
		Hence, 
		we have shown that finding the PULSE estimator, i.e., 
		finding a solution to the non-convex PULSE problem, is equivalent to solving the convex QCQP primal PULSE for a data dependent choice of $t_n^\star(p_{\min})$.\footnote{Given that value, we can use a numerical QCQP solver to calculate the PULSE estimate.
		} However,
		$t_n^\star(p_{\min})$ is still unknown. \Cref{fig:LevelsetsTestAndOLSAndIV} 
		shows
		an example of the equivalence in \Cref{thm:pPULSESolvesPULSE}.  \Cref{fig:LevelsetsTestAndOLSAndIV}(right) shows that the level set of $l_{\text{IV}}(\alpha) = t^{\star}(p_{\min})$ intersects the optimal level curve of $l_{\text{OLS}}^n(\alpha)$ in the same point given by minimizing over the constraint $T_n(\alpha) \leq Q_{\chi^2_q}(1-p_{\min})$.

		The set of solutions to the primal problem
		$\{\hat{\alpha}_{\text{Pr}}^n(t)  : t \in D_{\text{Pr} }\}$ can in the just- and over-identified setup be visualized as an (in general) non-linear path in $\R^{d_1+q_1}$ between the TSLS estimator $(t= l_{\mathrm{IV}}^n(\hat{\alpha}_{\text{TSLS}}^n))$ and the OLS estimator $(t= l_{\text{IV}}^n(\hat{\alpha}_{\text{OLS}}^n))$ \citep[see also][]{AnchorRegression}. 
		Theorem~\ref{thm:pPULSESolvesPULSE} yields that the PULSE estimator ($t=t_n^\star(p_{\min})$) then seeks the estimator 'closest' to the OLS estimator along this path that does not yield a rejected test of simultaneous vanishing correlation between the resulting prediction residuals and the exogenous variables $A$, see  \Cref{fig:LevelsetsTestAndOLSAndIV}. 
		The path of possible solutions is not necessarily a straight line (see black line); thus, in general, the PULSE estimator is different from the affine combination of OLS and TSLS estimators studied by e.g.\ \cite{judge2012minimum}.

		In the under-identified setup, the TSLS end point corresponding to $t= \min_\alpha l_{\mathrm{IV}}^n(\alpha)$ is instead given by the point in the IV solution space $\{\alpha \in \R^{d_1+q_1}: l_{\text{IV}}^n(\alpha)=0\}$ with the smallest mean squared prediction residuals.

		\subsection{Dual representation of PULSE} \label{sec:DualPULSE}
		In this section, we derive a dual representation of the primal PULSE problem which we will denote the dual PULSE problem. 
		This specific dual representation allows for the construction of a binary search algorithm for the PULSE estimator and yields that PULSE is a member of the K-class estimators with stochastic $\kappa$-parameter. 
		
		For any penalty parameter $\lambda \geq 0$ %
		we define the dual problem %
		(Dual$.\lambda.n$)  by
		\begin{align} \label{K.lambda.n} 
			\begin{array}{ll}
				\text{minimize} & l_{\mathrm{OLS}}^n(\alpha) + \lambda l_{\mathrm{IV}}^n(\alpha).
			\end{array}
		\end{align} 
		Whenever Assumption \Cref{ass:ZtZfullrank} holds, i.e., $\fZ^\t \fZ$ is of full rank, then for any $\lambda\geq 0$ the solution to (Dual$.\lambda.n$)  coincides with the 
		K-class estimator with $\kappa = \lambda/(1+\lambda)\in[0,1)$, see \Cref{lm:PenalizedKClassSolutionUniqueAndExists}. That is, 
		\begin{align*}
			\hat{\alpha}_{\text{K}}^n(\kappa)
			= (\fZ^\t (\fI+\lambda P_\fA)\fZ)^{-1} \fZ^\t(\fI+\lambda P_\fA)\fY 
		\end{align*}
		solves (Dual$.\lambda.n$). Henceforth,  let $\hat{\alpha}_{\mathrm{K}}^n (\lambda)$ denote the solution to (Dual$.\lambda.n$), i.e.,
		in a slight abuse of notation we will 
		denote 
		the solution to~(Dual$.\lambda.n$)  by
		$\hat{\alpha}_{\text{K}}^n(\lambda)$, such that
		$\hat{\alpha}_{\text{K}}^n(\lambda)=
		\hat{\alpha}_{\text{K}}^n(\kappa)$
		for
		$\kappa = \lambda/(1+\lambda)$. We refer to these two representations as the K-class estimator with penalty parameter $\lambda$ and parameter $\kappa$, respectively. The usage of 
		$\kappa$ or $\lambda$ as argument should clarify which notation we refer to.
		
		Under Assumption \Cref{ass:AtZfullrank} we have that the minimum of $l_{\text{IV}}^n(\alpha)$ is attainable (see the proof of \Cref{lm:PrimalUniqueSolAndSlatersConditions}). Hence, let the solution space for the minimization problem $\min_\alpha l_{\text{IV}}^n(\alpha)$ be given by
		\begin{align}\label{eq:Miv}
			\cM_{\mathrm{IV}} := \argmin_\alpha l_{\mathrm{IV}}^n(\alpha) =  \{\alpha \in \R^{d_1+q_1} : l_{\mathrm{IV}}^n(\alpha)= \min_{\alpha'}l_{\mathrm{IV}}^n(\alpha')\}.
		\end{align}
		In the under-identified setup $(q_2<d_1)$, %
		$\cM_{\text{IV}}$ is a  $(d_1-q_2)$-dimensional subspace of $\R^{d_1+q_1}$ and in the just- and over-identified setup it holds that $\cM_{\text{IV}}= \{\hat{\alpha}_{\text{TSLS}}^n\}$.

		We now prove that, in the generic case, K-class 
		estimators 
		for $\lambda \in [0,\infty)$
		are different from the TSLS estimator. This result may not come as a surprise, but we include it as we need the result
		later and have not found it elsewhere.
		\begin{restatable}[K-class estimators and TSLS differ]{lemma}{KclassNotEqualToTwoSLS}
			\label{lm:KclassNotEqualToTwoSLS}
			Assume that we are in the just- or over-identified setup and $n>q$. Furthermore, assume that 	$\ep$ has density with respect to Lebesgue measure and that the coefficient matrix $B$ of the SEM in \Cref{ARModel} is lower triangular. If the rank conditions of \Cref{ass:ZtZfullrankandAtZfullrank} hold almost surely, then it  %
			almost surely holds, that all K-class estimators with penalty parameter $\lambda\in[0,\infty)$ differ from the %
			TSLS estimator, i.e., $\hat{\alpha}_{\mathrm{TSLS}}^n  \not\in \{\hat{\alpha}_{\mathrm{K}}^n(\lambda): \lambda \geq 0 \}$. 
		\end{restatable}
		We conjecture that the corresponding statement 
		holds in the under-identified setup and without the lower triangular assumption on B, too. That is,  $\cM_{\text{IV}}\cap \{\hat{\alpha}_{\mathrm{K}}^n(\lambda): \lambda \geq 0 \}=\emptyset$ holds almost surely. We therefore introduce this as an assumption.
		\begin{assumption}
			No K-class estimator $\hat{\alpha}_{\mathrm{K}}^n(\kappa)$ with $\kappa\in[0,1)$, is a member of $\cM_{\text{IV}}$. \label{ass:KclassNotInIV}
		\end{assumption}
		
		Furthermore, when imposing that \Cref{ass:KclassNotInIV} holds we also have that the K-class estimators differ from each other. 
		\begin{restatable}[K-class estimators differ]{corollary}{KclassSolutionsDistinct}
			\label{cor:KclassSolutionsDistinct}
			Let %
			\Cref{ass:ZtZfullrankandAtZfullrank,ass:KclassNotInIV} hold.
			If $\lambda_1,\lambda_2\geq 0$ with $\lambda_1\not = \lambda_2$, then $\hat{\alpha}_{\mathrm{K}}^n(\lambda_1) \not = \hat{\alpha}_{\mathrm{K}}^n(\lambda_2)$.
		\end{restatable}
		
		The above corollary is proven as  \Cref{cor:KclassSolutionsDistinctApp} in \Cref{sec:SomeProofsOfSecPULSE}.
		We now show that the class of K-class estimators with penalty parameter $\lambda \geq 0$ , i.e., $\kappa\in[0,1)$, coincides with the class of constrained minimization-estimators that minimize the primal problems with constraint bounds $t> \min_{\alpha}l_{\mathrm{IV}}^n(\alpha)$. 
		\begin{restatable}[Connecting the primal and dual]{lemma}{EquivalenceBetweenKlikeAndPrimal}	\label{lm:EquivalenceBetweenKlikeAndPrimal}
			If %
			\Cref{ass:ZtZfullrankandAtZfullrank,ass:ZYfullcolrank,ass:KclassNotInIV} hold, then both of the following statements hold.  \textit{(a)} For any $t \in D_{\mathrm{Pr}}$, there exists a unique $\lambda(t) \geq 0$ such that (Primal$.t.n$) and (Dual$.\lambda(t).n$) have the same unique solution. \textit{(b)} For any $\lambda \geq 0$, there exists a unique $t(\lambda) \in D_{\mathrm{Pr}}$ such that (Primal$.t(\lambda).n$) and (Dual$.\lambda.n$) have the same unique solution.
		\end{restatable}
		\Cref{lm:EquivalenceBetweenKlikeAndPrimal} tells us that, under appropriate assumptions, $
		\{\hat{\alpha}_{\text{K}}^n(\kappa ): \kappa \in[0,1)\}= 	\{\hat{\alpha}_{\text{K}}^n(\lambda):\lambda \geq 0\}  = \{\hat{\alpha}_{\text{Pr}}^n(t): t\in D_{\text{Pr}} \}.
		$
		In words, we have recast the K-class estimators with $\kappa \in [0,1)$ as the class of solutions to the primal problems previously introduced. That the minimizers of $l_{\text{IV}}^n(\alpha)$ are different from all the K-class estimators with penalty $\lambda \geq 0$ (or $\kappa\in[0,1)$) guarantees that when representing a K-class  problem in terms of a constrained optimization problem it satisfies Slater's condition.%

		We are now able to show the main result of this section.
		The PULSE estimator $\hat{\alpha}^n_{\text{PULSE}}(p_{\min})$ 
		solves a K-class problem
		(Dual$.\lambda.n$)
		and can therefore be seen as a K-class estimator with 
		a data-dependent parameter. To see this, let us define 
		the dual PULSE penalty parameter, i.e., the dual analogue of the primal PULSE constraint $t_n^\star(p_{\min})$ as
		\begin{align} \label{eq.Def.lambda.star.p}
			\lambda_n^\star(p_{\min}) := \inf\{\lambda \geq 0 : T_n(\hat{\alpha}_{\mathrm{K}}^n (\lambda))\leq Q_{\chi^2_{q}}(1-p_{\min}) \}.
		\end{align}
		If $\lambda^\star_n(p_{\min})<\i$, we define 
		the
		dual PULSE problem by (Dual.$\lambda^\star_n(p_{\min}).n)$ with solution
		$
		\hat{\alpha}_{\text{K}}^n (\lambda_n^\star(p)) = \argmin_{\alpha\in \R^{d_1+q_1}} l_{\text{OLS}}^n(\alpha) + \lambda_n^\star(p_{\min}) l_{\text{IV}}^n(\alpha).
		$
		\begin{restatable}[Dual representation of PULSE]{theorem}{PrimalDualConnectionPvalConstraint}	\label{thm:PULSEpPULSEdPULSEEequivalent}
			Let $p_{\min}\in(0,1)$ and \Cref{ass:ZtZfullrankandAtZfullrank,ass:ZYfullcolrank,ass:KclassNotInIV} hold. %
			If $\lambda_n^\star(p_{\min}) <\i $, then it holds that $t_n^\star(p_{\min})>-\i$ and $		\hat{\alpha}_{\mathrm{K}}^n(\lambda_n^\star(p_{\min})) = \hat{\alpha}_{\mathrm{Pr}}^n(t_n^\star(p_{\min})) =	\hat{\alpha}_{\mathrm{PULSE}}^n(p_{\min})$. %
		\end{restatable}
		Thus, the PULSE estimator seeks to minimize the K-class penalty $\lambda$, i.e., to pull the estimator along the K-class path $\{\hat{\alpha}_{\text{K}}^n(\lambda):\lambda \geq 0\}$ as close to the ordinary least square estimator as possible.
		Furthermore, the statement implies that the PULSE estimator is a K-class estimator with data-driven penalty $\lambda_n^\star(p_{\min})$ or, equivalently, parameter $\kappa = \lambda_n^\star(p_{\min})/(1+\lambda_n^\star(p_{\min}))$.
		Given a finite dual PULSE penalty parameter $\lambda_n^\star(p_{\min})$ we can, by utilizing the closed form solution of the K-class problem, represent the PULSE estimator in the following form:
		\begin{align*}
			\hat{\alpha}_{\mathrm{PULSE}}^n(p_{\min}) &= \hat{\alpha}_{\mathrm{K}}^n(\lambda_n^\star(p_{\min})) = (\fZ^\t (\fI+\lambda_n^\star(p_{\min}) P_\fA)\fZ)^{-1} \fZ^\t(\fI+\lambda_n^\star(p_{\min}) P_\fA)\fY.
		\end{align*}
		However, to the best of our knowledge,
		$\lambda_n^\star(p_{\min})$ 
		has no known closed form, so the above expression cannot be computed in closed-form either.
		In \Cref{sec:BinarySearch}, we prove that the PULSE penalty parameter $\lambda_n^\star(p_{\min})$ can be approximated with arbitrary precision by a 
		simple binary search procedure.

		The following lemma provides a necessary and sufficient (in practice checkable) condition
		for when the PULSE penalty parameter $\lambda_n^\star(p_{\min})$ 
		is finite. 
		
		\begin{restatable}[Infeasibility of the dual representation]{lemma}{LambdaStarFinite}
			\label{lm:LamdaStarFiniteIFF}
			Let $p_{\min}\in(0,1)$ and \Cref{ass:ZtZfullrankandAtZfullrank,ass:ZYfullcolrank,ass:KclassNotInIV} hold. In the under- and just-identified setup we have that $\lambda_n^\star(p_{\min})<\i$. In the over-identified setup it holds that
			$
			\lambda^\star_n(p_{\min}) < \i \iff T_n(\hat{\alpha}_{\mathrm{TSLS}}^n)< Q_{\chi^2_q}(1-p_{\min}).
			$
			This is not guaranteed to hold as the event that $\cA_n(1-p_{\min})= \emptyset$ can have positive probability.
		\end{restatable}
		
		Thus, under suitable regularity assumptions \Cref{lm:LamdaStarFiniteIFF} yields that our dual representation of the PULSE estimator always holds in the under- and just-identified setup. It furthermore yields a sufficient and necessary condition for the dual representation to be valid in the over-identified setup, namely that the TSLS is in the interior of the acceptance region. Furthermore, this condition is possibly violated in the over-identified setup with non-negligible probability. 
		\subsubsection{Binary search for the dual parameter} \label{sec:BinarySearch}
		The key insight allowing for a binary search procedure for $\lambda_n^\star(p_{\min})$ is
		that
		the mapping
		$\lambda \mapsto T_n(\hat{\alpha}_{\text{K}}^n(\lambda))$
		is monotonically decreasing.

		\begin{restatable}[Monotonicity of the losses and the test statistic]{lemma}{MonotonicityOfTestOfLambda}
			\label{lm:OLSandIV_Monotonicity_FnctOfPenaltyParameterLambda}
			When Assumption \Cref{ass:ZtZfullrank} holds the maps $
			[0,\i)\ni \lambda \mapsto  l_{\mathrm{OLS}}^n(\hat{\alpha}_{\mathrm{K}}^n (\lambda) )$ and $ [0,\i)\ni  \lambda \mapsto l_{\mathrm{IV}}^n(\hat{\alpha}_{\mathrm{K}}^n (\lambda) ) $
			are monotonically increasing and monotonically decreasing, respectively. Consequently, if \Cref{ass:ZYfullcolrank} holds, we have that the map 
			$
			[0,\i)\ni \lambda \mapsto T_n (\hat{\alpha}_{\mathrm{K}}^n (\lambda) ) %
			$
			is monotonically decreasing. Furthermore, if \Cref{ass:KclassNotInIV} also holds, these monotonicity statements can be strengthened to strictly decreasing and strictly increasing.
		\end{restatable}

		The above lemma is proven as  \Cref{lm:OLSandIV_Monotonicity_FnctOfPenaltyParameterLambdaApp} in \Cref{sec:SomeProofsOfSecPULSE}. If the OLS solution is not strictly feasible in the PULSE problem, then $\lambda_n^\star(p_{\min})$ indeed is the smallest penalty parameter for which the test-statistic reaches a p-value of exactly $p_{\min}$; see \Cref{lm:TestInAlphaLambdaStarEqualsQuantileApp} in \Cref{sec:SomeProofsOfSecPULSE}. 
		
		We propose the binary search algorithm
		presented in Algorithm~\ref{Binary.Search.Lambda.Star} in \Cref{app:algo},
		that can approximate a finite $\lambda_n^\star(p_{\min})$ with arbitrary precision.
		We terminate the binary search (see line 2) if $\lambda^\star_n(p_{\min})$ is not finite, in which case we have no computable representation of the PULSE estimator. It is possible to improve this algorithm in the under- and just-identified setup, by initializing $\ell_{\max}$ as the quantity given by \Cref{eq:LambdaEquality} in the proof of \Cref{lm:LamdaStarFiniteIFF}. This initialization removes the need for the first while loop in (lines 4--6).
		We now prove that Algorithm~\ref{Binary.Search.Lambda.Star} 
		achieves the required precision and 
		is asymptotically correct.
		
		\begin{restatable}[]{lemma}{BinarySearchLambdaStarConverges}
			\label{lm:BinarySearchLambdaStarConverges} 
			Let $p_{\min}\in(0,1)$ and \Cref{ass:ZtZfullrankandAtZfullrank,ass:ZYfullcolrank} hold. If $\lambda_n^\star(p_{\min})<\i$, then $\lambda_n^\star(p_{\min})$ can be approximated with arbitrary precision by the binary search \Cref{Binary.Search.Lambda.Star}, that is, $	\mathrm{Binary.Search}(N,p_{\min}) - \lambda_n^\star(p_{\min}) \to 0,$ as $N\to\i$.
		\end{restatable}

		\subsection{Algorithm and consistency} \label{sec:AlgoAndConsistency}
		The dual representation of the PULSE estimator is not guaranteed to be well-defined in the over-identified setup.
		In particular, it is not well-defined if the TSLS is outside the interior of the acceptance region (which corresponds to a
		p-value of less than or equal to $p_{\min}$). In this case, we propose to output a warning. 
		This can be 
		helpful information for the user since it may indicate a model misspecification.
		For example, if the true relationship is in fact nonlinear, and one considers
		an over-identified 
		case (e.g., by 
		constructing different 
		transformations of the  instrument),
		even the TSLS may be rejected when 
		erroneously considering a linear model; see \citet{Keane2010} and \citet{Mogstad2010}.
		For any $p_{\min}\in(0,1)$ we 
		can still define an always well-defined
		modified PULSE estimator 
		$\hat{\alpha}_{\text{PULSE}+}^n(p_{\min})$ 
		as
		$\hat{\alpha}_{\mathrm{PULSE}}^n(p_{\min})$ if the dual representation is feasible %
		and some other 
		consistent estimator $\hat{\alpha}^n_{\text{ALT}}$ (such as  
		the TSLS, LIML or Fuller estimator) otherwise. 
		That is, we define
		\begin{align*}
			\hat{\alpha}_{\text{PULSE}+}^n(p_{\min}) := \left\{\begin{array}{ll}
				\hat{\alpha}_{\mathrm{PULSE}}^n(p_{\min}), & \text{if } T_n(\hat{\alpha}_{\text{TSLS}}^n)<Q_{\chi^2_{q}}(1-p_{\min}) \\
				\hat{\alpha}^n_{\text{ALT}}, & \text{otherwise}.
			\end{array} \right. 
		\end{align*}
		
		Similarly to the case of an empty rejection region, 
		we also output a warning 
		for the case when the OLS estimator is accepted. This may, but does not have to, indicate weak instruments. 
		Thus, we have the algorithm presented as \Cref{alg:2} in \Cref{algo:pulseplus} for computing the PULSE$+$ estimator.

		We now prove that the PULSE$+$ estimator consistently estimates the causal parameter in the just- and over-identified setting. 
		Assume that we choose 
		a consistent
		estimator
		$\hat{\alpha}_{\text{ALT}}^n$
		(under standard regularity assumptions, this is satisfied for the TSLS).\footnote{This holds as 
			$
			\hat{\alpha}_{\text{TSLS}}^n 
			= \alpha_0 +  (n^{-1}\fZ^\t \fA (n^{-1}\fA^\t \fA)^{-1} n^{-1}\fA^\t \fZ)^{-1} n^{-1}\fZ^\t \fA (n^{-1}\fA^\t \fA)^{-1} n^{-1}\fA^\t \fU_Y 
			$.}
		We can then show that, under mild conditions, the PULSE$+$ estimator, too, is a consistent estimator of $\alpha_0$.

		\begin{restatable}[Consistency of PULSE$+$]{theorem}{ConsistencyOfPULSE}
			\label{thm:ConsistencyOfPULSE}
			Consider the just- or over-identified setup and let $p_{\min}\in (0,1)$.  If \Cref{ass:AIndepUYandMeanZeroA,ass:ZtZfullrankandAtZfullrank,ass:ZYfullcolrank,ass:EAZtfullrank,ass:KclassNotInIV} hold almost surely for all $n\in \N$ and $\hat{\alpha}_{\mathrm{ALT}}^n$ consistently estimates $\alpha_0$, then $\hat{\alpha}_{\mathrm{PULSE}+}^n(p_{\min})\convp \alpha_0$, when $n\to \i$.
		\end{restatable}
		We believe that a similar statement also holds in the under-identified setting, see \Cref{app:underidentifiedexperiment}.
		
		\section{Simulation experiments}
		In \Cref{sec:Experiments} we conduct an extensive simulation study investigating the finite sample behaviour of the PULSE estimator. The concept of weak instruments is central to our analysis. An introduction to weak instruments can be found in \Cref{sec:WeakInst}. Here we give a brief overview of the study and the observations. 
		
		\subsection{Distributional robustness}
		The 
		theoretical results on distributional robustness 
		proved in \Cref{SEC:ROBUSTNESS}
		translate to finite data. The experiments of \Cref{app:DistributionalRobustness} shows that 
		even for small sample sizes, K-class estimators outperform 
		both OLS and TSLS for a certain range of interventions, matching the theoretical predictions with increasing sample size. In \cref{app:underidentifiedexperiment}, we furthermore consider an under-identified setting.
		
		\subsection{Estimating causal effects}
		When focusing on the estimation of a 
		causal effect in an identified setting,
		our simulations show 
		that there are several settings where PULSE outperforms the Fuller and TSLS estimators in terms of mean squared error (MSE).
		In univariate simulation experiments, such settings are 
		characterized by
		weakness of instruments and weak confounding (endogeneity). 
		The characterization becomes more  involved 
		in multivariate settings, 
		but is similar in that PULSE outperforms all other methods for small confounding strengths, an effect amplified by the weakness of instruments. 
		Below we detail the univariate simulation setup and refer the reader to \Cref{sec:Experiments} for further details and the multivariate simulation experiments mentioned above.
		
		\subsubsection{Univariate model} \label{sec:SimUnivariate}
		We first
		compared performance measures of the estimators in a univariate instrumental variable model. As seen in \citet{hahn2002new} and \citet{hahn2004estimation}, we consider structural equation models of the form
		\begin{align*}
			A := A \in \mathbb{R}^q,\quad 
			X :=  A^\t  \bar \xi  +  U_X \in \R, \quad
			Y :=  X \gamma + U_Y \in \R,
		\end{align*}
		where $A\sim \mathcal{N}(0,I)$ and $A\independent (U_X,U_Y) $ with 
		$
		\begin{psmallmatrix}
			U_X \\ U_Y
		\end{psmallmatrix}\sim \mathcal{N}\left(\begin{psmallmatrix}
			0 \\ 0
		\end{psmallmatrix}, \begin{psmallmatrix}
			1 & \rho \\
			\rho & 1
		\end{psmallmatrix}\right).$
		Furthermore, we let $\gamma =1$ and $\bar{\xi}^\t=(\xi,....,\xi)\in \R^q$, where $\xi>0$ is chosen according to the theoretical $R^2$-coefficient. We consider the following simulation scheme: for each $q\in \{1,2,3,4,5,10,20,30\}$, $\rho\in\{0.1,0.2,...,0.9\}$, $R^2\in\{0.0001,0.001,0.01,0.1,0.3\}$ and $n\in\{50,100,150\}$, we simulate $n$-samples from the above system and calculate the OLS, TSLS, Fuller(1), Fuller(4) and PULSE ($p_{\min}= 0.05$) estimates; see \Cref{sec:ExpResultsEstCausPerformanceMeasures}.
		
\begin{figure}[h] 
			\centering\includegraphics[width=\linewidth]{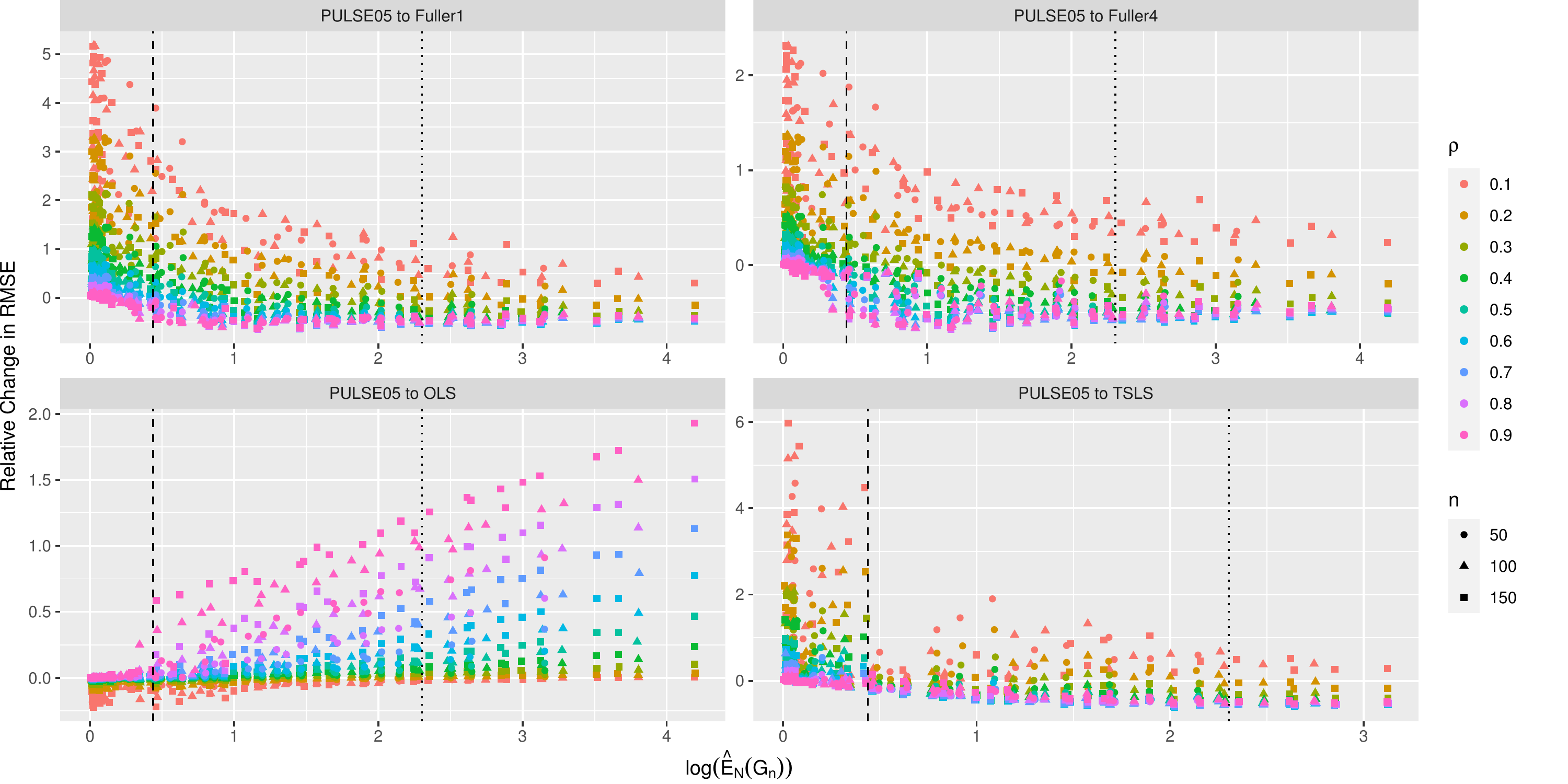}
			\caption{  \normalsize Illustrations of the relative change in  RMSE.}\label{fig:HahnExpRMSE}
		\end{figure}

		\Cref{fig:HahnExpRMSE} contains illustrations of the relative change in square-root mean squared error (RMSE) estimated from $15 000$ repetitions. On the horizontal axis we have plotted the average first stage F-test as a measure of weakness of instruments; see \Cref{sec:WeakInst} for further details. A test for $H_0:\bar \xi=0$, i.e.,  for the relevancy of instruments, 
		at a significance level of 5\%,	
		has different rejection thresholds in the range $[1.55,4.04]$ depending on $n$ and $q$. The vertical dashed line corresponds to the smallest rejection threshold of 1.55 and the dotted line corresponds to the `rule of thump' threshold of 10.  Note that the lowest possible negative relative change is $-1$ and a positive relative change means that PULSE is better.

		In \Cref{sec:AppFigs}, further illustrations of e.g. the relative change in mean bias and variance of the estimators are presented. 
		We also conducted the simulations for setups with combinations of $\gamma\in\{-1,0\}$, components of $\bar \xi$ chosen negatively, with random flipped sign in each coordinate and for negative $\rho$ (not shown but available in the folder 'Plots' in the code repository). The results with respect to MSE are similar to those shown in  \Cref{fig:HahnExpRMSE}, while the bias comparison changes depending on the setup.

		We observe that there are settings, in which the PULSE is superior to TSLS, Fuller(1) and Fuller(4) in terms of MSE.
		This is particularly often the case in weak instrument settings ($\hat E_N(G_n)<10$) for low confounding strength $(\rho\leq 0.2)$. Furthermore, as we tend towards the weakest instrument setting considered, we also see a gradual shift in favour of PULSE for higher confounding strengths. In these settings with weak instruments and low confounding we also see that OLS is superior to the PULSE in terms of MSE. However, for large confounding setups PULSE is superior to OLS in terms of both bias and MSE and this superiority increases as the instrument strength increases. 
		The PULSE is generally more biased than the Fuller and TSLS estimators but less biased than OLS. However, in the settings with weak instruments and low confounding the bias of PULSE and OLS is comparable.
		In summary, the PULSE is in these settings more biased but its variance is so small that it is MSE superior to the Fuller and TSLS estimators. 
		
		\section{Empirical applications}

		We now consider three classical 
		instrumental variable applications (see \citet{albouy2012colonial} and \citet{buckles2013season} for discussions on the underlying assumptions). 
		\begin{itemize}
			\item[\textit{(i)}] ``Does compulsory school attendance affect schooling and earnings?'' by \cite{angrist1991does}. This paper investigates the effects of education on wages. The endogenous effect of education on wages are remedied by instrumenting education on quarter of birth indicators.
			\item[\textit{(ii)}] ``Using geographic variation in college proximity to estimate the return to schooling'' by \cite{card1993using}. This paper also investigates the effects of education on wages. In this paper education is instrumented by proximity to college indicator.
			\item[\textit{(iii)}] ``The colonial origins of comparative development: An empirical investigation'' by \cite{acemoglu2001colonial}. This paper investigates the effects of extractive institutions (proxied by protection against expropriation) on the gross domestic product (GDP) per capita. The endogeneity of the explanatory variables are remedied by instrumenting protection against expropriation on early European settler mortality rates.
		\end{itemize}

		We have
		applied the different estimators OLS, TSLS, PULSE, and Fuller to the classical data sets
		\cite{acemoglu2001colonial}, \cite{angrist1991does} and  \cite{card1993using}.
		All models considered in \cite{angrist1991does} and  \cite{card1993using}, where we estimate the effect on years of education on wages, using quarter of birth and proximity to colleges as instruments, respectively, 
		the OLS estimates are not rejected by our test statistic and 
		PULSE outputs the OLS estimates; see \Cref{app:EmpricalApp} for futher details.
		This may be either due to weak endogeneity (weak confounding), or that the test has insufficient power to reject the OLS estimates due to either weak instruments or severe over-identification.

		\subsection{\cite{acemoglu2001colonial}} \label{sec:mainColonialApplication}
		The dataset of \cite{acemoglu2001colonial} consists of 64 
		observations,
		each corresponding to a different country for which mortality rate estimates encountered by the first European settlers are available. The endogenous target of interest is log GDP per capita (in 1995). The main endogenous regressor in the dataset is an  index of expropriation protection (averaged over 1985--1995), i.e., protection against expropriation of private investment by the respective governments. The average expropriation protection is instrumented by the settler mortality rates. We consider eight models M1--M8 which correspond to the models presented in column (1)--(8) in Table 4 of \cite{acemoglu2001colonial}. Model M1 is given by the reduced form structural equations
		\begin{align*}
			\log \mathrm{GDP} = \text{avexpr}\cdot \gamma + \mu_1 + U_1, \quad \text{avexpr} = \log \mathrm{em4}\cdot \delta + \mu_2 + U_2,
		\end{align*}
		where avexpr is the average expropriation protection, em4 is the settler mortality rates, $\mu_1$ and $\mu_2$ are intercepts and $U_1$ and $U_2$ are possibly correlated, unobserved noise variables. In model M2 we additionally introduce an included exogenous regressor describing the country latitude. In model M3 and M4 we fit model M1 and M2, respectively, on a dataset where we have removed Neo-European countries, Australia, Canada, New Zealand and the United States. In model M5 and M6 we fit model M1 and M2, respectively, on a dataset where we have removed observations from the continent of Africa. In model M7 and M8 we again fit model M1 and M2, respectively, but now also include three exogenous indicators for the continents Africa, Asia and other.
		
		\Cref{tbl:SettlerMortalityCoefficients} shows the OLS and TSLS estimates 
		(which replicate the values from the study), 
		as well as the Fuller(4) and PULSE estimates for the linear effect of the average expropriation protection on log GDP. In model M1, for example, we see that the PULSE estimate suggests that the average expropriation risk linear effect on log GDP is 0.6583 which is 26\% larger than the OLS estimate but 34\% smaller than TSLS estimate. In models M5--M8, the  OLS estimates are not rejected by the Anderson-Rubin test, so the PULSE estimates coincide with the OLS estimates.

		\begin{table}[htp]
			\caption{\label{tbl:SettlerMortalityCoefficients}The estimated return of expropriation protection on log GDP per capita. }
			\begin{center}
				\begin{tabu}to \textwidth {r c c c c c c c}
					\toprule \toprule
					Model & OLS & TSLS & FUL & PULSE & Message & Test & Threshold \\
					\midrule
					M1 &0.5221 & 0.9443 & 0.8584 & 0.6583 & --  & 5.9915 & 5.9915\\
					M2 &0.4679 & 0.9957 & 0.8457 & 0.5834 & -- & 7.8147 & 7.8147\\
					M3 &0.4868 & 1.2812 & 0.9925 & 0.7429 & -- & 5.9914 & 5.9915\\
					M4 &0.4709 & 1.2118 & 0.9268 & 0.6292 & -- & 7.8147 & 7.8147\\
					M5 &0.4824 & 0.5780 & 0.5573 & 0.4824 & OLS Accepted & 1.1798 & 5.9915\\
					M6 &0.4658 & 0.5757 & 0.5476 & 0.4658 & OLS Accepted & 1.1554 & 7.8147\\
					M7 &0.4238 & 0.9822 & 0.7409 & 0.4238 & OLS Accepted & 10.7722 & 11.0705\\
					M8 &0.4013 & 1.1071 & 0.7059 & 0.4013 & OLS Accepted & 9.7546 & 12.5916\\
					\bottomrule\bottomrule
				\end{tabu}
			\end{center}
			\footnotesize
			\renewcommand{\baselineskip}{11pt}
			\textbf{Note:} 
			Point estimates for the return of expropriation protection on log GDP per capita. 
			The OLS and TSLS values coincide with the ones shown in
			\cite{acemoglu2001colonial}.
			The right columns show the values of the test statistic (evaluated in the PULSE estimates) and the test rejection thresholds.
			The `--' indicates that OLS is not accepted and TSLS is not rejected.
		\end{table}
		
			We can also use this example to
		illustrate the robustness property of K-class estimators; 
		see \Cref{sthm:TheoremIntRobustKclas}. 
		Even though interventional data are not available, 
		we can consider the mean squared prediction error when 
		holding out the observations with the most extreme values of the instrument.
		Depending on the degree of generalization, we indeed see that 
		the PULSE and Fuller tend to outperform OLS or TSLS in terms of mean squared prediction error on the held out data; see \Cref{sec:Colonial} for further details.

		\section{Summary and future work}
		We have proved that a distributional robustness property similar to the one shown for anchor regression \citep[][]{AnchorRegression} 
		fully extends to general K-class estimators of possibly non-identifiable structural parameters in a general linear structural equation model that allows for latent endogenous variables.
		We have further proposed a novel estimator for structural parameters in linear structural equation models. This estimator, called PULSE, is derived as the solution to a minimization problem, where we seek to minimize mean squared prediction error constrained to a 
		confidence region 
		for the causal parameter. Even though this region is non-convex, we have shown 
		that the corresponding optimization problem allows for a computationally efficient algorithm 
		that approximates the above parameter with arbitrary precision using a simple binary search procedure.
		In the under-identified setting, this estimator extends existing work in the machine learning literature that considers invariant subsets or the best predictive sets among them: PULSE is applicable even in situations when no invariant subsets exist.
		We have proved that this estimator can also be written as a K-class estimator with data-driven 
		$\kappa$-parameter, which lies between zero and one. 
		Simulation experiments show that 
		in various settings 
		with weak instruments and weak confounding,
		PULSE  
		outperforms other estimators such as the Fuller(4) estimator. 
		We thus regard PULSE as an 
		interesting alternative for estimating causal effects
		in instrumental variable settings.
		It is easy to interpret and automatically provides the user feedback in case that the OLS is accepted (which may be an indication that the instruments are too weak)
		or that the TSLS is outside the acceptance region (which may indicate a model misspecification).
		We have applied the different estimators
		to classical data sets and have seen that, indeed, K-class estimators tend to be more distributionally robust than OLS or TSLS.

		There are several further directions that we consider worthwhile investigating. 
		This 
		includes better understanding of finite sample properties and for the identified setups, the study of loss functions other than MSE. It would be helpful, in particular with respect to real world applications, to understand to which extent similar principles can be applied to models allowing for a time structure of the error terms.
		We believe that the simple primal form of PULSE could make it applicable for model classes that are more complex than linear models \citep[see also][]{ChristiansenTPAMI}.
		Our procedure can be combined with other tests and it could furthermore be interesting to find efficient optimization procedures for tests that are robust with respect to weak instruments, such as  
		Kleibergen's K-statistic \citep[][]{Kleibergen2002},
		for example. 
		In an under-identified setting, the causal parameters are not identified but the solutions obtained by optimizing predictability under invariance might be promising candidates for models that generalize well to distributional shifts.

		\section*{Acknowledgements}
		We are grateful to 
		Trine %
		Boomsma,
		Peter %
		B\"uhlmann, 
		Rune Christiansen, 
		Steffen %
		Lauritzen, 
		Nicolai %
		Meinshausen, 
		Whitney %
		Newey,  
		Cosma %
		Shalizi, and
		Nikolaj %
		Thams for helpful 
		discussions.
		We thank the editor and two anonymous referees for 
		helpful and constructive comments.  MEJ and JP were supported by the Carlsberg Foundation; JP was, in addition, supported by a research grant
		(18968) from VILLUM FONDEN.

\appendix
\begin{appendices}
	
	\crefalias{section}{appendix}
	\crefalias{subsection}{appendix}
	\crefalias{subsubsection}{appendix}
	\crefname{appendix}{appendix}{appendices}
	\Crefname{appendix}{Appendix}{Appendices}
	\renewcommand{\theequation}{\thesection.\arabic{equation}}
	\renewcommand{\thefigure}{\thesection.\arabic{figure}}
	\renewcommand{\thetable}{\thesection.\arabic{table}}

	\section{Structural equation models and interventions} 
	\label{sec:simsem}
	\setcounter{figure}{0}
	\setcounter{table}{0}
	\setcounter{equation}{0}
	\renewcommand{\theremark}{\thesection.\arabic{remark}}
	Structural equation models and simultaneous equation models are causal models. That is, they contain more information than the description of an observational distribution. 
	We first
	introduce the notion of structural equation models (also called structural causal models)
	and use an example to show how they can be written as in the form of
	simultaneous equation models (SIM) commonly used in econometrics, see \Cref{sec:LinearModels}.
	
	\subsection{Structural equation models and interventions}
	A structural equation model (SEM)
	(e.g.\ \citealp{Bollen1989}, and \citealp{Pearl2009})
	over variables $X_1, \ldots, X_p$ consists of
	$p$ assignments of the form
	$$
	X_j := f_j(X_{\PA{j}}, \ep_j), \qquad j = 1, \ldots, p,
	$$
	where $\PA{j} \subseteq \{1, \ldots, p\}$
	are called the parents of $j$, 
	together with a distribution over 
	the noise variables $(\ep_1, \ldots, \ep_p)$, which is assumed to have jointly independent marginals. 
	The corresponding graph 
	over $X_1, \ldots, X_p$ 
	is obtained by drawing directed edges from the variables on the right-hand side to 
	the variables on the left-hand side. 
	If the corresponding 
	graph is acyclic, the SEM induces a unique distribution over 
	$(X_1, \ldots, X_p)$, which is often called the observational distribution.
	Section~\ref{sec:LinearModels} below discusses an example of linear assignments,
	which also allows for a cyclic graph structure.
	The framework of SEMs also models the effect of 
	interventions:
	An intervention on variable $j$ corresponds to 
	replacing the $j$th assignment.
	For example, 
	replacing it by
	$X_j = 4$, called a hard intervention, or, more generally, by
	$X_j = g(X_{\tPA{j}}, \tilde{\ep}_j)$
	induces yet another distribution over $X$ that is called an interventional distribution 
	and that we denote by
	$P^{\mathrm{do}(X_j = 4)}$ or 
	$P^{\mathrm{do}(X_j = g(X_{\tPA{j}}, \tilde{\ep}_j))}$, respectively.
	A formal introduction to SEMs, 
	in the general case of cyclic assignments is provided by \citet{Bongers2016b}, for example.
	In an SEM, 
	we call all $X$ variables endogenous and,
	in addition, all variables $X_j$, for which we have
	$\PA{j} = \emptyset$, will be called exogenous.
	A subset of variables is called exogenous relative to another subset
	if it does not contain a variable 
	that has a parent belonging to the other set. 
	
	In the  paper, 
	we are mostly interested in one of these equations and we denote the corresponding target variable as $Y$.
	Furthermore, some of the other $X$ variables  may be unobserved, which we indicate by using the notation $H$ (denoting a vector of variables). In linear models, hidden variables can 
	equivalently
	be represented 
	as correlation in the noise variables; see e.g.\ \citet{Bongers2016b}, and \citet{Hyttinen2012}.
	Finally, we let $A$ denote a collection of variables that are known 
	to enter the system as exogenous variables, relative to $(Y,X,H)$.
	
	\subsection{Example of a linear structural equation model} \label{sec:LinearModels}
	Let the distribution of $(Y,X,H,A)$ be generated according to the possibly cyclic 
	SEM,
	\begin{align} \label{ARModel}
		\begin{bmatrix}
			Y & X^\t & H^\t \end{bmatrix} 
		:=  \begin{bmatrix}
			Y &X^\t  & H^\t  \end{bmatrix} B   +  A^\t M+ \ep^\t.
	\end{align}
	Here, $B$ is a square matrix
	with eigenvalues whose absolute value is strictly smaller than one. This implies that $I-B$ is invertible ensuring that 
	the distribution of
	$(Y,X,H)$ is well-defined since 
	$(Y,X,H)$
	can be expressed in terms of $B,M,A$ and $\ep$ as $(I-B^\t)^{-1}(M^\t A+ \ep)$. 
	We denote the random vectors $Y\in \R, X\in \R^{d}, A\in \R^{q},H\in \R^r$  and $\ep\in \R^{d+1+r}$ 
	by target, endogenous regressor, anchor, hidden and noise variables, respectively. 
	We assume that $\ep\independent A$ rendering the so-called anchors as exogenous variables but the coordinate components of $A$ may be dependent on each other. 
	As above,
	we assume joint independence of the noises 
	$\ep_1, \ldots, \ep_{1+d+r}$.
	Let 
	$\fY \in \R^{n\times 1},\fX \in \R^{n\times d},\fA \in \R^{n\times q}, \fH \in \R^{n\times r}$ and $\bm{\ep}\in \R^{n\times(1+d+r)}$ be data-matrices with $n\in \N$ row-wise i.i.d.\ copies of the variables solving the system in \Cref{ARModel}. Transposing the structural equations and stacking them vertically by row-wise observations, we can represent all structural equations by $$[
	\fY  \, \, \fX  \, \,  \fH
	] := [
	\fY  \,  \, \fX \,  \,  \fH
	] B + \fA M + \bm{\ep}.$$ We can solve the structural equations for the endogenous variables and get the so-called structural and reduced form equations, commonly seen in econometrics, %
	\begin{align} \label{Eq:FullStructuralAndReducedFormOfSEM}
		\begin{bmatrix}
			\fY  & \fX &  \fH
		\end{bmatrix} \Gamma  = \fA M + \bm{\ep} \quad \text{and} \quad \begin{bmatrix}
			\fY  & \fX &  \fH
		\end{bmatrix}   = \fA \Pi + \bm{\ep}\Gamma^{-1},
	\end{align}
	respectively, where $\Gamma := I-B$ and  $\Pi := M \Gamma^{-1}$.
	Note that the equations in \Cref{Eq:FullStructuralAndReducedFormOfSEM} differ from the standard representations of simultaneous equation models as we have unobserved endogenous variables $\fH$ in the system. In this setup,
	identifiability of the full system parameters $\Gamma$ and $M$ in general breaks down due to the dependencies generated by the unobserved endogenous variables. 
	We now assume without loss of generality that $\Gamma$ has a unity diagonal, such that the target equation of interest, corresponding to the first column of \Cref{Eq:FullStructuralAndReducedFormOfSEM}, is given by
	\begin{align} \label{eq:StructuralEquationOfInterest}
		\fY = \fX \gamma_0 + \fA \beta_0  + \fH \eta_0 + \bm{\ep}_Y = %
		\fZ \alpha_0 + \tilde{\fU}_Y,
	\end{align}
	where $(1, -\gamma_0,- \eta_0)\in \R^{(1+d+r)}$, $ \beta_0\in \R^q$ and $\bm{\ep}_Y$  are the first columns of $\Gamma$, $M$ and $\bm{\ep}$ respectively, $\fZ := [
	\fX \, \, \fA]$, $\alpha_0 = (\gamma_0, \beta_0)\in \R^{d+q}$ and $\tilde{\fU}_Y := \fH\eta_0 + \bm{\ep}_Y$.

	The parameter of interest, $\alpha_0$, 
	can be derived directly from the corresponding entries in the matrices $B$ and $M$.
	It carries causal information in that, for example,
	after intervening on all variables except for $Y$, that 
	is, considering an intervention 
	$Z := z$, 
	and 
	$H := h$, 
	$Y$ has the mean 
	$z \alpha_0 + h \eta_0 + E \epsilon_1$, see \Cref{ARModel}.

	In \Cref{eq:StructuralEquationOfInterest} we have represented the target variable in terms of a linear combination of the observable variables $Z=(X^\t,A^\t)^\t$ and some unobservable noise term $\tilde{U}_Y$. 
	In contrast to \Cref{ARModel},   
	\Cref{eq:StructuralEquationOfInterest}, 
	which is more commonly used in the econometrics literature, 
	models 
	the influence of the latent variables using a dependence between endogenous 
	variables and the noise term $\tilde U_Y$;
	this equivalence is well-known and described by \citet{Bongers2016b} and \citet{Hyttinen2012}, for example.
	The construction in \Cref{ARModel} can be seen as a manifestation of Reichenbach's common cause principle (\citealp{reichenbach1956direction}). This principle stipulates that if two random variables are dependent then either one causally influences the other or there exists a third variable which causally influences both.

	\section{Algorithms} \label{app:algo}
	In this section we present two algorithms. Algorithm 1 details a binary search procedure for the dual PULSE parameter $\lambda^\star_n(p_{\min})$ and Algorithm 2 
	details the algorithmic construction and output messages of the PULSE estimator.
	\begin{algorithm}[H]
		\caption{Binary.Search with precision $1/N$.\label{Binary.Search.Lambda.Star}}
		\begin{algorithmic}[1]
			\State \textbf{input} $p_{\min}, N$
			\State {\bf if} {$T_n(\hat{\alpha}_{\text{TSLS}}^n) \geq  Q_{\chi^2_{q}}(1-p_{\min})$} %
			{\bf then} terminate procedure {\bf end if}
			\State $\ell_{\min}\leftarrow 0$; 
			$\ell_{\max} \leftarrow 2$
			\While{$T_n(\hat{\alpha}_\text{K}^n(\ell_{\max} )) >  Q_{\chi^2_{q}}(1-p_{\min}) $}
			\State $\ell_{\min} \leftarrow \ell_{\max}$; 
			$\ell_{\max} \leftarrow \ell_{\max}^2$
			\EndWhile
			\State $\Delta \leftarrow \ell_{\max}-\ell_{\min}$
			\While{$\Delta > 1/N$ } 
			\State $\ell \leftarrow (\ell_{\min}+ \ell_{\max})/2$
			\State {\bf if} $T_n(\hat{\alpha}_\text{K}^n (\ell)) >  Q_{\chi^2_{q}}(1-p_{\min}) $ {\bf then}
			$\ell_{\min} \leftarrow \ell$
			{\bf else}
			$\ell_{\max} \leftarrow \ell$
			{\bf end if}
			\State $\Delta \leftarrow \ell_{\max}-\ell_{\min}$
			\EndWhile
			\State return($\ell_{\max}$)
		\end{algorithmic}
	\end{algorithm}

	\begin{algorithm}[H] \label{algo:pulseplus}
		
		\caption{PULSE$+$}
		\begin{algorithmic}[1]
			\State \textbf{input} $p_{\min}$, precision $1/N$, $\hat{\alpha}_{\text{ALT}}^n$
			\If{$T_n(\hat{\alpha}_{\text{TSLS}}^n) \geq  Q_{\chi^2_{q}}(1-p_{\min})$}
			\State Warning: TSLS outside interior of acceptance region.
			\State $\hat{\alpha}_{\text{PULSE}+}^n(p_{\min}) \leftarrow \hat{\alpha}_{\text{ALT}}^n$
			\Else 
			\If{$T_n(\hat{\alpha}_{\text{OLS}}^n) \leq Q_{\chi^2_{q}}(1-p_{\min})$}
			\State Warning: The OLS is accepted.
			\State $\lambda^\star_n(p_{\min}) \leftarrow 0$
			
			\Else
			\State $\lambda^\star_n(p_{\min}) \leftarrow \text{Binary.Search}(N,p_{\min})$
			\EndIf
			\State $\hat{\alpha}_{\text{PULSE}+}^n(p_{\min}) \leftarrow (\fZ^\t (\fI+\lambda_n^\star(p_{\min}) P_\fA)\fZ)^{-1} \fZ^\t(\fI+\lambda_n^\star(p_{\min}) P_\fA)\fY$
			\EndIf
			\State return($\hat{\alpha}_{\text{PULSE}+}^n(p_{\min})$)
		\end{algorithmic}
		\label{alg:2}
	\end{algorithm} 

	\section{Proofs of results in Section~\ref{SEC:ROBUSTNESS}}
	\label{sec:RobustnessProofs}
	\setcounter{figure}{0}
	\setcounter{table}{0}
	\setcounter{equation}{0}
	\medskip

	\begin{proofenv}{\textbf{\Cref{lm:PenalizedKClassSolutionUniqueAndExists}}}
		The minimizations of \Cref{KclassLossFunctionPop}  and \Cref{KclassLossFunctionEmp} are unconstrained optimization problems. We  know that there exists a unique solution if the problems are strictly convex. Thus, it suffices to verify the second order condition for strict convexity of the objective functions, i.e., $D^2	l_{\mathrm{K}}^n(\alpha;\kappa)\succ 0$. To this end, note that
		$
		D l_{\mathrm{OLS}}^n(\alpha;\fZ_*,\fX) =  2(\alpha^\t \fZ_*^\t \fZ_* - \fY^\t\fZ_*)/n$  and $ 
		D l_{\mathrm{IV}}^n(\alpha;\fY,\fZ_*,\fA) =  2(\alpha^\t \fZ_*^\t P_\fA \fZ_* -  \fY^\t P_\fA \fZ_*)/n$. 
		Thus, the first order derivative of the K-class regression loss function is given by the $\kappa$-weighted affine combination of these two, that is,
		\begin{align*}
			D l_{\mathrm{K}}^n(\alpha;\kappa,\fY,\fZ_*,\fA) &= 2n^{-1}\lp (1-\kappa)\lp \alpha^\t \fZ_*^\t \fZ_* - \fY^\t\fZ_*\rp + \kappa \lp \alpha^\t \fZ_*^\t P_\fA \fZ_* -  \fY^\t P_\fA \fZ_*  \rp \rp \\
			&= 2n^{-1} \lp \alpha^\t \lp \fZ_*^\t \lp (1-\kappa)\fI +\kappa   P_\fA \rp \fZ_*  \rp - \lp \fY^\t \lp (1-\kappa) \fI + \kappa P_\fA\rp  \fZ_* \rp \rp  \\
			&= 2n^{-1} \lp \alpha^\t \lp \fZ_*^\t \lp \fI - \kappa\lp \fI- P_\fA \rp  \rp \fZ_*  \rp - \lp \fY^\t \lp \fI - \kappa\lp \fI- P_\fA \rp \rp  \fZ_* \rp \rp \\
			&= 2n^{-1} \lp \alpha^\t \lp \fZ_*^\t \lp \fI - \kappa P_\fA^\perp   \rp \fZ_*  \rp - \lp \fY^\t \lp \fI - \kappa P_\fA^\perp \rp  \fZ_* \rp \rp ,
		\end{align*}
		where $P_\fA^\perp = \fI - P_\fA$. 	The second order derivative is given by
		\begin{align*}
			D^2 l_{\mathrm{K}}^n(\alpha;\kappa,\fY,\fZ_*,\fA) %
			= 2n^{-1}\fZ_*^\t(I-\kappa P_\fA^\perp)\fZ_* ,
		\end{align*}
		The second derivative is
		and is proportional to the matrix we need to invert in order to solve the normal equation that yields the K-class estimator. As a consequence, we have that the K-class estimator is guaranteed to exist and 
		be unique if the second derivative is strictly positive definite, i.e., invertible.

		Let us first consider $\kappa < 1$.	
		To see that $D^2 l_{\mathrm{K}}^n(\alpha;\kappa,\fY,\fZ_*,\fA)\succ0$, 
		take any $y \in \R^{d_1+q_1}\setminus \{0\}$ and assume that Assumption \Cref{ass:ZtZfullRank} holds. That is, we assume that $\text{rank}(\fZ_*^\t \fZ_*)=\text{rank}(\fZ_*)=d_1+q_1$ almost surely such that  $ z= \fZ_* y \in \R^{n}\setminus \{0\}$ almost surely. 
		Without Assumption \Cref{ass:ZtZfullRank}, choosing $y\in \ker(\fZ_*)\setminus \{0\}$ yields a zero in the following quadratic form with positive probability. However, with this assumption (disregarding $2n^{-1}$) we get that
		\begin{align*}
			y^\t 	D^2 l_{\mathrm{K}}^n(\alpha;\kappa) y & \propto (1-\kappa)y^\t \fZ_*^\t \fZ_* y +  \kappa y^\t \fZ_*^\t P_\fA \fZ_* y  = (1-\kappa)\| z \|_2^2 + \kappa \|P_\fA z\|_2^2 \\
			&\geq \left\{\begin{array}{ll}
				(1-\kappa)\|z\|_2^2+\kappa \|z\|_2^2 = \|z\|_2^2, & \text{if } \kappa \in (-\infty,0) , \\
				(1-\kappa)\|z\|_2^2, & \text{if } \kappa \in [0,1), 
			\end{array}\right. >0.	\end{align*}
		Here, we used that $P_\fA = P_\fA^\t = \fA(\fA^\t\fA)^{-1} \fA^\t$ is an orthogonal projection matrix, hence $P_\fA=P_\fA^\t P_\fA$ and $0 \leq \|P_\fA w\|_2^2 \leq \|w\|_2^2$ for any $w\in \R^q$.

		Let us now consider the case $\kappa = 1$. The quadratic form is now given by
		$$
		y^\t 	D^2 l_{\mathrm{K}}^n(\alpha;\kappa)  y =\|P_\fA z\|_2^2 = y^\t \fZ_*^\t \fA (\fA^\t \fA)^{-1} \fA^\t \fZ_* y.
		$$
		If $\text{rank}(\fA^\t \fZ_*)< d_1+q_1$ with positive probability, then any $y\in \ker(\fA^\t \fZ_*)\setminus \{0\}\not = \emptyset$ yields a zero quadratic value, 	showing that $l_{\mathrm{K}}^n(\alpha;\kappa)$ is not strictly convex with positive probability. However, if Assumption \Cref{ass:AtZfullRank} holds, i.e., that $\fA^\t \fZ_*\in \R^{q\times(d_1+q_1)}$ satisfies $\text{rank}(\fA^\t \fZ_*)=d_1+q_1$ almost surely, then $D^2 l_{\mathrm{K}}^n(\alpha;\kappa)$ is also guaranteed to be positive definite almost surely. 
		
		Thus, we have shown sufficient conditions for $D^2 l_{\mathrm{K}}^n(\alpha;\kappa)$ to be almost surely positive definite, ensuring strict convexity of the $l_{\mathrm{K}}^n(\alpha;\kappa)$, hence almost sure uniqueness of a global minimum. The unique global minimum is then found as a solution to the normal equation $
			Dl_{\text{K}}^n(\alpha;\kappa) =0  %
		$
		which is given by $\hat{\alpha}_{\text{K}}^n (\kappa) = (\fZ_*^\t (\fI-\kappa P_\fA^\perp)\fZ_*)^{-1} \fZ_*^\t(\fI-\kappa P_\fA^\perp)\fY$. We conclude that under the above conditions the K-class estimator $\hat{\alpha}_{\text{K}}^n (\kappa)$ solves the unconstrained minimization problem $
		\argmin_{\alpha\in \R^{d_1+q_1}} l_{\mathrm{K}}^n(\alpha;\kappa)$ 
		almost surely.
	\end{proofenv}
	
	\noindent 
	\begin{proofenv}{\textbf{\Cref{lm:PopulationPenalizedKClassSolutionUniqueAndExists}}}
		We first prove that the population estimand that minimizes the population loss function is well-defined. It suffices to show strict convexity of the population loss function. Let Assumption \Cref{ass:VarianceOfZPositiveDefinite} hold, i.e., that $\text{Var}(Z_*)$ is positive definite, and consider $\kappa \in[0,1)$. For any $y\in \R^{d_1+q_1}\setminus \{0\}$ we see that
		\begin{align}
			y^\t 	D^2 l_{\mathrm{K}}(\alpha;\kappa) y &= (1-\kappa)y^\t E(Z_* Z_*^ \t) y +  \kappa y^\t E(Z_* A^\t) E(A A^\t)^{-1} E(A Z_*^\t) y \notag \\
			&\geq (1-\kappa)y^\t E(Z_* Z_*^\t) y = (1-\kappa) \lp y^\t  \text{Var}(Z_*)y + y^\t E(Z_*)E(Z_*)^\t y \rp  \notag\\
			& \geq  (1-\kappa)  y^\t  \text{Var}(Z_*)y   > 0,\label{eq:EZZtIsPositiveDefinite} 
		\end{align}
		proving strict convexity of the K-class penalized loss function. Now let $\kappa=1$ and let Assumption  \Cref{ass:VarianceOfAPositiveDefinite} and Assumption \Cref{ass:EAZtFullColumnRank} hold, i.e., $\text{Var}(A)$ is positive definite and  $E(AZ_*^\t)$ is of full column rank (which implicitly assumes we are in the just- or over-identified case). First note that by the above considerations this implies that $E(AA^\t)$ and its inverse $E(AA^\t)^{-1}$ are positive definite. For any $y\in \R^{d_1+q_1}\setminus \{0\}$ we note that $z:=E(A Z_*^\t) y \not = 0 $ by injectivity of $E(A Z_*^\t)$, and hence $$
		y^\t 	D^2 l_{\mathrm{K}}(\alpha;\kappa) y =   z^\t E(A A^\t)^{-1} z > 0,
		$$
		by the positive definiteness of $E(AA^\t)^{-1}$. Proving strict convexity.
		
		In both setups the minimization estimator of the population loss function solves the normal equation $0 = Dl_{\text{K}}(\alpha;\kappa) = (1-\kappa)Dl_{\text{OLS}}(\alpha) + \kappa Dl_{\text{IV}}(\alpha) $ which by rearranging the terms
		yields that
		\begin{align*}
			\alpha_{\text{K}}(\kappa) =&  \lp (1-\kappa)E(Z_*Z_*^\t) + \kappa E(Z_*A^\t)E(AA^\t)^{-1}E(AZ_*^\t)\rp^{-1} \\
			&\cdot \lp (1-\kappa) E(Z_*Y) + \kappa E(Z_*A^\t)E(AA^{\t})^{-1}E(AY)\rp.
		\end{align*}
		We now prove that the estimators are asymptotically well-defined if
		the population conditions of Assumption \Cref{ass:VarianceOfZPositiveDefinite} and Assumption \Cref{ass:EAZtFullColumnRank}	
		hold. 
		For $\kappa\in[0,1)$, we know from \Cref{lm:PenalizedKClassSolutionUniqueAndExists} that
		\begin{align*}
			P\left[ \argmin_{\alpha\in \R^{d_1+q_1}} l_{\mathrm{K}}^n(\alpha;\kappa,\fY,\fZ_*,\fA) \textit{ is well-defined} \right] \geq  P\left[ \fZ_*^\t \fZ_* \textit{ is positive definite} \right],
		\end{align*}
		So it suffices to show that the lower converges to one in probability. By the weak law of large numbers we have, for any $\ep >0$ that $
		P(\|\fZ_*^\t \fZ_* - E(Z_*Z_*^\t)\| < \ep )  \to 1 \label{eq:ZtZConvp}$. 
		Note that by Assumption \Cref{ass:VarianceOfZPositiveDefinite}, i.e., that $\text{Var}(Z_*)$ is positive definite, we also have that $E(Z_*Z_*^\t)$ is positive definite; see \Cref{eq:EZZtIsPositiveDefinite} above. Note that the set of positive definite matrices $S_+$ is an open set in the space of symmetric matrices $S$ of the same dimensions. Hence, there must exist an open ball $B(E(Z_*Z_*^\t),c)\subseteq S_+$ with center $E(Z_*Z_*^\t)$ and radius $c>0$, fully contained in the set of positive definite matrices. By virtue of the above convergence in probability, we have that
		\begin{align*}
			P\left[ \fZ_*^\t \fZ_* \textit{ is positive definite} \right] &\geq  P \lp \fZ_*^\t \fZ_* \in B(E(Z_*Z_*^\t),c)\rp \\& 
			\geq  P(\|\fZ_*^\t \fZ_* - E(Z_*Z_*^\t)\| < c )  \to 1,
		\end{align*}
		proving that the estimator minimizing the K-class penalized regression function is asymptotically well-defined. In the case of $\kappa=1$ the argument for asymptotic well-definedness follows by almost the same arguments. Arguing that $\fA^\t\fA$ is positive definite with probability converging to one since $\text{Var}(A)$ is assumed positive definite follows from the same arguments as above.
		To see that $\fA^\t\fZ_*$ is of full column rank with probability converging to one, we use that $E(AZ_*^\t)$ is assumed full column rank. If $q=d_1+q_2$, then follows from the above arguments. Otherwise, if $q> d_1+q_1$, then we modify the above arguments using that the set of injective linear maps from $\R^{d_1+q_1}$ to $\R^q$ is an open set of all linear maps from $\R^{d_1+q_1}$ to $\R^q$. 
		
		Finally, by the law of large numbers, Slutsky's theorem and the continuous mapping theorem, one can easily realize that $\hat{\alpha}_{\text{K}}^n (\kappa) \convp \alpha_{\text{K}}(\kappa)$.
	\end{proofenv}
	
	\noindent
	\begin{proofenv}{\textbf{\Cref{sthm:TheoremIntRobustKclas}}}
		Let $(Y,X,H,A)$ be generated by the SEM given by
		\begin{align} \label{KclassRobustnessEquationSEM}
			[
			Y \, \,  X^\t \, \, H^\t 
			]^\t := B [
			Y \, \, X^\t \, \, H^\t 
			]^\t + MA + \ep,
		\end{align}
		where $\ep$ satisfies $\ep\independent A$ 
		and has jointly independent marginals $\ep_1 \independent \cdots \independent \ep_{d+1+r}$ with finite second moment $E\|\ep\|_2^2<\i$ and mean zero $E(\ep)=0$. The distribution of $A$ is determined independently of \Cref{KclassRobustnessEquationSEM} and with the only requirement that $E\|A\|_2^2<\i$. Note that we have transposed $B$ and $M$ for ease of notation. This implies that $(Y,X,H)$ satisfies the  reduced form equations given by $
		[
		Y \, \, X^\t \, \, H^\t 
		]^\t = \Pi A + \Gamma^{-1}\ep$, where $\Gamma = I-B$ and $\Pi = \Gamma^{-1} M$. 
		
		Now let $X_*\subset X$ and $A_* \subset A$ be our candidate predictors of $Y$, regardless of which variables directly affect $Y$ and let $Z_* = [
		X_*^\t \, \, A_*^\t
		]^\t $. By the reduced form structural equations we  derive the marginal reduced forms as
		\begin{align} \label{Eq:MarginalReducedFormOfX}
			Y = \Pi_Y A + \Gamma^{-1}_Y \ep \quad \text{and} \quad X_* = \Pi_{X_*} A + \Gamma^{-1}_{X_*} \ep ,
		\end{align}
		where $\Pi_Y,\Pi_{X_*},\Gamma^{-1}_Y,\Gamma^{-1}_{X_*}$ are the relevant sub-matrices of rows from $\Pi$ and $\Gamma^{-1}$.  Furthermore, let $(Y^v,X^v,H^v)$ 
		be generated as a solution to the SEM of \Cref{KclassRobustnessEquationSEM} under the intervention $\text{do}(A:=v)$, where $v\in \mathcal{L}^2(\Omega,\cF,P)$ is any fixed stochastic element uncorrelated with $\ep$. Under the intervention and by  similar manipulations as above, we arrive at the following marginal reduced forms $
		Y^v = \Pi_Y v + \Gamma^{-1}_Y \ep$ and $X_*^v = \Pi_{X_*} v + \Gamma^{-1}_{X_*} \ep.$
		For a fixed $\gamma$ and $\beta$, with $A_{-*}$ being $A\setminus A_{*}$, we have that
		\begin{align*}
			Y-\gamma^\t X_* - \beta^\t A_* &= (\Pi_Y - \gamma^\t \Pi_{X_*})A +(\Gamma_Y^{-1}-\Gamma_{X_*}^{-1})\ep - \beta^\t A_* \\
			&= (\delta_1^\t-\beta^\t)A_* + \delta_2^\t A_{-*} + w^\t \ep = \xi^\t A + w^\t \ep,
		\end{align*}
		where $\delta_1$, $\delta_2$ are such that
		$(\Pi_Y - \gamma^\t \Pi_{X_*})A = \delta_1^\t A_* + \delta_2^\t A_{-*}$, $\xi$  is such that
		$\xi^\t A = (\delta_1^\t-\beta^\t)A_* + \delta_2^\t A_{-*} $  and $w^\t := (\Gamma_Y^{-1}-\Gamma_{X_*}^{-1})$.  Similar manipulations yield that 
		the regression residuals 
		under the intervention are given by
		$
		Y^v-\gamma^\t X_*^v - \beta^\t v_* = \xi^\t v + w^\t \ep. %
		$
		Since $A \independent \ep$ and $\ep$ has mean zero, we have that
		\begin{align}
			E(Y-\gamma^\t X_* - \beta^\t A_*| A) &= \xi^\t A + w^\t E(\ep) = \xi^\t A, \label{KclassRobustnessEquationRegResidualsCondMeanGivenA} \\
			Y-\gamma^\t X_* - \beta^\t A_* - E(Y-\gamma^\t X_* - \beta^\t A_*| A) &= w^\t \ep. \label{KclassRobustnessEquationRegResidualsMinusRegResidualsCondMeanOnA}
		\end{align}
		By construction $E(v\ep^\t)=0$, so
		\begin{align}
			E^{\text{do}(A:=v)}\left[ \lp Y-\gamma^\t X_* - \beta^\t A_* \rp^2  \right] &= E\left[ \lp \xi^\t v + w^\t \ep \rp^2  \right] \notag \\
			&= E\left[ \lp\xi^\t v \rp^2 \right] + E\left[ \lp w^\t \ep \rp^2 \right] + \xi^\t E(v\ep^\t)w \notag \\
			&= E\left[ \lp\xi^\t v \rp^2 \right] + E\left[ \lp w^\t \ep \rp^2 \right]. \label{KclassRobustnessEquationEqExpectedResidualsUnderInt}
		\end{align}
		We investigate the terms of \Cref{KclassRobustnessEquationEqExpectedResidualsUnderInt} and note by \Cref{KclassRobustnessEquationRegResidualsMinusRegResidualsCondMeanOnA} that
		\begin{align}
			E\left[ \lp w^\t \ep \rp^2 \right]  &=  E\left[ \lp Y-\gamma^\t X_* - \beta^\t A_* - E(Y-\gamma^\t X_* - \beta^\t A_*| A)\rp^2 \right]  \notag \\
			& =  E\left[ \lp Y-\gamma^\t X_* - \beta^\t A_* \rp^2 \right] +E\left[   E \lp Y-\gamma^\t X_* - \beta^\t A_*| A\rp^2 \right] \label{KclassRobustnessEquationSecondTerm}\\
			&\qquad - 2 E\left[ \lp Y-\gamma^\t X_* - \beta^\t A_*\rp  E(Y-\gamma^\t X_* - \beta^\t A_*| A) \right]. \notag 
		\end{align}
		In \Cref{KclassRobustnessEquationRegResidualsCondMeanGivenA} we established that $E(Y-\gamma^\t X_* - \beta^\t A_*| A)$ is a linear function of $A$, so it must hold that
		\begin{align*}
			E(Y-\gamma^\t X_* - \beta^\t A_*| A) &= \argmin_{Z\in\sigma(A)}\|Y-\gamma^\t X_* - \beta^\t A_*-Z\|_{L^2(P)}^2 \\
			&=A^\t \argmin_{c \in \R^{q}}\|Y-\gamma^\t X_* - \beta^\t A_*-  A^\t c\|_{L^2(P)}^2  \\
			&= A^\t E(AA^\t)^{-1} E\left[ A\lp Y-\gamma^\t X_* - \beta^\t A_*\rp \right],
		\end{align*}
		almost surely. In the first equality we used that the conditional expectation is the best predictor under the $L^2(P)$-norm and in the third equality we used that the minimizer is given by the population ordinary least square estimate. 
		An immediate consequence of this is that the second term of \Cref{KclassRobustnessEquationSecondTerm} equals 
		\begin{align*}
			E\left[  E \lp Y-\gamma^\t X_* - \beta^\t A_*|A  \rp^2 \right] &= E[(Y-\gamma^\t X_* - \beta^\t A_*)A^\t]E(AA^\t)^{-1}\\
			&\quad \quad \cdot  E[A(Y-\gamma^\t X_*- \beta^\t A_*)],
		\end{align*}
		which is seen to be of the same form of the third term in \Cref{KclassRobustnessEquationSecondTerm},
		\begin{align*}
			&E \left[ \lp Y-\gamma^\t X_* - \beta^\t A_*\rp  E(Y-\gamma^\t X_* - \beta^\t A_*|A)  \right] \\	= &E \left[ \lp Y-\gamma^\t X_* - \beta^\t A_*\rp A^\t \right] E(AA^\t)^{-1}  E\left[A(Y-\gamma^\t X_* - \beta^\t A_*)\right].
		\end{align*}
		Thus, we conclude that the second term of \Cref{KclassRobustnessEquationEqExpectedResidualsUnderInt} is given by
		\begin{align*}
			E \left[\lp w^\t \ep \rp^2 \right] &= E \left[\lp Y-\gamma^\t X_* - \beta^\t A_* \rp^2 \right]  - E\left[  E \lp Y-\gamma^\t X_* - \beta^\t A_*|A \rp^2 \right] \\
			&= l_{\mathrm{OLS}}(\alpha;Y,Z_*)- l_{\mathrm{IV}}(\alpha;Y,Z_*,A). 
		\end{align*}
		Taking the supremum over all $v\in C(\kappa)$ of the first term of \Cref{KclassRobustnessEquationEqExpectedResidualsUnderInt} we obtain
		\begin{align*}
			\sup_{v\in C(\kappa)}E \left[\lp \xi^\t v \rp^2 \right] &= \sup_{v\in C(\kappa)}\xi^\t E \left[   vv^\t  \right] \xi = \frac{1}{1-\kappa}\xi^\t E \left[  AA^\t \right] \xi = \frac{1}{1-\kappa} E \left[    \lp \xi^\t A \rp^2 \right] \\
			&= \frac{1}{1-\kappa} E \left[    E \lp Y-\gamma^\t X_* - \beta^\t A_*|A \rp^2 \right] = \frac{1}{1-\kappa} l_{\mathrm{IV}}(\alpha;Y,Z_*,A),
		\end{align*}
		where the 
		second last equation follows from 
		\Cref{KclassRobustnessEquationRegResidualsCondMeanGivenA}
		and the
		second equation follows from the following argument.	For any $v\in C(\kappa)$ we have that $E(vv^\t) \preceq \frac{1}{1-\kappa} E(AA^\t)$, that is,  for all $x\in \R^q$ it holds that $\frac{1}{1-\kappa} x^\t E(AA^\t) x \geq x^\t E(vv^\t)x$, which implies that the upper bound is attained for any $v$ such that $E(vv^\t) = \frac{1}{1-\kappa}E(AA^\t)$.
		Thus, we have that 
		\begin{align*}
			\sup_{v\in C(\kappa)}  E^{\mathrm{do}(A:=v)}\left[ (Y - \gamma^\t X_*  - \beta^\t A_*  )^2 \right] =&  \sup_{v\in C(\kappa)}  E\left[ \lp\xi^\t v \rp^2 \right] + E\left[ \lp w^\t \ep \rp^2 \right]  \\
			=& l_{\mathrm{OLS}}(\alpha;Y,Z_*)+ \frac{\kappa }{1-\kappa} l_{\mathrm{IV}}(\alpha;Y,Z_*,A).
		\end{align*}
		By the representation in  \Cref{eq:KclassLossFunctionAsPenalizedOLS} it therefore follows that the population K-class estimate with parameter $\kappa\not = 1$ is given as the estimate that minimizes the worst case mean squared prediction error over all interventions contained in $C(\kappa)$, that is,
		\begin{align*}
			\alpha_{\mathrm{K}}(\kappa;Z_*,A) &= 	\argmin_{\gamma\in \R^d,\beta\in \R^{q_1}}\sup_{v\in C(\kappa)}  E^{\mathrm{do}(A:=v)}\left[ (Y - \gamma^\t X_*  - \beta^\t A_*  )^2 \right] .%
		\end{align*}
	\end{proofenv}
	
	\section{Proofs of selected results in Section~\ref{SEC:PULSE}}
	\label{sec:SomeProofsOfSecPULSE}
	\setcounter{figure}{0}
	\setcounter{table}{0}
	\setcounter{equation}{0}
	\setcounter{corollary}{0}
	\medskip
	\renewcommand{\thecorollary}{\thesection.\arabic{corollary}}
	\renewcommand{\thelemma}{\thesection.\arabic{lemma}}
	\renewcommand{\thetheorem}{\thesection.\arabic{theorem}}
	
	\begin{restatable}[K-class estimators differ]{corollary}{KclassSolutionsDistinct}
		\label{cor:KclassSolutionsDistinctApp}
		Let %
		\Cref{ass:KclassNotInIV,ass:ZtZfullrankandAtZfullrank} hold.
		If $\lambda_1,\lambda_2\geq 0$ with $\lambda_1\not = \lambda_2$, then $\hat{\alpha}_{\mathrm{K}}^n(\lambda_1) \not = \hat{\alpha}_{\mathrm{K}}^n(\lambda_2)$.
	\end{restatable}
	\noindent	\textbf{Proof:}
	Let \Cref{ass:ZtZfullrankandAtZfullrank,ass:KclassNotInIV} hold. $\hat{\alpha}_{\text{K}}^n(\lambda)$ is well-defined for all $\lambda \geq 0$ by \Cref{lm:PenalizedKClassSolutionUniqueAndExists}. Let $\lambda_1,\lambda_2\geq 0$
	with $\lambda_1 \not = \lambda_2$ and note that the orthogonality condition derived in the proof of \Cref{lm:KclassNotEqualToTwoSLS} also applies here. That is, $
	\la \fY- \fZ \hat{\alpha}_{\text{K}}^n(\lambda_i), (\fI + \lambda_i P_\fA) z\ra =0, $ for all $z\in \cR(\fZ)$ and  $i=1,2$. Assume for contradiction that $\hat{\alpha}_{\text{K}}^n(\lambda_1) = \hat{\alpha}_{\text{K}}^n(\lambda_2)$. This implies that
	\begin{align*}
		0&=\la \fY - \fZ \hat{\alpha}_{\mathrm{K}}^n(\lambda_1),  (\fI + \lambda_1 P_\fA) z - (\fI + \lambda_2 P_\fA) z\ra \\
		&= \la \fY - \fZ \hat{\alpha}_{\mathrm{K}}^n(\lambda_1),   (\lambda_1-\lambda_2) P_\fA  z\ra = (\lambda_1-\lambda_2) \la \fY - \fZ \hat{\alpha}_{\mathrm{K}}^n(\lambda_1),   P_\fA  z\ra,
	\end{align*}
	for any $z\in \cR(\fZ)$. Thus, by symmetry and idempotency of $P_\fA$ we have that for all $z\in \cR(\fZ)$, $$
	\la P_\fA \fY - P_\fA \fZ \hat{\alpha}_{\mathrm{K}}^n(\lambda_1),   P_\fA  z\ra = \la \fY - \fZ \hat{\alpha}_{\mathrm{K}}^n(\lambda_1),   P_\fA  z\ra =0.$$
	That is, $P_\fA \fZ \hat{\alpha}_{\mathrm{K}}^n(\lambda_1)$ is the orthogonal projection of $P_\fA \fY$ onto $\cR(P_\fA \fZ)$.
	This is equivalent with saying that $\hat{\alpha}_{\mathrm{K}}^n(\lambda_1) \in \cM_{\text{IV}}$ as the space of minimizers of $l_{\text{IV}}^n$ are exactly the coefficients in $\R^{d_1+q_1}$ which mapped through $P_\fA \fZ$ yields this orthogonal projection. See the proof of \Cref{lm:KclassNotEqualToTwoSLS} for further elaboration on this equivalence.
	This is a contradiction to \Cref{ass:KclassNotInIV}, hence $\hat{\alpha}_{\text{K}}^n(\lambda_1) \not =  \hat{\alpha}_{\text{K}}^n(\lambda_2)$.
	\hfill$\square$\bigskip

	\begin{restatable}[Monotonicity of the losses and the test statistic]{lemma}{MonotonicityOfTestOfLambda} \textcolor{white}{,}\label{lm:OLSandIV_Monotonicity_FnctOfPenaltyParameterLambdaApp} \\When Assumption \Cref{ass:ZtZfullrank} holds the maps $
		[0,\i)\ni \lambda \mapsto  l_{\mathrm{OLS}}^n(\hat{\alpha}_{\mathrm{K}}^n (\lambda) )$ and $ [0,\i)\ni  \lambda \mapsto l_{\mathrm{IV}}^n(\hat{\alpha}_{\mathrm{K}}^n (\lambda) ) $
		are monotonically increasing and monotonically decreasing, respectively. Consequently, if \Cref{ass:ZYfullcolrank} holds, we have that the map 
		$
		[0,\i)\ni \lambda \longmapsto T_n (\hat{\alpha}_{\mathrm{K}}^n (\lambda) ) %
		$
		is monotonically decreasing. Furthermore, if \Cref{ass:KclassNotInIV} also holds, these monotonicity statements can be strengthened to strictly decreasing and strictly increasing.
	\end{restatable}
	\noindent	\textbf{Proof:}
	Let Assumption \Cref{ass:ZtZfullrank} hold, such that $\hat{\alpha}_{\text{K}}^n(\lambda)$ is well-defined for all $\lambda\geq 0$; see \Cref{lm:PenalizedKClassSolutionUniqueAndExists}.
	Let $\lambda_2 > \lambda_1 \geq 0$ and note that
	\begin{align*}
		l_{\mathrm{OLS}}^n(\hat{\alpha}_{\mathrm{K}}^n (\lambda_1) ) &+ \lambda_1 l_{\mathrm{IV}}^n(\hat{\alpha}_{\mathrm{K}}^n (\lambda_1) )  \leq l_{\mathrm{OLS}}^n(\hat{\alpha}_{\mathrm{K}}^n (\lambda_2) ) + \lambda_1 l_{\mathrm{IV}}^n(\hat{\alpha}_{\mathrm{K}}^n (\lambda_2) ) \\
		&= l_{\mathrm{OLS}}^n(\hat{\alpha}_{\mathrm{K}}^n (\lambda_2) ) + \lambda_2 l_{\mathrm{IV}}^n(\hat{\alpha}_{\mathrm{K}}^n (\lambda_2) ) + (\lambda_1-\lambda_2)l_{\mathrm{IV}}^n(\hat{\alpha}_{\mathrm{K}}^n (\lambda_2) ) \\
		&\leq l_{\mathrm{OLS}}^n(\hat{\alpha}_{\mathrm{K}}^n (\lambda_1) ) + \lambda_2 l_{\mathrm{IV}}^n(\hat{\alpha}_{\mathrm{K}}^n (\lambda_1) ) + (\lambda_1-\lambda_2)l_{\mathrm{IV}}^n(\hat{\alpha}_{\mathrm{K}}^n (\lambda_2) ),
	\end{align*}
	where we used that $\hat{\alpha}_{\mathrm{K}}^n (\lambda)$ minimizes the expressions with penalty factor $\lambda$. Thus, $$
	(\lambda_1-\lambda_2)l_{\mathrm{IV}}^n(\hat{\alpha}_{\mathrm{K}}^n (\lambda_1) ) \leq (\lambda_1-\lambda_2)l_{\mathrm{IV}}^n(\hat{\alpha}_{\mathrm{K}}^n (\lambda_2) ),$$ which is equivalent with $$ l_{\mathrm{IV}}^n(\hat{\alpha}_{\mathrm{K}}^n (\lambda_1) ) \geq l_{\mathrm{IV}}^n(\hat{\alpha}_{\mathrm{K}}^n (\lambda_2) ),$$ 
	proving that $\lambda \mapsto l_{\mathrm{IV}}^n(\hat{\alpha}_{\mathrm{K}}^n (\lambda) )$ is monotonically decreasing. 
	
	If $\lambda_2 > \lambda_1 =0$, then we note that $$
	l_{\mathrm{OLS}}^n(\hat{\alpha}_{\mathrm{K}}^n (\lambda_1) ) = \min_{\alpha} \{l_{\mathrm{OLS}}^n( \alpha) \}  \leq l_{\mathrm{OLS}}^n(\hat{\alpha}_{\mathrm{K}}^n (\lambda_2) ).$$	For any $\lambda >0$, $$
	\hat{\alpha}_{\mathrm{K}}^n (\lambda) = \argmin_{\alpha} \{ l_{\mathrm{OLS}}^n(\alpha) + \lambda l_{\mathrm{IV}}^n(\alpha) \} 
	= \argmin_{\alpha} \{ \lambda^{-1}l_{\mathrm{OLS}}^n(\alpha) +  l_{\mathrm{IV}}^n(\alpha) \}.$$ 
	Thus, if $\lambda_2> \lambda_1 > 0,$ we have that
	\begin{align*}
		\lambda_1^{-1}l_{\mathrm{OLS}}^n(\hat{\alpha}_{\mathrm{K}}^n (\lambda_1) ) +&  l_{\mathrm{IV}}^n(\hat{\alpha}_{\mathrm{K}}^n (\lambda_1) ) 
		\leq \lambda_1^{-1}l_{\mathrm{OLS}}^n(\hat{\alpha}_{\mathrm{K}}^n (\lambda_2) ) +  l_{\mathrm{IV}}^n(\hat{\alpha}_{\mathrm{K}}^n (\lambda_2) ) \\
		= &\lambda_2^{-1}l_{\mathrm{OLS}}^n(\hat{\alpha}_{\mathrm{K}}^n (\lambda_2) ) +  l_{\mathrm{IV}}^n(\hat{\alpha}_{\mathrm{K}}^n (\lambda_2) ) + (\lambda_1^{-1}-\lambda_2^{-1})l_{\mathrm{OLS}}^n(\hat{\alpha}_{\mathrm{K}}^n (\lambda_2) ) \\
		\leq &\lambda_2^{-1}l_{\mathrm{OLS}}^n(\hat{\alpha}_{\mathrm{K}}^n (\lambda_1) ) +  l_{\mathrm{IV}}^n(\hat{\alpha}_{\mathrm{K}}^n (\lambda_1) ) + (\lambda_1^{-1}-\lambda_2^{-1})l_{\mathrm{OLS}}^n(\hat{\alpha}_{\mathrm{K}}^n (\lambda_2)),
	\end{align*}
	hence $
	l_{\mathrm{OLS}}^n(\hat{\alpha}_{\mathrm{K}}^n (\lambda_1) ) \leq l_{\mathrm{OLS}}^n(\hat{\alpha}_{\mathrm{K}}^n (\lambda_2) ),$
	so $\lambda \mapsto l_{\mathrm{OLS}}^n(\hat{\alpha}_{\mathrm{K}}^n (\lambda) )$ is monotonically increasing. 	
	
	When \Cref{ass:ZYfullcolrank} holds, the map $$
	\lambda \mapsto T_n (\hat{\alpha}_{\mathrm{K}}^n (\lambda) ) = n  \frac{l_{\text{IV}}^n(\hat{\alpha}_{\mathrm{K}}^n (\lambda))}{l_{\mathrm{OLS}}^n(\hat{\alpha}_{\mathrm{K}}^n (\lambda))} ,$$
	is well-defined and monotonically decreasing, as it is given by a positive,  monotonically decreasing function over a strictly positive and monotonically increasing function.
	
	Furthermore, when \Cref{ass:KclassNotInIV} holds, 
	\Cref{cor:KclassSolutionsDistinct} yields that for $\lambda_1,\lambda_2\geq0 $ with $\lambda_1 \not = \lambda_2$ it holds that $\hat{\alpha}_{\text{K}}^n(\lambda_1) \not = \hat{\alpha}_{\text{K}}^n(\lambda_2)$. 
	As a consequence, the above inequalities become strict, 
	since
	otherwise (Dual.$\lambda.n$) has two distinct solutions which contradicts
	\Cref{lm:PenalizedKClassSolutionUniqueAndExists}. Replacing the above inequalities with strict inequalities yields that the functions are strictly increasing and decreasing, respectively.
	\hfill$\square$\bigskip
	
	\begin{restatable}[]{lemma}{TestInAlphaLambdaStarEqualsQuantile}
		\label{lm:TestInAlphaLambdaStarEqualsQuantileApp}
		Let $p_{\min}\in(0,1)$ and let Assumption \Cref{ass:ZtZfullrank} and \Cref{ass:ZYfullcolrank} hold. If $\lambda_n^\star(p_{\min})<\infty$, it holds that \begin{align} \label{eq:TestInAlphaLambdaStarLessQ}
			T_n(\hat{\alpha}_{\mathrm{K}}^n (\lambda_n^\star(p_{\min}))) \leq  Q_{\chi^2_{q}}(1-p_{\min}).
		\end{align}
		If the ordinary least square estimator  satisfies
		$T_n(\hat{\alpha}_{\mathrm{OLS}}^n)< Q_{\chi^2_{q}}(1-p_{\min})$, then \Cref{eq:TestInAlphaLambdaStarLessQ} holds with strict inequality, otherwise it holds with equality.
	\end{restatable}
\noindent	\textbf{Proof:}
	Let $p_{\min}\in(0,1)$ and let let Assumption \Cref{ass:ZtZfullrank} and \Cref{ass:ZYfullcolrank} hold, such that $\hat{\alpha}_{\text{K}}^n(\lambda)$ for all $\lambda\geq 0$ and $T_n(\alpha)$ for all $\alpha\in \R^{d_1+q_1}$ are well-defined, by \Cref{lm:PenalizedKClassSolutionUniqueAndExists}.

	Assume that $\lambda_n^\star(p_{\min})< \i$, so we know that $T_n(\hat{\alpha}_{\text{K}}^n(\lambda)) \leq Q_{\chi^2_{q}}(1-p)$ for all $\lambda > \lambda_n^\star(p_{\min})$ by the monotonicity of  \Cref{lm:OLSandIV_Monotonicity_FnctOfPenaltyParameterLambda}. 
	Thus, the first statement follows
	if we can show that $\lambda \mapsto T_n(\hat{\alpha}_{\text{K}}^n(\lambda))$ is a continuous function. Since $\alpha \mapsto T_n(\alpha)$ is continuous it suffices to show that $[0,\infty) \ni \lambda \mapsto \hat{\alpha}_{\text{K}}^n(\lambda)$ is continuous. Recall that $
	\hat{\alpha}_{\text{K}}^n(\lambda) =(\fZ^\t (\fI+\lambda P_\fA)\fZ)^{-1} \fZ^\t(\fI+\lambda P_\fA)\fY, $	for any $\lambda \geq 0$. Note that the functions
	$\text{Inv}:S_{++}^{d_1+q_1} \to S_{++}^{d_1+q_1}$ given by $\fM\stackrel{\text{Inv}}{\mapsto} \fM^{-1}$,  $\lambda \mapsto \fZ^\t (\fI+\lambda P_\fA)\fZ$, $\lambda \mapsto  \fZ^\t(\fI+\lambda P_\fA)\fY$ and $(\fB,\fC) \mapsto \fB \fC$ 
	are all continuous maps, where $S_{++}^{d_1+q_1}$ is the set of all positive definite $(d_1+q_1)\times(d_1+q_1)$ matrices. We have that $\lambda \mapsto \hat{\alpha}_{\text{K}}^n(\lambda)$ is a composition of these continuous maps, hence it itself is continuous. This proves the first statement.
	
	In the case that OLS is strictly feasible in the PULSE problem, $T_n(\hat{\alpha}_{\text{OLS}}^n)<  Q_{\chi^2_{q}}(1-p_{\min}) $, we have that
	$$
	\lambda^\star(p_{\min})=\inf \left\{  \lambda \geq 0  :   T_n ( \hat{\alpha}_{\text{K}}^n (\lambda) )\leq Q_{\chi^2_{q}}(1-p_{\min}) \right\} = 0,
	$$
	since $\hat{\alpha}_{\text{K}}^n(0) = \hat{\alpha}_{\text{OLS}}^n$,	hence
	$$
	T_n(\hat{\alpha}_{\mathrm{K}}^n (\lambda_n^\star(p_{\min}))) =T_n(\hat{\alpha}_{\mathrm{K}}^n (0)) = T_n(\hat{\alpha}_{\text{OLS}}^n) <  Q_{\chi^2_{q}}(1-p_{\min}).
	$$
	Similar arguments show that, if the OLS is just-feasible in the PULSE problem, $T_n(\hat{\alpha}_{\text{OLS}}^n)= Q_{\chi^2_{q}}(1-p_{\min})$, then $T_n(\hat{\alpha}_{\mathrm{K}}^n (\lambda_n^\star(p_{\min})))= Q_{\chi^2_{q}}(1-p_{\min})$.

	In the case that the OLS estimator is infeasible in the PULSE problem, $Q_{\chi^2_{q}}(1-p_{\min}) < T_n(\hat{\alpha}_{\text{OLS}}^n)$, continuity and monotonicity of $\lambda \mapsto T_n(\hat{\alpha}_{\text{K}}^n(\lambda))$ 
	entail it must hold that $T_n(\hat{\alpha}_{\text{K}}^n(\lambda_n^\star(p_{\min})))=Q_{\chi^2_{q}}(1-p_{\min})$, 
	as otherwise $$T_n(\hat{\alpha}_{\text{K}}^n(\lambda_n^\star(p_{\min})))<Q_{\chi^2_{q}}(1-p_{\min}) < T_n(\hat{\alpha}_{\text{K}}^n(0)),$$
	implying that there exists $\tilde{\lambda} < \lambda_n^\star(p_{\min})$ such that
	$T_n(\hat{\alpha}_{\text{K}}^n(\tilde{\lambda}))\leq Q_{\chi^2_{q}}(1-p_{\min})$, contradicting $\lambda_n^\star(p_{\min}) = \inf\{\lambda \geq 0 : T_n(\hat{\alpha}_{\mathrm{K}}^n (\lambda))\leq Q_{\chi^2_{q}}(1-p_{\min}) \}$.
	\hfill$\square$\bigskip
	
	\section{Proofs of remaining results in Section~\ref{SEC:PULSE}}
	\label{sec:RemainingProofsOfSecPULSE}
	\setcounter{figure}{0}
	\setcounter{table}{0}
	\setcounter{equation}{0}
	\medskip
	
	\noindent
	\begin{proofenv}{\textbf{\Cref{prop:TestingVanishingCorr}}}
		We want to show an asymptotic guarantee that type I errors (rejecting a true hypothesis) occur with probability $p$. That is,  if $\cH_0(\alpha)$ is true, then $P(T_n^c(\alpha) >  Q_{\chi^2_{q}}(1-p) ) \stackrel{n\to\i }{\longrightarrow} p$. Furthermore, we want to show that for any fixed alternative, the probability of	type II errors (failure to reject) converges to zero. That is, if $P$ is such that	
		$\cH_0(\alpha)$ is false, then $P(T_n^c(\alpha) \leq  Q_{\chi^2_{q}}(1-p)) \stackrel{n\to\i }{\longrightarrow} 0$.

		Fix any $\alpha\in \R^{d_1+q_1}$. It suffices to show that under the null-hypothesis
		$T_n^c(\alpha)$ is asymptotically Chi-squared distributed with $q$ degrees of freedom and that $T_n^c(\alpha)$ tends to infinity under any fixed alternative. Without loss of generality assume that $c(n)=n$ for all $n\in \N$ and recall that
		$$
		T_n^c(n)= T_n(\alpha) = n \frac{l_{\text{IV}}^n(\alpha)}{l_{\text{OLS}}^n(\alpha)} 
		=  n \frac{\|P_\fA (\fY-  \fZ \alpha) \|_2^2}{ \|\fY- \fZ \alpha\|_2^2} .
		$$
		By the idempotency of $P_\fA$ the 
		numerator can be rewritten as $$
		\|P_\fA (\fY-  \fZ \alpha) \|_2^2 =   \|(\fA^\t \fA)^{-1/2} \fA^\t  \fR(\alpha) \|_2^2,
		$$
		while the denominator takes the form $\|\fR(\alpha)\|_2^2$. Here, $\fR(\alpha) := \fY - \fZ \alpha$ and $R(\alpha):= Y-Z^\t\alpha$ denotes the empirical and population regression residuals, respectively. \Cref{ass:ZYfullcolrank} ensures that $T_n$ is well-defined on the entire domain of $\R^{d_1+q_1}$ as the denominator is never zero. Furthermore, note that both $R(\alpha)$ for any $\alpha\in \R^{d_1+q_1}$ and $A_i$ for any $i=1,...,q$ have finite second moments
		by virtue of Assumption \Cref{ass:SecondMomentA}. %
		
		Assume that the null hypothesis of zero correlation between the components of
		$A$ and the regression residuals $R(\alpha)$ holds. First we show that the null hypothesis, under the stated assumptions, implies independence between the exogenous variables $A$ and the regression residuals $R(\alpha)$.
		It holds that $E(AR(\alpha)) = E(A)E(R(\alpha))=0$ by Assumption \Cref{ass:MeanZeroA}, the mean zero assumption of $A$. Assumption \Cref{ass:AIndepUy}, i.e., $A\independent U_Y$, yields that
		\begin{align} \label{Eq:TempMomentOfProductAandResiduals}
			0 &= E(AR(\alpha)) %
			= E(AZ^\t)(\alpha_0-\alpha) + E(AU_Y) = E(AZ^\t)(\alpha_0-\alpha),
		\end{align}
		proving that $\alpha - \alpha_0 = w$ 
		for some $w \in \text{kern}(E(AZ^\t))$. Recall that
		the marginal structural equation of \Cref{Eq:MarginalReducedFormOfX} states that $X_* = \Pi_{X_*} A + \Gamma^{-1}_{X_*} \ep$. Thus, $Z$ has 
		the following representation
		\begin{align*}
			Z = \begin{bmatrix}
				X_* \\ A_*
			\end{bmatrix} =\begin{bmatrix}
				\Pi_{X_*} A + \Gamma^{-1}_{X_*} \ep\\ A_*
			\end{bmatrix}  = \begin{bmatrix}
				\Pi_{X_*}^{(*)} & \Pi_{X_*}^{(-*)} \\
				I & 0
			\end{bmatrix} \begin{bmatrix}
				A_* \\ A_{-*}
			\end{bmatrix} + \begin{bmatrix}
				\Gamma_{X_{*}}^{-1} \\ 0 
			\end{bmatrix} \ep =: \Lambda A + \Psi \ep,
		\end{align*}
		where $\Pi_{X_*} = [
		\Pi_{X_*}^{(*)} \, \, \Pi_{X_*}^{(-*)} ] \in \R^{d_1 \times(q_1+q_2)}$ and $\Lambda$, $\Psi$ are the 
		conformable block-matrices. Since $A\independent \ep$ by Assumption \Cref{ass:AindepEp} we have that $E(A\ep^\t)=0$, hence $$
		0=E(AZ^\t) w = E(AA^\t)\Lambda^\t w + E(A\ep^\t)\Psi^\t w = E(AA^\t)\Lambda^\t w.$$ 
		This proves that $\Lambda^\t w =0$  as $E(AA^\t)$ is of full rank by Assumption \Cref{ass:VarianceOfAPositiveDefinite}. Hence, 
		\begin{align*}
			R(\alpha) &= Y-Z^\t \alpha = Z^\t(\alpha_0-\alpha)+ U_Y 
			= Z^\t w +U_Y \\
			&= A^\t \Lambda^\t w + \ep^\t \Psi^\t w +U_Y =\ep^\t \Psi^\t w +U_Y.
		\end{align*}
		Furthermore, $U_Y=  \alpha_{0,-*}^\t Z_{-*}+\eta^\t_0  H + \ep_Y$ can be written as a linear function of $A$ plus a linear function of $\ep$. To realize this, simply express $Z_{-*}$ and $H$ by their marginal reduced form structural equations. Hence, the assumptions that $A \independent U_Y$ must entail that $A$ vanishes from the expression of $U_Y$. As a consequence we have that $R(\alpha)$ is a linear function only of $\ep$, from which the assumption that $A\independent \ep$ yields that $A \independent R(\alpha)$. That is, the null hypothesis of zero correlation implies independence in the linear structural equation model, 
		under the given assumptions. Thus, $E\|AR(\alpha)\|_2^2 = E\|A\|_2^2 E\|R(\alpha)\|_2^2<\i$, so the covariance matrix of $AR(\alpha)$ is well-defined.

		By the established independence and \Cref{Eq:TempMomentOfProductAandResiduals}, the covariance matrix of $AR(\alpha)$ has the following representation
		$$
		\Cov( AR(\alpha) )=  
		E(AA^\t)E(R(\alpha)^2) \succ 0.$$
		The positive definiteness follows from the facts that $E(AA^\t)\succ 0$ and $E(R(\alpha)^2)>0$ for any $\alpha\in \R^{d_1+q_1}$. $E(AA^\t)\succ 0$ follows by Assumption \Cref{ass:VarianceOfAPositiveDefinite} and $E(R(\alpha)^2)>0$ for any $\alpha\in \R^{d_1+q_1}$ follows by  Assumption \Cref{ass:SpectralRadiusOfBLessThanOne}, Assumption \Cref{ass:epIndependentMarginals} and \Cref{ass:NonDegenYNoise};  non-degeneracy and mutual independence of the marginal noise variables in $\ep$. To see this, expand $R(\alpha)$ in terms of the marginal reduced form structural equations of $Y$ and $Z$ and use that $(I-B^\t)$ is invertible to see that $\ep$ does not vanish in the expression $R(\alpha)$. The multi-dimensional Central Limit Theorem yields that
		$$
		\frac{1}{\sqrt{n}}\fA^\t \fR(\alpha) = \sqrt{n} \lp \frac{1}{n}\sum_{i=1}^n \begin{pmatrix}	A_{i,1}R(\alpha)_{i} \\ 	\vdots \\	A_{i,q} R(\alpha)_{i}	\end{pmatrix}\rp  
		\convd \cN(0,\Cov(AR(\alpha))).$$
		Furthermore, note that regardless of whether or not the null-hypothesis is true, we have that $$
		\sqrt{n}(\fA^\t \fA)^{-1/2} %
		\convp E(AA^\t)^{-1/2},
		$$
		and
		$$
		\frac{1}{\sqrt{n}}\|\fR(\alpha)\|_2 = \sqrt{\frac{1}{n}\sum_{i=1}^n R(\alpha)_i^2} 
		\convp \sqrt{E(R(\alpha)^2)}>0,
		$$
		by the law of large numbers and the continuity of the matrix square root operation on the cone of symmetric positive-definite matrices.	
		We can represent the test-statistic as $
		T_n(\alpha):= \| \sqrt{n}W_n(\alpha)\|_2^2$ with $$W_n(\alpha):= (\fA^\t \fA)^{-1/2}\fA^\t \fR(\alpha)/\|\fR(\alpha)\|_2,$$ 
		and have that $
		\sqrt{n}W_n(\alpha) %
		\convd W \sim  \cN\lp 0, I \rp$, 
		by Slutsky's theorem and linear transformation rules of multivariate normal distributions.	Hence, %
		the continuous mapping theorem yields that $$
		T_n(\alpha) = \left\|\sqrt{n}W_n(\alpha) \right\|_2^2 \convd \|W\|_2^2 = \sum_{i=1}^qW_i^2 \sim  \chi^2_{q},$$ 
		where $\chi^2_{q}$ is the Chi-squared distribution with $q$ degrees of freedom, since $W_1\independent \cdots \independent W_q$. This proves that the test-statistic $T_n$ has the correct asymptotic distribution under the null-hypothesis.

		Now fix a distribution $P$, for which the null hypothesis of simultaneous zero correlation between the components of $A$ and the residuals $R(\alpha)$ does not hold. That is, there exists an $j\in\{1,...,q\}$ such that $E(A_{j}R(\alpha)) \not = E(A_j)E(R(\alpha))=0$.  Note that 
		\begin{align*}
			\left\| n^{-1/2}(\fA^\t \fA)^{1/2} \right \|_{\text{op}}^2 T_n(\alpha) &= \left\| n^{-1/2}(\fA^\t \fA)^{1/2} \right \|_{\text{op}}^2 \left\| \frac{\sqrt{n}(\fA^\t \fA)^{-1/2}\frac{1}{\sqrt{n}}\fA^\t \fR(\alpha)}{\frac{1}{\sqrt{n}}\|\fR(\alpha)\|_2} \right\|_2^2 \\
			&\geq \left\| \frac{\frac{1}{\sqrt{n}}\fA^\t \fR(\alpha)}{\frac{1}{\sqrt{n}}\|\fR(\alpha)\|_2} \right\|_2^2 \geq \left| \frac{\frac{1}{\sqrt{n}}\fA_j^\t \fR(\alpha)}{\frac{1}{\sqrt{n}}\|\fR(\alpha)\|_2} \right|^2= n \left| \frac{ \frac{1}{n} \fA_j^\t \fR(\alpha)}{\frac{1}{\sqrt{n}}\|\fR(\alpha)\|_2} \right|^2, %
		\end{align*}
		where $\fA_j^\t := (\fA_j)^\t$ and $\fA_j$ is the \textit{j}'th column
		of $\fA$ corresponding to the i.i.d.\ vector consisting of $n$ copies of the \textit{j}'th exogenous variable $A_j$ and $\|\cdot\|_{\text{op}}$ is the operator norm.
		The lower bound diverges to infinity in probability
		as the latter factor tends to $|E(A_jR(\alpha)) / \sqrt{E(R(\alpha)^2)}|^2 > 0$ in probability by the law of large numbers and Slutsky's theorem. Hence, it holds that $T_n(\alpha) \convp \i,$
		as $$\left\| n^{-1/2}(\fA^\t \fA)^{1/2} \right \|_{\text{op}} \to \left\| E(A A^\t )^{1/2} \right\|_{\text{op}} \in(0,\i).$$ %
		This concludes the proof.
	\end{proofenv}

	\noindent
	\begin{proofenv}{\textbf{\Cref{lm:PrimalUniqueSolAndSlatersConditions}}}
		Let \Cref{ass:ZtZfullrankandAtZfullrank} hold, i.e., that $\fZ^\t \fZ$ and $\fA^\t \fZ$ are of full rank.
		That  $\alpha \mapsto l_{\mathrm{IV}}^n(\alpha;\fY,\fZ,\fA)$ is a convex function and $\alpha \mapsto l_{\text{OLS}}^n(\alpha)$ is a strictly convex function can be seen from the quadratic forms of their second derivatives, i.e., $$
		y^\t D^2l_{\text{IV}}^n(\alpha)y = 2n^{-1} y^\t \fZ^\t \fA (\fA^\t \fA)^{-1} \fA^\t \fZ y = 
		2n^{-1}\|(\fA^\t \fA)^{-1/2} \fA^\t \fZ y\|_2^2\geq 0,$$ and $$	y^\t D^2l_{\text{OLS}}^n(\alpha) y = 2n^{-1} y^\t \fZ^\t \fZ y =2
		n^{-1}\|\fZ y\|_2^2 >0,$$ 
		for any $y\in \R^{d_1+q_1}\setminus \{0\}$.  Here, we also used that $\fA^\t \fA$ is of full rank by Assumption \Cref{ass:AtAfullRank} and that $\fZ\in \R^{n\times(d_1+q_1)}$ is an injective linear transformation as $d_1+q_1=\text{rank}(\fZ^\t \fZ)=\text{rank}(\fZ)$.

		Suppose that there exists two optimal solutions $\alpha_1,\alpha_2$ to the (Primal$.t.n$) problem. By the convexity of the feasibility set any convex combination is also feasible. However, $$l_{\mathrm{OLS}}^n \lp \alpha_1/2 + \alpha_2/2\rp < l_{\mathrm{OLS}}^n(\alpha_1)/2 + l_{\mathrm{OLS}}^n(\alpha_2)/2 = l_{\mathrm{OLS}}^n(\alpha_1),$$ since $l_{\mathrm{OLS}}^n(\alpha_1)=l_{\mathrm{OLS}}^n(\alpha_2)$.	This means that $\alpha_1/2+\alpha_2/2$ has a strictly better objective value than the optimal point $\alpha_1$, which is a contradiction. 
		Hence, there cannot exist multiple solutions to the optimization problem (Primal$.t.n$).
		
		Regarding the claim of solvability, note that $\fZ^\t \fZ$ is positive definite and as a consequence the smallest eigenvalue $\lambda_{\min}(\fZ^\t \fZ)$ is strictly positive. Thus, using the lower bound of the symmetric quadratic form $\alpha^\t \fZ^\t \fZ \alpha \geq\lambda_{\min}(\fZ^\t \fZ) \|\alpha\|_2^2$, we get that
		\begin{align}
			l_{\text{OLS}}^n(\alpha) &= \fY^\t \fY + \alpha^\t \fZ^\t  \fZ \alpha - 2\fY^\t \fZ \alpha \notag \geq \fY^\t \fY + \lambda_{\min}(\fZ^\t \fZ)\| \alpha\|_2^2 - 2|\fY^\t \fZ \alpha|  \notag\\
			& \geq  \fY^\t \fY + \lambda_{\min}(\fZ^\t \fZ)\| \alpha\|_2^2 - 2\|\fY^\t \fZ\|_{\text{op}} \|\alpha\|_2  \to \i, \label{eq:lOLSTendsToInfinityAsAlphaGrowsLarge}
		\end{align}
		as $\|\alpha\|_2\to\i$, where we used that for the linear operator $\fY^\t \fZ:\R^{d_1+q_1}\to \R$ the operator
		norm is given by $\|\fY^\t \fZ\|_{\text{op}} := \inf\{c\geq 0 : |\fY \fZ v|\leq c\|v\|_2, \forall v\in \R^{d_1+q_1}\}$, obviously satisfying $|\fY^\t \fZ v|\leq  \|\fY^\t \fZ\|_{\text{op}}\|v\|_2$ for any $v\in \R^{d_1+q_1}$. 
		
		Now assume that $t> \inf_\alpha l_{\text{IV}}^n(\alpha)$.
		This implies that there exists at least one point $\tilde{\alpha}\in\R^{d_1+q_1}$ such that $l_{\text{IV}}^n (\tilde{\alpha}) \leq t$, hence 
		we only need to consider points $\alpha$ such that $l_{\text{OLS}}^n(\alpha) \leq l_{\text{OLS}}^n(\tilde{\alpha})$ 
		as possible solutions of the optimization problem.	
		By the considerations 
		in \Cref{eq:lOLSTendsToInfinityAsAlphaGrowsLarge}
		above, there exists
		$c\geq 0$ such that is suffices to search over the closed ball $\overline{B(0,c)}$. 
		Indeed, 
		for a sufficiently large $c\geq 0$ we know that $\alpha \not \in \overline{B(0,c)}$	 
		implies that $l_{\text{OLS}}^n(\alpha)> l_{\text{OLS}}^n(\tilde{\alpha})$ by \Cref{eq:lOLSTendsToInfinityAsAlphaGrowsLarge}. Furthermore, as the inequality constraint function $\alpha \mapsto l_{\text{IV}}^n(\alpha)$ is continuous, the set of feasible points $(l_{\text{IV}}^n)^{-1}((-\infty,t])$ is closed. Hence, our minimization problem is equivalent with the minimization of the continuous function $\alpha\mapsto l_{\text{OLS}}^n(\alpha)$ over the convex and compact set $\overline{B(0,c)} \cap (l_{\text{IV}}^n)^{-1}((-\infty,t])$. By the extreme value theorem, the minimum exist and is attainable. We conclude that the primal problem is solvable if $t > \inf_\alpha l_{\text{IV}}^n(\alpha)$.  
		
		By definition, Slater's condition is satisfied if there exists a point in the relative interior of the problem domain where the constraint inequality is strict \citep[][]{boyd2004convex}. Since the problem domain is 
		$\R^{d_1+q_1}$, we need the existence of $\alpha\in \R^{d_1+q_1}$ such that $l^n_{\text{IV}}(\alpha)<t$. This is clearly satisfied if $t > \inf_{\alpha} l_{\text{IV}}^n(\alpha)$. Let us now specify the exact lower bound for the constraint bound as a function of the over-identifying restrictions. 	\textit{Under- and just-identified case: $q_2 \leq d_1$ ($q\leq d_1+q_1$).}  %
		Assumption \Cref{ass:AtZfullrank} yields that $\fA^\t \fZ\in \R^{q\times (d_1+q_1)}$  satisfies $\text{rank}(\fA^\t \fZ) = q$. That is, $\fA^\t \fZ $ is of full row rank, hence surjective. Thus, we are guaranteed the existence of a $\tilde{\alpha}\in \R^{d_1+q_1}$ such that $	\fA^\t \fZ \tilde{\alpha} = \fA ^\t \fY$, implying that  $l_{\mathrm{IV}}^n(\tilde{\alpha}) =0$. \textit{Over-identified case: $d_1 < q_2$ ($d_1+q_1< q$).} Note that the constraint function $l_{\mathrm{IV}}^n(\alpha):\R^{d_1+q_1} \to \R$  is strictly  convex as the  second derivative $D^2l_{\mathrm{IV}}^n(\alpha;\fY,\fZ,\fA) \propto \fZ^\t \fA(\fA^\t \fA)^{-1}\fA^\t \fZ$ is 
		positive definite by the assumption that $\fA^\t \fZ\in \R^{q\times (d_1+q_1)}$ has full (column)
		rank. 	The global minimum of $l_{\text{IV}}$ is 
		therefore	 
		attained in the unique stationary point. Furthermore,  the stationary point is found by solving the normal equation $
		Dl_{\mathrm{IV}}^n(\alpha;\fY,\fZ,\fA) =0$. The solution to the normal equation is given by $\hat{\alpha}_{\text{TSLS}}^n = (\fZ^\t P_\fA \fZ)^\t \fZ^\t P_\fA\fY,
		$
		which is the standard TSLS estimator. 
	\end{proofenv}

	\noindent
	\begin{proofenv}{\textbf{\Cref{thm:pPULSESolvesPULSE}}}
		Let $p_{\min}\in(0,1)$ and let  \Cref{ass:ZtZfullrankandAtZfullrank} and \Cref{ass:ZYfullcolrank} hold. That is, $\fA^\t\fZ$ and $\fZ^\t \fZ$ are of full rank and $[
		\fZ \, \, \fY]$ is of full column rank. Furthermore, assume that $t_n^\star(p_{\min})>-\i$ and  $T_n(\hat{\alpha}_{\text{Pr}}^n(t_n^\star(p_{\min})))\leq Q_{\chi^2_q}(1-p_{\min})$. 
		First assume that $\hat{\alpha}_\text{Pr}^n(t_n^\star(p_{\min})) = \hat{\alpha}_{\text{OLS}}^n$. We note that
		$$
		T_n(\hat{\alpha}_{\text{OLS}}^n) = T_n(\hat{\alpha}_\text{Pr}^n(t_n^\star(p_{\min})) ) %
		\leq Q_{\chi^2_q}(1-p_{\min}),
		$$
		hence the global minimizer $\hat{\alpha}_{\text{OLS}}^n$ of $\alpha \mapsto l_{\text{OLS}}^n(\alpha)$ is unique, feasible and necessarily optimal in the PULSE problem, so $\hat{\alpha}_\text{Pr}^n(t_n^\star(p_{\min}))  = \hat{\alpha}_{\text{OLS}}^n = \hat{\alpha}^n_{\text{PULSE}}$ and we are done.

		Now assume that $\hat{\alpha}_\text{Pr}^n(t_n^\star(p)) \not = \hat{\alpha}_{\text{OLS}}^n$. Consider the PULSE problem of interest
		\begin{align} \tag{PULSE}
			\begin{array}{ll}
				\mathrm{min}_\alpha & l_{\mathrm{OLS}}^n(\alpha)  \\
				\mathrm{subject \, to} & T_n(\alpha) \leq Q_{\chi^2_{q}}(1-p_{\min}),
			\end{array} %
		\end{align}
		which is, in general, a non-convex quadratically constrained quadratic program. First we argue that the problem is solvable, i.e., the optimum is attainable.
		\begin{quote} \normalsize
			To see this, let $p=p_{\min}$, $Q=Q_{\chi^2_{q}}(1-p_{\min})$ and note that by the assumption  $t_n^\star(p_{\min})>-\i$ we have that the feasible set of the PULSE problem is non-empty. By the assumptions that $[ \fZ \, \, \fY]$ is of full column rank we have that $T_n(\alpha)$ is well-defined for any $\alpha \in \R^{d_1+q_1}$, as the denominator is never zero. By continuity of $\R^{d_1+q_1} \ni \alpha \mapsto T_n(\alpha)$ we have that the feasible set $
			\cF := T_n^{-1} \lp (-\i,Q] \rp,$
			is closed and non-empty, since it is the continuous preimage of a closed set. Applying the same arguments as seen earlier in the proof of \Cref{lm:PrimalUniqueSolAndSlatersConditions}, we know that $l_{\text{OLS}}^n(\alpha)\to \i$ when $\|\alpha\| \to \i$. Hence, for a sufficiently large $c>0$ we know that if $\alpha \not \in \overline{B(0,c)}$, where $\overline{B(0,c)}\subset \R^{d_1+q_1}$ is the closed ball with centre 0 and radius $c$, then we only get suboptimal objective values $l_{\text{OLS}}^n(\alpha) > l_{\text{OLS}}^n(\hat{\alpha}_{\text{Pr}}^n (t_n^\star(p)))$. That is, we can without loss of optimality or loss of solutions restrict the feasible set to $\cF'= T_n^{-1} ( (-\i, Q] ) \cap \overline{B(0,c)}$ a closed and bounded set in $\R^{d_1+q_1}$. Hence, by the extreme value theorem the minimum over $\cF'$ is guaranteed to be attained. That is, the PULSE problem is solvable. 	
		\end{quote}	
		However, by the non-convexity of $T_n$, the preimage $T_n^{-1} ( (-\i, Q] )$ is in general not convex, so the minimum is not yet guaranteed to be attained in a unique point.	
		We will show that the minimum of the PULSE problem is attained in a unique point, that exactly coincides with the primal PULSE solution. Fix any solution $\hat{\alpha}$ to 
		the
		PULSE problem and realize that the PULSE constraint is active in  $\hat{\alpha}$,  \begin{align}
			T_n(\hat{\alpha})= Q. \label{Eq:PULSEsolActiveInConstraint}
		\end{align}
		\begin{quote} \normalsize
			This is seen by noting that $\hat{\alpha}_\text{Pr}^n(t_n^\star(p)) \not = \hat{\alpha}_{\text{OLS}}^n$ by assumption, so $\hat{\alpha}_{\text{OLS}}^n$ is not feasible in the PULSE problem, that is,  $
			\hat{\alpha}_{\text{OLS}}^n\not \in \cF$. 
			If $\hat{\alpha}_{\text{OLS}}^n$ was feasible, then $t_n^\star(p) =  \sup \{ t \in D_{\text{Pr}} : T_n(\hat{\alpha}_{\mathrm{Pr}}^n(t))\leq Q_{\chi^2_{q}}(1-p_{\min})\} = l_{\text{IV}}^n(\hat{\alpha}^n_{\text{OLS}}),$	since $T_n(\hat{\alpha}_{\text{Pr}}^n(l_{\text{IV}}^n(\hat{\alpha}^n_{\text{OLS}})) = T_n(\hat{\alpha}_{\text{OLS}}^n)\leq Q$,  hence  $$
			\hat{\alpha}_\text{Pr}^n(t_n^\star(p))  =   \argmin_{\alpha: l_{\text{IV}}^n(\alpha) \leq l_{\text{IV}}^n(\hat{\alpha}_{\text{OLS}}^n)} 
			l_{\text{OLS}}^n(\alpha)   =  \hat{\alpha}_{\text{OLS}}^n,
			$$
			which is a contradiction. That the optimum must be attained in a point, where the PULSE inequality constraint is active then	 
			follows from \Cref{lm:SolutionIsTightIfNotStationary} of \Cref{sec:AuxLemmas}
			and the conclusion above that the only stationary point of $l_{\text{OLS}}^n$, $\hat{\alpha}_{\text{OLS}}^n$, is not feasible.
		\end{quote}	 
		Thus,
		\begin{align} \label{eq:PULSEsolutionTIGHTconstraint}
			T_n(\hat{\alpha}) = n \frac{l_{\text{IV}}^n(\hat{\alpha})}{l_{\text{OLS}}^n(\hat{\alpha})} = Q \iff   l_{\text{IV}}^n(\hat{\alpha}) = \frac{Q}{n} l_{\text{OLS}}^n(\hat{\alpha}).
		\end{align}
		Furthermore, the assumption that $T_n(\hat{\alpha}_{\text{Pr}}^n(t_n^\star(p)))\leq Q$ means that the solution to the primal PULSE, $\hat{\alpha}_{\text{Pr}}^n(t_n^\star(p))$, is feasible in the PULSE problem. That is, $\hat{\alpha}_\text{Pr}^n(t_n^\star(p)) \in \cF$. As a consequence of this we have that
		\begin{align} \label{eq:TheoremPULSEequalpPULSE_pPULSEfeasibleInPULSE}
			l_{\text{OLS}}^n(\hat{\alpha}_\text{Pr}^n(t_n^\star(p)))  \geq \min_{\alpha\in \cF} l_{\text{OLS}}^n(\alpha) = l_{\text{OLS}}^n(\hat{\alpha}).
		\end{align}
		Now we show that the PULSE solution $\hat{\alpha}$ is feasible in the primal PULSE problem (Primal$.t_n^*(p).n$).   
		\begin{quote} \normalsize
			To see this, Note that the feasibility set of the PULSE problem can be shrunk in the following manner
			\begin{align*}
				\cF &= \lb \alpha \in \R^{d_1+q_1} :  l_{\text{IV}}^n(\alpha )\leq \frac{Q}{n}l_{\text{OLS}}^n(\alpha) \rb \\
				&=\lb \alpha \in \R^{d_1+q_1} :  l_{\text{IV}}^n(\alpha )\leq \frac{Q}{n}l_{\text{OLS}}^n(\alpha) , l_{\text{OLS}}^n(\alpha) \geq  l_{\text{OLS}}^n(\hat{\alpha})\rb \\
				&\supseteq \lb \alpha \in \R^{d_1+q_1} :  l_{\text{IV}}^n(\alpha )\leq \frac{Q}{n}l_{\text{OLS}}^n(\hat{\alpha}) , l_{\text{OLS}}^n(\alpha) \geq  l_{\text{OLS}}^n(\hat{\alpha})\rb \\ 
				&= \lb \alpha \in \R^{d_1+q_1} :  l_{\text{IV}}^n(\alpha )\leq  l_{\text{IV}}^n(\hat{\alpha} ) , l_{\text{OLS}}^n(\alpha) \geq  l_{\text{OLS}}^n(\hat{\alpha})\rb  \\
				&= \lb \alpha \in \R^{d_1+q_1} :  l_{\text{IV}}^n(\alpha )\leq  l_{\text{IV}}^n(\hat{\alpha} ) \rb  =: \hat{\cF}(\hat{\alpha}),
			\end{align*}
			where the third equality follows from \Cref{eq:PULSEsolutionTIGHTconstraint}. 	The only claim above that needs justification is that: 
			\begin{equation} \label{eq:toprove}
				l_{\text{IV}}^n(\alpha )\leq  l_{\text{IV}}^n(\hat{\alpha} ) \implies l_{\text{OLS}}^n(\alpha) \geq  l_{\text{OLS}}^n(\hat{\alpha}).
			\end{equation}
			For now we assume that this claim holds and provide a proof later. Thus, we have that $\hat{\cF}(\hat{\alpha}) \subseteq \cF$ and we note that $\hat{\alpha}\in \hat{\cF}(\hat{\alpha})$.
			An important consequence of this is that the PULSE solution $\hat{\alpha}$ is also the unique solution to the primal problem (Primal$.l_{\text{IV}}^{n}(\hat{\alpha}).n$). That is,
			\begin{align*}
				\hat{\alpha}  = \hat{\alpha}_{\text{Pr}}^n(l_{\text{IV}}^n(\hat{\alpha})) = \begin{array}{ll}
					\mathrm{argmin}_\alpha & l_{\mathrm{OLS}}^n(\alpha)  \\
					\mathrm{subject \, to} & l_{\text{IV}}^n(\alpha) \leq l_{\text{IV}}^n(\hat{\alpha}).
				\end{array}
			\end{align*}
			We will now prove that
			$
			l_\text{IV}^n(\hat{\alpha}) \in \cE:=
			\{ t\in [ \min_{\alpha}l_{\mathrm{IV}}^n(\alpha),l_{\text{IV}}^n(\hat{\alpha}_{\text{OLS}}^n) ] : T_n(\hat{\alpha}_{\text{Pr}}^n(t)) \leq  Q_{\chi^2_{q}}(1-p)\}.
			$
			This follows from the following two observations: 
			(1)
			$\min_{\alpha}l_{\mathrm{IV}}^n(\alpha) \leq l_{\text{IV}}^n(\hat{\alpha}) < 
			l_{\text{IV}}^n(\hat{\alpha}_{\text{OLS}}^n)$ 
			and (2) $T_n(\hat{\alpha}_{\text{Pr}}^n(l_\text{IV}^n(\hat{\alpha}))) \leq  Q_{\chi^2_{q}}(1-p)$.
			(1) follows 	
			because 
			$\hat{\alpha}_{\text{OLS}}^n\not \in \cF$, 
			which implies, by the above inclusion, that $\hat{\alpha}_{\text{OLS}}^n\not \in \hat{\cF}(\hat{\alpha})$. 
			(2)	follows because $\hat \alpha$ solves
			(Primal$.l_{\text{IV}}^{n}(\hat{\alpha}).n$) and thus
			$\hat{\alpha}_{\text{Pr}}^n(l_\text{IV}^n(\hat{\alpha})) = \hat{\alpha}$;
			$T_n(\hat{\alpha}) \leq Q_{\chi^2_q}(1-p)$
			holds because $\hat{\alpha}$ is feasible for the PULSE problem.

			Now, since $t_n^\star(p)=\sup(\cE\setminus \{\min_{\alpha}l_{\mathrm{IV}}^n(\alpha)\})\in \R$ implies $t_n^\star(p)= \sup(\cE)$, it follows that	 $l^n_{\text{IV}}(\hat{\alpha}) \leq t_n^\star(p)$. In other words, any solution $\hat{\alpha}$ to the PULSE problem is feasible in the primal PULSE problem (Primal$.t_n^*(p).n$). 
		\end{quote}		
		Hence,
		\begin{align} \label{eq:TheoremPULSEequalpPULSE_PULSEfeasibleInpPULSE}
			l_{\text{OLS}}^n (\hat{\alpha}) \geq l_{\text{OLS}}(\hat{\alpha}_{\text{Pr}}^n(t_n^\star(p))).
		\end{align}
		\Cref{eq:TheoremPULSEequalpPULSE_pPULSEfeasibleInPULSE} and \Cref{eq:TheoremPULSEequalpPULSE_PULSEfeasibleInpPULSE}  now yield that $l_{\text{OLS}}^n(\hat{\alpha}_{\text{Pr}}^n(t_n^\star(p))) =  l_{\text{OLS}}^n (\hat{\alpha})$ for any PULSE solution $\hat{\alpha}$. 
		Thus, any solution $\hat{\alpha}$ to the PULSE problem  is feasible in the primal PULSE problem (Primal$.t_n^*(p).n$) and it attains the optimal primal PULSE objective value. We conclude that $\hat{\alpha}$ solves the primal PULSE problem. Furthermore, it must hold that $
		\hat{\alpha} = \hat{\alpha}_{\text{Pr}}^n(t_n^\star(p))$, 
		by uniqueness of solutions to the primal PULSE problem(see \Cref{lm:PrimalUniqueSolAndSlatersConditions}). This implies two things: solutions to the PULSE problem are unique and the PULSE solution coincides with the primal PULSE solution.
		
		It only remains to prove the claim of \Cref{eq:toprove}, which ensures
		$\hat{\cF}(\hat{\alpha}) \subseteq \cF$. Assume for contradiction that there exists an
		$\alpha$ such that $l_{\text{IV}}^n(\alpha )\leq  l_{\text{IV}}^n(\hat{\alpha} )$ and $l_{\text{OLS}}^n(\alpha) <  l_{\text{OLS}}^n(\hat{\alpha})$, that is, we assume that
		\begin{align*}
			\cA := \underset{=:\cB}{\underbrace{\{\alpha \in \R^{d_1+q_1} : l_{\text{OLS}}^n(\alpha) < l_{\text{OLS}}^n(\hat{\alpha})\}}} \cap \underset{=:\cC}{\underbrace{\{\alpha \in \R^{d_1+q_1}: l_{\text{IV}}^n(\alpha) \leq l_{\text{IV}}^n(\hat{\alpha})\}}}   \not = \emptyset.
		\end{align*}
		Define $\cM_{\text{IV}} := \{\alpha : l_{\text{IV}}^n(\alpha) = \min_{\alpha'} l_{\text{IV}}^n(\alpha')\}$ as the solution space to the generalized method of moments formulation of the instrumental variable minimization problem.
		We now prove that  $
		\cM_{\text{IV}} \cap \cA = \emptyset$. 
		\begin{quote} \normalsize
			That is, we claim that
			in the just- and over-identified setup  $\hat{\alpha}_{\text{TSLS}}^n\not \in \cA$ and in the under-identified setup none of the infinitely many solutions in the solution space of the instrumental variable minimization problem  lies in $\cA$. These statements follow by first noting that $\cM_{\text{IV}} \subset \cF$ in any identification setting. In the under- and -just identified setup this is seen by noting that $l_{\text{IV}}^n(\alpha)=0$ for any $\alpha \in \cM_{\text{IV}}$, which implies $T_n(\alpha)=0\leq  Q$, hence $\cM_{\text{IV}}\subset \cF$. 
			In the over-identified setup, where $\cM_{\mathrm{IV}}=\{\hat{\alpha}_{\text{TSLS}}^n\}$, we will now  argue that $\cM_{\text{IV}} \subset \cF$
			follows from the assumption that $t_n^\star(p) <\infty$. 
			We first prove
			that $D_{\text{Pr}}\ni t \mapsto T_n(\hat{\alpha}^n_{\text{Pr}}(t))$ is weakly increasing. If $t_1<t_2$ are two constraint bounds for which the primal problem is solvable, then $l_{\text{OLS}}^n(\hat{\alpha}^n_{\text{Pr}}(t_1)) \geq l_{\text{OLS}}^n(\hat{\alpha}^n_{\text{Pr}}(t_2))$ as the feasibility set
			for $t_2$ is larger than the 
			one for $t_1$.
			Furthermore, the solution $\hat{\alpha}^n_{\text{Pr}}(t_2)$ either equals
			$\hat{\alpha}^n_{\text{Pr}}(t_1)$ or is contained in the set $\{\alpha\in \R^{d_1+q_1}: t_1< l_{\text{IV}}^n(\alpha) \leq t_2 \}$; in the latter case we have
			$l_{\text{IV}}^n (\hat{\alpha}^n_{\text
				{Pr}}(t_1)) \leq t_1 < l_{\text{IV}}^n (\hat{\alpha}^n_{\text
				{Pr}}(t_2))  \leq t_2$. Thus, we have in both cases that
			$l_{\text{IV}}^n (\hat{\alpha}^n_{\text
				{Pr}}(t_1)) \leq  l_{\text{IV}}^n (\hat{\alpha}^n_{\text
				{Pr}}(t_2))$. Combining 
			the
			two observations above we have that $$
			T_n(\hat{\alpha}^n_{\text{Pr}}(t_1)) = n \frac{l_{\text{IV}}^n(\hat{\alpha}^n_{\text{Pr}}(t_1))}{l_{\text{OLS}}^n(\hat{\alpha}^n_{\text{Pr}}(t_1))} \leq n \frac{l_{\text{IV}}^n(\hat{\alpha}^n_{\text{Pr}}(t_2))}{l_{\text{OLS}}^n(\hat{\alpha}^n_{\text{Pr}}(t_2))} = T_n(\hat{\alpha}^n_{\text{Pr}}(t_2)).$$ 
			Hence, as $-\infty<\min_{\alpha}l_{\text{IV}}^n(\alpha) = l_{\text{IV}}^n(\hat{\alpha}_{\text{TSLS}}^n) < t_n^\star(p) < \i$ are two points for which the primal problem is solvable we get that $$
			T_n(\hat{\alpha}_{\text{TSLS}}^n) = T_n(\hat{\alpha}^n_{\text{Pr}}(l_{\text{IV}}^n(\hat{\alpha}_{\text{TSLS}}^n))) \leq T_n(\hat{\alpha}^n_{\text{Pr}}(t_n^\star(p))) \leq Q.$$ 
			This proves that $\cM_{\text{IV}} \subset \cF$ in the over-identified setup.
			Now, if $\cM_{\text{IV}}\cap \cA \not = \emptyset$, there exists an $\alpha\in \cM_{\text{IV}}\cap \cA\subseteq \cF\cap \cA$ such that $\alpha$ is feasible in the PULSE problem $(\alpha\in \cF)$ and $\alpha$ is super-optimal compared to $\hat{\alpha}$, $l_{\text{OLS}}^n(\alpha) < l_{\text{IV}}^n(\hat{\alpha})$ ($\alpha\in \cA$), contradicting that $\hat{\alpha}$ is an solution to the PULSE problem. We can thus conclude that
			$\cM_{\mathrm{IV}} \cap \cA = \emptyset$. 
		\end{quote}
		This allows us to fix two distinct points $\bar{\alpha}\not = \alpha'$ such that  $\bar{\alpha} \in \cA $ and $\alpha'\in \cM_{\text{IV}}$. Consider the proper line segment function between $\bar{\alpha}$ and  $\alpha'$, $f(t):[0,1]\to \R^{d_1+q_1}$ given by
		$
		f(t) := t\alpha' + (1-t) \bar{\alpha}.
		$
		A multivariate convex function is convex in any direction from any given starting point in its domain, so both $l_{\text{IV}}^n\circ f:[0,1]\to \R_+$ and $l_{\text{OLS}}^n\circ f:[0,1]\to \R_+$ are convex. Since $\cM_{\mathrm{IV}} \cap \cA = \emptyset$ it is obvious that the function $f$ will for sufficiently large $t$ 'leave' the set $\cA$.
		We will now prove that $f$ actually leaves the superset $ \cB\supset \cA$.
		More precisely, we will prove that
		there exists a $t_1 \in (0,1]$ such that 
		for all $t'\in[0,t_1)$ it holds that $f(t') \in \cB$
		and
		for all $t'\in[t_1,1]$ it holds that 
		$f(t') \notin \cB$
		(which implies
		$f(t') \notin \cA$).
		\begin{quote} \normalsize
			Because $l_{\text{IV}}^n(\alpha')=\min_{\alpha}l_{\text{IV}}^n(\alpha)$ we have that $\alpha'\in \cC=\{\alpha : l_{\text{IV}}^n(\alpha) \leq l_{\text{IV}}^n(\hat{\alpha})\}$. 
			By convexity of $l_{\text{IV}}^n$ 	(Lemma~\ref{lm:PrimalUniqueSolAndSlatersConditions}) the sublevel set $\cC$ is convex and thus contains the entire line segment between $\bar{\alpha}$ and $\alpha'$. As a consequence $a'\not\in \cB$.
			It therefore suffices to
			construct a $t_1 \in (0,1]$ such that 
			for all $t'\in[0,t_1)$ it holds that $f(t') \in \cB$
			and
			for all $t'\in[t_1,1]$ it holds that 
			$f(t') \notin \cB$. 
			We now consider the set
			$
			\{t \in [0,1] \,:\,
			l_{\text{OLS}}^n(f(t)) < l_{\text{OLS}}^n(\hat{\alpha})
			\} = f^{-1}(\cB).
			$
			This set 
			contains $0$ 
			because $\bar{\alpha} \in \cA\subset \cB$;
			it does not contain $1$ because
			$\alpha'\not \in \cB$;
			it
			is convex, as it is a sublevel set of a convex function ($l_{\text{OLS}}^n \circ f$);
			it is relatively open in $[0,1]$ because 
			it is a pre-image of an open set under a continuous function ($l_{\text{OLS}}^n \circ f$).
			Thus, the set must be of the form
			$[0,t_1)$ for some $t_1 \in (0,1]$. This $t_1$ satisfies the desired criteria.
		\end{quote}
		We constructed $t_1$ above such that for all $t'\in[0,t_1)$ it holds that 
		$l_{\text{OLS}}^n(f(t'))< l_{\text{OLS}}^n(\hat{\alpha})$
		and for all $t'\in[t_1,1]$ it holds that 
		$l_{\text{OLS}}^n(f(t'))\geq l_{\text{OLS}}^n(\hat{\alpha})$. By continuity of  $l_{\text{OLS}}^n \circ f$ we must therefore have that $l_{\text{OLS}}^n(f(t_1))= l_{\text{OLS}}^n(\hat{\alpha})$. 
		Since $f(1)= \alpha'$ is a global minimum for $l_{\text{IV}}^n$, we have that 1 must also be a global minimum for $l_{\text{IV}}^n\circ f$, implying that the convex the function $l_{\text{IV}}^n\circ f:[0,1]\to \R_+$ is monotonically decreasing. 
		It must therefore hold that $$
		l_{\text{IV}}^n(f(t_1)) <  l_{\text{IV}}^n(f(0)) =  l_{\text{IV}}^n(\bar{a}) \leq l_{\text{IV}}^n(\hat{\alpha})
		.$$
		The first inequality is strict because if $l_{\text{IV}}^n(f(t_1)) = l_{\text{IV}}^n(f(0)) =l_{\text{IV}}^n(\bar{\alpha})$, then convexity of $l_{\text{IV}}^n$ implies that
		$$
		l_{\text{IV}}^n(f(t_1)) = l_{\text{IV}}^n(t_1 \alpha' + (1-t_1) \bar\alpha)  \leq t_1 l_{\text{IV}}^n(\alpha') + (1-t_1)l_{\text{IV}}^n(\bar{\alpha}),$$ which happens if and only if $
		l_{\text{IV}}^n(\bar{\alpha})\leq l_{\text{IV}}^n(\alpha')
		$
		contradicting the already established fact that $l_{\text{IV}}^n(\bar{\alpha})> l_{\text{IV}}^n(\alpha')$, which holds since $\alpha'\in \cM_{\text{IV}}$ but $\bar{\alpha} \not \in \cM_{\mathrm{IV}}$. We conclude that $l_{\text{IV}}^n(f(t_1))< l_{\text{IV}}^n(\hat{\alpha})$.

		Thus, we have argued that $\cM_{\mathrm{IV}} \cap \cA = \emptyset$ implies the existence of an $\tilde{\alpha} := f(t_1) = t_1 \alpha ' + (1-t_1) \bar{\alpha}$ such that $l_{\text{IV}}^n(\tilde{\alpha}) < l_{\text{IV}}^n(\hat{\alpha})$ and $l_{\text{OLS}}^n( \tilde{\alpha}) = l_{\text{OLS}}^n(\hat{\alpha})$. We have illustrated the above considerations  in \Cref{fig:pPULSEequalsPULSEProofInsersectionPlot}.
		\begin{figure}[htp] 
			\centering		\includegraphics[width=\textwidth-150pt]{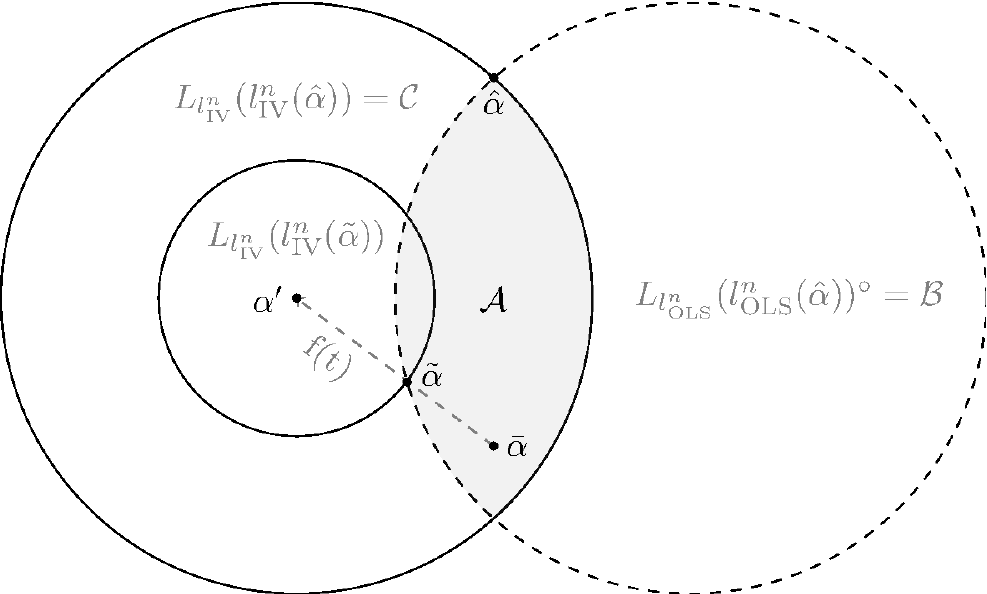}
			\caption{	
				Illustration of the described procedure in the
				just- or over-identified setup with $d_1+q_1=2$, where we show that  $\cA \not = \emptyset$ leads to a contradiction. Here, $L_g(c) := \{\alpha: g(\alpha) \leq c\}$ 
				is the $c$ sublevel set of the function $g$ and $A^\circ$ denotes the interior of a set $A$. 
				The illustration is simplified, e.g., because 
				the sublevel sets are convex but not necessarily Euclidean balls.		
				Note that the position of $\hat{\alpha}_{\text{OLS}}^n$ is not specified, as it can possibly be in either $\cA$ or $L_{l_{\text{OLS}}^n}(l_{\text{OLS}}^n(\hat{\alpha}))\setminus \cA$. In the under-indentified setup $\alpha'$ would lie in the $d-q_2=1$ dimensional subspace $\cM_{\text{IV}}$ and the level sets would be slabs around this line. } \label{fig:pPULSEequalsPULSEProofInsersectionPlot}
		\end{figure}
	
		It follows that $$
		T_n(\tilde{\alpha}) = n \frac{l_{\text{IV}}^n(\tilde{\alpha})}{l_{\text{OLS}}^n(\tilde{\alpha})} =  n \frac{l_{\text{IV}}^n(\tilde{\alpha})}{l_{\text{OLS}}^n(\hat{\alpha})} < n \frac{l_{\text{IV}}^n(\hat{\alpha})}{l_{\text{OLS}}^n(\hat{\alpha})}  = Q,$$ 
		implying that $\tilde{\alpha}$ is 	
		strictly feasible in the PULSE problem and, in fact, a solution  
		as the objective value is optimal.
		We argued earlier in \Cref{Eq:PULSEsolActiveInConstraint} that any solution to the PULSE problem must be tight in the inequality constraint, hence we have arrived at a contradiction.  We conclude that $\cA =\emptyset$, which implies 
		that \Cref{eq:toprove} must hold.
	\end{proofenv}

	\noindent
	\begin{proofenv}{\textbf{\Cref{lm:KclassNotEqualToTwoSLS}}}
		Assume that we are in the just- or over-identified setup and that \Cref{ass:ZtZfullrankandAtZfullrank} are satisfied. That is, $\fZ^\t\fZ$, $\fA^\t \fZ$ and $\fA^\t \fA$ are almost surely of full rank. In particular we have that 	
		$\fZ$, $\fA^\t \fZ$ and $P_\fA \fZ$ are almost surely of full column rank (injective linear maps).  Furthermore, let $\ep$ have density with respect to the Lebesgue measure and let $B$ be lower triangular.
		Fix $\lambda \geq 0$ and $\omega\in W_\lambda$, where 
		\begin{align*}
			W_\lambda:=& (\hat{\alpha}_{\mathrm{K}}^n(\lambda)= \hat{\alpha}_{\text{TSLS}}^n) \cap (\text{rank}(\fZ^\t \fZ)=d_1+q_1) \\&\quad \cap  (\text{rank}(\fA^\t \fZ)=d_1+q_1) \cap (\text{rank}(\fA^\t \fA)=q),
		\end{align*}
		satisfying
		$P(\hat{\alpha}_{\mathrm{K}}^n(\lambda)= \hat{\alpha}_{\text{TSLS}}^n) = P(W_\lambda)$. 	By \Cref{eq:KclassLossFunctionAsPenalizedOLS} we have that 
		\begin{align*}
			\hat{\alpha}_{\mathrm{K}}^n(\lambda) &= \argmin_{\alpha} \{l_{\mathrm{OLS}}^n(\alpha) + \lambda l_{\mathrm{IV}}^n(\alpha)\}  
			\\
			&=\argmin_{\alpha} \{(\fY-\fZ\alpha)^\t (\fY-\fZ\alpha) + \lambda (\fY-\fZ\alpha)^\t P_\fA (\fY-\fZ\alpha)\} \\
			&=\argmin_{\alpha}  (\fY-\fZ\alpha)^\t(\fI+  \lambda P_\fA) (\fY-\fZ\alpha) \\
			&= \argmin_{\alpha} \| (\fI + \lambda P_\fA)^{1/2} (\fY - \fZ \alpha ) \|_2^2 \\
			&= \argmin_{\alpha} \| \fY - \fZ \alpha \|_{(\fI+\lambda P_\fA)}^2 ,
		\end{align*}
		where $\| \cdot \|_{(\fI+\lambda P_\fA)}$ is the norm induced by the inner product $
		\la x , y \ra_{(\fI+\lambda P_\fA)} = x^\t (\fI+\lambda P_\fA) y$.
		The solution $\fZ \hat{\alpha}_{\mathrm{K}}^n(\lambda)$ is well-known to coincide with the orthogonal projection of $\fY$ onto $\cR(\fZ)$, the range of $\fZ$, with respect to the inner product $\la \cdot , \cdot \ra_{(\fI+\lambda P_\fA)}$. Hence, $\fZ \hat{\alpha}_{\mathrm{K}}^n(\lambda)$ is the unique element in this closed linear subspace such that for all $z\in \cR(\fZ)$ it holds that
		$$
		\la \fY - \fZ \hat{\alpha}_{\mathrm{K}}^n(\lambda),  z \ra_{(\fI+\lambda P_\fA)} = \la \fY - \fZ \hat{\alpha}_{\mathrm{K}}^n(\lambda), (\fI+\lambda P_\fA) z \ra = 0  ,$$ or equivalently,  
		\begin{align} \label{eq:LemmaKclassDifferentFromIVSolutionsKclassPredictionUniqueEquality}
			\la \fY - \fZ \hat{\alpha}_{\mathrm{K}}^n(\lambda),  z \ra = - \lambda \la \fY - \fZ \hat{\alpha}_{\mathrm{K}}^n(\lambda),  P_\fA z \ra, \quad \forall z\in \cR(\fZ).
		\end{align}
		We note that if $\lambda = 0$ then $\hat{\alpha}_{\mathrm{K}}^n(\lambda) = \hat{\alpha}_{\text{OSL}}^n$, seen either by directly inspecting the closed form solution of $\hat{\alpha}_{\mathrm{K}}^n(\lambda)$ or concluding the same from \Cref{eq:LemmaKclassDifferentFromIVSolutionsKclassPredictionUniqueEquality}. 
		
		Furthermore, when $\lambda > 0$ 
		we have that  $\hat{\alpha}_{\mathrm{K}}^n(\lambda)=\hat{\alpha}_{\text{TSLS}}^n$ implies that, again, 	$\hat{\alpha}_{\mathrm{K}}^n(\lambda) = \hat{\alpha}_{\text{OSL}}^n$. 
			To see this, we note that $$
			\hat{\alpha}_{\text{TSLS}}^n =\argmin_\alpha l_{\text{IV}}^n(\alpha) =  \argmin_\alpha \| P_\fA \fY - P_\fA \fZ \alpha \|_2^2,$$
			so $P_\fA \fZ\hat{\alpha}_{\text{TSLS}}^n$ is the orthogonal projection of $P_\fA \fY$ onto $\cR(P_\fA\fZ)$. That is, $P_\fA \fZ\hat{\alpha}_{\text{TSLS}}^n$ is the unique element in $\cR(P_\fA \fZ )$ such that $\la P_\fA \fY - P_\fA \fZ \hat{\alpha}_{\text{TSLS}}^n , s \ra = 0$ for all $s\in \cR(P_\fA \fZ )$, i.e.,  %
			$$\la P_\fA \fY - P_\fA \fZ \hat{\alpha}_{\text{TSLS}}^n, P_\fA z \ra  = 0,$$ for all  $z\in \cR(\fZ)$. Thus, if $\hat{\alpha}_{\mathrm{K}}^n(\lambda)= \hat{\alpha}_{\text{TSLS}}^n$ for some $\lambda >0$ we have that
			\begin{align*}
				0 &= 	\la P_\fA \fY - P_\fA \fZ \hat{\alpha}_{\text{TSLS}}^n , P_\fA z \ra  \\
				&= 	\la P_\fA \fY - P_\fA \fZ \hat{\alpha}_{\mathrm{K}}^n(\lambda) , P_\fA z \ra 
				\\
				&= \la \fY -  \fZ \hat{\alpha}_{\mathrm{K}}^n(\lambda), P_\fA z \ra \\
				&= -\lambda^{-1}\la \fY - \fZ \hat{\alpha}_{\mathrm{K}}^n(\lambda),  z \ra,  %
			\end{align*}
			hence $\la \fY - \fZ \hat{\alpha}_{\mathrm{K}}^n(\lambda),  z \ra = 0 $ for all $z\in \cR(\fZ)$, where we used \Cref{eq:LemmaKclassDifferentFromIVSolutionsKclassPredictionUniqueEquality} and in the third equality we used	
			that $P_\fA$ is idempotent and symmetric. This implies that $\hat{\alpha}_{\mathrm{K}}^n(\lambda) = \hat{\alpha}_{\text{OLS}}^n$, as it satisfies the uniquely determining condition for the ordinary least square estimator.

		Hence, for any $\lambda \geq 0$, whenever $\hat{\alpha}_{\mathrm{K}}^n(\lambda)= \hat{\alpha}_{\text{TSLS}}^n$ we know that $
		\hat{\alpha}_{\text{TSLS}}^n =\hat{\alpha}_{\mathrm{K}}^n(\lambda) = \hat{\alpha}_{\text{OLS}}^n$. 
		Thus, for any $\lambda \geq 0$ it holds that $$
		P(\hat{\alpha}_{\mathrm{K}}^n(\lambda)= \hat{\alpha}_{\text{TSLS}}^n) \leq P(\hat{\alpha}_{\text{TSLS}}^n = \hat{\alpha}_{\text{OLS}}^n).$$
		Recall that the reduced form equations of our system are given by
			$	[\fY  \, \, \fX \, \,  \fH ]
		= \fA \Pi + \bm{\ep}\Gamma^{-1}$ where $\Gamma := I-B$. When $B$ is lower triangular, so is $\Gamma$ and $\Gamma^{-1}$. %
		By selecting the relevant columns of $\Pi$ and $\Gamma^{-1}$ we may express the marginal reduced form structural equations of $\fS$  that consist of any collection of columns from $[
		\fY \, \,  \fX \, \, \fH
		]$ by $\fS = \fA \Pi_{S} + \bm{\ep} \Gamma^{-1}_S$ for conformable matrices $\Pi_{S}$ and $\Gamma^{-1}_S$. In particular, we have that the marginal reduced form structural equations for $\fY$ and $\fX_{*}$ are given by $
		\fY = \fA \Pi_Y + \bm{\ep}\Gamma^{-1}_{Y}$ and $
		\fX_{*} = \fA \Pi_{X_*} + \bm{\ep}\Gamma^{-1}_{X_*}$, 
		where $\Pi_Y, \Pi_{X_*},\Gamma^{-1}_{Y}$ and $\Gamma^{-1}_{X_{*}}$ are matrices conformable with the following block representation
		\begin{align*}
			\Pi &= [ \, 
			\underbrace{\Pi_{Y_{}}}_{q\times 1} \,\,  \underbrace{\Pi_{X_*}}_{q\times d_1} \, \, \underbrace{\Pi_{X_{-*}}}_{d\times q_2} \, \, \underbrace{\Pi_{H_{}}}_{q\times r} \,
			] \in \R^{q \times l}, 
			\quad \text{and} \quad	
			\Gamma^{-1} = 
			[ \,\underbrace{\Gamma^{-1}_{Y_{}}}_{l\times 1}  \,\,
			\underbrace{\Gamma^{-1}_{X_*}}_{l\times d_1} \,\,
			\underbrace{\Gamma^{-1}_{X_{-*}}}_{l \times d_2} \,\,
			\underbrace{\Gamma^{-1}_{H_{}}}_{l \times r}  \,]  \in \R^{l\times l},
		\end{align*}
		where $l:=1+d+r$. 	Note that by the lower triangular structure of $\Gamma^{-1}$ we have that the only matrix among $ \Gamma^{-1}_{Y}$, $\Gamma^{-1}_{X_*}$, $\Gamma^{-1}_{X_{-*}}$ and $\Gamma^{-1}_{H}$ that has a non-zero first row is $\Gamma^{-1}_Y$.
		
		Now assume without loss of generality that the first row of $\Gamma^{-1}$ is given by the first canonical Euclidean basis vector $(1,0,...,0)\in \R^{1\times l}$ such that we have the following partitionings
		\begin{align*}
			\bm{\ep} &= [ \,
			\underbrace{\bm{\ep}_Y}_{n\times 1}  \, \,  \underbrace{\bm{\ep}_{-Y}}_{n\times(d+r)}
			\, ]\in \R^{n\times l}, &
			\Gamma_{Y}^{-\t} &= [ \, 1  \, \, 
			\underbrace{\Gamma_{-Y,Y}^{-\t}}_{1\times (d+r)}
			\,] \in \R^{1 \times l},\\
			\Gamma_{X_{*}}^{-\t} &= [ \,
			\bm{0}_{d_1\times 1}  \, \, 
			\underbrace{\Gamma_{-Y,X_{*}}^{-\t}}_{d_1\times(d+r)}
			\, ] \in \R^{d_1 \times l}, &
			\Gamma_{X_{1}}^{-\t} &= [ \,
			0  \, \,
			\underbrace{\Gamma_{-Y,X_{1}}^{-\t}}_{1\times (d+r)}
			\,] \in \R^{1 \times l},
		\end{align*}
		where $\fX_1$ is the first column of $\fX$.
		Hence, we note that $\bm{\ep}\Gamma^{-1}_{Y} =\bm{\ep}_Y + \bm{\ep}_{-Y}\Gamma_{-Y,Y}^{-1}$, such that the marginal reduced form structural equation for $\fY$ has the following representation $$
		\fY = \fA \Pi_Y + \bm{\ep}\Gamma^{-1}_{Y} 
		= \fA \Pi_Y + \bm{\ep}_{-Y}\Gamma_{-Y,Y}^{-1}+ \bm{\ep}_Y  =: f_y(\fA,\bm{\ep}_{-Y}) + \bm{\ep}_{Y}.$$
		We can also represent $\fZ$ in terms of these structural coefficient block matrices by
		\begin{align*}
			\fZ &=  \begin{bmatrix}
				\fX_{*} & \fA_{*} 
			\end{bmatrix} =  \begin{bmatrix}
				\fA \Pi_{X_{*}} + \bm{\ep}\Gamma^{-1}_{X_{*}} & \fA_{*}
			\end{bmatrix}  = \begin{bmatrix}
				\fA \Pi_{X_{*}}  & \fA_{*}
			\end{bmatrix} + \bm{\ep} \begin{bmatrix}
				\Gamma^{-1}_{X_{*}} & \bm{0}_{l \times q_1}
			\end{bmatrix}  \\
			&= \fA \begin{bmatrix}
				\Pi_{X_{*}}  & \begin{bmatrix}
					\fI_{q_1\times q_1} \\ \bm{0}_{q_2\times q_1}
				\end{bmatrix}
			\end{bmatrix} +  \begin{bmatrix}
				\bm{\ep}_{-Y} \Gamma_{-Y,X_{*}}^{-1} & \bm{0}_{l \times q_1}
			\end{bmatrix} =:f_z(\fA,\bm{\ep}_{-Y}).
		\end{align*}
		Assumption \Cref{ass:AindepEp} and Assumption \Cref{ass:epIndependentMarginals} together  with the assumption that the data matrices consist of row-wise i.i.d.\ copies of the system variables, yield that $\fA \independent \bm{\ep}_{Y}$ and $\bm{\ep}_{-Y}\independent \bm{\ep}_{Y} $. This implies that the conditional distribution of $\bm{\ep}_{Y}$ given $\fA$ and $\bm{\ep}_{-Y}$ satisfies $P_{\bm{\ep}_{Y} |\fA = A,\bm{\ep}_{-Y}=e} = P_{\bm{\ep}_{Y}}$ for $P_{\fA,\bm{\ep}_{-Y}}$-almost all $(A,e)\in \R^{n\times q}\times \R^{n\times (d+r)}$. Hence, conditional on $\fA = A$ and $\bm{\ep}_{-Y} =e$ we have that
		$
		\fY|(\fA = A,\bm{\ep}_{-Y} =e) \stackrel{a.s.}{=} f_y(A,e) + \bm{\ep}_Y,$ 
		and 
		$ \fZ |(\fA = A,\bm{\ep}_{-Y} =e) \stackrel{a.s.}{=} f_z(A,e).$
		Now let $(P_\fA\fZ)^+ = (\fZ^\t P_\fA \fZ)^{-1}\fZ^\t P_\fA$ and $
		\fZ^+ =(\fZ^\t \fZ)^{-1}\fZ^\t$ denote the pseudo-inverse matrices of the almost surely full column rank matrices $P_\fA \fZ$ and $\fZ$. Furthermore, note that the pseudo-inverses are unique for all matrices, 	
		i.e., if $P_\fA \fZ \not = \fZ$, then $(P_\fA\fZ)^+ \not = \fZ^+$.   We realize that $
		\hat{\alpha}^n_{\text{TSLS}} =(\fZ^\t P_\fA \fZ)^{-1}\fZ^\t P_\fA \fY = (P_\fA \fZ)^+\fY $ and $ \hat{\alpha}^n_{\text{OLS}} = (\fZ^\t\fZ)^{-1} \fZ^\t \fY =\fZ^+\fY$. 
		Thus, with slight abuse of notation we let $Z:=f_z(A,e)$ for any $A,e$, and note that
		\begin{align} \notag
			&P(\hat{\alpha}^n_{\text{TSLS}}= \hat{\alpha}^n_{\text{OLS}})\\ \notag
			&= P((P_\fA\fZ)^+ \fY =\fZ^+\fY) \\\notag
			&= \int P\lp [(P_\fA \fZ)^+- \fZ^+] \fY =0 | \fA = A,\bm{\ep}_{-Y} =e \rp \, \mathrm{d} P_{\fA,\bm{\ep}_{-Y}} (A,e) \\\notag
			&= \int P\lp [(P_A Z)^+- Z^+](f_y(A,e) + \bm{\ep}_Y)  =0 \rp \, \mathrm{d} P_{\fA,\bm{\ep}_{-Y}} (A,e) \\ \label{Eq:KclassNotEqTSLSTempEq1}
			&= \int \mathbbm{1}_{(P_A Z \not = Z)}P\lp [(P_A Z)^+- Z^+](f_y(A,e) + \bm{\ep}_Y)  =0 \rp \, \mathrm{d} P_{\fA,\bm{\ep}_{-Y}} (A,e),
		\end{align}
		where $P_A = A(A^\t A)^{-1}A^\t\in \R^{n\times n}$. The last equality is due to the claim that $\mathbbm{1}_{(P_A Z \not = Z)}=1$ for $P_{\fA,\bm{\ep}_{-Y}}$ almost all $(A,e)$, or equivalently  $$
		\int \mathbbm{1}_{(P_A Z  = Z)} \, \mathrm{d} P_{\fA,\bm{\ep}_{-Y}} (A,e) 
		= \int \mathbbm{1}_{(P_\fA \fZ  = \fZ)} \, \mathrm{d} P 
		=  P(P_\fA \fZ =\fZ)=0.$$
		We prove this claim now.
		\begin{quote} \normalsize
			We now prove that $P(P_\fA \fZ =\fZ)=0$. First we note that $P_\fA \fZ =\fZ$ implies that $\cR(\fZ) \subseteq \cR(\fA)$.  
			Since $ \fZ = [ \fX_* \, \, \fA_* ]$ with $\fA = [\fA_*\,\,  \fA_{-*} ]$ it must hold that $\cR(\fX_*) \subseteq \cR(\fA)$. 
			Assume without loss of generality that $\fX_1$, the first column of $\fX$, is also a column of $\fX_*$. 
			Note that $\cR(\fX_*)\subseteq \cR(\fA)$ implies that $\fX_1$ can be written as a linear combination of the columns in $\fA$, i.e., there exists a $b=(b_1,...,b_q)\in \R^{q}$ such that $
			\fX_1 = b_1 \fA_1 + \cdots + b_{q} \fA_q = \fA b$, 
			namely $b=(\fA^\t \fA)^{-1}\fA \fX_1$. The marginal reduced form structural equation for $\fX_1$ is given by $		\fX_1 = \fA \Pi_{X_1} + \bm{\ep}\Gamma^{-1}_{X_1} = \fA \Pi_{X_1} + \bm{\ep}_{-Y} \Gamma_{-Y,X_{1}}^{-1}=\fA \Pi_{X_1} + \tilde{\bm{\ep}}$, 
			where $\tilde{\bm{\ep}}:= \bm{\ep}_{-Y}\Gamma_{-Y,X_{1}}^{-1}$. These two equalities are only possible if $\tilde{\bm{\ep}}\in \cR(\fA)$. 
			Note that $\tilde{\bm{\ep}}$ has jointly independent marginals (i.i.d.\ observations). Each coordinate is an independent copy of a linear combination of $1+d+r$ independent random variables $\ep_1,...,\ep_{1+d+r}$ all with density with respect to Lebesgue measure. 
			We conclude that $\tilde{\bm{\ep}}$ has density with respect to the $n$-dimensional Lebesgue measure as the linear 
			combination is non-vanishing. This holds because
			$\Gamma_{-Y,X_{1}}^{-1}\not=0$ by virtue of being a column of the invertible matrix $
			\Gamma^{-1}$, where we have removed the first entry (which was 
			a zero element). Furthermore, since $\fA \independent \bm{\ep}$, we also have that $\fA 
			\independent \tilde{\bm{\ep}}$. Hence, the conditional distribution of $\tilde{\bm{\ep}}$ given $\fA$ satisfies 
			$P_{\tilde{\bm{\ep}} |\fA = A} = P_{\tilde{\bm{\ep}}}$ for $P_{\fA}$-almost all $A\in \R^{n\times q}$. We conclude that 
			\begin{align*}
				P(P_\fA \fZ =\fZ)  &\leq P(\tilde{\bm{\ep}} \in \cR(\fA)) \\
				&= \int P(\tilde{\bm{\ep}} \in \cR(\fA)|\fA=A) \, \mathrm{d} P_{\fA}(A) 
				\\
				&= \int P(\tilde{\bm{\ep}} \in \cR(A)) \, \mathrm{d} P_{\fA}(A) \\
				&=0.
			\end{align*}
			The last equality follows from the fact that $q= \text{rank}(\fA^\t\fA) = \text{rank}(\fA)< n$ implies that $\cR(\fA)$ is a $q$-dimensional subspace of $\R^n$. Hence, for $P_\fA$-almost all $A\in \R^{n\times q}$ it
			holds that $\cR(A)$ is a $q$-dimensional subspace of $\R^n$. The probability that $\tilde{\bm{\ep}}$ lies in a $q$-dimensional subspace of $\R^n$ is zero, since it has density with respect to the $n$-dimensional Lebesgue measure.
		\end{quote}		 
		Thus, it suffices to show that $$ \label{Eq:EpYInaffineSubspaceWithProbZero}
		P\lp [(P_A Z)^+- Z^+](f_y(A,e) + \bm{\ep}_Y)  =0 \rp =0,$$
		for any $A\in \R^{n\times q}$
		and $Z=f_z(A,e)\in \R^{n\times (d_1+q_1)}$ with $P_AZ \not = Z$. 
			Therefore, let $A\in \R^{n\times q}$ 
			and $Z=f_z(A,e)\in \R^{n\times (d_1+q_1)}$ with $P_AZ \not = Z$. It holds that $(P_AZ)^+\not = Z^+$, which implies that $(P_AZ)^+-Z^+\not = 0$. Furthermore, we have that $$
			[(P_A Z)^+- Z^+](f_y(A,e) + \bm{\ep}_Y)  =0 ,$$ if and only if $$\bm{\ep}_\fY \in \text{ker}((P_A Z)^+- Z^+)-[(P_A Z)^+- Z^+]f_y(A,e),$$
			so it suffices to show that $\bm{\ep}_Y$ has zero probability to be in the affine (translated) subspace $$\text{ker}((P_A Z)^+- Z^+)-[(P_A Z)^+- Z^+]f_y(A,e)\subset \R^n.$$ This affine subspace has dimension $n$ if and only if $(P_AZ)^+-Z^+ = 0$, which we know is false. Hence, the dimension of the affine subspace is strictly less than $n$. As $\bm{\ep}_Y$ has density with respect to the $n$-dimensional Lebesgue measure, we know that the probability of being in a $N<n$ dimensional affine subspace is zero.

		Hence, we have shown that $
		P(\hat{\alpha}^n_{\text{TSLS}}= \hat{\alpha}^n_{\text{OLS}}) = 0$. 
		Combining all of our observations we get that $
		P(\hat{\alpha}_{\mathrm{K}}^n(\lambda)= \hat{\alpha}_{\text{TSLS}}^n) \leq P(\hat{\alpha}_{\text{TSLS}}^n = \hat{\alpha}_{\text{OLS}}^n) = 0.$
		We conclude that $$
		P \lp   \hat{\alpha}_{\text{TSLS}}^n \not = \hat{\alpha}_{\mathrm{K}}^n(\lambda) \rp = 1,\quad \text{for all }\lambda \geq 0.$$ However, we can easily strengthen this to  
		$P ( \cap_{\lambda \geq 0}   (\hat{\alpha}_{\text{TSLS}}^n \not = \hat{\alpha}_{\mathrm{K}}^n(\lambda) ) ) = 1.
		$
		To this end, let $\omega$ be a realization in the almost sure set $\cap_{\lambda \in \bQ_+}	W_\lambda.
		$
		Then, 
		$\omega \in \bigcap_{\lambda \geq 0 } \lp \hat{\alpha}_{\text{TSLS}}^n \not = \hat{\alpha}_{\mathrm{K}}^n(\lambda) \rp$.
		Otherwise,
		there exists an $\tilde{\lambda} \in \R_+ \setminus \bQ_+$ such that $\hat{\alpha}_{\text{TSLS}}^n = \hat{\alpha}_{\mathrm{K}}^n(\tilde{\lambda})$. By \Cref{lm:OLSandIV_Monotonicity_FnctOfPenaltyParameterLambda} we have that $\lambda \mapsto l_{\text{IV}}^n( \hat{\alpha}_{\text{IV}}^n (\lambda))$ is monotonically decreasing, but since  $\hat{\alpha}_{\mathrm{K}}^n(\tilde{\lambda})$ already minimizes the $l_{\text{IV}}^n$ function, so will all $\hat{\alpha}_{\mathrm{K}}^n(\lambda )$ for all $\lambda \geq \tilde{\lambda}$. As $\hat{\alpha}_{\text{TSLS}}^n$ is the unique point that minimizes $l_{\text{IV}}^n$ we conclude that $\hat{\alpha}_{\text{TSLS}}^n = \hat{\alpha}_{\mathrm{K}}^n(\lambda )$ for all $\lambda \geq \tilde{\lambda}$, which yields a contradiction.
		We conclude that $
		P \lp \cap_{\lambda \geq 0 } \lp \hat{\alpha}_{\text{TSLS}}^n \not = \hat{\alpha}_{\mathrm{K}}^n(\lambda) \rp \rp =1$.

	\end{proofenv}

	\noindent
	\begin{proofenv}{\textbf{\Cref{lm:EquivalenceBetweenKlikeAndPrimal}}}
		Let \Cref{ass:ZtZfullrankandAtZfullrank} and \Cref{ass:ZYfullcolrank} hold, i.e., that $\fZ^\t \fZ$ and $\fA^\t \fZ$ are  of full rank and  $[ \fZ \,\, \fY]$ is of full column rank. Furthermore, let \Cref{ass:KclassNotInIV} hold, i.e., that $\hat{\alpha}_{\mathrm{K}}^n(\lambda)\not \in \cM_{\mathrm{IV}}$ for all $\lambda\geq 0$. It holds that (Primal$.t.n$) has a unique solution and satisfies Slater's condition for all  $t>  \min_{\alpha}l_{\mathrm{IV}}^n(\alpha)$ (\Cref{lm:PrimalUniqueSolAndSlatersConditions}). Furthermore,  (Dual$.\lambda.n$) has a unique solution for all $\lambda \geq 0$ (\Cref{lm:PenalizedKClassSolutionUniqueAndExists}).
		
		First consider an arbitrary $t\in D_{\text{Pr}}$ and note that the dual problem of (Primal$.t.n$), 	not to be confused with the problem (Dual$.\lambda.n$), is given by
		\begin{align} \label{Eq:DualOfPrimal}
			\begin{array}{ll}
				\text{maximize}_\lambda & g_t(\lambda)  \\
				\text{subject to} & \lambda \geq 0.
			\end{array}
		\end{align}
		However, (Dual$.\lambda.n$) is equivalent with the infimum problem in the definition of  $g_t:\R_+ \to \R$ given by $$
		g_t(\lambda) := \inf_\alpha \{l_{\text{OLS}}^n(\alpha)+ \lambda (l_{\text{IV}}^n(\alpha) - t)\}.$$ 
		Now consider $\hat{\alpha}_{\text{Pr}}^n(t)$ solving the primal (Primal$.t.n$). 
		Slater's condition is satisfied, so there exists a $\lambda(t)\geq 0$ solving the dual problem and strong duality holds, $l_{\text{OLS}}^n(\hat{\alpha}_{\text{Pr}}^n(t)) = g_t(\lambda(t))$. 
		We will now show that $\hat{\alpha}_{\text{Pr}}^n(t)$ also solves to the K-class penalized regression problem (Dual.$\lambda(t).n$). That is, we will show that $
		\hat{\alpha}_{\text{Pr}}^n(t) = \begin{array}{ll}
			\underset{\alpha}{\text{argmin}} & l_{\text{OLS}}^n(\alpha) + \lambda(t) l_{\text{IV}}^n(\alpha).
		\end{array}
		$
		To that end, note that
		\begin{align*}
			g_t(\lambda(t)) &=  \inf_\alpha \{ {l_{\text{OLS}}^n(\alpha)+ \lambda(t) (l_{\text{IV}}^n(\alpha) - t)} \} = \inf_\alpha \{ {l_{\text{OLS}}^n(\alpha)+ \lambda(t)l_{\text{IV}}^n(\alpha) } \}- \lambda(t) t\\
			&\leq l_{\text{OLS}}^n(\hat{\alpha}_{\text{Pr}}^n(t))+ \lambda(t) (l_{\text{IV}}^n(\hat{\alpha}_{\text{Pr}}^n(t)) - t) 
			=  l_{\text{OLS}}^n(\hat{\alpha}_{\text{Pr}}^n(t))  =g_t(\lambda(t)),
		\end{align*}
		where in the last equality
		we used strong duality 
		and  the second last equality we used that for any constraint bound $t\in D_{\text{Pr}}$ the inequality constraint of (Primal$.t.n$) is active in the solution $\hat{\alpha}_{\text{Pr}}^n(t)$, i.e., $l_{\text{IV}}^n(\hat{\alpha}_{\text{Pr}}^n(t))=t$; see \Cref{lm:SolutionIsTightIfNotStationary} of \Cref{sec:AuxLemmas}.
		Thus, it holds that 
		\begin{align*}
			& \inf_\alpha \{ {l_{\text{OLS}}^n(\alpha)+ \lambda(t) (l_{\text{IV}}^n(\alpha) - t)} \} = l_{\text{OLS}}^n(\hat{\alpha}_{\text{Pr}}^n(t))+ \lambda(t) (l_{\text{IV}}^n(\hat{\alpha}_{\text{Pr}}^n(t)) - t) \\
			\iff & 
			\inf_\alpha \{ l_{\text{OLS}}^n(\alpha)+ \lambda(t) l_{\text{IV}}^n(\alpha) \} = l_{\text{OLS}}^n(\hat{\alpha}_{\text{Pr}}^n(t))+ \lambda(t) l_{\text{IV}}^n(\hat{\alpha}_{\text{Pr}}^n(t))
			,\end{align*}
		proving that $\hat{\alpha}_{\text{Pr}}^n(t)$ coincides with the unique solution $\hat{\alpha}_{\text{K}}^n(\lambda(t))$ to the K-class problem (Dual.$\lambda(t).n$) as it attains the same objective. Furthermore, there can only be one $\lambda(t)$ solving the dual problem in \Cref{Eq:DualOfPrimal}.
		If there are two distinct solutions $\lambda',\lambda''\geq 0$ with $\lambda'\not = \lambda''$, then by the above observations we get that $
		\hat{\alpha}_{\text{Pr}}^n(t) = \hat{\alpha}_{\text{K}}^n(\lambda') = \hat{\alpha}_{\text{K}}^n(\lambda''),$
		in contradiction to \Cref{cor:KclassSolutionsDistinct}.

		Conversely, fix $\lambda \geq 0$ and recall that $\hat{\alpha}_{\mathrm{K}}^n(\lambda)$ solves the
		penalized K-class regression problem (Dual.$\lambda.n$), that is, $
		\hat{\alpha}_{\mathrm{K}}^n(\lambda) = \argmin_{\alpha} \{ l_{\text{OLS}}^n(\alpha) + \lambda l_{\text{IV}}^n(\alpha) \}$.
		Now consider a primal constraint bound $t(\lambda) :=  l_{\text{IV}}^n(\hat{\alpha}_{\text{K}}^n(\lambda)) $ and consider the corresponding primal optimization problem (Primal.$t(\lambda).n$) and its dual form given by
		\begin{align} \label{Eq:PrimalKclassEquivalence_PrimalDualSpecificToTheorem}
			\textit{Primal}:\begin{array}{lr}
				\begin{array}{ll}
					\text{minimize} & l_{\text{OLS}}^n(\alpha)  \\
					\text{subject to} &  l_{\text{IV}}^n(\alpha) \leq  t(\lambda)
				\end{array}&  \quad \quad \textit{Dual}:	\begin{array}{ll}
					\text{maximize} & g_{t(\lambda)}(\gamma)  \\
					\text{subject to} & \gamma \geq 0,
				\end{array}
			\end{array}
		\end{align}
		where  $g_{t(\lambda)}:[0,\i) \to \R$ is given by
		$
		g_{t(\lambda)}(\gamma) = \inf_\alpha \{ l_{\text{OLS}}^n(\alpha) + \gamma [l_{\text{IV}}^n(\alpha)-t(\lambda)] \} .
		$
		Here we note that the proposed primal problem satisfies Slater's condition. To see this note that $\hat{\alpha}_{\text{K}}^n(\lambda)\not \in \cM_{\text{IV}}$, by \Cref{ass:KclassNotInIV}, 
		hence $\inf_{\alpha}l_{\text{IV}}^n(\alpha)=\min_{\alpha}l_{\text{IV}}^n(\alpha) < t(\lambda)= l_{\text{IV}}^n( \hat{\alpha}_{\text{K}}^n(\lambda))$. 
		Furthermore, we conclude that $t(\lambda) \in (\min_{\alpha}l_{\text{IV}}^n(\alpha),l_{\text{IV}}(\hat{\alpha}_{\text{OLS}}^n)]=D_{\text{Pr}}$ as $\lambda \mapsto l_{\text{IV}}(\hat{\alpha}_{\text{K}}^n(\lambda))$ 
		is monotonically decreasing and $\hat{\alpha}_{\text{K}}^n(0) = \hat{\alpha}_{\text{OLS}}^n$; see \Cref{lm:OLSandIV_Monotonicity_FnctOfPenaltyParameterLambda}.

		Let $p^\star$ and $d^\star$ denote the optimal objective values for the above primal and dual problem in \Cref{Eq:PrimalKclassEquivalence_PrimalDualSpecificToTheorem}, respectively. It 
		holds that $\hat{\alpha}_{\text{K}}^n(\lambda)$  is primal feasible since it satisfies the inequality constraint of the primal problem in	 \Cref{Eq:PrimalKclassEquivalence_PrimalDualSpecificToTheorem}. This implies that $p^\star \leq l_{\text{OLS}}^n(\hat{\alpha}_{\text{K}}^n(\lambda))$ 
		since $p^\star$ is the infimum of all attainable objective values. By the non-negative duality gap we also have that
		\begin{align*}
			p^\star & \geq d^\star = \sup_{\gamma\geq 0} g_{t(\lambda)}(\gamma)  \geq  g_{t(\lambda)}(\lambda) =\inf_\alpha \{ l_{\text{OLS}}^n(\alpha) + \lambda [l_{\text{IV}}^n(\alpha)-t(\lambda)] \} \\
			&=\inf_\alpha \{ l_{\text{OLS}}^n(\alpha) + \lambda l_{\text{IV}}^n(\alpha) \} - \lambda t(\lambda)=   l_{\text{OLS}}^n(\hat{\alpha}_{\text{K}}^n(\lambda)) + \lambda [l_{\text{IV}}^n(\hat{\alpha}_{\text{K}}^n(\lambda))-l_{\text{IV}}^n(\hat{\alpha}_{\text{K}}^n(\lambda))] \\
			&= l_{\text{OLS}}^n(\hat{\alpha}_{\text{K}}^n(\lambda)),
		\end{align*}
		implying that $l_{\text{OLS}}^n(\hat{\alpha}_{\text{K}}^n(\lambda))= p^\star$. This proves that strong duality holds and that the solution 
		to the K-class regression problem $\alpha_{\text{K}}^n(\lambda)$ solves the primal optimization problem (Primal.$t(\lambda).n$), since it attains the unique optimal objective value while also satisfying the inequality constraint. 
	\end{proofenv}

	\noindent
	\begin{proofenv}{\textbf{\Cref{thm:PULSEpPULSEdPULSEEequivalent}}}
		Fix any $p_{\min}\in(0,1)$ and let \Cref{ass:ZtZfullrankandAtZfullrank,ass:ZYfullcolrank} hold, 
		i.e., that $\fZ^\t \fZ$ and $\fA^\t \fZ$ are  of full rank and  $[ \fZ \,\, \fY]$ is of  full column rank. Furthermore, let \Cref{ass:KclassNotInIV} hold, i.e., that $\hat{\alpha}_{\mathrm{K}}^n(\lambda)\not \in \cM_{\mathrm{IV}}$ for all $\lambda \geq 0$. It holds that (Primal$.t.n$) has a unique solution and satisfies Slater's condition for all  $t>  \min_{\alpha}l_{\mathrm{IV}}^n(\alpha)$ (\Cref{lm:PrimalUniqueSolAndSlatersConditions}), that (Dual$.\lambda.n$) has a unique solution for all $\lambda \geq 0$ (\Cref{lm:PenalizedKClassSolutionUniqueAndExists}) and that $\{\hat{\alpha}_{\text{Pr}}^n(t):t\in D_{\text{Pr}}\}=\{\hat{\alpha}_{\text{K}}^n(\lambda):\lambda \geq 0\}$ (\Cref{lm:EquivalenceBetweenKlikeAndPrimal}). 
		Finally, we assume that $\lambda_n^\star(p_{\min})<\i$. 
		To simplify notation, we write $Q=Q_{\chi^2_{q}}(1-p_{\min})$.	
		
		We claim that the PULSE estimator can be represented in the dual form of the primal PULSE problem. That is, as a K-class estimator $\hat{\alpha}_{\text{PULSE}}^n(p_{\min}) = \hat{\alpha}^n_{\text{K}}(\lambda_n^\star(p_{\min}))$ with stochastic penalty parameter given by $
		\lambda_n^\star(p_{\min}) := \inf\{\lambda \geq 0 : T_n(\hat{\alpha}_{\text{K}}^n (\lambda))\leq Q_{\chi^2_{q}}(1-p_{\min}) \}.
		$
		We show this by proving that $\hat{\alpha}_{\text{K}}^n(\lambda_n^\star(p_{\min}))= \hat{\alpha}_{\text{Pr}}^n(t_n^\star(p_{\min})))$, which by \Cref{thm:pPULSESolvesPULSE} implies that the claim is true, if the conditions $t_n^\star(p_{\min})>-\i$ and 
		$T_n(\hat{\alpha}_{\text{Pr}}^n(t_n^\star(p_{\min})))\leq Q$ can be verified from the assumption that $\lambda_n^\star(p_{\min})<\i$. First, we note that if $\lambda_n^\star(p_{\min})<\i$, then $t_n^\star(p_{\min})>-\i$. 
		\begin{quote} \normalsize
			This follows by noting that, with $t(\lambda) := l_{\text{IV}}^n(\hat{\alpha}_{\text{K}}^n(\lambda))$,  proof of \Cref{lm:EquivalenceBetweenKlikeAndPrimal} \textit{ii)} yields that $\hat{\alpha}_{\text{K}}^n(\lambda) = \hat{\alpha}_{\text{Pr}}^n(t(\lambda))$ for any $\lambda \geq 0$ 
			which yields $
			\lambda_n^\star(p_{\min}) = \inf \left\{  \lambda  \geq 0:   T_n ( \hat{\alpha}_{\text{Pr}}^n \circ t(\lambda) )\leq Q \right\}.
			$
			Hence, if $\lambda_n^\star(p_{\min}) <\i$ we know there exists a $\lambda' \geq 0$ such that $T_n ( \hat{\alpha}_{\text{Pr}}^n \circ t(\lambda') )\leq Q$, i.e., there exists a $t'=t(\lambda')\in(\min_{\alpha'} l_{\text{IV}}^n(\alpha'),\i)$ such that $T_n ( \hat{\alpha}_{\text{Pr}}^n (t') )\leq Q$. 
			We have excluded that $t'=\min_{\alpha'} l_{\text{IV}}^n(\alpha')$ as $t'=l_{\text{IV}}^n(\hat{\alpha}_{\text{K}}^n(\lambda')) > \min_{\alpha'} l_{\text{IV}}^n(\alpha')$ since $\hat{\alpha}_{\text{K}}(\lambda') \not \in \cM_{\text{IV}}$. 
			Furthermore, we can without loss of generality assume that $t' \in (\min_{\alpha'} l_{\text{IV}}^n(\alpha'),l_{\text{IV}}^n(\hat{\alpha}^n_{\text{OLS}})]$ because if $t' > l_{\text{IV}}^n(\hat{\alpha}^n_{\text{OLS}})$, then it holds that $T_n ( \hat{\alpha}_{\text{Pr}}^n (l_{\text{IV}}^n(\hat{\alpha}^n_{\text{OLS}})) )\leq Q$ as $\hat{\alpha}_{\text{Pr}}^n (l_{\text{IV}}^n(\hat{\alpha}^n_{\text{OLS}})) = \hat{\alpha}_{\text{Pr}}^n (t')$ since the ordinary least square solution solves 
			all (Primal$.t.n$) with constraints bounds larger than
			$l_{\text{IV}}^n(\hat{\alpha}^n_{\text{OLS}})$. As a consequence, the set for which we take the supremum over in the definition of $t_n^\star(p_{\min})$ is non-empty, such that $t_n^\star(p_{\min}) >-\i$.
		\end{quote}
		Next we show that $\hat{\alpha}_{\text{K}}^n(\lambda_n^\star(p_{\min}))= \hat{\alpha}_{\text{Pr}}^n(t_n^\star(p_{\min})))$. 
		When this equality is shown, then the remaining condition that 
		$T_n(\hat{\alpha}_{\text{Pr}}^n(t_n^\star(p_{\min})))\leq Q$ 
		follows by \Cref{lm:TestInAlphaLambdaStarEqualsQuantileApp} 
		and we are done. For any constraint bound $t \in D_{\text{Pr}}= ( \min_{\alpha}l_{\mathrm{IV}}^n(\alpha),l_{\text{IV}}^n(\hat{\alpha}_{\text{OLS}}^n) ]$, consider the primal and corresponding dual optimization problems
		\begin{align} \label{eq:dPULSEequalpPULSEPrimalDualProblems}
			\textit{Primal}:\begin{array}{lr}
				\begin{array}{ll}
					\text{minimize} & l_{\mathrm{OLS}}^n(\alpha)  \\
					\text{subject to} & l_{\mathrm{IV}}^n(\alpha)\leq t
				\end{array} &  \quad \quad \textit{Dual}:	\begin{array}{ll}
					\text{maximize} & g_t(\lambda)  \\
					\text{subject to} & \lambda \geq 0,
				\end{array}
			\end{array}
		\end{align}
		with dual function $g_t:\R_{\geq 0} \to \R$ given by $
		g_t(\lambda) := \inf_\alpha\{l_{\mathrm{OLS}}^n(\alpha)+ \lambda (l_{\mathrm{IV}}^n(\alpha) - t)\}.$
		The proof of \Cref{lm:EquivalenceBetweenKlikeAndPrimal} yields that there exists a unique $\tilde{\lambda}(t)\geq 0$ solving the dual problem of \Cref{eq:dPULSEequalpPULSEPrimalDualProblems} such that   $ \label{eq:TheoremPrimalDualConnectionPvalConstraintEqThatPrimalEqualKclass}
		\hat{\alpha}_{\text{Pr}}^n(t) = \hat{\alpha}_{\text{K}}^n(\tilde{\lambda}(t)).
		$
		We now prove that $D_{\text{Pr}}\ni t \mapsto \tilde{\lambda}(t)$ is strictly decreasing.
		\begin{quote} \normalsize
			Note that by the definition of $g_t$ and \Cref{lm:PenalizedKClassSolutionUniqueAndExists} (or  equivalently the discussion in the beginning of \Cref{sec:DualPULSE}) we have that
			\begin{align} 
				g_t(\lambda) &= \inf_\alpha\{l_{\mathrm{OLS}}^n(\alpha)+ \lambda l_{\mathrm{IV}}^n(\alpha) \} -\lambda t  \label{Eq:RepresentationOfGtInTermsOfKclass}
				= l_{\mathrm{OLS}}^n(\hat{\alpha}_{\text{K}}^n (\lambda))+ \lambda l_{\mathrm{IV}}^n(\hat{\alpha}_{\text{K}}^n (\lambda)) -\lambda t.
			\end{align}
			For any $t_1$, $t_2$ with
			$0\leq  \min_{\alpha}l_{\mathrm{IV}}^n(\alpha)<t_1<t_2 \leq l_{\text{IV}}^n(\hat{\alpha}_{\text{OLS}}^n) $ we have that $g_{t_1}(\tilde{\lambda}(t_1)) \geq g_{t_1}(\tilde{\lambda}(t_2))$ and $g_{t_2}(\tilde{\lambda}(t_2)) \geq g_{t_2}(\tilde{\lambda}(t_1))$ as $\tilde{\lambda}(t)$ maximizes $g_t$.
			Hence, by bounding the first term we get that
			\begin{align} \label{eq:TheoremLambdaStarFiniteTStarFinite_Ineq1}
				g_{t_1}(\tilde{\lambda}(t_1)) - g_{t_2}(\tilde{\lambda}(t_2)) & \geq g_{t_1}(\tilde{\lambda}(t_2)) - g_{t_2}(\tilde{\lambda}(t_2)) 
				= \tilde{\lambda}(t_2)(t_2-t_1), 
			\end{align}
			where the last equality follows from the representation in \Cref{Eq:RepresentationOfGtInTermsOfKclass}.
			Similarly, by bounding the other term we get that
			\begin{align} \label{eq:TheoremLambdaStarFiniteTStarFinite_Ineq2}
				g_{t_1}(\tilde{\lambda}(t_1)) - g_{t_2}(\tilde{\lambda}(t_2)) & \leq  g_{t_1}(\tilde{\lambda}(t_1)) - g_{t_2}(\tilde{\lambda}(t_1)) 
				= \tilde{\lambda}(t_1)(t_2-t_1). 
			\end{align}
			Combining the inequalities from \Cref{eq:TheoremLambdaStarFiniteTStarFinite_Ineq1,eq:TheoremLambdaStarFiniteTStarFinite_Ineq2} we conclude that $
			\tilde{\lambda}(t_2) (t_2-t_1) \leq \tilde{\lambda}(t_1) (t_2-t_1)$ which implies $\tilde{\lambda}(t_2) \leq \tilde{\lambda}(t_1),
			$
			proving that $D_{\text{Pr}} \ni t \mapsto \tilde{\lambda}(t)$, the dual solution as a function of the primal problem constraint bound, is weakly decreasing. We now  strengthen this statement to strictly decreasing. 
			For any constraint bound $t\in D_{\text{Pr}}= ( \min_{\alpha}l_{\mathrm{IV}}^n(\alpha),l_{\text{IV}}^n(\hat{\alpha}_{\text{OLS}}^n)]$ we have that the solution $\hat{\alpha}^n_{\text{Pr}}(t)$ yields an active inequality constraint in the (Primal.$t.n$) problem, i.e., $l_{\text{IV}}^n(\hat{\alpha}_{\text{Pr}}(t))=t$; see \Cref{lm:SolutionIsTightIfNotStationary} of \Cref{sec:AuxLemmas}.
			Therefore, for any $\min_{\alpha}l_{\mathrm{IV}}^n(\alpha) < t_1 < t_2 \leq  l_{\text{IV}}^n(\hat{\alpha}_{\text{OLS}}^n)$ we get that
			$
			l_{\text{IV}}^n(\hat{\alpha}_{\text{K}}^n(\tilde{\lambda}(t_1))) = l_{\text{IV}}^n(\hat{\alpha}_{\text{Pr}}^n(t_1)) = t_1 < t_2 = l_{\text{IV}}^n(\hat{\alpha}_{\text{Pr}}^n(t_2)) = l_{\text{IV}}^n(\hat{\alpha}_{\text{K}}^n(\tilde{\lambda}(t_2))),
			$
			proving that $\tilde{\lambda}(t_1) \not = \tilde{\lambda}(t_2) $, which implies that $D_{\text{Pr}} \ni t\mapsto \tilde{\lambda}(t)$ is strictly increasing.
		\end{quote}
		Recall, by \Cref{lm:EquivalenceBetweenKlikeAndPrimal} that the K-class estimators for $\kappa\in[0,1)$ coincides with the collection of solutions to every primal problem satisfying Slater's condition. 
		That is,
		\begin{align} \label{Eq:EqualityOfPrimalSoltuionsAndKclassSolutions}
			\{\hat{\alpha}_{\text{K}}^n(\lambda):\lambda \geq 0\} 	
			= \{\hat{\alpha}_{\text{Pr}}^n(t): t \in D_{\text{Pr}} \}	
			= \{\hat{\alpha}_{\text{K}}^n(\tilde{\lambda}(t)):t \in D_{\text{Pr}}\},
		\end{align}
		where $\tilde{\lambda}$ is as introduced above.

		It now only remains to show	
		that $\tilde{\lambda}( t_n^\star(p_{\min}))= \lambda_n^\star(p_{\min})$, which implies the 
		wanted 
		conclusion as $\hat{\alpha}_{\text{Pr}}^n(t_n^\star(p_{\min})) = \hat{\alpha}_{\text{K}}^n(\tilde{\lambda}(t_n^\star(p_{\min}))) = \hat{\alpha}_{\text{K}}^n(\lambda_n^\star(p_{\min}))$. 
		We know that $\hat{\alpha}_{\text{K}}^n \circ \tilde{\lambda} (t) = \hat{\alpha}_{\text{Pr}}^n (t)$, hence for all $t\in D_{\text{Pr}}$, $
		(T_n \circ \hat{\alpha}_{\text{K}}^n \circ \tilde{\lambda}) (t) = (T_n \circ \hat{\alpha}_{\text{Pr}}^n) (t)$,
		and that for any $A\subset [0,\infty)$ it holds that  $\tilde{\lambda} ( \tilde{\lambda}^{-1}(A)) = A \cap \cR (\tilde{\lambda})$, 
		where $\cR (\tilde{\lambda}) = \{\tilde{\lambda}(t): t\in D_{\text{Pr}}\}\subseteq [0,\i)$ is the range of the reparametrization function $\tilde{\lambda}:D_{\text{Pr}} \to [0,\i)$. In fact, 
		$\tilde{\lambda}$ is surjective. To see this, note that $[0,\i) \ni \lambda \mapsto \hat{\alpha}_{\text{K}}^n(\lambda)$ is injective by \Cref{cor:KclassSolutionsDistinct}. Thus, $\cR(\tilde{\lambda})=[0,\i)$ must hold, for otherwise 	\Cref{Eq:EqualityOfPrimalSoltuionsAndKclassSolutions} would not hold. Hence, by surjectivity of $\tilde{\lambda}$ we get that for $A\subset [0,\infty)$ it holds that  $\tilde{\lambda} ( \tilde{\lambda}^{-1}(A)) = A$.

		Now consider $\hat{\alpha}_{\text{Pr}}^n: D_{\text{Pr}} \to \R^{d_1+q_1}$, $\hat{\alpha}_{\text{K}}^n: [0,\infty) \to \R^{d_1+q_1}$ and $\tilde{\lambda}:D_{\text{Pr}} \to [0,\infty)$  as measurable (which follows by continuity and monotonicity) mappings  such that $$
		t_n^\star(p_{\min}) = \sup \{ (T_n \circ \hat{\alpha}_{\text{Pr}}^n)^{-1} (-\i,  Q] \} 
		= \sup \{  (T_n \circ \hat{\alpha}_{\text{K}}^n \circ \tilde{\lambda})^{-1} (-\i,  Q] \}.$$ 
		Since $t\mapsto \tilde{\lambda}(t)$ is strictly decreasing, we get that
		\begin{align*}
			\tilde{\lambda}(t_n^\star(p_{\min})) &= \tilde{\lambda} (   \sup \{   (T_n \circ \hat{\alpha}_{\text{K}}^n \circ \tilde{\lambda})^{-1} (-\i,  Q] \} ) \\
			&= \inf \left\{ \tilde{\lambda}\lp  (T_n \circ \hat{\alpha}_{\text{K}}^n \circ \tilde{\lambda})^{-1} (-\i,  Q]  \rp \right\} \\
			&= \inf \{ \tilde{\lambda} ( \tilde{\lambda}^{-1} (  (T_n \circ \hat{\alpha}_{\text{K}}^n )^{-1} (-\i,  Q] )) \} \\
			&= \inf \left\{   (T_n \circ \hat{\alpha}_{\text{K}}^n )^{-1} (-\i,  Q]  \right\} \\
			&=  \inf \left\{  \lambda \geq 0  :   T_n ( \hat{\alpha}_{\text{K}}^n (\lambda) )\leq Q \right\} 
			\\&= \lambda_n^\star(p_{\min}),
		\end{align*}

\end{proofenv}

\noindent
\begin{proofenv}{\textbf{\Cref{lm:LamdaStarFiniteIFF}}}
	Let $p_{\min}\in(0,1)$ and let \Cref{ass:ZtZfullrankandAtZfullrank,ass:ZYfullcolrank,ass:KclassNotInIV} hold. We have that $$
	l_{\mathrm{OLS}}^n(\hat{\alpha}_{\mathrm{K}}^n (\lambda)) \geq l_{\mathrm{OLS}}^n(\hat{\alpha}_{\text{OLS}}^n) 
	= n^{-1}\|\fY-\fZ\hat{\alpha}_{\text{OLS}}^n\|_2^2 
	= n^{-1}\| \fY - P_\fZ \fY \|_2^2 
	>0,$$
	as $P_\fZ \fY \not = \fY$ (by \Cref{ass:ZYfullcolrank} we have that $\fY \not \in \text{span}(\fZ)$, such that the projection of $\fY$ onto the column space of $\fZ$ does not coincide with $\fY$ itself). Hence,
	$T_n:\R^{d_1+q_1}\to \R$ is well-defined, and the following upper bound
	$$
	T_n (\hat{\alpha}_{\text{K}}^n (\lambda) ) = n \frac{l_{\text{IV}}^n(\hat{\alpha}_{\text{K}}^n (\lambda))}{l_{\mathrm{OLS}}^n(\hat{\alpha}_{\text{K}}^n (\lambda))}  \leq n \frac{l_{\text{IV}}^n(\hat{\alpha}_{\text{K}}^n (\lambda)) }{l_{\mathrm{OLS}}^n(\hat{\alpha}_{\text{OLS}}^n)}
	,$$
	is valid for every $\lambda \geq 0$. In the under- and just-identified setup we know that there exists an $\tilde{\alpha}\in \cM_{\text{IV}} \subset \R^{d_1+q_1}$ such that $
	0=l_{\mathrm{IV}}^n(\tilde{\alpha})$. 
	Now let $\Lambda>0$ be given by
	\begin{align} \label{eq:LambdaEquality}
		\Lambda 
		:= 
		n \frac{  l_{\mathrm{OLS}}^n(\tilde{\alpha}) }{l_{\mathrm{OLS}}^n(\hat{\alpha}_{\text{OLS}}^n) Q_{\chi^2_{q}}(1-p_{\min})}.
	\end{align}
	For any $\lambda > \Lambda$ we have by the non-negativity of $l_{\text{OLS}}^n(\alpha)/\lambda$ that
	\begin{align*}
		l_{\text{IV}}^n(\hat{\alpha}_{\text{K}}^n (\lambda))  &\leq  \lambda^{-1}l_{\mathrm{OLS}}^n(\hat{\alpha}_{\text{K}}^n (\lambda)) + l_{\text{IV}}^n(\hat{\alpha}_{\text{K}}^n (\lambda)) = \min_\alpha \{ \lambda^{-1}l_{\mathrm{OLS}}^n(\alpha) + l_{\text{IV}}^n(\alpha)\} \\
		&\leq  \lambda^{-1}l_{\mathrm{OLS}}^n(\tilde{\alpha}) + l_{\text{IV}}^n(\tilde{\alpha})  <  \frac{l_{\text{OLS}}^n(\tilde{\alpha})}{\Lambda} = \frac{l_{\mathrm{OLS}}^n(\hat{\alpha}_{\text{OLS}}^n)Q_{\chi^2_{q}}(1-p_{\min})}{n},
	\end{align*}
	This implies $$
	T_n (\hat{\alpha}_{\mathrm{K}}^n (\lambda) )  \leq n \frac{l_{\text{IV}}^n(\hat{\alpha}_{\text{K}}^n (\lambda))}{l_{\mathrm{OLS}}^n(\hat{\alpha}_{\text{OLS}}^n)}  < Q_{\chi^2_{q}}(1-p_{\min}),
	$$
	whenever 
	$\lambda > \Lambda$, proving that $
	\lambda_n^\star(p_{\min}) = \inf \{\lambda \geq 0 : T_n(\hat{\alpha}_{\text{K}}^n(\lambda)) \leq Q_{\chi^2_{q}}(1-p_{\min})\}<\i.
	$
	Now consider the over-identified setup $(q> d_1+q_1)$. We claim that $
	\lambda^\star_n(p_{\text{min}}) < \i$ if and only if $T_n(\hat{\alpha}_{\text{TSLS}}^n)< Q_{\chi^2_q}(1-p_{\min})$. 
	If $T_n(\hat{\alpha}_{\text{TSLS}}^n)< Q_{\chi^2_q}(1-p_{\min})$, then by continuity of $\lambda \mapsto \hat{\alpha}_{\text{K}}^n(\lambda)$ and $\alpha \mapsto T_n(\alpha)$ it must hold that $\lambda^\star_n(p_{\text{min}}) < \i$. This follows by noting that $$ %
	T_n(\hat{\alpha}_{\text{K}}^n(\lambda)) \downarrow T_n(\lim_{\lambda \to \i }\hat{\alpha}_{\text{K}}^n(\lambda))= T_n(\hat{\alpha}_{\text{TSLS}}^n) < Q_{\chi^2_q}(1-p_{\min}),$$
	when $\lambda \to \i$, as $\lambda \mapsto T_n(\hat{\alpha}_{\text{K}}^n(\lambda))$ 
	is strictly decreasing (\Cref{lm:OLSandIV_Monotonicity_FnctOfPenaltyParameterLambda})
	Here, we also used that 
	\begin{align*}
		\lim_{\lambda \to \i } \hat{\alpha}_\text{K}^n(\lambda) &= \lim_{\lambda \to \i } (\fZ^\t (\fI+\lambda P_\fA)\fZ)^{-1} \fZ^\t(\fI+\lambda P_\fA)\fY  \\
		&=  \lim_{\lambda \to \i } (\fZ^\t (\lambda^{-1}\fI+P_\fA)\fZ)^{-1} \fZ^\t(\lambda^{-1}\fI+ P_\fA)\fY \\ \notag
		&= (\fZ^\t P_\fA \fZ)^{-1} \fZ^\t  P_\fA\fY  \\
		&= \hat{\alpha}_{\text{TSLS}}^n.
	\end{align*}
	Hence, there must exist a $\lambda \in [0,\i)$ such that $T_n(\hat{\alpha}_{\text{K}}^n(\lambda))  <Q_{\chi^2_q}(1-p_{\min})$, proving that $\lambda^\star_n(p_{\text{min}}) < \i$.  Furthermore, note that the above arguments also imply that $\label{Eq:TSLSyieldsSmallestTestAmongKclass}
	T_n(\hat{\alpha}_{\text{K}}^n(\lambda)) > T_n(\hat{\alpha}_{\text{TSLS}}^n)
	$,
	for any $\lambda \geq 0$, as $\lambda \mapsto T_n(\hat{\alpha}_{\text{K}}^n(\lambda))$ is strictly decreasing and $T_n(\hat{\alpha}_{\text{TSLS}}^n)$ is the limit as $\lambda \to\i$.

	Conversely, assume that $\lambda^\star_n(p_{\text{min}}) < \i$, which implies that there exists a $\lambda'\in[0,\i)$ such that $T_n(\hat{\alpha}_{\text{K}}^n(\lambda'))\leq Q_{\chi^2_q}(1-p_{\min} )$. Thus, %
	$$
	T_n(\hat{\alpha}_{\text{TSLS}}^n) < T_n(\hat{\alpha}_{\text{K}}^n(\lambda'))\leq Q_{\chi^2_q}(1-p_{\min}),
	$$
	proving that the converse implication also holds. 
	
	We furthermore note that, if the acceptance region is empty, that is $$
	\cA_n(1-p_{\min}) := \{\alpha\in \R^{d_1+q_1}:T_n(\alpha) \leq Q_{\chi^2_q}(1-p_{\min})\} = \emptyset,$$ 
	then it obviously holds that $\lambda^\star_n(p_{\min})=\{\lambda \geq 0 :T_n(\hat{\alpha}_{\text{K}}^n(\lambda) )\leq Q_{\chi^2_q}(1-p_{\min}) \} = \i$.
	The possibility of the acceptance region being empty, follows from the fact that the Anderson-Rubin confidence region can be empty; see \Cref{rm:ConnectionToAndersonRubinCI}. To realize that the Anderson-Rubin confidence region can be empty we refer to the discussions and Monte-Carlo simulations of \citet{davidson2014confidence}.

\end{proofenv}

\noindent
\begin{proofenv}{\textbf{\Cref{lm:BinarySearchLambdaStarConverges}}}
	Assume that $\lambda_n^\star(p_{\min})<\i$ and that Assumption  \Cref{ass:ZtZfullrank} and \Cref{ass:ZYfullcolrank} hold. Consider \Cref{Binary.Search.Lambda.Star} for any fixed $N\in \N$. 
	The first `while loop' guarantees that $\lambda_{\min}$ and $\lambda_{\max}$ are such that
	$\lambda^{\star}\in(\lambda_{\min},\lambda_{\max}]$. This is seen by noting that $\lambda \mapsto T_n(\hat{\alpha}_\text{K}^n(\lambda))$ is monotonically decreasing
	(\Cref{lm:OLSandIV_Monotonicity_FnctOfPenaltyParameterLambda}) and that $\lambda_n^\star(p_{\min})<\i$. Hence, $T_n(\hat{\alpha}_\text{K}^n(\lambda_{\max}))$
	eventually drops below $Q_{\chi^2_{q}}(1-p)$. 
	The second `while loop' keeps iterating until the 
	interval $(\lambda_{\min},\lambda_{\max}]$,  which  is guaranteed to contain $\lambda_n^\star(p_{\min})$, has a length less than or equal to $1/N$. Let $\lambda_{\min}$ and $\lambda_{\max}$ denote the last boundaries achieved 
	before the procedure terminates. Then  $
	0\leq \mathrm{Binary.Search}(N,p) - \lambda_n^\star(p_{\min}) = 
	\lambda_{\max} - \lambda_n^\star(p_{\min})  \leq 
	\lambda_{\max}- \lambda_{\min} 	\leq 1/N$. 
	Hence,  $
	\mathrm{Binary.Search}(N,p) - \lambda_n^\star(p_{\min}) \to 0$, as $N\to\i$.
\end{proofenv}

\noindent
\begin{proofenv}{\textbf{\Cref{thm:ConsistencyOfPULSE}}}
	Consider the just- or over-identified setup $(q\geq d_1+q_1)$, let  \Cref{ass:AIndepUYandMeanZeroA} hold. We furthermore assume that the population rank condition, \Cref{ass:EAZtfullrank}, i.e., $E(AZ^\t)$ is of full rank, are satisfied. We furthermore work under the finite-sample conditions of  \Cref{ass:ZtZfullrankandAtZfullrank,ass:ZYfullcolrank}, i.e., that $\fA^\t \fA$ and  $\fZ^\t\fA$ are of full rank and $\fY\not\in \text{span}(\fZ)$ for all sample-sizes $n\in \N$ almost surely. 
	The  first two 
	of these are not strictly necessary as the population version of these rank assumptions guarantee that $\fA^\t \fA$ and  $\fZ^\t\fA$ are of full rank with probability tending to one; see 
	proof of
	\Cref{lm:PopulationPenalizedKClassSolutionUniqueAndExists}.
	Likewise, we can drop the last finite-sample assumption as it is almost surely guaranteed if we assume that the distribution of $\ep_Y$ has density with respect to Lebesgue measure.
	The proof below is easily modified	 
	to accommodate these more relaxed assumptions, but for notational simplicity we 
	prove the statement
	under the stronger finite-sample assumptions. We also let \Cref{ass:KclassNotInIV} hold which in addition with the previous assumptions guarantees that the dual representation of the PULSE holds whenever $\lambda^\star_n(p_{\min})<\i$; see \Cref{thm:PULSEpPULSEdPULSEEequivalent}.
	Furthermore, many of the previous theorems and lemmas were shown for a specific realization that satisfies the finite sample assumptions. Hence, we may only invoke the conclusions of these theorems almost surely.	 Note that the assumptions guarantee that the TSLS estimator is consistent, i.e., $\hat \alpha_{\mathrm{TSLS}}^n \convp \alpha_0$.
	
	Fix any $p_{\min}\in(0,1)$ and let an arbitrary $\ep>0$ be given. We want to prove  that  $
	P(\|\hat{\alpha}_{\text{PULSE}+}^n(p_{\min})-\alpha_0\|>\ep)  \to 0.$
	To that end, define the events $(A_n)_{n\in \N}$ by $
	A_n := (T_n(\hat{\alpha}_{\text{TSLS}}^n)< Q_{\chi^2_{q}}(1-p_{\min})),
	$
	such that
	\begin{align} \label{eq:ConsistencyFirstTerm}
		P(\|\hat{\alpha}_{\text{PULSE}+}^n(p_{\min})-\alpha_0\|>\ep) =& P((\|\hat{\alpha}_{\text{PULSE}}^n(p_{\min})-\alpha_0\|>\ep)\cap A_n) \\
		&+P((\|\hat{\alpha}_{\text{ALT}}^n-\alpha_0\|>\ep)\cap A_n^c), \label{eq:ConsistencySecondTerm}
	\end{align}
	for all $n\in\N$. The last term,  \Cref{eq:ConsistencySecondTerm}, tends to zero as $n\to\i$, $$
	P((\|\hat{\alpha}_{\text{ALT}}^n-\alpha_0\|>\ep)\cap A_n^c) \leq P(\|\hat{\alpha}_{\text{ALT}}^n-\alpha_0\|>\ep) \to 0,
	$$
	by the assumption that $\hat{\alpha}_{\text{ALT}}^n\convp \alpha_0$ as $n\to\i$. In regards to the first term, the right-hand side of \Cref{eq:ConsistencyFirstTerm}, we note that $A_n = (\lambda^\star_n(p_{\min})<\i)$, by \Cref{lm:LamdaStarFiniteIFF}. Formally, this event equality only holds when intersecting both sides with the almost sure event that the finite sample rank condition holds. However, we suppress this intersection for ease of notation. Thus, on $A_n$, it holds that $\hat{\alpha}_{\text{PULSE}}^n(p_{\min}) = \hat{\alpha}_{\text{K}}^n(\lambda^\star_n(p_{\min})),$ by \Cref{thm:PULSEpPULSEdPULSEEequivalent}, implying that $$
	P((\|\hat{\alpha}_{\text{PULSE}}^n(p_{\min})-\alpha_0\|>\ep)\cap A_n) = P((\|\hat{\alpha}_{\text{K}}^n(\lambda^\star_n(p_{\min}))-\alpha_0\|>\ep)\cap A_n).$$
	Furthermore, \Cref{lm:TestInAlphaLambdaStarEqualsQuantileApp} yields that on $A_n$, it holds that $$
	T_n(\hat{\alpha}_{\text{K}}^n(\lambda^\star_n(p_{\min}))) \leq Q_{\chi^2_{q}}(1-p_{\min}),$$ or equivalently $$l_{\text{IV}}^n(\hat{\alpha}_{\text{K}}^n(\lambda^\star_n(p_{\min})) \leq n^{-1}Q_{\chi^2_{q}}(1-p_{\min}) l_{\text{OLS}}^n(\hat{\alpha}_{\text{K}}^n(\lambda^\star_n(p_{\min})).$$ 
	On $A_n$, the stochastic factor in the upper bound above, is further bounded from above by 
	\begin{align*}
		l_{\mathrm{OLS}}^n(\hat{\alpha}_{\mathrm{K}}^n (\lambda_n^\star(p_{\min}))) &\leq \sup_{\lambda\geq 0 }l_{\mathrm{OLS}}^n(\hat{\alpha}_{\mathrm{K}}^n (\lambda ))  \\
		&= \lim_{\lambda \to \i }l_{\mathrm{OLS}}^n(\hat{\alpha}_{\mathrm{K}}^n (\lambda )) \\
		&= l_{\mathrm{OLS}}^n(\lim_{\lambda \to \i }\hat{\alpha}_{\mathrm{K}}^n (\lambda )) \\
		&= l_{\mathrm{OLS}}^n(\hat{\alpha}_{\text{TSLS}}^n), 
	\end{align*}
	where we used continuity of $\alpha \mapsto l_{\mathrm{OLS}}^n(\alpha)$, that $\lambda \mapsto l_{\mathrm{OLS}}^n(\hat{\alpha}_{\mathrm{K}}^n (\lambda ))$ is weakly increasing (\Cref{lm:OLSandIV_Monotonicity_FnctOfPenaltyParameterLambda}) and that	$\lim_{\lambda \to \i } \hat{\alpha}_\text{K}^n(\lambda) = \hat{\alpha}_{\text{TSLS}}^n$.
	Recall that the TSLS estimator is consistent 	$\hat{\alpha}_{\text{TSLS}}^n \convp \alpha_0$, where $\alpha_0$ is the causal coefficient of $Z$ onto $Y$. Hence, Slutsky's theorem and the weak law of large numbers yield that 
	\begin{align*}
		l_{\mathrm{OLS}}^n(\hat{\alpha}_{\text{TSLS}}^n)  &= n^{-1} (\fY - \fZ \hat{\alpha}_{\text{TSLS}}^n)^\t (\fY - \fZ \hat{\alpha}_{\text{TSLS}}^n) \\&= 
		n^{-1} \fY^\t \fY + (\hat{\alpha}_{\text{TSLS}}^n)^\t n^{-1} \fZ^\t \fZ\hat{\alpha}_{\text{TSLS}}^n - 2n^{-1}\fY^\t \fZ\hat{\alpha}_{\text{TSLS}}^n  \\&\convp E(Y^2) + \alpha_0^\t E(Z Z^\t ) \alpha_0 - 2 E(YZ^\t) \alpha_0 \\&= E[(Y-Z\alpha_0)^2]. 
	\end{align*}
	Thus, on the event $A_n$, we have that $$
	0\leq l_{\text{IV}}^n(\hat{\alpha}_{\text{K}}^n(\lambda^\star_n(p_{\min})) \leq n^{-1}Q_{\chi^2_{q}}(1-p_{\min}) l_{\text{OLS}}^n(\hat{\alpha}_{\text{TSLS}}^n)=:H_n,$$
	where the upper bound $H_n$ converges to zero in probability by Slutsky's theorem. Furthermore, note that
	\begin{align*}
		\notag
		l_\text{IV}^n(\alpha_0) &= \|n^{-1/2} (\fA^\t \fA)^{-1/2} \fA^\t(\fY - \fZ\alpha_0)\|_2^2 \\ \notag
		&=\| (n^{-1}\fA^\t \fA)^{-1/2} n^{-1}\fA^\t\fU_Y\|_2^2 \\
		&\convp \| E(AA^\t)^{-1/2} E(AU_Y)\|_2^2 \\
		&= \| E(AA^\t)^{-1/2} E(A)E(U_Y)\|_2^2 \\\notag 
		& = 
		0,	
	\end{align*}
	where we used that $Y=Z^\t \alpha_0 +U_Y$, Assumption \Cref{ass:AIndepUy}: $A\independent U_Y$, and Assumption \Cref{ass:MeanZeroA}: $E(A)=0$ (Alternatively,  $E(U_Y|A)=0$).

	Now define a sequence of (everywhere) well-defined estimators $(\tilde{\alpha}_n)_{n\in\N}$ by $$
	\tilde{\alpha}_n := 
	\mathbbm{1}_{A_n} \hat{\alpha}_{\text{K}}^n(\lambda^\star_n(p_{\min})) 
	+ \mathbbm{1}_{A_n^c} \alpha_0,$$
	for each $n\in \N$. We claim that the loss function $l_{\text{IV}}^n$ evaluated in this estimator tends to zero in probability, i.e., as $n\to\i$ it holds that
	\begin{align} \label{eq:lIVinAlphatildeConvPZero}
		l_{\text{IV}}^n(\tilde{\alpha}_n) = \| (n^{-1}\fA^\t \fA)^{-1/2} n^{-1}\fA^\t(\fY - \fZ\tilde{\alpha}_n)\|_2^2 \convp 0.
	\end{align}
	This holds by the above observations as for any $\ep'>0$ we have that
	\begin{align*}
		P(|l_{\text{IV}}^n(\tilde{\alpha}_n)| > \ep') 
		&= P((|l_{\text{IV}}^n(\hat{\alpha}_{\text{K}}^n(\lambda^\star_n(p_{\min})) )| > \ep')\cap A_n) +P((|l_{\text{IV}}^n(\alpha_0)| > \ep')\cap A_n^c) \\
		&\leq    P((|H_n| > \ep')\cap A_n) +P((|l_{\text{IV}}^n(\alpha_0)| > \ep')\cap A_n^c) \\
		&\leq P(|H_n| > \ep') +P(|l_{\text{IV}}^n(\alpha_0)| > \ep') \to 0,
	\end{align*}
	when $n\to\i$. 	Now define the random linear maps $g_n:\Omega\times \R^{d_1+q_1}\to \R^q$ by $$
	g_n(\omega,\alpha) := (n^{-1}\fA^\t(\omega) \fA(\omega))^{-1/2}n^{-1} \fA^\t(\omega)  \fZ(\omega) \alpha,$$
	for all $n\in \N$. 	
	The maps $(g_n)$ converge point-wise, that is, for each $\alpha$, in probability to $g:\R^{d_1+q_1}\to \R^q$, given by $
	g(\alpha) := E(AA^\t)^{-1/2}E(AZ^\t)\alpha$, 
	as $n\to \i$. The map $g$ is injective. This follows by  Assumption \Cref{ass:VarianceOfAPositiveDefinite} and  \Cref{ass:EAZtfullrank}, which implies and state that $E(AA^\t)\in \R^{q\times q}$ and $E(AZ^\t)\in \R^{q\times(d_1+q_1)}$ are of full rank, respectively, hence 
	$
	\text{rank}(E(AA^\t)^{-1/2}E(AZ^\t))=\text{rank}(E(AZ^\t))= d_1+q_1,
	$
	since we are in the just- and over-identified setup, where $q\geq d_1+q_1$. We conclude that $g$ is injective, as its matrix representation is of full column rank.
	Furthermore, by \Cref{eq:lIVinAlphatildeConvPZero} it holds that $
	g_n(\tilde{\alpha}_n) \convp E(AA^\t)^{-1/2}E(AY)$. 
	Hence, we have that
	\begin{align*}	
		g_n(\tilde{\alpha}_n)-g_n(\alpha_{0}) &\convp E(AA^\t)^{-1/2}E(AY) -E(AA^\t)^{-1/2}E(AZ^\t)\alpha_{0} \\
		&= E(AA^\t)^{-1/2}E(AU_Y) \\
		&= 0,
	\end{align*}
	
	as $n\to \i$. \Cref{lm:LemmaRandomFunctionOfRandomArgument} of \Cref{sec:AuxLemmas} now yields that $\tilde{\alpha}_n \convp \alpha_0$. Finally, note that as $\hat{\alpha}_{\text{K}}^n(\lambda^\star_n(p_{\min}))= \tilde{\alpha}_n$ on $A_n$ we have that
	\begin{align*}
		P((\|\hat{\alpha}_{\text{K}}^n(\lambda^\star_n(p_{\min}))-\alpha_0\|>\ep)\cap A_n) &=P((\|\tilde{\alpha}_n-\alpha_0\|>\ep)\cap A_n)\\ 
		&\leq P(\|\tilde{\alpha}_n-\alpha_0\|>\ep) \\
		&\to 0,
	\end{align*}
	proving that  $\hat{\alpha}_{\text{PULSE}+}^n(p_{\min}) \convp \alpha_0$, as $n\to\i$.
\end{proofenv}

\section{Auxiliary lemmas} 
\label{sec:AuxLemmas}
\setcounter{figure}{0}
\setcounter{table}{0}
\setcounter{equation}{0}
\medskip

\begin{lemma} \label{lm:LemmaRandomFunctionOfRandomArgument}
	Suppose that $g_n: \mathbb{R}^G \to \mathbb{R}^K$ are random linear maps
	converging point-wise in probability to a non-random  linear map $g:\R^G\to \R^K$ that is injective. If
	$$
	g_n(\hat{\beta}_n - \beta_0) \underset{n\to\i}{\convp}0,
	$$
	then $\hat{\beta}_n$ is a consistent estimator of $\beta_0$. That  is, $	\hat{\beta}_n \underset{n\to\i}{\convp}\beta_0$.
\end{lemma}
\textbf{Proof:}
As $g$ is injective, we have that $\text{rank}(g) = G$, and as such $\text{rank}(g^\t g) = \text(g) = G$ which implies that $g^\t g$ is invertible. Furthermore, by Slutsky's theorem we get that $
g_n\convp g \implies g_n^\t g_n \convp g^\t g$,
as $n\to\i$, that is, for any $\ep>0$, $$
P(||g_n^\t g_n-g^\t g\| \leq \ep) =P(g_n^\t g_n \in \overline{B(g^\t g,\ep)})  \underset{n\to\i}{\to }  1.$$ 
Here $\|\cdot\|$ is any norm on the set of $G\times G$ matrices and $\overline{B(g^\t g,\ep)}$ is the closed ball around $g^\t g$ with radius $\ep$ with respect the this norm.
Now note that the set $\text{NS}_G$ of all non-singular $G\times G$ matrices  is an open subset of all $G\times G$ matrices, which implies that there exists an $\ep>0$, such that  $\overline{B(g^\t g,\ep)} \subset \text{NS}_G.$
Hence, $g_n^\t g_n$ is invertible with probability tending towards 1, that is, $
P(g_n^\t g_n \in \text{NS}_G) \underset{n\to\i}{\to } 1$.
Let $h_n:\Omega\to\text{NS}_G$ be given by $$
h_n(\omega)  := \mathbbm{1}_{(g_n^\t g_n \in \text{NS}_G)} g_n^\t(\omega) g_n(\omega) + \mathbbm{1}_{(g_n^\t g_n \in \text{NS}_G)^c} I.$$
Then $h_n \underset{n\to\i}{\convp} g^\t g$, since for any $\ep >0$
\begin{align*}
	P(\|h_n-g^\t g\| >\ep ) &= P((\|g_n^\t g_n-g^\t g\|> \ep) \cap (g_n^\t g_n \in \text{NS}_G)) \\
	& \qquad + P((\|I-g^\t g\|> \ep )\cap (g_n^\t g_n \in \text{NS}_G)^c) \\
	&\leq  P(\|g_n^\t g_n-g^\t g\|> \ep) + P(g_n^\t g_n \in \text{NS}_G)^c) \\
	&\underset{n\to\i}{\to}  0,
\end{align*}
Continuity of the inverse operator and the continuous mapping theorem, yield that $\|h_n^{-1}\|_{\text{op}}\convp \|(g^\t g)^{-1}\|_{\text{op}} \in \R$ and $\|g_n^\t \|_{\text{op}} \convp \|g^\t\|_{\text{op}}\in\R$ as $n$ tends to infinity, where $\|\cdot\|_{\text{op}}$ is the operator norm induced by the Euclidean norm $\|\cdot\|_2$. Furthermore, 
\begin{align*}
	\|g_n^\t g_n(\hat{\beta}_n-\beta_0) \|_2 &\leq \|g_n^\t \|_{\text{op}} \| g_n(\hat{\beta}_n-\beta_0) \|_2 \\ &\underset{n\to\i}{\convp} \|g^\t \|_{\text{op}}  \cdot  0\\
	&= 0,
\end{align*}

by the assumptions and Slutsky's theorem. Hence, for any  $\ep>0$ 
\begin{align*}
	P(	\|h_n(\hat{\beta}_n-\beta_0) \|_2 > \ep ) &= P(	(\|g_n^\t g_n(\hat{\beta}_n-\beta_0) \|_2 > \ep ) \cap (g_n^\t g_n \in \text{NS}_G) ) \\
	&\qquad + P(	(\|\hat{\beta}_n-\beta_0\|_2 > \ep) \cap  (g_n^\t g_n \in \text{NS}_G)^c) \\
	&\leq  P(	(\|g_n^\t g_n(\hat{\beta}_n-\beta_0) \|_2 > \ep ) ) + P((g_n^\t g_n \in \text{NS}_G)^c)\\
	&\underset{n\to\i}{\to}  0.
\end{align*}
Thus,
\begin{align*}
	\|\hat{\beta}_n-\beta_0\|_2 &= \| h_n^{-1} h_n(\hat{\beta}_n-\beta_0) \|_2 \\& \leq \|h_n^{-1}\|_{\text{op}} \| h_n(\hat{\beta}_n-\beta_0) \|_2 \\ &\underset{n\to\i}{\convp} \|(g^\t g)^{-1}\|_{\text{op}} \cdot 0\\ &=0,
\end{align*}

by Slutsky's theorem, yielding that $\hat{\beta}_n$ is an consistent estimator of $\beta_0$.
\hfill$\square$\bigskip

\begin{lemma} \label{lm:SolutionIsTightIfNotStationary}
	Let $\hat{\alpha}$ be a solution to a constrained optimization problem of the form
	\begin{align*}
		\begin{array}{ll}
			\underset{\alpha \in \R^{k}}{\mathrm{minimize}} & f(\alpha) \\
			\mathrm{subject \, to} & g(\alpha) \leq c,
		\end{array}
	\end{align*}
	where $f$ is an everywhere differentiable strictly convex function on $\R^k$ for which a stationary point exists, $g$ is continuous and $c\in \R$.  If the stationary point of $f$ is not feasible, then the constraint inequality is tight (active)
	in the solution $\hat{\alpha}$, that is, $
	g(\hat{\alpha}) = c$.
\end{lemma}
\textbf{Proof:}
Since	
$\hat{\alpha}$ feasible and the stationary point of $f$ is not feasible,	
we know that $\hat{\alpha}$ is not a stationary point of $f$, hence $Df(\hat{\alpha}) \not = 0$. Now assume that the constraint bound is not tight (active) in the solution $\hat{\alpha}$, that is $g(\hat{\alpha})<c$. By continuity of $g$, we know that there exists an $\epsilon >0$, such that for all $\alpha \in B(\hat{\alpha}, \ep)$, it holds that $g(\alpha)<c$. Furthermore, since $Df(\hat{\alpha}) =0$, we can look at the line segment going through $\hat{\alpha}$ in the direction of the negative gradient of $f$ in $\hat{\alpha}$. That is, $l:\R \to \R^k$ defined by $l(t) = \hat{\alpha} - tDf(\hat{\alpha})$. Note that $$D (f\circ l)(0) = Df(l(0))Dl(0) = -Df(\hat{\alpha}) Df(\hat{\alpha})^\t = -\|Df(\hat{\alpha})\|<0,$$ meaning that the derivative of $f\circ l:\R \to \R$ is negative in zero. Therefore, there exists a $\delta>0$, such that for all $t\in(0,\delta)$ it holds that $f\circ l (t) < f \circ l(0)$, i.e., $$ f(\hat{\alpha}-tDf(\hat{\alpha})) < f(\hat{\alpha}).$$ Thus, for sufficiently small $t'$, is it holds that $t'<\delta$ and $\hat{\alpha}-t'Df(\hat{\alpha}) \in B(\hat{\alpha},\ep)$. We conclude that $\tilde{\alpha}:= \hat{\alpha}-tDf(\hat{\alpha})$ is feasible, $g(\tilde{\alpha})<c$, and super-optimal compared to $\hat{\alpha}$, $f(\tilde{\alpha})< f(\hat{\alpha})$, which contradicts that $\hat{\alpha}$ solves the optimization problem. In words, if the solution is not tight we can take a small step in the negative gradient direction of the objective function and get a better objective value while still being feasible. 
\hfill$\square$\bigskip

\section{Additional remarks} 
\label{app:AddRemarks}
\setcounter{figure}{0}
\setcounter{table}{0}
\setcounter{equation}{0}
\renewcommand{\theremark}{\thesection.\arabic{remark}}

\medskip

\begin{remark}[Model misspecification] \label{rm:ModelMispecification}
	\textnormal{Theorem~\ref{sthm:TheoremIntRobustKclas} still holds under
		the following three model misspecifications, which may arise from erroneous non-sample information and unobserved endogenous variables
		(these violations 
		may break the identification of $\alpha_{0,*}$ and 
		generally render 
		the K-class estimators inconsistent even when $\plim\kappa  =1$).}
	\begin{itemize}
		\item[\textnormal{(a)}] 	\textnormal{Exclude included endogenous variables. Consider the setup where no hidden variable enters the target equation given by $Y = \gamma_0^\t X + \beta_0^\t A + \ep_Y$, with $\ep_Y \independent A$.  If we 
			erroneously 	
			exclude an  endogenous variable that directly affects $Y$, i.e., $\gamma_{0,-*} \not =0$, this is equivalent to drawing inference from the model $	Y= \gamma_{0,*}^\t X_{*} + \beta_{0,*}^\t A_{*} +U $,
			where $U= \ep_Y + \gamma_{0,-*}^\t X_{-*}$. If $E(A_{-*}U)= E(A_{-*}X_{-*}^\t)\gamma_{0,-*}\not = 0$, we have introduced dependence that renders at least some of the instruments $A_{-*}$ invalid, breaking identifiability.} %
		\item[\textnormal{(b)}] \textnormal{Exclude included exogenous variables. Consider again the setup from $(a)$ where there is no hidden variables entering the target equation. If we 
			erroneously 				
			exclude a exogenous variable that directly affects $Y$, i.e., $\beta_{0,-*} \not =0$, then this is equivalent with drawing inference from the model $	Y= \gamma_{0,*}^\t X_{*} + \beta_{0,*}^\t A_{*} +U $,
			where $U= \ep_Y + \beta_{0,-*}^\t A_{-*}$. It holds that $E(A_{-*}U)= E(A_{-*}A_{-*}^\t)\beta_{0,-*}\not =0$, again rendering the instruments invalid. }%
		\item[\textnormal{(c)}] \textnormal{Possibility of hidden endogenous variables. 
			Consider the case with
			included hidden variables that are directly influenced by the excluded exogenous variables, i.e., $A_{-*}\to H\to Y$. 
			This implies that the excluded exogenous variables $A_{-*}$ are correlated with the collapsed noise variable in the structural equation $Y =  \alpha^\t_{0,*} Z_*  + U$, where $U = \ep_Y + \eta_0^\t H$ with $\eta_0\not =0$. In the case that $E(A_{-*}U) = E(A_{-*}H^\t)\eta_0\not =0$ the instruments are invalid.} 
	\end{itemize}
	
\end{remark}

\begin{remark}[Connection to the Anderson-Rubin Test] \label{rm:ConnectionToAndersonRubinCI}
	\textnormal{Our acceptance region $
		\cA_n^c(1-p_{\min}) := \{\alpha \in \R^{d_1+q_1}: T_n^c(\alpha) \leq Q_{\chi^2_{q}}(1-p_{\min})\}$, 
		is closely related to the Anderson-Rubin \citep[][]{anderson1949estimation} confidence region of the simultaneous causal parameter $\alpha_0 = (\gamma_0,\alpha_0)$ in an identified model. When the causal parameter $\alpha_0$ is identifiable, i.e., in a just- or over-identified setup $(q\geq d_1+q_2)$ and \Cref{ass:EAZtfullrank} 
		holds, only the causal parameter yields regression residuals $Y-\alpha_0^\t Z$ that are uncorrelated with the exogenous variables $A$. In this restricted setup, our null hypothesis is equivalent with $\tilde{H}_0(\alpha): \alpha = \alpha_0$. 
		The hypothesis $\tilde{H}_0(\alpha)$	
		can be tested by the Anderson-Rubin test and all non-rejected coefficients constitute 
		the Anderson-Rubin confidence region of $\alpha_0$, which is given by $
		\text{CR}_{\text{AR}}^{ex,n}(1-p_{\min}) := \left\{\alpha \in \R^{d_1+q_1}:T^{\text{AR}}_n(\alpha) \leq Q_{F(q,n-q)}(1-p_{\min}) \right\}$, 
		where $Q_{F(q,n-q)}(1-p_{\min})$ is the $1-p_{\min}$ quantile of the $F$ distribution with $q$ and $n-q$ degrees of freedom and the 
		Anderson-Rubin test-statistic $T^{\text{AR}}_n(\alpha)$ is  given by
		\begin{align*}
			T^{\text{AR}}_n(\alpha) :=  \frac{n-q}{q}\frac{(\fY-\fZ\alpha)^\t P_\fA(\fY-\fZ\alpha)}{(\fY-\fZ\alpha)^\t P_\fA^\perp(\fY-\fZ\alpha)} = \frac{n-q}{q}\frac{l_{\text{IV}}^n(\alpha)}{l_{\text{OLS}}^n(\alpha)-l_{\text{IV}}^n(\alpha)}.
		\end{align*}
		The confidence region $\text{CR}_{\text{AR}}^{ex,n}$ is exact whenever several regularity conditions are satisfied, such as
		deterministic exogenous variables and normal distributed errors
		\citep[][Theorem 3]{anderson1949estimation}.
		In a general SEM model the regularity conditions are not fulfilled, but changing the rejection threshold to $Q_{\chi^2_q/q}(1-p_{\min})$, we obtain an asymptotically valid confidence region. That is,
		\begin{align*}
			\text{CR}_{\text{AR}}^{as,n}(1-p_{\min}) := \left\{\alpha \in \R^{d_1+q_1}:T^{\text{AR}}_n(\alpha) \leq Q_{\chi^2_q/q}(1-p_{\min}) \right\},
		\end{align*}
		is an asymptotically valid approximate confidence region
		\citep[][Theorem 6]{anderson1950asymptotic}.	
		This relies on the fact that $T^{\text{AR}}_n(\alpha) \convd \chi^2_q/q$ under the null and $T^{\text{AR}}_n$ diverges to infinity under the general alternative. The test-statistic $T_n^c(\alpha)$ can be seen as a scaled coefficient of determination ($R^2$-statistic) for which $T_n^{\text{AR}}(\alpha)$ is the  corresponding $F$-statistic. That is, one can realize that
		\begin{align*}
			T_n^{\text{AR}}(\alpha) = \frac{n-q}{q} \frac{T_n^c(\alpha)/c(n)}{1-T_n^c(\alpha)/c(n)} \leq Q_{\chi^2_q/q}(1-p_{\min}),
		\end{align*}
		is equivalent to 
		\begin{align*}
			\frac{n-q+ Q_{\chi^2_q}(1-p_{\min})}{c(n)} T_n^c(\alpha) \leq Q_{\chi^2_q}(1-p_{\min}).
		\end{align*}
		Thus, if
		$Q_{\chi^2_q}(1-p_{\min}) \geq q$, then  $\cA_n(1-p_{\min}) \supseteq \text{CR}_{\text{AR}}^{as,n}(1-p_{\min})$ and $\cA_n(1-p_{\min}) \subset \text{CR}_{\text{AR}}^{as,n}(1-p_{\min})$ otherwise, where $\cA_n(1-p_{\min})$ is the acceptance region when using the scaling scheme $c(n)=n$. 
		Furthermore, $\cA_n^c(1-p_{\min})$, the acceptance region under a general scaling scheme $c(n) \sim n$,  is asymptotically equivalent to the Anderson-Rubin approximate confidence region $\text{CR}_{\text{AR}}^{as,n}(1-p_{\min})$. If we choose the specific scaling to be 
		$c(n)=n-q+ Q_{\chi^2_q}(1-p_{\min})$, then 
		they coincide, 
		$\text{CR}_{\text{AR}}^{as,n}(1-p_{\min})=\cA_n^c(1-p_{\min})$ for each $n\in \N$. Whenever the Anderson-Rubin confidence region is exact, we could change the rejection threshold from $Q_{\chi^2_{q}}(1-p_{\min})$ to $c(n)Q_{B(q/2,(n-q)/2)}(1-p_{\min})$ and also get an exact acceptance region, where $B(q/2,(n-q)/2)$ is the Beta distribution with shape and scale parameter $q/2$ and $(n-q)/2$ respectively.}
\end{remark}

\begin{remark}[Connections to pre-test estimators] \label{rm:Pretest}
	\textnormal{
		It has been suggested to use pre-test for choosing	
		between the TSLS and OLS estimator. When using the Hausman test for endogeneity \citep[][]{hausman1978} one considers the pre-test estimator studied by, e.g., \citet{chmelarova2010} and \citet{guggenberger2010}. If $H$ denotes the Hausman test-statistic that rejects the hypothesis of endogeneity when $H\leq Q$, the pre-test estimator is given by $\alpha_{\mathrm{pretest}}^n=1_{(H\leq Q)}\alpha_{\mathrm{OLS}}^n+1_{(H>Q)}\alpha_{\mathrm{TSLS}}^n$.  The PULSE estimator can  be seen as a 
		pre-test estimator using the Anderson-Rubin test as a test for endogeneity.  However, PULSE differs from the above in the sense that when endogeneity is not rejected we do not revert to the TSLS estimate but rather to the coefficient within the Anderson-Rubin confidence region that minimizes the mean squared prediction error.}
\end{remark}

\section{Simulation study} 
\label{sec:Experiments}
\setcounter{figure}{0}
\setcounter{table}{0}
\setcounter{equation}{0}

\medskip

\subsection{Distributional robustness} \label{app:DistributionalRobustness}
\begin{figure}[H]
	\centering
	\includegraphics[width=\linewidth]{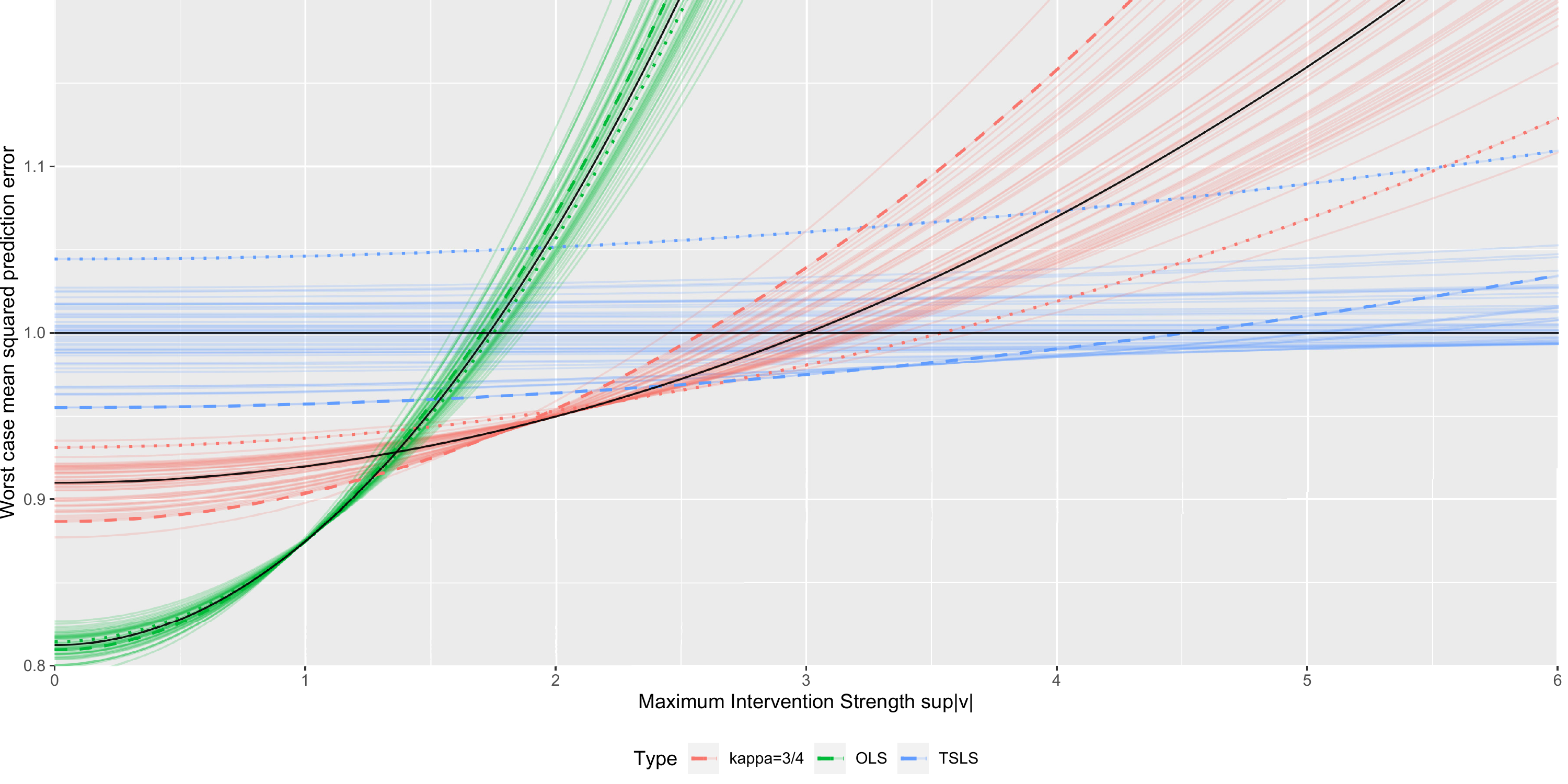}
	\caption{  \normalsize Distributional Robustness of K-class estimators. The plot shows
		the worst case MSPE against the maximum intervention strength condsidered. Each of the 50 repetitions corresponds 
		to three lines (green, red, blue), corresponding to the three
		estimates using $\kappa \in \{0,3/4,1\}$, respectively. The solid black line corresponds to the population coefficients. The OLS is optimal for small interventions but yields a large loss for strong interventions; the TSLS is optimal 
		for large interventions but yields a relatively large loss for small interventions. Choosing a $\kappa$ different from zero and allows us to trade off these two regimes.
		The dashed and the dotted lines correspond
		to the two samples,
		for which 
		the interval on which the 
		$\kappa=3/4$ estimator 
		outperforms 
		TSLS and OLS in terms of worst case MSPE
		is shortest and longest, respectively.
	}
	\label{fig:DistRobustness}
\end{figure}

We first illustrate the distributional robustness property of K-class estimators 
discussed in Section~\ref{sec:intervrobustnessofKclass} in  
a finite sample setting.
We consider the model given by
\begin{align*}
	X := A + U_X, \qquad 
	Y := \gamma  X + U_Y,
\end{align*}
where $\gamma=1$ and $A\sim N(0,1)$ independent of $
\begin{psmallmatrix}
	U_X \\U_Y
\end{psmallmatrix} \sim \mathcal{N} \left( \begin{psmallmatrix}
	0 \\ 0 
\end{psmallmatrix}, \begin{psmallmatrix}
	1 & 0.5 \\
	0.5 & 1
\end{psmallmatrix}\right)$.
We estimate $\gamma$ from $n=2000$  observations generated by the above system and estimate $\hat \gamma^n_{\text{K}}(\kappa)$ for all $\kappa \in \{0,3/4,1\}$ for which the corresponding population coefficients are given by $\gamma_{\text{K}}(0)=\gamma_{\text{OLS}}=1.25$, $\gamma_{\text{K}}(3/4)=1.1$ and $\gamma_{\text{K}}(1)=\gamma_{\text{TSLS}}=1$.  
We repeat the simulation 50 times and save the estimated coefficients. 
\Cref{fig:DistRobustness} 
illustrates the distributional robustness property of \Cref{sthm:TheoremIntRobustKclas}. For all estimated coefficients $\hat \gamma$ of $\gamma$ we have plotted the analytically computed worst case mean squared prediction error (MSPE) under all hard interventions of absolute strength up to $x$ given by \begin{align} \label{eq:analyticDerivationOfWCMSPE}
	\sup_{|v| \leq x} E^{\mathrm{do}(A:=v)} [ \lp Y-\hat \gamma X\rp^2 ] %
	&=  x^2(1-\hat \gamma)^2 +\hat \gamma^2+3(1-\hat \gamma)
\end{align}
against the maximum intervention strength $x$ for the range $x\in[0,6]$. The plot also shows results for the population coefficient as seen in  \citet[Figure~2]{AnchorRegression}.

In all 50 repetitions the K-class estimator for $\kappa=3/4$  outperforms both OLS and TSLS in terms of worst case MSPE for maximum intervention strength of $2$.
This is in line with the theory presented in Section~\ref{sec:intervrobustnessofKclass}.
In terms of population coefficients  our theoretical results predict
that $\kappa=3/4$ is worst case MSPE superior, relative to OLS and TSLS, for all maximum intervention strengths in the range $[1.37,3]$. 
Among the 50 repetitions we find the outcomes for which the superiority range of $\kappa = 3/4$ has the shortest and longest superiority range length. The shortest superiority range is $[1.27, 2.15]$ and the longest is $[1.46,5.54]$. Clearly, these numbers vary with changing sample size and number of repetitions. For example, with $50$, $200$, $500$, $2000$, $5000$ and $10000$ observations and 50 repetitions, the median lengths of the MSPE superiority range for $\kappa=3/4$ equal 0.82, 1.16, 1.44, 1.74, 1.58 and 1.63, respectively (1.63 is also the length of the theoretically computed interval $[1.37, 3]$).

\subsection{Estimating causal effects}
In this subsection we 
investigate the finite sample behaviour of the PULSE estimator by simulation experiments. We look at how the PULSE estimator fairs in comparison to other well-known single equation estimators in terms of different performance measures.  We generate $n \in \N$ realizations of the SEM in question and construct the estimators of interest based on these $n$ observations. This is repeated $N\in \N$ times, allowing us to estimate different finite sample performance measures of the estimators of interest. The characterization of weak instruments through the minimum eigenvalue of $G_n$, a multivariate analogue to the first stage \textit{F}-statistic, as introduced in \citet{stock2002testing} is important for some of our experimental findings. We refer the reader to \Cref{sec:WeakInst} for a brief introduction.
\subsubsection{Benchmark Estimators and Performance Measures}
\label{sec:ExpResultsEstCausPerformanceMeasures}
We compare the PULSE(5)  estimator, that is PULSE with $p_{\min}=0.05$, to four specific K-class estimators that are well-known to have second moments (in sufficiently over-identified setups). This will allow us to conduct both bias and mean squared error analysis of estimators. %
Most importantly, we benchmark against  Fuller estimators. 
The $\kappa$-parameter of the Fuller estimators are given by 
$\kappa_{\text{FUL}}^n(a) =\kappa_{\text{LIML}}^n -\frac{a}{n-q},$
where $n-q$ is the degrees of freedom in the first stage regression, $a>0$ is a hyper parameter 
and $\kappa_{\text{LIML}}^n$ is the stochastic $\kappa$-parameter corresponding the to LIML estimator. One way to represent the $\kappa$-parameter of the LIML estimator is
$
\kappa_{\text{LIML}} = \lambda_{\min}(W_1W^{-1})  %
$
where $\lambda_{\min}$ denotes the smallest eigenvalue, $W_1$ and $W$ are defined as $
W = [
\fY  \, \, \fX
]^\t P_\fA^\perp [
\fY  \, \, \fX
]$ and $W_1 =[
\fY  \, \, \fX
]^\t P_{\fA_*}^\perp [
\fY  \, \, \fX
],$
and $P_\fA^\perp = \fI-\fA (\fA^\t \fA)^{-1}\fA^\t$; see, e.g.,  \citet{amemiya1985advanced}.  We choose to benchmark the PULSE estimator against the
following K-class estimators: OLS ($\kappa=0$), TSLS ($\kappa=1$), Fuller(1) ($\kappa =\kappa_{\text{FUL}}^n(1)$) and Fuller(4) ($\kappa =\kappa_{\text{FUL}}^n(4)$).

The Fuller(1) estimator is approximately unbiased in that the mean bias is zero up to $\mathcal{O}(n^{-2})$ \citep[][Theorem 1]{fuller1977some} and  Fuller(4) exhibits approximate superiority in terms of MSE compared to all other Fuller estimators \citep[][Corollary 2]{fuller1977some}. As we shall see below the PULSE estimator has good MSE performance when instruments are weak and therefore we especially benchmark against Fuller(4) which  has shown better MSE performance than TSLS in simulation studies when instruments are weak; see e.g.\  \citet{hahn2004estimation}. In the over-identified setup we let the PULSE estimator revert to Fuller(4) whenever the dual representation is infeasible.

We compare the estimators in terms of bias and mean squared error (MSE), which for an $n$-sample estimator $\hat{\alpha}_n$ with target $\alpha\in \R^{d_1+q_1}$ 
are given by $
\text{Bias}(\hat \alpha_n) = E(\hat \alpha_n)-\alpha\in \R^{d_1+q_1}$, $
\text{MSE}(\hat \alpha_n) = E[(\hat \alpha_n-\alpha) (\hat \alpha_n-\alpha)^\t]\in \R^{(d_1+q_1)\times (d_1+q_1)}$. 
The empirical quantities, estimated 
from $N$ independent repetitions are denoted by 
$\widehat{\text{Bias}}(\hat{\alpha}_n)$
and
$\widehat{\text{MSE}}(\hat \alpha_n)$.
In the multivariate setting,
we compare biases by comparing their Euclidean norms %
When comparing MSEs, we 
call
$\hat{\alpha}_n$ 
MSE superior to 
$\tilde{\alpha}_n$ 
if they are ordered in the partial ordering generated by the proper cone of positive semi-definite matrices (that is, $\widehat{\text{MSE}}(\tilde{\alpha}_n) - \widehat{\text{MSE}}(\hat\alpha_n)$ is
positive semi-definite).
We also consider 
the ordering of its scalarizations given by the determinant and trace (the latter satisfies
$\text{trace}(\widehat{\text{MSE}}(\hat{\alpha}_n)) = \text{trace}(\widehat{\text{Var}}(\hat{\alpha}_n))+ \|\widehat{\text{Bias}}(\hat{\alpha}_n)\|_2^2$).

We conduct the simulation experiments 
even though it is not proved that the
PULSE estimator has finite second moments. 
In the simulations, the empirical estimates of the mean squared error were stable, possibly
even more so than for the Fuller estimators for which we know that second moments exists in settings where the noise is Gaussian; see e.g.,\ \citet{fuller1977some,chao2012expository}.

Below we describe two multivariate simulation experiments and refer the reader to \Cref{sec:SimUnivariate} in the main paper for a univariate simulation experiment.

\subsubsection{Varying confounding multivariate experiment.} \label{sec:ExpResultsEstCausalMultidim}
\begin{figure}[h] 
	\centering
	\begin{minipage}{0.40\textwidth}
		\vspace*{\fill}
		
		\begin{align*}
			A &:= N_A \in \mathbb{R}^2, \\[7pt]
			H &:= N_H \in \mathbb{R}^2, \\[7pt]
			X &:=  \xi^\t A +  \delta^\t H + N_X \in \mathbb{R}^2 , \\[7pt]
			Y &:=  \gamma^\t X +  \mu^\t H + N_Y  \in \mathbb{R}.
		\end{align*}
		\vspace*{\fill}
	\end{minipage}	
	\begin{minipage}{0.4\textwidth}
		\centering
		\includegraphics[width=\textwidth]{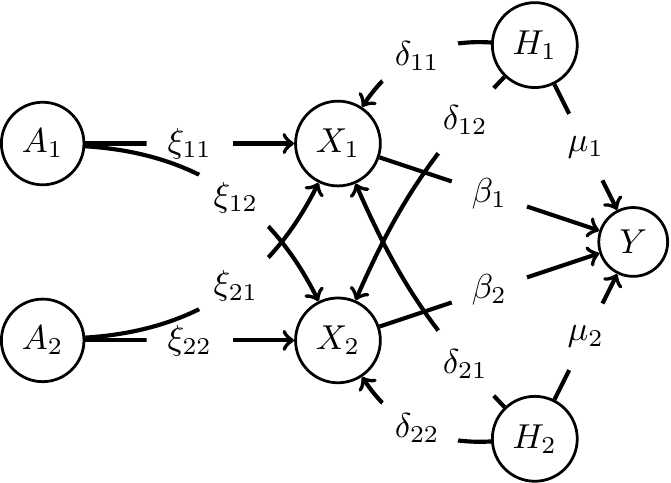}
	\end{minipage}	
	\caption{ The SEM and graph representation used for simulating data in the experiments described in Section~\ref{sec:ExpResultsEstCausalMultidim}.} \label{fig:IV2d_crossdelta}
\end{figure}

In this simulation scheme we consider just-identified two-dimensional instrumental variable models with the 
SEM and causal graph illustrated in \Cref{fig:IV2d_crossdelta}.  Since we want to compare MSE statistics that require estimators with second moments we drop comparisons with the TSLS estimator. 

Here, $\xi, \delta\in \R^{2\times 2}$, $\mu\in \R^2$ and $(N_A,N_X,N_X,N_Y)$ are independent noise innovations. We let $\gamma=(0,0)$ and let the noise innovations for $A,H,Y$ have distribution $(N_A ,N_H, N_Y) \sim  \mathcal{N} (0, I )$. 

We randomly generate 10000 models by letting $
N_X \sim \mathcal{N} \left( \begin{psmallmatrix}
	0 \\ 0 
\end{psmallmatrix}, \begin{psmallmatrix}
	\sigma_{1}^2 & 0 \\
	0 & \sigma_{2}^2
\end{psmallmatrix} \right)$, 
where the standard deviations is drawn by $\sigma_{1}^2,\sigma_{2}^2\sim \text{Unif}(0.1,1)$ and all other model coefficients are drawn according to $
\xi_{11},\xi_{12},\xi_{21},\xi_{22},\delta_{11},\delta_{12},\delta_{21},\delta_{22},\mu_1,\mu_2 \sim \text{Unif}(-2,2)$. 
The hidden confounding induces dependence between the collapsed noise variables $U_X = \delta^\t H + N_X$ and $U_Y = \mu^\t H + N_Y$, which we capture by a normalized cross covariance vector $\rho := \Sigma_{U_X}^{-1/2} \Sigma_{U_XU_Y} \Sigma_{U_Y}^{-1/2}\in \R^2$, where $\Sigma_{U_X} = \text{Var}(U_X)$, $\Sigma_{U_XU_Y}=\text{Cov}(U_X,U_Y)$ and $\Sigma_{U_Y}= \text{Var}(U_Y)$. As such, the degree of confounding can be explained by the norm of $\rho$ given by $
\| \rho\|_2^2 =  \Sigma_{U_YU_X} \Sigma_{U_X}^{-1}\Sigma_{U_XU_Y}/\Sigma_{U_Y} = \mu^\t \delta (\delta^\t \delta + \mathrm{diag}(\sigma_{1}^2,\sigma_{2}^2))^{-1}{\delta}^\t \mu /(\mu^\t \mu + 1)$. 
For each of the 10000 generated models we simulate $n=50$ observations and compute the PULSE and benchmark estimators and repeat this $N=5000$ times to estimate the performance measures. 

\begin{figure}[t] 
	\centering\includegraphics[width=\linewidth-50pt]{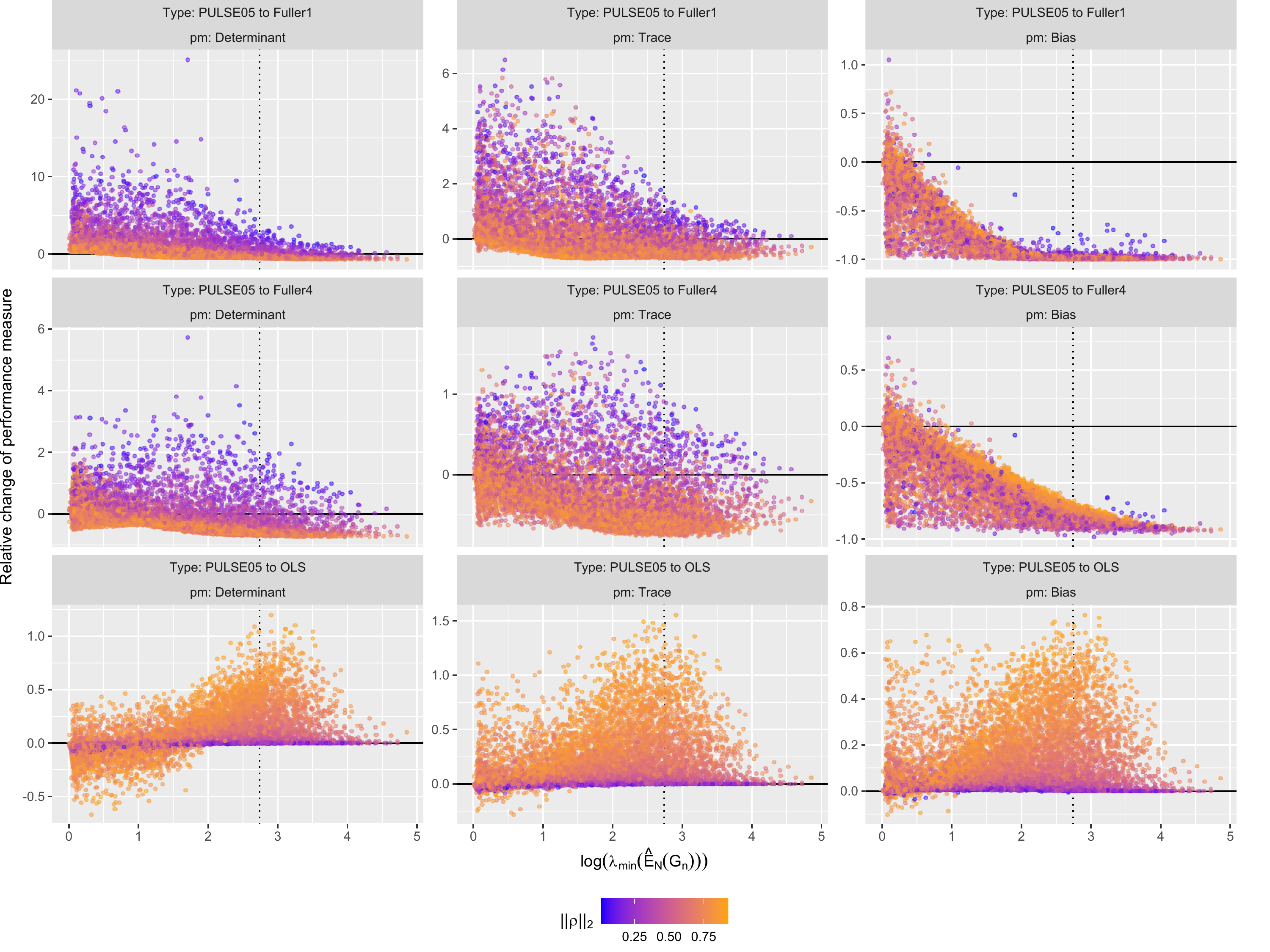}
	\caption{  \normalsize Illustrations of the relative change in the determinant (left) and trace (middle) of the MSE matrix and the Euclidean norm of the bias vector (right) (a positive relative change means that PULSE is better).  Each of the 10000 models corresponds to a point which is color-graded according the the value of $\|\rho\|_2$ (which indicates the strength of confounding), see \Cref{sec:ExpResultsEstCausalMultidim}. PULSE tends to outperform the Fuller estimators for weak instruments and weak confounding.	
		The vertical dotted line at $\log(15.5)$ corresponds to a rejection threshold for weak instruments based on relative change in bias for Fuller estimators \citep[][Table 5.3]{stock2002testing}. Note that the lowest possible negative relative change is $-1$.  }\label{fig:AllRandom_Beta00}
\end{figure}

\Cref{fig:AllRandom_Beta00} shows 
the relative change in the determinant and trace of the MSE matrix and the Euclidean norm of the bias vector.  
Similarly to the univariate setup, 
PULSE seems to perform better than Fuller(1) and Fuller(4) in terms of the determinant and trace for 
settings with weak
confounding (small $\|\rho\|_2$)
and weak instruments (small $\lambda_{\min}(\hat E_NG_n)$).
Most of the MSE matrices do not allow for an ordering:
PULSE is MSE superior to Fuller(1), 
Fuller(4), and  
OLS 
in 
9.2\%, 4.6\% and $1\%$ of the cases, while 
the MSE matrices are not ordered  
in 
90.8\%, 95.4\% and 95.8\% of the cases. Note that both Fuller(1) and Fuller(4) is never MSE superior to PULSE.
In contrast to the univariate setup, there are  models with very  weak instruments for which Fuller outperforms PULSE; these models seems to be exclusively with strong confounding.
We also see models with strong confounding and moderate to strong instrument strength where the PULSE is superior and models with weak confounding where PULSE is inferior. Hence, the degree of confounding $\|\rho\|_2$  does not completely characterize whether or not PULSE is superior to the Fuller estimators in terms of MSE performance measures in the multi-dimensional setting.
In regards to the bias we see that both Fuller estimators are less biased than PULSE for all but a few models with very weak instruments. Furthermore, PULSE is for models with strong confounding less biased than OLS but has comparable bias for models with small to moderate confounding.

We also conducted the above simulation experiment for $\gamma=(1,1)$ and $\gamma=(-1,1)$. The results (not shown but available in the folder 'Plots' in the code repository) are similar to the case $\beta = (0,0)$ and the above observations still apply. \Cref{sec:AppFigs} shows the results of additional experiments,
where we consider, e.g., PULSE with $p_{\min} = 0.1$.

\subsubsection{Fixed confounding multivariate experiment.}
In the varying confounding experiment, we saw that when  $\|\rho\|_2$ is small then the majority of the simulated models had PULSE superior to Fuller(1) and Fuller(4) in terms of the determinant and trace of MSE. However, we also saw models with large $\|\rho\|_2$ where PULSE was still superior and models with small $\|\rho\|_2$ where PULSE was inferior. 
In this experiment, we will investigate this further by fixing the confounding strength $\|\rho\|_2$ and investigating other model aspects that affect which estimator is superior. That is, we consider models with structural assignments given by
\begin{align*}
	A := N_A \in \mathbb{R}^2, \quad 
	X :=  \xi^\t A +U_X\in \mathbb{R}^2 ,  \quad 
	Y :=  \gamma^\t X + U_Y \in \mathbb{R},
\end{align*}
for some $\xi = \in \R^{2\times 2}$ and independent noise innovations $(N_A,(U_X,U_Y))$. We let $\gamma=(0,0)$ and fix the noise innovations for $A$ with distribution $N_A \sim  \mathcal{N} (0, I )$. We let
\begin{align*}
	\begin{pmatrix}
		U_X \\ U_Y 
	\end{pmatrix} \sim  \cN \lp \begin{pmatrix}
		0 \\ 0 \\ 0 
	\end{pmatrix}, \begin{pmatrix}
		1 & \eta & \phi_1 \\
		\eta & 1 & \phi_2\\
		\phi_1 & \phi_2 & 1
	\end{pmatrix}\rp,
\end{align*}
for some $\eta,\phi_1,\phi_2\in[0,1)$. With this noise structure we have that
$
\|\rho \|_2^2 = (\phi_1^2 +\phi_2^2-2\eta \phi_1\phi_2)/(1-\eta^2),
$
and when $\phi=\phi_1=\phi_2$ it holds that $\|\rho \|_2^2  =  2\phi^2/(1+\eta)$.
We randomly generate 5000 copies of $\xi$ with each entry drawn by $\text{Unif}(-2,2)$ distribution. For each model, that is, each combination of selected noise-parameter values and $\xi$, we simulate $n=50$ observations and compute the estimators. This is repeated $N=5000$ times to estimate the performance measures.

\begin{figure}[htp] 
	\centering\includegraphics[width=\linewidth]{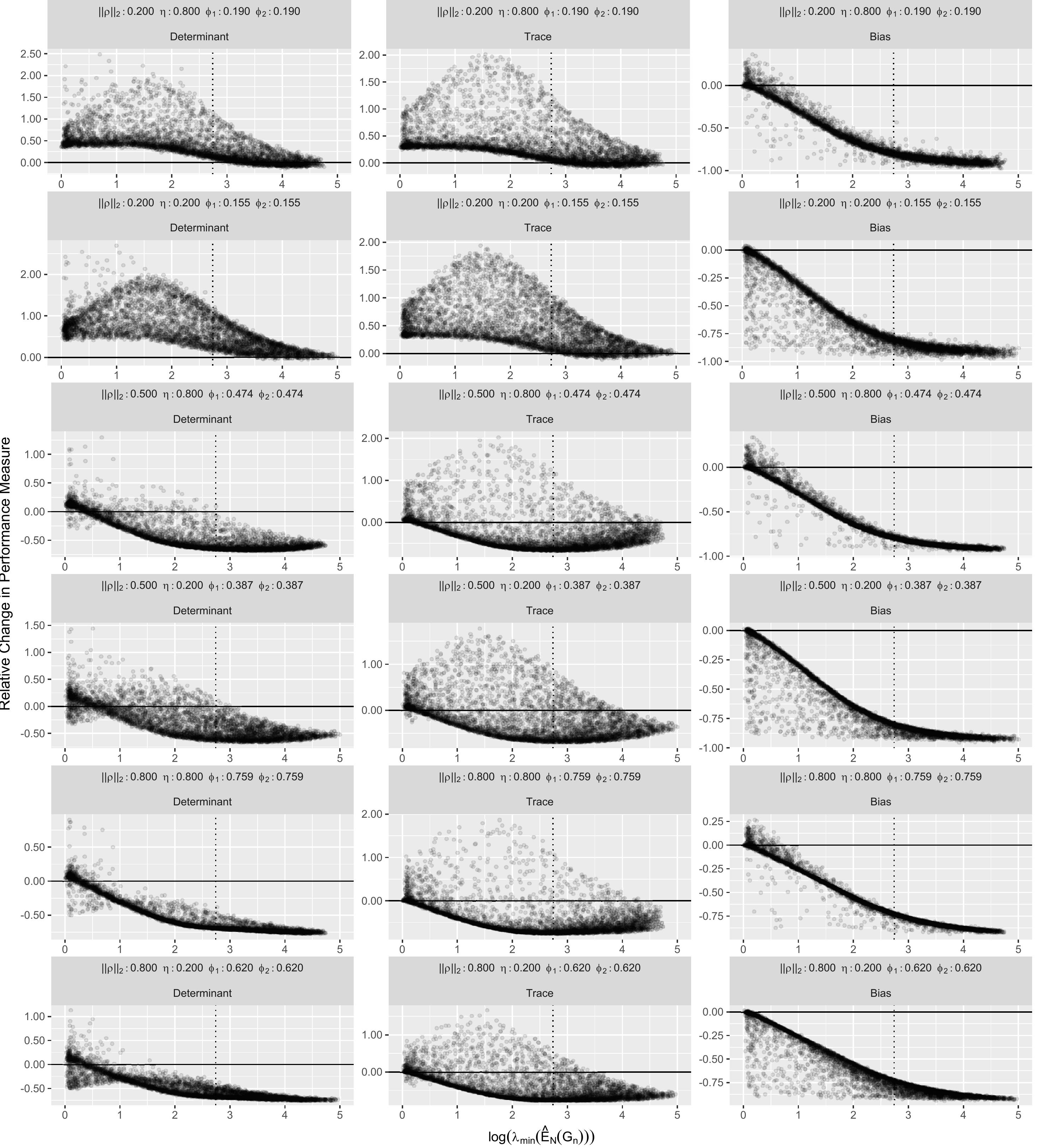}
	\caption{  \normalsize Illustrations of the relative change from PULSE to Fuller(4) in the determinant and trace of the MSE matrix and the Euclidean norm of the bias vector. 	 The vertical dotted line at $\log(15.5)$ corresponds to a rejection threshold for weak instruments based on relative change in bias for Fuller estimators \citep[][Table 5.3]{stock2002testing}.  }\label{fig:AllRandomFixedNoiseCorr}
\end{figure}

In \Cref{fig:AllRandomFixedNoiseCorr} we have illustrated the relative change in the performance measures when comparing 
PULSE to Fuller(4). For setups with weak confounding ($\|\rho\|_2=0.2$), it is seen that if instruments are sufficiently weak ($\lambda_{\min}(\hat{E}_N(G_n))\leq 15.5$), then PULSE is superior to Fuller(4) in terms of both the determinant and trace performance measures. For setups with larger confounding there are still models where PULSE is superior but the characterization of superiority by weakness of instruments is no longer valid.  

In \Cref{tbl:parametervalues} the percentage of models for which PULSE is superior to Fuller(4) in terms of the MSE partial ordering, determinant and trace performance measures is presented. It is seen that setups with identical $\|\rho\|_2$ does 
not yield
similar comparisons between PULSE and Fuller(4). 

\begin{table}[ht]
	\caption{MSE superiority} \label{tbl:parametervalues}
	\begin{center}
		\begin{tabu}to \textwidth { X[c] X[c] X[c] X[c] | X[r] X[r] X[c]}
			\toprule \toprule
			\multicolumn{4}{c}{Model Parameters}  & \multicolumn{3}{c}{PULSE Superiority (\%)}  \\ 
			$\|\rho\|_2$ & 	$\eta$ & $\phi_1$ & $\phi_2$ & MSE & determinant & trace\\ \midrule
			0.20 & 0.80 & 0.19 & 0.19 & 48.46 & 85.52 & 86.74 \\ 
			0.20 & 0.20 & 0.15 & 0.15 & 32.34 & 98.66 & 92.16 \\ 
			0.50 & 0.80 & 0.47 & 0.47 & 1.60 & 13.04 & 19.80 \\ 
			0.50 & 0.20 & 0.39 & 0.39 & 0.76 & 19.86 & 27.86 \\ 
			0.80 & 0.80 & 0.76 & 0.76 & 0.14 & 7.48 & 12.80 \\ 
			0.80 & 0.20 & 0.62 & 0.62 & 0.06 & 7.64 & 15.50 \\
			\bottomrule\bottomrule
		\end{tabu}
	\end{center}
	\footnotesize
	\renewcommand{\baselineskip}{11pt}
	\textbf{Note:} The rows show different  noise-parameter values for the different experimental setups. The last three columns describe the percentage of models (out of the 5000 randomly generated models) for which PULSE (with $p_{\min}=0.05$) is superior to Fuller(4) in terms of the MSE partial ordering, determinant and trace performance measures. Whenever PULSE is not superior to Fuller(4) in terms of the MSE partial ordering the MSE matrices are not comparable.
\end{table}

For any two setups with identical confounding strength $\|\rho\|_2$ we see that decreasing $\eta$ yields a larger percentage of models for which PULSE is superior in terms of the determinant and trace. Furthermore, we see that decreasing $\|\rho\|_2$ (for fixed $\eta$) has a similar effect. Thus, it seems that both $\rho$ and $\eta$ negatively influences the size of the parameter space of $\xi$ for which PULSE is superior to Fuller(4) in terms of both the determinant and trace performance measures. However, superiority with respect to the partial ordering of the MSE matrices does not exhibit similar behaviour. Decreasing $\|\rho\|_2$ (for fixed $\eta$) still leads to a percentage increase but decreasing $\eta$ (for fixed $\|\rho\|_2$) leads to a percentage decrease, of models for which PULSE is superior to Fuller(4).

\newpage	
\subsection{Under-identified setup} \label{app:underidentifiedexperiment}
In an under-identified setup the causal parameter is not identified by instrumental variable methods. Instead the usual two stage least square procedure, $\argmin_\alpha l_\mathrm{IV}(\alpha)$, yields an entire linear solution space of coefficients that renders the regression residuals uncorrelated with the instruments. The causal coefficient lies within this solution space but we are unable to identify it. In the under-identified setup, the population PULSE coefficient is the point in the solution space which provides the best mean squared prediction error. That is, the population PULSE coefficient is given by
\begin{align*}
	\alpha^*= \argmin_{\alpha: E[A(Y-Z\alpha)]=0} E[(Y-Z\alpha)^2] = \argmin_{\alpha:l_{\mathrm{IV}}(\alpha)=0}l_{\mathrm{OLS}}(\alpha).
\end{align*}
The PULSE estimator in the under-identified setup remains unchanged from the exposition in the main paper. 
	Here, the function $l_{\mathrm{IV}}^n$ does not have a unique solution but we can define a modified TSLS estimator 
\begin{align*}
	\hat \alpha_{\mathrm{TSLS.mod}}^n := \lim_{\kappa\uparrow 1}\alpha_{\mathrm{K}}^n(\kappa) = \argmin_{\alpha: l_{\mathrm{IV}}^n(\alpha)=0} l_{\mathrm{OLS}}^n (\alpha).
\end{align*}
The modified TSLS estimator is the minimum of a quadratic  
function
subject to a feasible linear constraint, and can be computed efficiently using QP solvers. 
\subsubsection{Underindentified example} \label{sec:underIDexample}
\begin{figure}[]
	\centering\includegraphics[scale=1]{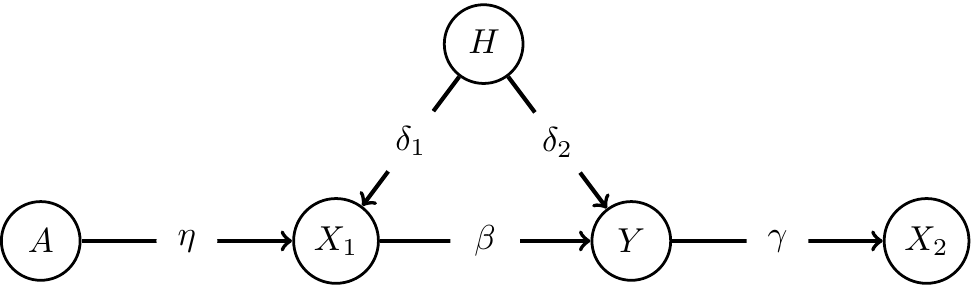}
	\caption{Causal graph of the under-indentified setup in \Cref{sec:underIDexample} Here, $H$ is hidden and the causal parameter $\beta$ is, in general, not identifiable from the distribution over $(A, X_1, X_2, Y)$. 
		Existing methods in machine learning try to find invariant sets of covariates (i.e., sets $S$ that, after regressing $Y$ on $X_S$, yield residuals which are uncorrelated with $A$). In this example, no such set exists. PULSE finds a solution and outputs a vector with non-zero coefficients for $X_1$ and $X_2$.}
	\label{fig:underidcausalgraph}
\end{figure}

Consider an under-identified setup with structural assignments given by
\begin{align*}
	A &:= \ep_A, \quad \quad H := \ep_H, \quad \quad X_1 := \eta A + \delta_1H + \ep_1,\\
	Y &:=	\beta X_1 + \delta_2 H + \ep_Y,\quad \quad X_2 := \gamma Y + \ep_2,
\end{align*}
with $(\ep_A,\ep_H,\ep_Y,\ep_1,\ep_2)\sim \cN(0,I_5)$. The causal graph of this structural equation model is illustrated in \Cref{fig:underidcausalgraph}. 
In general, the causal parameter $\beta$ is not identifiable.
Existing methods \citep[e.g.,][]{Peters2016jrssb, rojas2018invariant, Pfister2019stab} propose to look for invariant sets that yield residuals which are uncorrelated with $A$ after regressing $Y$ on that set. In general, because of the hidden variable $H$, no such sets exist either. 
The best predictive model under all invariant models, however, is still well-defined. 
To see this, let us derive the population PULSE coefficient 
\begin{align*}
	\alpha^* = \argmin_{\alpha: l_{\mathrm{IV}}(\alpha)=0} E[(Y-\alpha_1X_1-\alpha_2X_2)^2] .
\end{align*}
We know that a necessary and sufficient condition for $l_{\mathrm{IV}}(\alpha)=0$ is that $\Corr(Y-\alpha_1X_1 -\alpha_2X_2,A)=0$. We have
\begin{align*}
	Y-\alpha_1X_1-\alpha_2X_2 &=  Y-\alpha_1X_1-\alpha_2(\gamma Y + \ep_2) \\
	&= (1-\alpha_2 \gamma) (\beta X_1 + \delta_2 H + \ep_Y) -\alpha_1X_1-\alpha_2 \ep_2 \\
	&= (\beta-\alpha_1-\alpha_2\gamma \beta) X_1 + (1-\alpha_2\gamma)\delta_2 H + (1-\alpha_2\gamma)\ep_Y - \alpha_2 \ep_2.
\end{align*}
As $\eta\not = 0$, the regression residuals are uncorreleted with $A$ if and only if $  \alpha_1 = (1-\alpha_2\gamma) \beta$. Hence, 
\begin{align*}
	\alpha^* &= \argmin_{\alpha: \alpha_1 = (1-\alpha_2\gamma)\beta} E[((1-\alpha_2\gamma)\delta_2 H + (1-\alpha_2\gamma)\ep_Y - \alpha_2 \ep_2)^2] \\
	&= \argmin_{\alpha: \alpha_1 = (1-\alpha_2\gamma)\beta} (1-\alpha_2\gamma)^2\delta_2^2 \mathrm{Var}(H) + (1-\alpha_2\gamma)^2 \mathrm{Var}(\ep_Y) + \alpha_2^2 \mathrm{Var}(\ep_2).
\end{align*}
The latter function is convex in $\alpha_2$, so the minimum is attained in a stationary point. We have that
\begin{align*}
	&\frac{\partial}{\partial \alpha_2}   (1-\alpha_2\gamma)^2\delta_2^2 \mathrm{Var}(H) + (1-\alpha_2\gamma)^2 \mathrm{Var}(\ep_Y) +\alpha_2^2 \mathrm{Var}(\ep_2) \\
	&= 2\left[ \alpha_2(\mathrm{Var}(\ep_2)+\gamma^2\delta_2^2 \mathrm{Var}(H) + \gamma^2 \mathrm{Var}(\ep_Y)) - \gamma \delta_2^2 \mathrm{Var}(H) - \gamma\mathrm{Var}(\ep_Y)    \right]=0,
\end{align*}
if and only if 
\begin{align*}
	\alpha_2(\mathrm{Var}(\ep_2)+\gamma^2\delta_2^2 \mathrm{Var}(H) + \gamma^2 \mathrm{Var}(\ep_Y)=   \gamma \delta_2^2 \mathrm{Var}(H) + \gamma\mathrm{Var}(\ep_Y).
\end{align*}
Hence,
\begin{align} \label{eq:alst}
	\alpha_2^* &= \frac{(\mathrm{Var}(\ep_Y)+\delta_2^2\mathrm{Var}(H))\gamma }{\mathrm{Var}(\ep_2)+(\mathrm{Var}(\ep_Y)+\delta_2^2\mathrm{Var}(H))\gamma^2} = \frac{(1+\delta_2^2)\gamma}{1+(1+\delta_2^2)\gamma^2 };
	\quad  \alpha_1^* = (1-\alpha_2^*\gamma)\beta.
\end{align}

We now generate 
models by randomly drawing the model coefficients using $\alpha \sim \mathrm{Unif}(1,2)$, $\delta_1 \sim \mathrm{Unif}(1,2), \delta_2 \sim \mathrm{Unif}(1,2), \gamma \sim \mathrm{Unif}(1,2)$ and $\eta \sim \mathrm{Unif}(0.1,1)$
and compute the corresponding population quantities according to Equation~\eqref{eq:alst}.

\begin{figure}[t]
		\begin{center}
			\includegraphics[width=\linewidth]{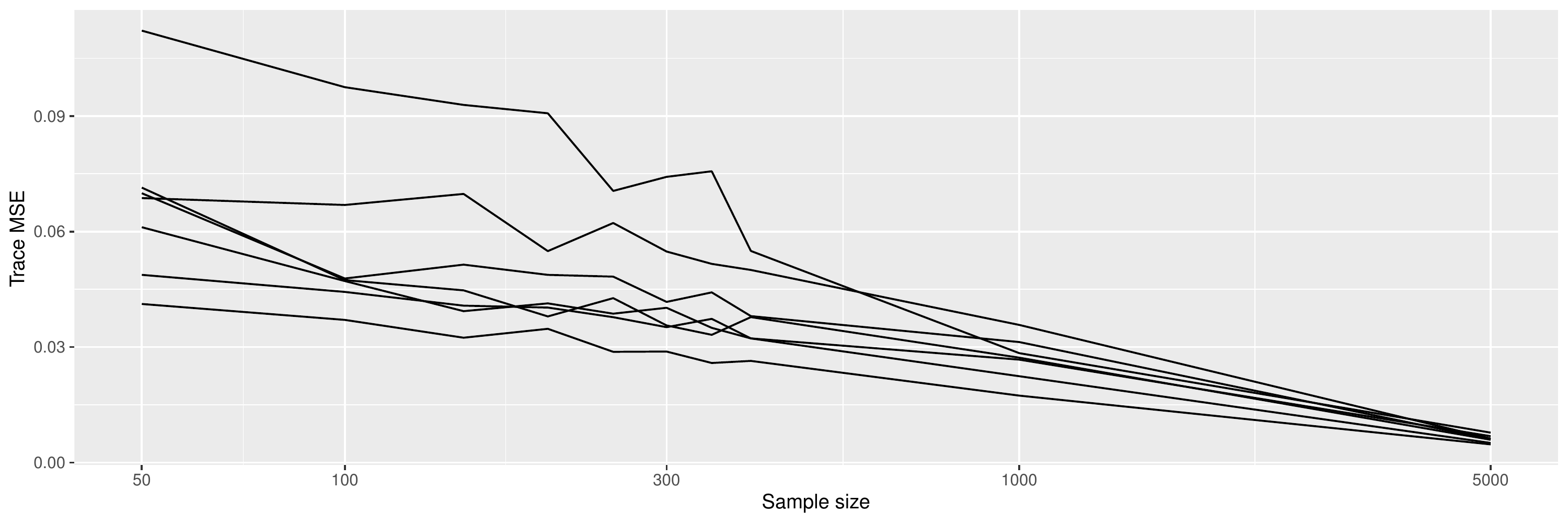}
		\end{center}
		\caption{Illustration of the trace of the estimated MSE matrix of the PULSE estimator in the under-identified setup based on 100 repetitions. PULSE converges towards the population quantities computed in Equation~\eqref{eq:alst}.} \label{fig:underidConvergence}
	\end{figure}

For different sample sizes, we then simulate data sets from such models and compute the PULSE estimator.
\Cref{fig:underidConvergence} shows the trace of the estimated MSE of the PULSE estimator (with $p_{\min}=0.05$) 
when comparing to the population quantity derived above.
For each model and sample size, the MSE is estimated based on 100 repetitions. As sample size increases, the MSE indeed approaches the population quantity.

	As a comparison, we also implemented the TSLS modification from Equation~\eqref{eq:alst}. %
	Similarly to the identified setups, the TSLS modification may come with poor finite sample properties, in particular for weak instruments and small sample size. Indeed, in this example
	we observe that PULSE has superior MSE properties for small sample sizes. For example, the trace MSE for the PULSE estimator is on average (over 1000 random models) 50\%  lower than the trace MSE of the modified TSLS estimator for a sample size of 50.
	
	\section{Empirical applications} 
	\label{app:EmpricalApp}
	\setcounter{figure}{0}
	\setcounter{table}{0}
	\setcounter{equation}{0}
	\medskip
	
	We now consider three classical 
	instrumental variable applications (see \citet{albouy2012colonial} and \citet{buckles2013season} for discussions on the underlying assumptions). 
	\begin{itemize}
		\item[\ref{sec:QOB}] ``Does compulsory school attendance affect schooling and earnings?'' by \cite{angrist1991does}. This paper investigates the effects of education on wages. The endogenous effect of education on wages are remedied by instrumenting education on quarter of birth indicators.
		\item[\ref{sec:Proximity}] ``Using geographic variation in college proximity to estimate the return to schooling'' by \cite{card1993using}. This paper also investigates the effects of education on wages. In this paper education is instrumented by proximity to college indicator.
		\item[\ref{sec:Colonial}] ``The colonial origins of comparative development: An empirical investigation'' by \cite{acemoglu2001colonial}. This paper investigates the effects of extractive institutions (proxied by protection against expropriation) on the gross domestic product (GDP) per capita. The endogeneity of the explanatory variables are remedied by instrumenting protection against expropriation on early European settler mortality rates.
	\end{itemize}

	For each study, we replicate the OLS and TSLS estimates of these studies and provide in addition the corresponding Fuller(4) (see
	\Cref{sec:ExpResultsEstCausPerformanceMeasures}) 
	and PULSE estimates. 
	Since we do not have access to interventional data,
	we cannot directly test the distributional 
	robustness properties discussed in Section~\ref{sec:RobustnessOfKclass}.
	For the third study, however, the exogenous variable is continuous, which allows us to investigate 
	distributional robustness empirically 
	by holding out data points with extreme values of the exogenous variable and predict on these held-out data.

	For the remainder of this section we use the PULSE estimator with $p_{\min}=0.05$ and the test scaling-scheme that renders the test equivalent to the asymptotic version of the Anderson-Rubin test (see \Cref{sec:VanishCorr}). 
	Code replicating this analysis is available on GitHub.\footnote{\url{https://github.com/MartinEmilJakobsen/PULSE/tree/master/Empirical_Applications}}

	\subsection{\cite{angrist1991does}} \label{sec:QOB}
	The dataset of \cite{angrist1991does} consists, in part, of 1980 US census data of 329,509 men born between 1930--1939. The endogenous target of interest is log weakly wages and the main endogenous regressor is years of education is instrumented on year and quarter of birth indicators. We consider four models M1--M4 corresponding to the models presented in column (1)--(8) in Table 5 of \cite{angrist1991does}. Model M1 is given by the structural reduced form equations
	\begin{align*}
		\log \mathrm{weakly \, wage} &= \mathrm{educ}\cdot \gamma + \sum_{i} \mathrm{YR}_i \cdot \beta_i + U_1, \\
		\mathrm{educ} &= \sum_{i} \mathrm{YR}_i \cdot  \delta_i + \sum_{i,j} \mathrm{YR}_i\cdot \mathrm{QOB}_j \cdot \delta_{i,j} + U_2,
	\end{align*}
	where $\mathrm{educ}$ is years of education, $(\mathrm{YR}_i)$ is year of birth indicators and $(\mathrm{QOB}_j)$ is quarter of birth indicators. Model M2 is given by M1 with the additional included exogenous regressors of age and age-squared. Models M3 and M4 are given by model M1 and M2, respectively, with additional included exogenous indicators describing race, marital status, metropolitan area and eight regional indicators. All models are over-identified, instrumenting education on a total of 30 binary instruments.
	
	\Cref{tbl:QOB} shows the OLS and TSLS estimates, as well as the Fuller(4) and PULSE estimates for the linear effect of education on log weakly wages.  In all models the PULSE estimates coincide with the OLS estimates. 
	
	\begin{table}[h]
		\caption{\label{tbl:QOB}The estimated return of education on log weakly wage. }
		\begin{center}
			\begin{tabu}to \textwidth {r c c c c c c c}
				\toprule \toprule
				Model & OLS & TSLS & FUL & PULSE & Message & Test & Threshold \\
				\midrule
				M1 &0.0711 & 0.0891 & 0.0926 & 0.0711 & OLS Accepted & 26.9231 & 55.7585\\
				M2 &0.0711 & 0.0760 & 0.0739 & 0.0711 & OLS Accepted & 23.1512 & 55.7585\\
				M3 &0.0632 & 0.0806 & 0.0835 & 0.0632 & OLS Accepted & 23.7903 & 68.6693\\
				M4 & 0.0632 & 0.0600 & 0.0555 & 0.0632 & OLS Accepted & 19.5856 & 68.6693\\
				\bottomrule\bottomrule
			\end{tabu}
		\end{center}
		\footnotesize
		\renewcommand{\baselineskip}{11pt}
		\textbf{Note:} 
		Point estimates for the return of education on log weakly wage. The OLS and TSLS values coincide with the ones in
		Table V of \citet{angrist1991does}.
		The right columns show the values of the test statistic (evaluated in the PULSE estimates) and the test rejection thresholds.
		For all models, the OLS is accepted and the PULSE coincides with the OLS.
	\end{table}

	\subsection{\cite{card1993using}} \label{sec:Proximity}
	The dataset of \cite{card1993using} consists of a US National Longitudinal Survey of Young Men spanning from 1966 to 1981. The subset of interest consists of 3010 observations	
	for which there is recorded a valid wage and education level in a 1976 interview. 
	The endogenous target of interest is log hourly wages and the main endogenous regressor is years of education. 
	Proximity to a four year college, recorded in 1966, is used as an instrument. 
	We consider two models, M1 and M2, corresponding to models in Panel B, column (5) and (6) of Table 3 \citep[][]{card1993using}%
	, respectively. 
	Model M1 is given by regressing the target, log hourly wages, on included exogenous  indicators of race, metropolitan area and region; the included endogenous regressors are years of education, work-experience and work-experience-squared. The endogenous regressors are instrumented by the excluded exogenous variables age, age-squared and indicator of proximity to college.	In model M2, we have model M1 with the addition of several exogenous indicators of parents education level. %
	
	\Cref{tbl:Proximity} shows the OLS and TSLS estimates, as well as the Fuller(4) and PULSE estimates for the linear effect of education on log hourly wages. 
	Again, in all models the OLS estimates are not rejected by the Anderson-Rubin test. Hence, all PULSE estimates coincide with the OLS estimates. 
	\begin{table}[H]
		\caption{\label{tbl:Proximity} The estimated return of education on log hourly wages. }
		\begin{center}
			\begin{tabu}to \textwidth {r c c c c c c c}
				\toprule \toprule
				Model & OLS & TSLS & FUL & PULSE & Message & Test & Threshold \\
				\midrule
				M1 &0.0747 & 0.1224 & 0.1156   & 0.0747 & OLS Accepted & 1.2218 & 26.2962\\
				M2& 0.0726 & 0.1324 & 0.1283  & 0.0726 & OLS Accepted & 1.7095 & 43.7730\\
				\bottomrule\bottomrule
			\end{tabu}
		\end{center}
		\footnotesize
		\renewcommand{\baselineskip}{11pt}
		\textbf{Note:}  
		Point estimates for the return of education on log hourly wage. The OLS and TSLS values coincide with the ones shown in
		Table 3 of \citet{card1993using}.
		The right columns show the values of the test statistic (evaluated in the PULSE estimates) and the test rejection thresholds.
		For all models, the OLS is accepted and the PULSE coincides with the OLS.
	\end{table}

	\subsection{\cite{acemoglu2001colonial}} \label{sec:Colonial}
	In \Cref{sec:mainColonialApplication} of the main paper we describe the data and models of \cite{acemoglu2001colonial}. Furthermore, we replicate the OLS and TSLS estimates and presented the corresponding Fuller(4) and PULSE estimates.
	
	To investigate 
	distributional robustness,
	we conduct an out-of-sample 
	mean squared prediction error (MSPE) analysis on a mean-centered dataset of the just-identified identified model M1.  
	This is the simplest model proposed in \cite{acemoglu2001colonial} but the MSPE robustness property of \Cref{sthm:TheoremIntRobustKclas} 
	is robust to model misspecifications; see \Cref{rm:ModelMispecification}. 
	We do not have access to interventional data.
	Instead, 
	for different values of $n_{\mathrm{test}} \in \mathbb{N}$, that is, 
	for each $n_{\mathrm{test}}\in\{4,8,...,32\}$, 
	we remove the data points with the  $n_{\mathrm{test}}/2$ lowest and $n_{\mathrm{test}}/2$ highest settler mortality rates.
	We then fit the OLS, TSLS, PULSE and Fuller(4) on the remaining $64-n_{\mathrm{test}}$ observations and compute the out-of-sample MSPE on the $n_{\mathrm{test}}$ held-out observations,
	measuring the model's ability to generalize.
	
	The instrument has a larger variance on the held-out data and the population robustness property of K-class estimators (see \Cref{sthm:TheoremIntRobustKclas})
	suggests 
	that PULSE and Fuller(4) 
	might generalize slightly better than OLS or TSLS.\footnote{Here, we consider a just-identified model, so the Fuller(4) K-class parameter $\kappa$ lies in $(0,1)$.} 
	The results of this analysis is summarised in  \Cref{tbl:MSPECOLONIAL}. 
	Indeed, we see that the 
	OLS is optimal for a small number of held-out data points (when little generalization is required) 
	and that for an
	increasing 
	number of held-out data points, 
	PULSE and FULLER(4) outperform the other estimators in terms of MSPE.

	For comparison, we also consider
	random sample splits, i.e., taking out a random subset of the dataset rather.
	Here, no generalization is required and as expected, 
	OLS performs better than the other estimates, see \Cref{tbl:MSPECOLONIALrandomsplits}. 
	The MSPE is minimized by 
	OLS, PULSE, Fuller(4), and TSLS 
	in 
	65.9\%, 21.8\%, 6.1\%, and  6.2\% of the cases, respectively.

	\begin{table}[htp]
		\caption{\label{tbl:MSPECOLONIALrandomsplits} log GPD MSPE orderings on random sample splits. }
		\begin{center}
			\begin{tabu}to \textwidth {r | r r r r }
				\toprule \toprule
				MSPE&\multicolumn{4}{c}{Outperforms} \\ 
				&  OLS & PULSE & FUL  & TSLS  \\
				\midrule
				OLS & \xmark &65.9\% &  79.7\% & 85.3\% \\
				PULSE & 34.1\% &\xmark & 87.7\% & 90.5\% \\
				FUL & 20.3\% & 12.3\%&\xmark & 93.8\% \\
				TSLS & 14.7\% & 9.5\% & 6.2\% & \xmark \\
				\bottomrule\bottomrule
			\end{tabu}
		\end{center}
		\footnotesize
		\renewcommand{\baselineskip}{11pt}
		\textbf{Note:} 
		The table shows generalization performance for different estimators on model M1 of \cite{acemoglu2001colonial}. 
		The data set is split randomly into a subset of	
		90\% of the data (that is, 58 observations) and the MSPE for the OLS, PULSE, Fuller(4), and TSLS 
		are calculated on the remaining 10\% of the data. 
		This procedure is repeated 1000 times. 
		The table shows how often the estimators outperform each other. E.g., 
		OLS has lower MSPE than TSLS in 85.3\% of the cases. 
		Here, no generalziation is needed 
		and, as expected, the OLS performs best.
	\end{table}	
	\begin{landscape}
		\hfill
		\begin{table}[H]
			\caption{\label{tbl:MSPECOLONIAL} log GPD MSPE on extreme out-of-sample instrument observations. }
			\begin{center}
				\begin{tabu} to \textwidth {r | c c c c| c c  | c c c c } 
					\toprule \toprule 
					& \multicolumn{4}{c|}{Estimated coefficient} & \multicolumn{2}{c|}{K-class $\kappa$} & \multicolumn{4}{c}{MSPE}   \\ 
					$n_{\mathrm{test}}$&OLS&TSLS&PULSE&FUL&PULSE&FUL&OLS&TSLS&PULSE&FUL\\
					\midrule
					4 & 0.5015 & 1.1592 & 0.7852 & 0.9509 & 0.8286 & 0.9322  & \bt0.2072 & 2.0358 & 0.3211 & 0.8613 \\
					6 & 0.5113 & 0.9441 & 0.6590 & 0.8313 & 0.7075 & 0.9298  & \bt0.8282 & 1.5889 & 0.8692 & 1.2034 \\
					8 & 0.5017 & 0.9433 & 0.6287 & 0.8150 & 0.6781 & 0.9273  & 0.7800 & 1.5331 & \bt0.7796 & 1.0961 \\
					10 & 0.4978 & 0.8795 & 0.5810 & 0.7717 & 0.5733 & 0.9245 & 0.7018 & 1.0850 & \bt0.6769 & 0.8479 \\
					12 & 0.4901 & 0.8693 & 0.5390 & 0.7512 & 0.4407 & 0.9216 & 0.6605 & 1.0346 & \bt0.6357 & 0.7788 \\
					14 & 0.4748 & 0.8439 & 0.4748 & 0.7091 & 0.0000 & 0.9184 & \bt0.6562 & 0.8910 & \bt0.6562 & 0.6722 \\
					16 & 0.4581 & 0.7655 & 0.4581 & 0.6359 & 0.0000 & 0.9149 & 0.7290 & 0.7581 & 0.7290 & \bt0.6573 \\
					18 & 0.4247 & 0.6861 & 0.4247 & 0.5451 & 0.0000 & 0.9111 & 0.7476 & \bt0.6263 & 0.7476 & \bt0.6263 \\
					20 & 0.3883 & 0.8604 & 0.3883 & 0.6096 & 0.0000 & 0.9070 & 0.8886 & 0.8354 & 0.8886 & \bt0.6632 \\
					22 & 0.3789 & 0.8867 & 0.3789 & 0.6046 & 0.0000 & 0.9024 & 0.8285 & 0.8315 & 0.8285 & \bt0.6072 \\
					24 & 0.3784 & 0.7016 & 0.3784 & 0.5450 & 0.0000 & 0.8974 & 0.9152 & \bt0.7251 & 0.9152 & 0.7334 \\
					26 & 0.4156 & 0.8753 & 0.5240 & 0.6723 & 0.6682 & 0.8919 & 0.8794 & 1.0333 & \bt0.7957 & 0.8012 \\
					28 & 0.4155 & 0.7867 & 0.4676 & 0.6306 & 0.4789 & 0.8857 & 0.8340 & 0.8530 & 0.7880 & \bt0.7468 \\
					30 & 0.4016 & 0.8725 & 0.4710 & 0.6278 & 0.5754 & 0.8788 & 0.7989 & 0.9223 & 0.7370 & \bt0.6991 \\
					32 & 0.4087 & 0.9103 & 0.4893 & 0.6228 & 0.6344 & 0.8710 & 0.7823 & 0.9880 & 0.7225 & \bt0.7016 \\
					\bottomrule\bottomrule
				\end{tabu}
			\end{center}
			\footnotesize
			\renewcommand{\baselineskip}{11pt}
			\textbf{Note:} 
			The table shows generalization performance for different estimators on model M1 of \cite{acemoglu2001colonial}. 
			We remove the $n_{\mathrm{test}}$ 
			observations with the most extreme values of settler mortality, 
			fit 
			OLS, TSLS, PULSE and Fuller(4)
			on the $64-n_{\mathrm{test}}$ samples, and compute the MSPE on the $n_{\mathrm{test}}$ held-out samples (four right-most columns). 
			Indeed, in particular for larger values of $n_{\mathrm{test}}$, where more generalization is needed, PULSE and Fuller(4) outperform OLS and TSLS in the majority of cases. 
			The columns ``Estimated coefficient'' show the estimates for the linear effect of average expropriation risk on log GPD of each estimation method. 
			The column ``K-class $\kappa$'' shows K-class $\kappa$ parameters for both the PULSE and Fuller(4) estimates; EQ is computed  according to  $\mathrm{Var}(\mathrm{\text{out-of-sample}})=\mathrm{Var}(\mathrm{\text{in-sample}})/(1-\kappa_{\mathrm{EQ}})$.  
		\end{table}
		
	\end{landscape}
	\newpage

	\newpage
	\section{Weak instruments} 
	\label{sec:WeakInst}
	\setcounter{figure}{0}
	\setcounter{table}{0}
	\setcounter{equation}{0}
	\medskip	
	There is a wide variety of attempts to quantify weakness of instruments, see e.g.\ \citet{andrews2019weak}  and  \citet{stock2002survey} for an overview.  Heuristically, the presence of weak instruments in a instrumental variable setup refers to the notion that the causal effects of the instruments onto regressors are weak relative to the noise variance of the regressors. This strength of the instruments has direct effects on the finite sample behavior of instrumental variable estimators. For simplicity consider a mean zero collapsed causal structural model with no included exogenous variables entering the equation of interest, that is,
	\begin{align} \label{eq:weakInstrumentSetup}
		\begin{split}
			Y= \gamma^\t X + U_Y, \qquad X= \xi^\t A + U_X,
		\end{split}
	\end{align}
	where $A\in \R^q$ are the collection of exogenous variables and the noise variables $U_X$ and $U_Y$ are possibly correlated. Let $\fA,\fX,\fY$ be a  $n$-sample data matrices of i.i.d. realizations of the system in \Cref{eq:weakInstrumentSetup}. A key statistic used to quantify weakness of instruments is the concentration matrix  given by
	$ \mu_n = \Sigma_{U_X}^{-1/2} \xi^\t \fA^\t \fA \xi \Sigma_{U_X}^{-1/2}$,
	where $\Sigma_{U_X}$ is the variance matrix of $U_X$. This statistic turns up in numerous different aspect of the finite sample properties of the two-stage least square estimator. 
	\Citet{rothenberg1984approximating} argues
	that the one-dimensional analogue of $ \mu_n$ under deterministic instruments and normal distributed noise variables directly influences the goodness of approximating a finite sample standardized two-stage least square estimator by its Gaussian asymptotic distribution. 
	He argues that for large concentration parameters the Gaussian approximation is good. 
	The concentration parameter can also be connected to approximate bias of the two-stage least squares estimator. Under assumptions similar to the above,  \Citet{nagar1959bias} showed that an approximate (to the order of $\mathcal{O}(n^{-1})$) finite sample bias of the two-stage least square estimator is inversely proportional to $ \mu_n$. Note that the concentration matrix $\mu_n$ is not observable, but may be approximated by
	$\hat \mu_n = \hat{\Sigma}_{U_X}^{-1/2}\fX^\t P_{\fA} \fX \hat{\Sigma}_{U_X}^{-1/2}$,
	where  $\hat{\Sigma}_{U_X} =  \frac{1}{n-q} \fX^\t P_{\fA}^\perp \fX$ is an estimator of the variance matrix of $U_X$ and $P_\fA \fX$ is the ordinary least square prediction of $\fA \xi$.   Now define 
	\begin{align*}
		G_n := \frac{\hat \mu_n}{q}=\frac{ \hat{\Sigma}_{U_X}^{-1/2}\fX^\t P_{\fA} \fX \hat{\Sigma}_{U_X}^{-1/2}}{q},
	\end{align*}
	which can be seen
	as a multivariate
	first-stage $F$-statistic for testing the hypothesis $H_0:\xi=0$. That is, when $X\in \R$, then $G_n = \frac{n-q}{q} \frac{\fX^\t P_\fA \fX }{\fX^\t \fX - \fX^\t P_\fA \fX }$  
	is recognized as the F-test for testing $H_0$. \citet{stock2002testing} propose to reject the
	hypothesis of a presence of weak instruments if the test-statistic $\lambda_{\min}(G_n)$, the smallest eigenvalue of $G_n$, is larger than a critical value that, for example, depends on 
	how much bias you allow your estimator to have.  Prior to this $G_n$ had been  used to test under-identifiability in the sense that the concentration matrix is singular \citep[][]{cragg1993testing}, while the former uses a small minimum eigenvalue of $G_n$  as a proxy for the presence of weak instruments in identified models. From the work of \citet{staiger1997instrumental} a frequently appearing rule of thumb for instruments being non-weak is that the \textit{F}-statistic $G_n$ ($\lambda_{\min}(G_n)$ in higher dimensions) is larger than 10.  A more formal justification of this rule is due to  \citet{stock2002testing} who showed (under weak-instrument asymptotics) that it approximately (in several models) corresponds to a 5\% significance test that the bias of TSLS is at most 10\% of the bias of OLS.

	We can, under further model simplification, strengthen the intuition on how the concentration matrix $G_n$ and especially the minimum eigenvalue $\lambda_{\min}(G_n)$ governs the weakness of instruments. To this end assume that $\text{Var}(U_X)= \Sigma_{
		U_X}=I$  and note that $\hat \mu_n$ is approximately proportional to the Hessian of the two-stage least squares objective function. That is, $\hat \mu_n  \approx \Sigma_{U_X}^{-1/2} \fX^\t P_{\fA} \fX \Sigma_{U_X}^{-1/2} = \fX^\t \fA ( \fA^\t \fA)^{-1} \fA^\t \fX \propto H(l_{\text{IV}}^n)$. Hence, we have that $\lambda_{\min}(G_n)$ is approximately proportional to the curvature of two-stage least squares objective function in the direction of least curvature. Thus, if $\lambda_{\min}(G_n)$ is small, 
	then, heuristically, 
	the objective function $l_{\text{IV}}^n$ has weak identification in the direction of the corresponding eigenvector. That is, changes to the point estimate of $\beta$ away from the two-stage least square solution in this direction does not have a strong effect on the objective value.
	Finally, the weak instrument problem is a small sample problem. To this end note that 
	$
	n^{-1} G_n= n^{-1}\hat{\Sigma}_{U_X}^{-1/2} \fX^\t P_{\fA} \fX \hat{\Sigma}_{U_X}^{-1/2} \convp \text{Var}(U_X)^{-1/2} \xi^\t \text{Var}(A)^{-1} \xi  \text{Var}(U_X)^{-1/2}$, 
	hence
	by the continuity of the minimum eigenvalue operator, we have that $\lambda_{\min}(G_n) \convp \i$.	
	\section{Additional simulation experiments} 
	\label{sec:AppFigs}
	
	\setcounter{figure}{0}
	\setcounter{table}{0}
	\setcounter{equation}{0}
	
	\subsection{Additional illustrations for the univariate experiment}

	\begin{figure}[H] 
		\centering\includegraphics[width=\linewidth-70pt]{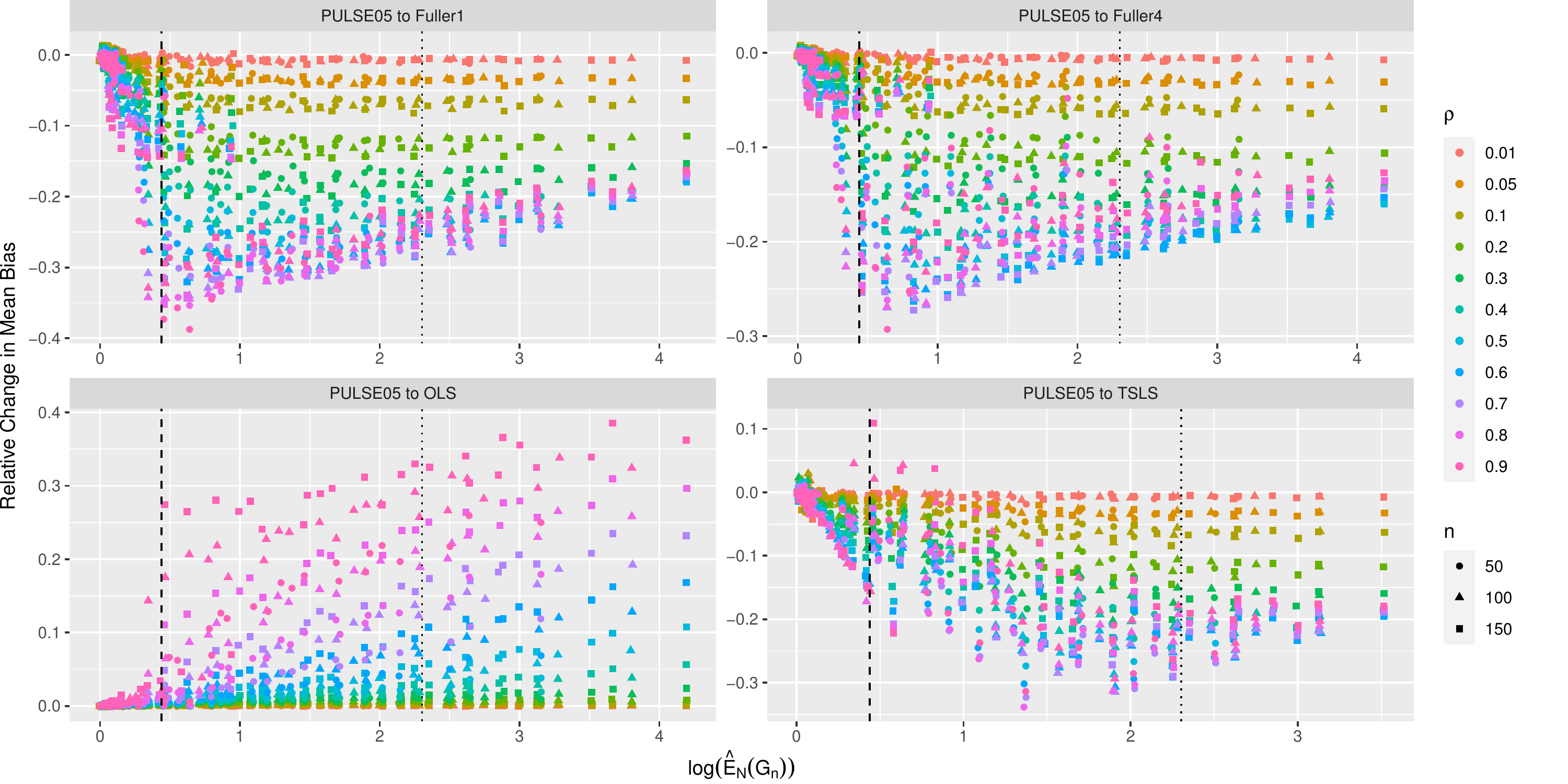}
		\caption{  \normalsize Illustrations of the relative change in the absolute value of the mean bias (a positive relative change means that PULSE is better). The vertical dotted line corresponds to the rule of thumb for classifying instruments as weak, i.e., an  F-test rejection threshold of 10. The first stage F-test for $H_0:\bar \xi=0$, i.e.,  for the relevancy of instruments, 
			at a significance level of 5\%,	
			has different rejection thresholds in the range $[1.55,4.04]$ depending on $n$ and $q$. The vertical dashed line corresponds to the smallest rejection threshold of 1.55.  Note that the lowest possible negative relative change is $-1$.  For the comparison with the TSLS estimator we have removed the case $q=1$ to ensure existence of first moments. TSLS, Fuller(1) and Fuller(4) outperforms PULSE while PULSE outperforms  OLS.  }\label{fig:HahnExpMeanBias}
	\end{figure} 
	
	\begin{figure}[H] 
		\centering\includegraphics[width=\linewidth-40pt]{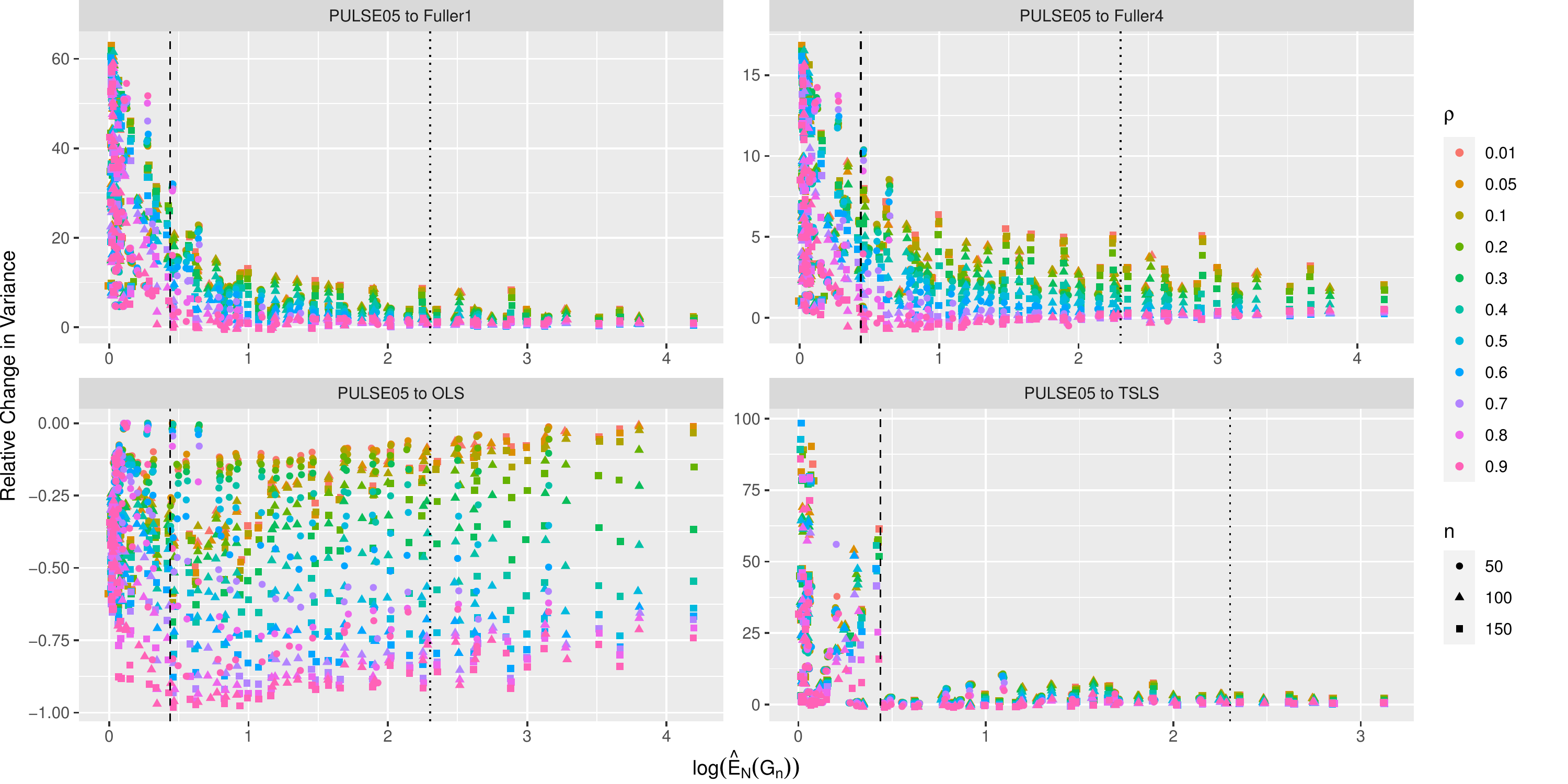}
		\caption{  \normalsize Illustrations of the relative change in variance (a positive relative change means that PULSE is better). The vertical lines are identical to those of \Cref{fig:HahnExpMeanBias}.
			For the comparison with the TSLS estimator we have removed the case $q\in \{1,2\}$ to ensure existence of second moments. We have removed two observations with relative change above 100, in the very weak instrument setting, for aesthetic reasons. PULSE outperforms TSLS, Fuller(1) and Fuller(4), especially for low confounding and weak instruments. We also see that OLS outperforms PULSE with the largest decrease in variance for the large confounding cases. }\label{fig:HahnExpVariance}
	\end{figure} 
	
	\begin{figure}[H] 
		\centering\includegraphics[width=\linewidth-40pt]{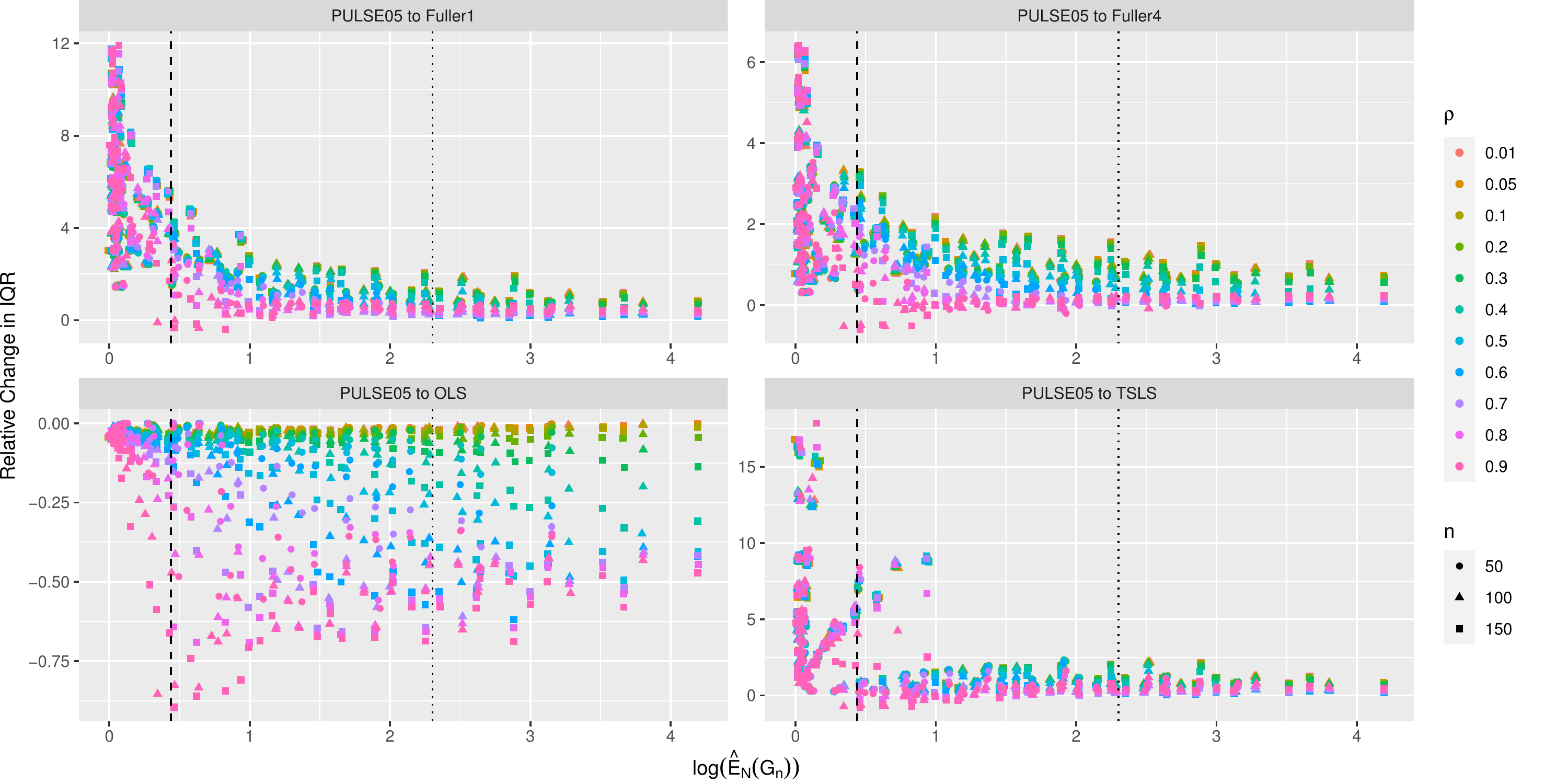}
		\caption{  \normalsize Illustrations of the relative change in interquartile range (a positive relative change means that PULSE is better). 
			The vertical lines are identical to those of \Cref{fig:HahnExpMeanBias}.
			We see that PULSE is superior to Fuller(1), Fuller(4) and TSLS except in very few cases with very large confounding. Furthermore, OLS outperforms PULSE with relatively small difference for low confounding and larger difference for large confounding. }\label{fig:HahnExpIQR}
	\end{figure} 
	\newpage
	\subsection{Additional illustrations for the multivariate experiment}
	
	\begin{figure}[H] 
		\centering\includegraphics[width=\linewidth-75pt]{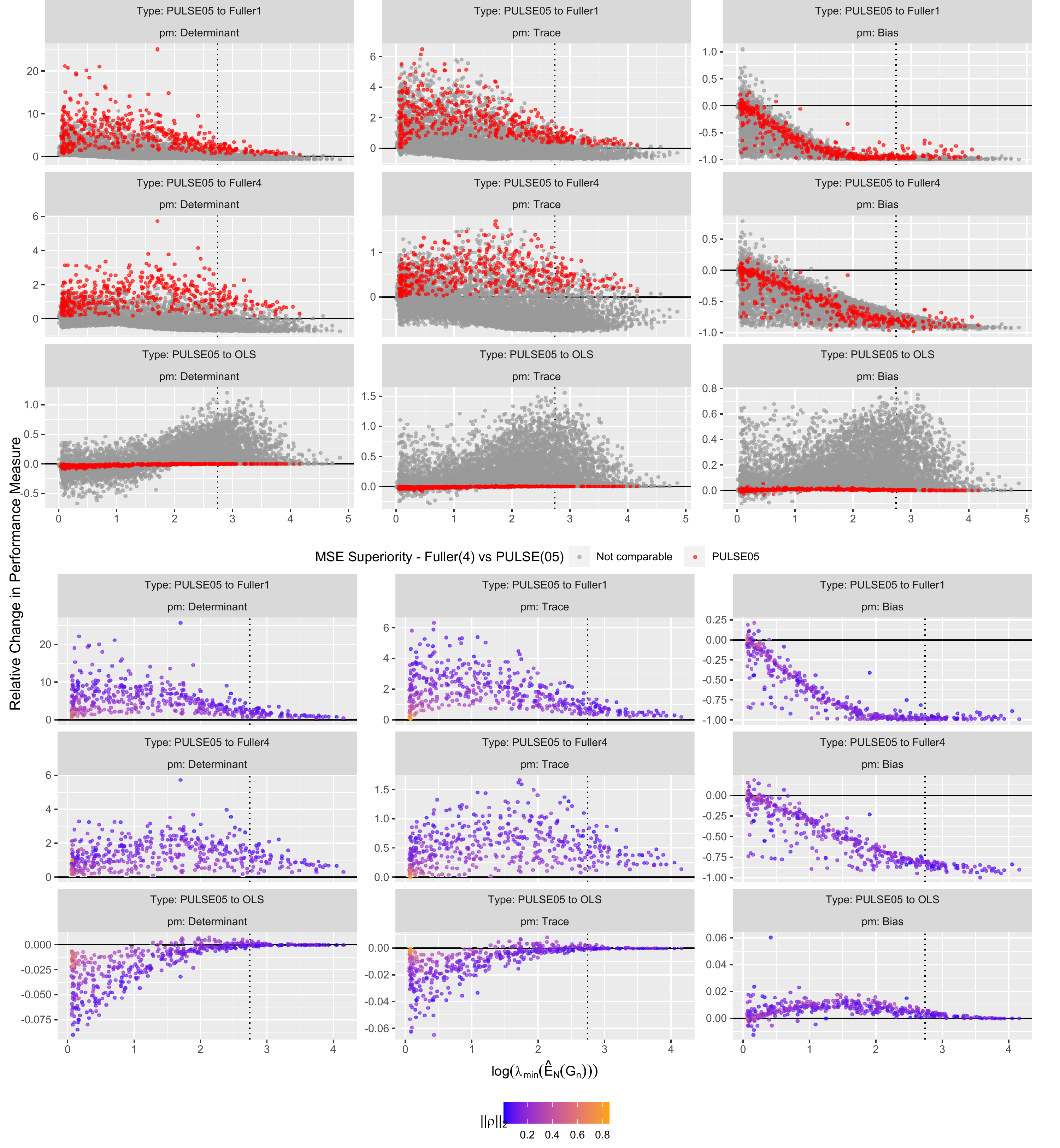}
		\caption{  \normalsize There are two illustrations, both illustrating relative changes in performance measures as in \Cref{fig:AllRandom_Beta00} except that the points are color-graded according to MSE superiority when comparing Fuller(4) and PULSE (top $3\times 3$) and confounding strength $\|\rho\|_2$ (bottom $3\times 3$) . Among the 10000 randomly generated models there are 461 models where PULSE is MSE superior to Fuller(4). In the remaining 9539 models the MSE matrices are not comparable. For the 461 models where PULSE was MSE superior the simulations were repeated with $N=25000$ repetitions to account for possible selection bias.  Of the 461 models 445 were still superior when increasing $N$ from $5000$ to $25000$. The bottom $3\times 3$ grid is an illustration of the relative change in performance measure for the 445 models that remained superior, each model color-graded according to confounding strength. We see that in almost all of these models there is weak to moderate confounding. The exception being a few models in the very weak instrument setting where the confounding is strong.}\label{fig:AllRandom_Beta00_TrueSuperior}
	\end{figure}

	\begin{figure}[H] 
		\centering\includegraphics[width=\linewidth]{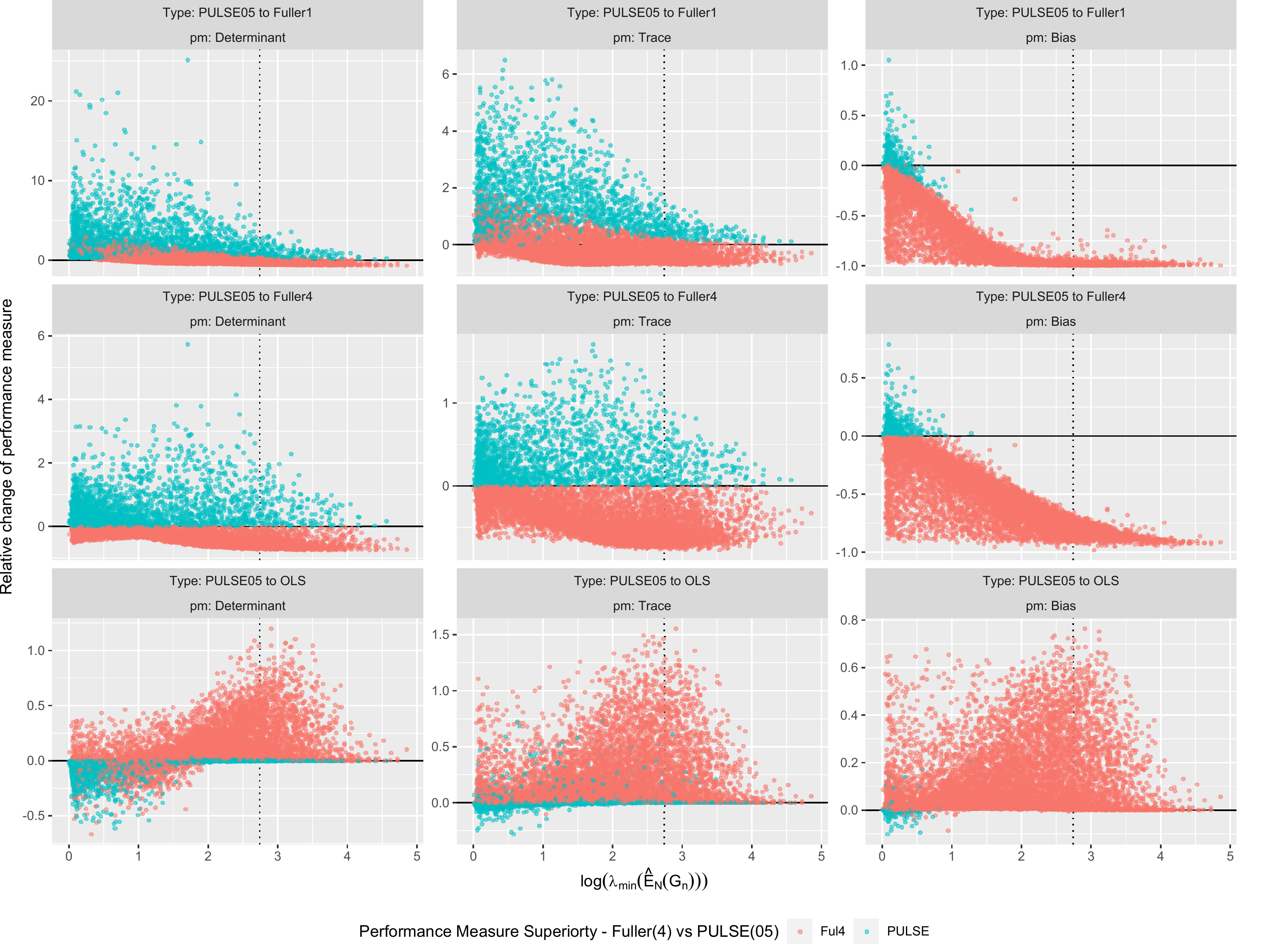}
		\caption{  \normalsize 
			This figure shows the same results as in \Cref{fig:AllRandom_Beta00} except that the points are color-graded according to performance measure superiority when comparing Fuller(4) and PULSE(05). That is, the models have fixed column-wise color-grading according to the comparison between Fuller(4) and PULSE(05).}\label{fig:AllRandom_Beta00_PmSuperior}
	\end{figure}

	\begin{figure}[H] 
		\centering\includegraphics[width=\linewidth]{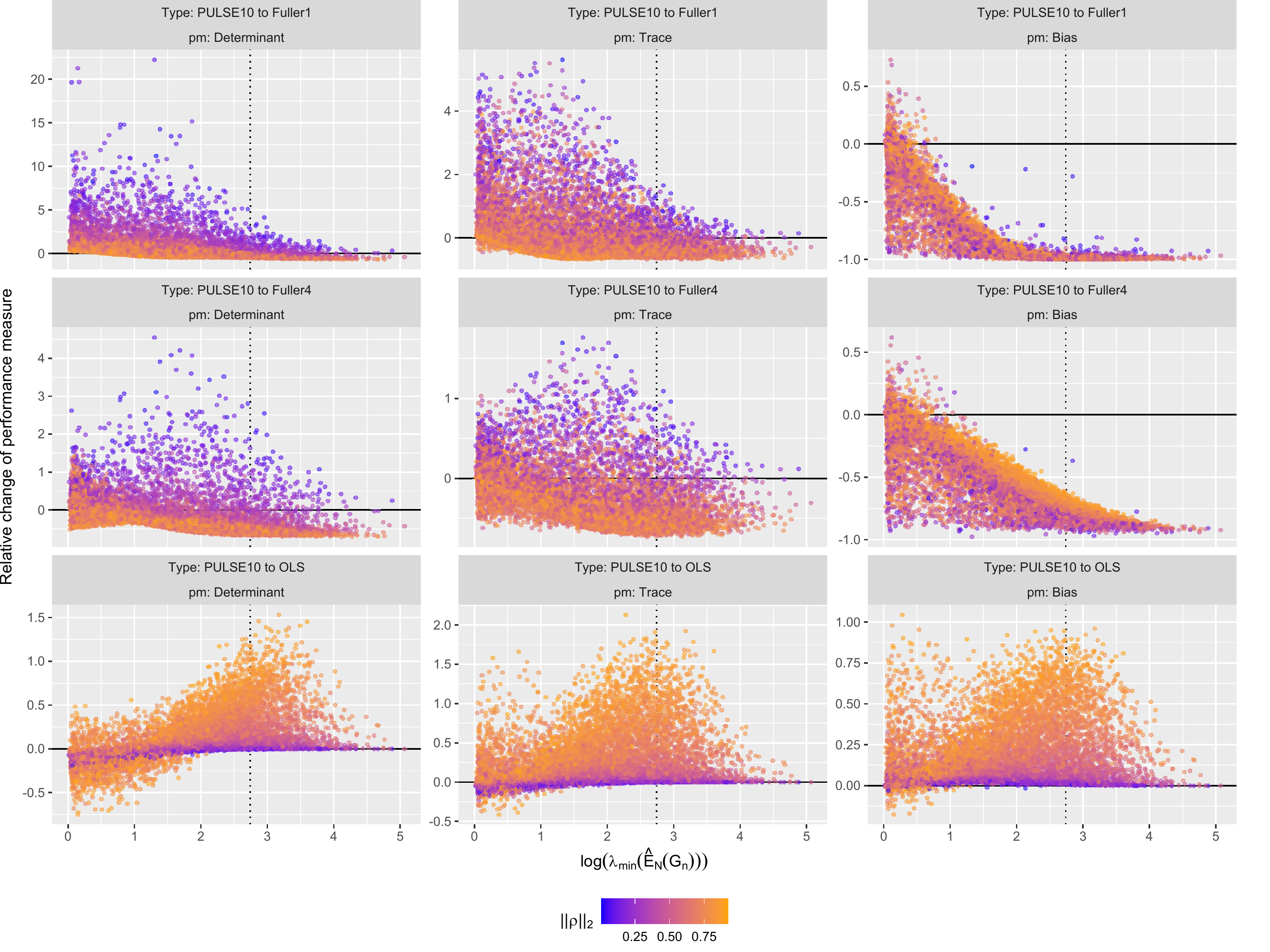}
		\caption{  \normalsize 	This figure shows the same results as in \Cref{fig:AllRandom_Beta00} except that we here compare PULSE with $p_{\min}=0.1$ to the benchmark estimators.}\label{fig:AllRandom_PULSE10}
	\end{figure} 
			
\end{appendices}		
		
		\setlength{\bibsep}{2pt plus 0ex}
		
		\bibliographystyle{chicago}
		\bibliography{bibfile}
\end{document}